\documentclass[
11pt, 
polish,
singlespacing, 
headsepline, 
]{MastersDoctoralThesis} 
\usepackage{amsmath, amsthm, amssymb, calrsfs, wasysym, verbatim, bbm, color, graphics, geometry}
\usepackage[utf8]{inputenc} 
\usepackage[T1]{fontenc} 
\usepackage{multirow}
\usepackage{cancel}
\usepackage{tensor}
\usepackage{comment}
\usepackage{bm}

\usepackage{empheq}
\newcommand*\widefbox[1]{\fbox{\hspace{2em}#1\hspace{2em}}}

\usepackage{tikz}
\usepackage{tikz-feynman}
\tikzfeynmanset{compat=1.0.0}
\usepackage{tikz-cd}

\usepackage{slashed}
\usepackage{mathbbol}
\usepackage[stable]{footmisc}
\usepackage{graphicx}

\usepackage{dcolumn}
\usepackage{rotating}
\usepackage{adjustbox,booktabs}

\usepackage{makecell}

\allowdisplaybreaks

\usepackage{natbib}
\setcitestyle{numbers}
\usepackage[colorlinks=true, pdfencoding=auto,psdextra]{hyperref}
\usepackage{url}

\usepackage[autostyle=true]{csquotes} 

\makeatletter
\def\blfootnote{\gdef\@thefnmark{}\@footnotetext}
\makeatother

\usepackage{mathtools}
\usepackage{lettrine}
\usepackage{amsmath,amssymb}
\usepackage{dsfont}
\usepackage{subcaption}
\usepackage{ stmaryrd }
\usepackage{pdfpages}
\usepackage{enumerate}

\thesistitle{$\kappa$-deformed scalar field} 
\supervisor{Jerzy \textsc{Kowalski-Glikman} \\ Wojciech \textsc{Wiślicki}} 
\examiner{} 
\degree{Doctor of Philosophy} 
\author{Andrea \textsc{Bevilacqua}} 
\addresses{} 

\subject{Physical Sciences} 
\university{\href{http://www.ncbj.gov.pl}{National Centre for Nuclear Research}} 
\department{\href{}{Department BP2}} 
\group{\href{}} 
\faculty{\href{}{Factulty of Physics}} 

\AtBeginDocument{
\hypersetup{pdftitle=\ttitle} 
\hypersetup{pdfauthor=\authorname} 
\hypersetup{pdfkeywords=\keywordnames} 
}

\begin{document}

\frontmatter 

\pagestyle{plain} 


\begin{titlepage}
\begin{center}

{\scshape\LARGE \univname\par}\vspace{1.5cm} 
\textsc{\Large Doctoral Thesis}\\[0.5cm] 

\HRule \\[0.4cm] 
{\huge \bfseries \ttitle\par}\vspace{0.4cm} 
\HRule \\[1.5cm] 
 
\begin{minipage}[t]{0.4\textwidth}
\begin{flushleft} \large
\emph{Author:}\\
\href{https://orcid.org/0000-0002-9801-4405}{\authorname} 
\end{flushleft}
\end{minipage}
\begin{minipage}[t]{0.4\textwidth}
\begin{flushright} \large
\emph{Supervisors:} \\
\href{https://www.ncbj.gov.pl/}{\supname}\\ 
\end{flushright}
\end{minipage}\\[2cm]
 
\vspace{-1 cm}
\large \textit{A thesis submitted in fulfillment of the requirements\\ for the degree of \degreename}\\[0.3cm] 
\textit{in the}\\[0.4cm]
\groupname  \deptname\\[-2cm] 
\includegraphics[width=0.5\textwidth]{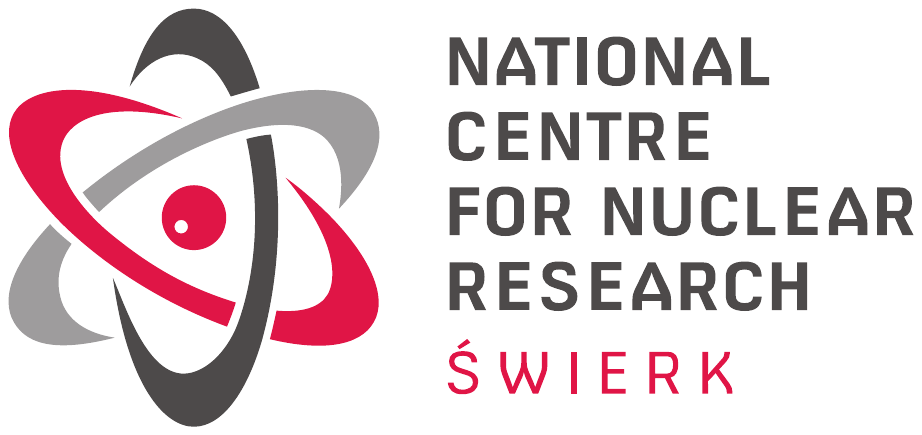}\\[-3cm]
\vfill
{\large \today} 
 
\end{center}
\end{titlepage}


\begin{declaration}
\addchaptertocentry{\authorshipname} 
\noindent I, \authorname, declare that this thesis titled, \enquote{\ttitle} and the work presented in it are my own. I confirm that:

\begin{itemize} 
\item This work was done wholly or mainly while in candidature for a research degree at the National Centre for Nuclear Research.
\item Where any part of this thesis has previously been submitted for a degree or any other qualification at the National Centre for Nuclear Research or any other institution, this has been clearly stated.
\item Where I have consulted the published work of others, this is always clearly attributed.
\item Where I have quoted from the work of others, the source is always given. With the exception of such quotations, this thesis is entirely my own work.
\item I have acknowledged all main sources of help.
\item Where the thesis is based on work done by myself jointly with others, I have made clear exactly what was done by others and what I have contributed myself.\\
\end{itemize}
 
\noindent Signed:\\
\rule[0.5em]{25em}{0.5pt} 
 
\noindent Date:\\
\rule[0.5em]{25em}{0.5pt} 
\end{declaration}

\cleardoublepage





\begin{abstract}
\addchaptertocentry{\abstractname} 
In the following work we will introduce and discuss in detail a particular model of complex $\kappa$-deformed scalar field, whose behaviour under $C$, $P$, $T$ transformation is particularly transparent from both a formal and phenomenological point of view. We will begin by introducing the key mathematical structure at the basis of our investigation, namely the $\kappa$-Poincar\'e (Hopf) algebra and the $\kappa$-Minkowski spacetime. We will then investigate the behaviour of general two-particle states under deformed boost. After this we will introduce the action of our $\kappa$-deformed complex scalar field. From it, we will derive the equations of motion, as well as the Noether charges due to the continuous symmetries. The peculiar features of $\kappa$-deformation in general, and of our model in particular, allow for very non-trivial interaction between discrete and continuous symmetries, of which we will investigate the phenomenological consequences (particularly in terms of difference of lifetime of decaying particles). To conclude, we will obtain the $\kappa$-deformed propagator of the $\kappa$-deformed complex scalar field, and the imaginary part of the 1-loop contribution to it, ending with additional phenomenological consequences. The third chapter is new, unpublished work.

\end{abstract}

\begin{extraAbstract}
\addchaptertocentry{\extraAbstractname}
W niniejszej pracy wprowadzimy i przedyskutujemy w szczegółach model zespolonego,
$\kappa$-zdeformowanego pola skalarnego, którego zachowanie pod wpływem transformacji C, P
i T można jasno przedstawić, zarówno z formalnego, jak z fenomenolo\-gicznego punktu
widzenia. Zaczniemy od wprowadzenia struktur matematycznych, leżących u podstaw
naszych badań, a mianowicie algebry $\kappa$-Poincarego (Hopfa) oraz czasoprzestrzeni $\kappa$-Minkowskiego. Następnie zbadamy zachowanie ogólnych stanów dwucząstkowych pod
wpływem zdeformowanego pchnięcia Lorentza. Potem wprowa\-dzimy działanie $\kappa$-zdeformowanego, zespolonego pola skalarnego, po czym wyprowa\-dzimy równania ruchu i
ładunki Noether odpowiadające symetriom ciągłym. Pewne szczególne własności $\kappa$-deformacji, zarówno te ogólne, jak i dotyczące tylko naszego modelu, umożliwiają na
istnienie nietrywialnch związków między symetriami dyskretnymi i ciągłymi, z których
wyprowadzimy konsekwencje fenomenologiczne i poddamy je badaniom (szczególnie
konsekwencje dotyczące różnic czasów życia niestabilnych cząstek i antycząstek).
Obliczymy także $\kappa$-zdeformowany propagator $\kappa$-zdeformowane\-go, zespolonego pola
skalarnego, jak również część urojoną jednopętlowej poprawki do propagatora, a
następnie zakończymy dyskusją konsekwencji fenomenologicznych. Trzeci rozdział to nowa, nieopublikowana praca.
\end{extraAbstract}

\begin{acknowledgements}
\addchaptertocentry{\acknowledgementname} 

I would like to thank my supervisors Jurek and Wojtek. Their deep knowledge and understanding of physics has been the source of many interesting discussions, and their guidance has made my Ph.D. studies a wonderful experience. For the same reasons, I would also like to thank Giulia, Michele, Giacomo, and Josua. I would like to thank Prof. Stanisław Mrówczyński for the interesting discussions, and for giving me the opportunity to assist him as a tutor in his Statistical Mechanics lectures. I am also deeply grateful to Przemek 
for his help in making these four years possible in the first place. I would also like to thank Monika for all the incredible help, and for the very interesting discussions during my frequent trips to her office, as well as for the enjoyable Polish lectures. Additionally, I would like to thank Ubaldo and Anatolii for the wonderful time spent together. A special thanks goes also to my family for their help, I wouldn't have been able to study for so many years without their support. 

Finally, my partner Alice has accompanied me throughout all my studies, in a voyage which has been made even more beautiful (and much more chaotic) by the birth of our son Levi. To both of them goes all my gratitude.


\bigskip

Part of this work has been supported by funds provided by the Polish National Science Center, the project number 2019/33/B/ST2/00050, and by the NAWA scholarship financed by the STER program. My participation to COST Action CA18108 activities has been financed by COST Action. Part of this work's bibliography has been collected using \cite{elicit}.

\end{acknowledgements}


\tableofcontents 

\dedicatory{To A.B. and L.B-B.} 

\mainmatter 

\pagestyle{thesis} 

\chapter{$\kappa$-Minkowski and $\kappa$-Poincar\'e} 

\label{Chapter1}

\section{Introduction to $\kappa$-deformation}\label{intro-ch1}

Immediately after both general relativity and quantum mechanics were formalized, physicists were already trying to understand what a quantum theory of gravity would look like \cite{Einstein}. Several approaches were tried, each with their pros and cons (an extensive discussion of each approach would be out of place here, but one can consult \cite{Rovelli:2000aw} for a review). From a theoretical point of view, the wealth of approaches was due to the vastly different nature of general relativity and quantum mechanics, both conceptually and formally. Moreover, from the more pragmatic point of view, effects of quantum gravitational origin were thought to appear when particles with Plank energy $E_p$ scatter with impact parameter given by the Plank length $l_p$. This means that very early on, there was significant doubt that any quantum gravitational effect could be phenomenologically relevant, even in the distant future \cite{Bronstein:2012zz}. Moreover, the measurement of phenomena happening at distances comparable or smaller than $l_p$ would require apparatuses so large that they inevitably collapse to a black hole \cite{Dyson:2013hbl}, which means there may be some limitation from first principles in the conventional definition of spacetime as a continuous structure (this was already noted in \cite{Bronstein:2012zz}, \cite{Jackiw:2003dw}, \cite{Snyder:1947}). Despite these advances, however, the lack of phenomenological feedback seemed to be destined to be a long lasting issue in quantum gravity (QG) research. It was therefore unexpected when it was suggested that, switching the focus to cumulative effects of QG, a phenomenology was not only possible, but within reach \cite{Amelino-Camelia:1999hpv}. Indeed, instead of focussing on high energy scattering processes with very small impact parameter, one could try to study the cumulative effects of QG, which can for example manifest themselves with a slightly different propagation speed for photons of different energies \cite{Ellis:1999uh}, \cite{Ellis:1999rz}, \cite{Ellis:1999sd}, \cite{Amelino-Camelia:2010pnk}, \cite{Amelino-Camelia:2011ebd}, \cite{Amelino-Camelia:2011uwb}, and may also affect neutrino properties \cite{Mavromatos:2007hv}, \cite{Amelino-Camelia:2016fuh}, \cite{Amelino-Camelia:2016wpo}, \cite{Amelino-Camelia:2016ohi}, \cite{Rosati:2017hod}, \cite{Amelino-Camelia:2022pja}. One of the most effective ways to go about studying potential QG effects is therefore not to study any model of QG directly, but a potential effective theory of QG. 

One of the most studied model in this sense is based on a deformation of the Poincar\'e algebra, called $\kappa$-Poincar\'e algebra, where $\kappa$ is a deformation parameter with dimensions of energy. Its origin can be traced back to the 1990's, particularly to \cite{Lukierski:1992dt}, \cite{Lukierski:1993wxa}, \cite{Lukierski:1993wx}, \cite{Kosinski:1994je}, \cite{Kosinski:1994br}, \cite{Lukierski:1996ib}. The idea is to deform the canonical Poincar\'e algebra to investigate how symmetries might change in some well defined physical limit of QG. 

At the same time, new models were being proposed where in addition to a constant speed $c$, there is also a length scale $l=\frac{\hbar}{\kappa}$ (or $l=\frac{1}{\kappa}$ if $\hbar=1$) which is independent of the observer \cite{Amelino-Camelia:2000stu}, \cite{Amelino-Camelia:2002cqb}, \cite{Amelino-Camelia:2002uql}, \cite{Kowalski-Glikman:2001vvk}. This length scale also implies a momentum scale in momentum space; the idea is therefore that such observer-independent scale can be used to define some non-linear dispersion relation while keeping the theory Lorentz invariant (where `Lorentz invariance' in this context is defined in the appropriate way). Intuitively speaking, such an additional scale would allow for a non-trivial geometry of momentum space, which is usually flat in non-deformed theories which lack such a scale. To give just a small example, one can start from the action of a classical particle with non-deformed dispersion relation
moving in a curved spacetime \cite{Arzano:2021scz}, \cite{Kowalski-Glikman:2022xtr}
\begin{align}\label{actioncurvedst}
	S = \int d\tau \,\, \dot{x}^\mu \tensor{e}{_\mu^a}(x) p_a - N(\eta^{ab} p_a p_b + m^2).
\end{align}
From this, one can by analogy write down the action of a particle moving in flat spacetime, but curved momentum space 
\begin{align}\label{actioncurvedms}
	S_\kappa = 
	-\int d\tau \,\, x^a \tensor{E}{_a^\mu}(k, \kappa) \dot{k}_\mu - N(\mathcal{C}_\kappa(k) + m^2).
\end{align}
where $\tensor{E}{_a^\mu}$ is the tetrad which is related to the momentum-space metric by
\begin{align}\label{momentumspacemetric}
	G^{\mu\nu}(k, \kappa) = \eta^{ab} \tensor{E}{_a^\mu}(k, \kappa) \tensor{E}{_b^\nu}(k, \kappa),
\end{align}
and $\mathcal{C}_\kappa(k)$ is the deformed dispersion relation (once again, here we are using a deformation parameter with dimensions of energy). For example, one could choose
\begin{align}
	\mathcal{C}_\kappa(k) = -4 \kappa^2 \sinh^2 
	\left(
	\frac{k_0}{2\kappa}
	\right)
	+
	\mathbf{k}^2 e^{\frac{k_0}{\kappa}}.
\end{align}
One can show that
\begin{align}\label{explicitdefaction}
	S_\kappa = 
	\int d\tau \,\, \dot{x}^0 k_0
	-
	e^{\frac{k_0}{\kappa}}
	\mathbf{x} \cdot \dot{\mathbf{k}}
	+ 
	N(\mathcal{C}_\kappa(k) + m^2)
\end{align}
which is invariant under a deformed set of transformations which satisfy the so called $\kappa$-Poincar\'e algebra (which will be discussed in detail in later sections).
The symplectic form is then given by
\begin{align}\label{symplecticform-intro}
	\Omega_\kappa
	=
	\int d\tau \,\,  \delta k_0 \wedge \delta x^0
	+
	e^{\frac{k_0}{\kappa}}
	\delta \mathbf{k}_i \wedge \delta \mathbf{x}^i
	-
	\frac{\mathbf{x}^i}{\kappa}
	e^{\frac{k_0}{\kappa}}
	\delta k_0 \wedge \delta \mathbf{k}_i
\end{align}
which implies 
\begin{align}\label{Poissonb-intro}
	\{x^0, p_0\} = 1
	\qquad
	\{\mathbf{x}^i, \mathbf{k}_j\} = e^{-\frac{k_0}{\kappa}}\delta^i_j
	\qquad
	\{x^0, \mathbf{x}^i\}
	=
	-\frac{1}{\kappa} \mathbf{x}^i.
\end{align}
We see that, upon quantization, spacetime coordinates do not commute. These type of non-commutativity defines the so called $\kappa$-Minkowski spacetime. This kind of models showcases interesting phenomena, such as relative locality \cite{Amelino-Camelia:2011lvm}, \cite{Amelino-Camelia:2011hjg}, \cite{Amelino-Camelia:2011uwb}, and are related to $\kappa$-Poincar\'e algebras \cite{Kowalski-Glikman:2002iba}, \cite{Pachol:2011tp}, \cite{Arzano:2022har}, \cite{Gubitosi:2019ymi}. 

Finally, in $2+1$ dimensions, one can show that $\kappa$-deformation naturally emerges when considering point-like particles coupled with gravity (see for example \cite{Freidel:2003sp},\cite{Kowalski-Glikman:2017ifs}, \cite{Arzano:2021scz} and references therein). This is due to the fact that the Newton constant $G_N$ is dimensionful in $2+1$ dimensions, and it has dimensions of energy. The procedure is fairly straightforward and can be summarized in the following steps \cite{Kowalski-Glikman:2022xtr},\cite{Arzano:2021scz}:
\begin{itemize}
	\item[1)] From the total action of gravity and point-like  particles coupled to it, one can get the explicit solutions of the (finitely many) gravitational degrees of freedom;
	\item[2)] Substitute these solutions back into the action, obtaining a new action which describes the motion of point-like particles `dressed' in their own gravitational field;
	\item[3)] This new action can then be shown to correspond to a $\kappa$-deformed model, much like the one described by eq. \eqref{explicitdefaction} (although in the case of $2+1$ particles with gravity we have a different kind of deformations).
\end{itemize}
Furthermore, one can argue that the deformed symmetry group $\mathcal{P}^3$ acting on the Hilbert space $\mathcal{H}^3$ of $2+1$ gravity with particles is expected to be a subgroup of the full group $\mathcal{P}^4$ acting on the Hilbert space $\mathcal{H}^4$ of particles coupled with gravity in $3+1$ dimensions \cite{Freidel:2003sp}. All this motivates the study of fields in the context of $\kappa$-Poincar\'e and $\kappa$-Minkowski. 

$\kappa$-deformation models have also proven to be interesting from the phenomenological point of view, ranging from symmetry violation, to string theory, astronomy, particle physics, and black holes \cite{Lobo:2020qoa}, \cite{Amelino-Camelia:1997ieq}, \cite{Ellis:1999sd}, \cite{Aschieri:2017ost}, \cite{Ellis:1999sf}, \cite{Amelino-Camelia:1996bln}, \cite{Amelino-Camelia:2023srg}, \cite{Barcaroli:2017gvg}.

In this work, we will investigate the kinematical properties of particles in $\kappa$-Minkowski spacetime (chapter \ref{Chapter1}), and we will then introduce the $\kappa$-deformed complex scalar field, studying its properties and computing the Noether charges (chapter \ref{Chapter2}). Finally, we will compute the propagator and the imaginary part of the 1-loop correction to it (chapter \ref{Chapter3}). Each of the chapters will contain a final section investigating possible experimental signature of the theoretical results obtained earlier in the chapter. There have been already several approaches building fields on top of $\kappa$-Minkowski spacetime, or using $\kappa$-Minkowski algebra in general, see for example 
\cite{Meljanac:2007xb}, 
\cite{Freidel:2007hk}, 
\cite{Daszkiewicz:2004xy}, \cite{Mercati:2018hlc}, \cite{Meljanac:2011ge}, \cite{Govindarajan:2009wt}, \cite{Kim:2009jk}, \cite{Poulain:2018two}, \cite{Daszkiewicz:2008bm}, \cite{Daszkiewicz:2007ru}, \cite{Arzano:2007ef}, \cite{Amelino-Camelia:2001rtw}, \cite{Kosinski:2001ii}, \cite{Kosinski:2003xx}. In contrast to these models, the construction which we will present in this work allows for a much simpler, canonical definition of deformed symmetries in the $\kappa$-deformed context. The particularly simple form of the discrete symmetries in turn allows for an interesting (and easy to interpret) phenomenology.

Of course, key ingredients in all our constructions are the $\kappa$-Minkowski spacetime, and the $\kappa$-Poincar\'e algebra, so we will dedicate the first two section of the current chapter to a brief exposition of their properties. It turns out, however, that it is not just the deformed $\kappa$-Poincar\'e algebra to be important, but the Hopf algebra structure of which the $\kappa$-Poincar\'e algebra is but a part \cite{Majid:1994cy}. The mathematical aspects of $\kappa$-deformations and the importance of the relevant Hopf algebra have been investigated in detail in \cite{Kosinski:1994ng}, \cite{Borowiec:2008uj}, \cite{Borowiec:2010yw}, \cite{Pachol:2011tp}, \cite{Borowiec:2013lca}, \cite{Borowiec:2013gca}, and the important physical ultra-relativistic and non-relativistic limits in \cite{Ballesteros:2020uxp}, \cite{Ballesteros:2022gep}, \cite{Ballesteros:2021dob}. The constructions of $\kappa$-Minkowski spacetime and the $\kappa$-Poincar\'e Hopf algebra, as well as the relations between them, have been extensively studied for decades. 
Here we will highlight how one can obtain $\kappa$-Minkowski spacetime starting from the $\kappa$-Poincar\'e Hopf algebra following \cite{Kowalski-Glikman:2002eyl}, \cite{Kowalski-Glikman:2017ifs} and references therein as a guide. We will then show the opposite direction, namely how one can obtain the $\kappa$-Poincar\'e Hopf algebra structure starting from $\kappa$-Minkowski spacetime, following the approach presented in \cite{Arzano:2022ewc}. Both the approaches can be useful when working in the context of $\kappa$-deformation, and each of them sheds light on some feature of $\kappa$-deformation in general. Since these notions are well known in the literature, we will not delve into too much details, referring the reader to the provided references for a deeper treatment. Nevertheless, we will be explicit in out computations, and show each important step.

\section{Formal definition of Hopf algebra}

Before beginning our discussion, we give a mathematical definition of a Hopf algebra. A Hopf algebra is technically defined as a bi-algebra with an antipode, satisfying several compatibility conditions. To understand what is a bi-algebra with antipode, we will need to introduce the notion of algebra, co-algebra, and antipode separately. Using the same notation as \cite{Mercati:2011fmm}, we can define a unital algebra as a vector field $\mathcal{A}$ over a field $\mathbb{K}$ together with two maps, the product map $\mu: \mathcal{A}\otimes \mathcal{A} \rightarrow \mathcal{A}$, and the unit map $\eta: \mathbb{K} \rightarrow \mathcal{A}$, which need to satisfy the associativity and unity axiom defined respectively by the following commutative diagrams \cite{Mercati:2011fmm}
\begin{figure}[h]
	\centering
	\includegraphics[width=1\textwidth]{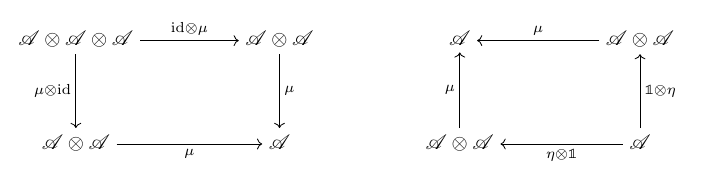}
\end{figure}

A co-algebra can basically be defined by reversing all the arrows. In particular, a co-algebra is a vector field $\mathcal{C}$ over a field $\mathbb{K}$ together with two maps, the co-product map $\Delta:\mathcal{C}\rightarrow\mathcal{C}\otimes\mathcal{C}$ and the co-unit map $\epsilon:\mathcal{C}\rightarrow\mathbb{K}$, which need to satisfy the co-associativity and co-unit axiom, which are defined respectively by the following commutative diagrams \cite{Mercati:2011fmm}
\begin{figure}[h]
	\centering
	\includegraphics[width=1\textwidth]{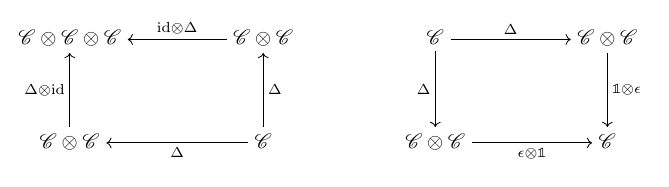}
\end{figure}

A bi-algebra $B$ is therefore both an algebra and a co-algebra, with additional compatibility conditions. In fact, we now have a structure with product $\mu$, unity $\eta$, co-product $\Delta$, and co-unity $\epsilon$, and they must satisfy \cite{Mercati:2011fmm}
\begin{align}
	\Delta(\mu(a,b)) = \mu(\Delta(a),\Delta(b)),
	\qquad
	\epsilon(\mu(a,b)) = \epsilon(a)\epsilon(b)
\end{align}
for all $a,b\in B$, and where 
\begin{align}
	\Delta(\text{id}) = \text{id}\otimes \text{id}
	\qquad
	\epsilon(\text{id}) = 1
\end{align}
where $1\in \mathbb{K}$. Finally, a Hopf algebra $\mathcal{H}$ is a bi-algebra with an additional map $S:\mathcal{H}\rightarrow \mathcal{H}$ such that 
\begin{align}
	S(\mu(a,b))=\mu(S(a),S(b))
	\qquad
	S(\mathbb{1})=\mathbb{1}
	\qquad
	(S\otimes S)\circ\Delta = \tau\circ\Delta\circ S
\end{align}
where $a,b\in \mathcal{H}$ and  $\tau:\mathcal{H}\otimes\mathcal{H}\rightarrow \mathcal{H}\otimes\mathcal{H}$ is the flip map defined by
\begin{align}
	\tau(a\otimes b) = b\otimes a
\end{align}
for all $a\in \mathcal{H}$. 

In what follows we will concentrate on the $\kappa$-Poincar\'e Hopf algebra. All the above equalities and commutative diagrams are satisfied by the product, co-product, unity, co-unity, and antipode, but we will not check them explicitly here. Our aim with the discussion in this section and the next is to highlight how the $\kappa$-Poincar\'e Hopf algebra, and more in particular its co-product sector, is related to the $\kappa$-Minkowski spacetime. 

\section{From $\kappa$-Poincar\'e to $\kappa$-Minkowski}

The algebra sector of the $\kappa$-Poincar\'e Hopf algebra in the so called \textit{bicrossproduct basis} is given by 
\begin{align}\label{Mk-comm}
	[M_i, k_j] = i\epsilon_{ijl} k_l,
	\qquad
	[M_i,k_0] = 0,
\end{align}
\begin{align}\label{Nk-comm}
	[N_i,k_j]
	=
	i\delta_{ij}
	\left(
	\frac{\kappa}{2}
	(1-e^{-2k_0/\kappa})
	+
	\frac{\mathbf{k}^2}{2\kappa}
	\right)
	-\frac{i}{\kappa}
	k_i k_j, 
	\qquad
	[N_i,k_0] =  ik_i,
\end{align}
\begin{align}\label{MN-comm}
	[M_i,M_j] = i \epsilon_{ijk} M_k,
	\qquad
	[M_i,N_j] = i\epsilon_{ijk}N_k
	\qquad
	[N_i,N_j] = -i\epsilon_{ijk} M_k.
\end{align}
The co-algebra sector is given by 
\begin{align}\label{momcopbi}
	\Delta k_i = k_i \otimes \mathbb{1} + e^{-k_0/\kappa} \otimes k_i,
	\qquad
	\Delta k_0 = k_0 \otimes \mathbb{1} + \mathbb{1} \otimes k_0,
\end{align}
\begin{align}\label{boostcopbi}
	\Delta N_i = N_i \otimes \mathbb{1} + e^{-k_0/\kappa} \otimes N_i + \frac{1}{\kappa} \epsilon_{ijk} k_j \otimes M_k,
\end{align}
\begin{align}\label{rotcopbi}
	\Delta M_i = M_i \otimes \mathbb{1} + \mathbb{1} \otimes M_i
\end{align}
and finally the antipode sector is given by
\begin{align}\label{antik}
	S(k_0) = -k_0, 
	\qquad
	S(k_i) = -k_i e^{k_0/\kappa},
\end{align}
\begin{align}\label{antiMN}
	S(M_i) = -M_i,
	\qquad
	S(N_i) = -e^{k_0/\kappa} \left(N_i - \frac{1}{\kappa} \epsilon_{ijk} k_j M_k\right).
\end{align}
The co-units $\epsilon(k)$, $\epsilon(N_i)$, $\epsilon(M_j)$ are all zero. Starting from this Hopf algebra, and in particular from the co-algebra sector, we can build the spacetime associated to the algebra. The construction is quite intuitive, because one first introduces the objects $x^\mu$ (interpreted as `coordinates', the dual objects of the momenta $k_\mu$), and then one defines the action of the algebra generators $k_\mu$, $N_i$, $M_j$ on them. Having done this, the co-algebra sector is used to define the action of $k_\mu$, $N_i$, $M_j$ on products $x^\mu x^\nu$, and one immediately gets the coordinates commutation relation defining $\kappa$-Minkowski spacetime from the action of $k_\alpha$ on $x^\mu x^\nu$ (or more in general on polynomials of $x^\mu$). Furthermore, one can also check that such commutation relations are invariant under the action of $N_i$, $M_j$, showing that the Poincar\'e algebra indeed describes the symmetries of $\kappa$-Minkowski spacetime. More in detail, using the symbol $\triangleright$ to indicate the action of one of the generators of the $\kappa$-Poincar\'e algebra on spacetime variables, we have the definitions \cite{Kowalski-Glikman:2017ifs}
\begin{align}\label{actiongenonx}
	k_\mu \triangleright x^\nu = -i\delta_\mu^\nu, 
	\qquad
	N_i \triangleright \hat{x}^j = i\hat{x}^0 \delta_i^j, 
	\qquad
	N_i \triangleright \hat{x}^0 = i x_i,
	\qquad
	k_\mu\triangleright 1 = \epsilon(k_\mu) = 0
\end{align}
We then use the co-products to define the actions on polynomials of $x^\mu$ using the so called \textit{Sweedler notation}. Writing everything explicitly, we have
\begin{align}
	\Delta f \triangleright g_1 g_2 =
	(\Delta f)(g_1, g_2) &= 
	\sum_\alpha (f_\alpha^{(1)}\triangleright g_1) (f_\alpha^{(2)}\triangleright g_2)
\end{align}
where $g_1, g_2$ are some coordinates. The objects with ${}^{(1)}$ and ${}^{(2)}$ represent respectively the `first' and `second' terms in each of the tensor products in the definition of co-product in eq. \eqref{momcopbi}, \eqref{boostcopbi}, \eqref{rotcopbi}. To make things more concrete, take for example $k_i$ whose co-product rule is given in eq. \eqref{momcopbi}. We have
\begin{align}
	k_i \triangleright \hat{x}^0 \hat{x}^j 
	&=
	\sum_\alpha ((k_i)_{(1)}^\alpha\triangleright \hat{x}^0) ((k_i)_{(2)}^\alpha\triangleright \hat{x}^j) \\
	&=
	((k_i)_{(1)}^1\triangleright \hat{x}^0)  ((k_i)_{(2)}^1\triangleright \hat{x}^j) + 
	((k_i)_{(1)}^2\triangleright \hat{x}^0)  ((k_i)_{(2)}^2\triangleright \hat{x}^j) \\
	&=
	(k_i\triangleright \hat{x}^0)(\mathbb{1}\triangleright \hat{x}^j) + (e^{-k_0/\kappa}\triangleright \hat{x}^0) (k_i\triangleright \hat{x}^j) \\
	&= 
	-i\delta_{i}^{0}  \hat{x}^j
	+
	\left[\left(1 - \frac{k_0}{ \kappa} + \frac{k_0^2}{2\kappa^2} - \dots\right)\triangleright \hat{x}^0 \right] (-i\delta_i^j) \\
	&=
	0
	-
	i x_0 \delta_i^j
	+
	\frac{(k_0\triangleright \hat{x}^0)}{\kappa} i\delta_i^j
	-
	\frac{k_0\triangleright(k_0\triangleright \hat{x}^0)}{2\kappa^2} i\delta_i^j
	+
	\dots \\
	&=
	-i \hat{x}^0 \delta_i^j
	+
	\frac{1}{\kappa}\delta_i^j 
	+
	0
	-
	\dots \\
	&=
	-i x_0 \delta_i^j
	+
	\frac{1}{\kappa}\delta_i^j 
\end{align}
Notice that we also used the relations $k^2 \triangleright x = k\triangleright (k\triangleright x)$, and recall that $\epsilon(k)=0$ in the $\kappa$-Poincar\'e Hopf algebra. In the same way, we also have
\begin{align}
	k_i \triangleright \hat{x}^j \hat{x}^0 
	&=
	\sum_\alpha ((k_i)_{(1)}^\alpha\triangleright \hat{x}^j) ((k_i)_{(2)}^\alpha\triangleright \hat{x}^0) \\
	&=
	((k_i)_{(1)}^1\triangleright \hat{x}^j) ((k_i)_{(2)}^1\triangleright \hat{x}^0) + 
	((k_i)_{(1)}^2\triangleright \hat{x}^j) ((k_i)_{(2)}^2\triangleright \hat{x}^0) \\
	&=
	(k_i\triangleright \hat{x}^j) (\mathbb{1}\triangleright \hat{x}^0) + (e^{-k_0/\kappa}\triangleright \hat{x}^j) (k_i\triangleright \hat{x}^0) \\
	&=
	-i\hat{x}^0 \delta_i^j.
\end{align}
\begin{align}
	k_0 \triangleright \hat{x}^0 \hat{x}^j 
	&=
	\sum_\alpha ((k_0)_{(1)}^\alpha\triangleright \hat{x}^0) ((k_0)_{(2)}^\alpha\triangleright \hat{x}^j) \\
	&=
	((k_0)_{(1)}^1\triangleright \hat{x}^0)  ((k_0)_{(2)}^1\triangleright \hat{x}^j) + 
	((k_0)_{(1)}^2\triangleright \hat{x}^0)  ((k_0)_{(2)}^2\triangleright \hat{x}^j) \\
	&=
	(k_0\triangleright \hat{x}^0)(\mathbb{1}\triangleright \hat{x}^j) + (\mathbb{1}\triangleright \hat{x}^0) (k_0\triangleright \hat{x}^j) \\
	&= 
	-i \hat{x}^j
\end{align}
\begin{align}
	k_0 \triangleright \hat{x}^j \hat{x}^0 
	&=
	\sum_\alpha ((k_0)_{(1)}^\alpha\triangleright \hat{x}^j) ((k_0)_{(2)}^\alpha\triangleright \hat{x}^0) \\
	&=
	((k_0)_{(1)}^1\triangleright \hat{x}^j) ((k_0)_{(2)}^1\triangleright \hat{x}^0) + 
	((k_0)_{(1)}^2\triangleright \hat{x}^j) ((k_0)_{(2)}^2\triangleright \hat{x}^0) \\
	&=
	(k_0\triangleright \hat{x}^j) (\mathbb{1}\triangleright \hat{x}^0) + (\mathbb{1}\triangleright \hat{x}^j) (k_0\triangleright \hat{x}^0) \\
	&=
	-i\hat{x}^j
\end{align}
Furthermore, one can also obtain the relation
\begin{align}
	k_i \triangleright \hat{x}^j \hat{x}^k 
	&=
	\sum_\alpha ((k_i)_{(1)}^\alpha\triangleright \hat{x}^j) ((k_i)_{(2)}^\alpha\triangleright \hat{x}^k) \\
	&=
	((k_i)_{(1)}^1\triangleright \hat{x}^j)  ((k_i)_{(2)}^1\triangleright \hat{x}^k) + 
	((k_i)_{(1)}^2\triangleright \hat{x}^j)  ((k_i)_{(2)}^2\triangleright \hat{x}^k) \\
	&=
	(k_i\triangleright \hat{x}^j)(\mathbb{1}\triangleright \hat{x}^k) + (e^{-k_0/\kappa}\triangleright \hat{x}^j) (k_i\triangleright \hat{x}^k) \\
	&= 
	-i\delta_{i}^{j}  \hat{x}^k
	+
	\left[\left(1 - \frac{k_0}{ \kappa} + \frac{k_0^2}{2\kappa^2} - \dots\right)\triangleright \hat{x}^j \right] (-i\delta_i^k) \\
	&=
	-i\delta_{i}^{j}  \hat{x}^k
	-i\delta_{i}^{k}  \hat{x}^j
\end{align}
One can therefore deduce 
\begin{align}
	k_i \triangleright \hat{x}^0 \hat{x}^j - k_i \triangleright \hat{x}^j \hat{x}^0 = \frac{\delta_i^j}{\kappa}
	\qquad
	k_0 \triangleright \hat{x}^0 \hat{x}^j - k_0 \triangleright \hat{x}^j \hat{x}^0 = 0
\end{align}
\begin{align}
	k_i \triangleright \hat{x}^j \hat{x}^k - k_i \triangleright \hat{x}^k \hat{x}^j = 0
\end{align}
and therefore 
\begin{align}\label{kminkcomm}
	[\hat{x}^0, \hat{x}^j] = \frac{i}{\kappa} \hat{x}^j
	\qquad
	[\hat{x}^i,\hat{x}^j] = 0
\end{align}
which are the commutation relations for the coordinates defining $\kappa$-Minkowski spacetime. For the second commutator, notice that we used $k_0\triangleright \hat{x}^i = 0$. This algebra is called $\mathfrak{an}(3)$ algebra. 

One can then show that the commutators \eqref{kminkcomm} is invariant under boosts and rotations. The invariance under rotations is trivial since, as can be seen from eq. \eqref{Mk-comm}, \eqref{MN-comm}, \eqref{rotcopbi}, \eqref{antiMN} the rotation sector is completely not deformed in the $\kappa$-Poincar\'e Hopf algebra, and so it acts as usual. For the boosts, using eq. \eqref{actiongenonx}, as well as the previous computations one gets \cite{Majid:1994cy}
\begin{align}
	N_i \triangleright \hat{x}^0\hat{x}^j
	&=
	(N_i\triangleright \hat{x}^0)(\mathbb{1}\triangleright \hat{x}^j)
	+
	(e^{-k_0/\kappa}\triangleright \hat{x}^0)(N_i\triangleright \hat{x}^j)
	+
	\frac{1}{\kappa}
	\epsilon_{ijk}
	(k_j\triangleright \hat{x}^0)(M_k \triangleright \hat{x}^j) \nonumber \\
	&=
	ix_i \hat{x}^j
	+
	\left[\left(1 - \frac{k_0}{ \kappa} + \frac{k_0^2}{2\kappa^2} - \dots\right)\triangleright \hat{x}^0 \right] (i \hat{x}^0 \delta_i^j) 
	+
	0\\
	&=
	ix_i \hat{x}^j
	+
	i(\hat{x}^0)^2 \delta_i^j
	-
	\frac{\hat{x}^0\delta_i^j}{\kappa}
\end{align} 
\begin{align}
	N_i \triangleright \hat{x}^j\hat{x}^0
	&=
	(N_i\triangleright \hat{x}^j)(\mathbb{1}\triangleright \hat{x}^0)
	+
	(e^{-k_0/\kappa}\triangleright \hat{x}^j)(N_i\triangleright \hat{x}^0)
	+
	\frac{1}{\kappa}
	\epsilon_{ijk}
	(k_j\triangleright \hat{x}^j)(M_k \triangleright \hat{x}^0) \nonumber \\
	&=
	i\hat{x}^0\delta_i^j \hat{x}^0
	+
	i\hat{x}^jx_i
	+
	0\\
	&=
	i(\hat{x}^0)^2\delta_i^j + i\hat{x}^jx_i
\end{align} 
and therefore
\begin{align}
	N_i\triangleright[\hat{x}^0,\hat{x}^j] = N_i\triangleright \frac{i \hat{x}^j}{\kappa} +i[x_i,\hat{x}^j] = N_i\triangleright \frac{i \hat{x}^j}{\kappa}
\end{align}
which shows that indeed the commutation relation \eqref{kminkcomm} are invariant under $\kappa$-deformed boosts. Notice that we used the fact that $M_k \triangleright \hat{x}^0 = 0$ (spatial rotations do not act on the time direction) and eq. \eqref{kminkcomm}.

Although we used the bicrossproduct basis for the computations up to now, this is by no means the only choice. In particular, another choice of coordinates for momentum space is the so called \textit{classical basis}, which is related to the bicrossproduct one by the following coordinate change
\begin{align}\label{classicalbasis1a}
	P_0(k_0, \mathbf{k})
	=
	\kappa \sinh \frac{k_0}{\kappa} + \frac{\mathbf{k}^2}{2\kappa} e^{\frac{k_0}{\kappa}}, 
	\qquad
	P_i(k_0, \mathbf{k})
	=
	k_i
	e^\frac{k_0}{\kappa}
\end{align}
\begin{align}\label{classicalbasis1b}
	P_4(k_0, \mathbf{k})
	=
	\kappa\cosh \frac{k_0}{\kappa} - \frac{\mathbf{k}^2}{2\kappa} e^{\frac{k_0}{\kappa}}.
\end{align}
One can check by direct computations \cite{Borowiec:2009vb}, \cite{Arzano:2021scz} that
\begin{align}\label{classicalconstraints}
	-P_0^2 + \mathbf{P}^2 + P_4^2 = \kappa^2
	\qquad
	P_+:= P_0+P_4 >0
	\qquad
	P_4>0.
\end{align}
In this basis, the algebra sector of the $\kappa$-Poincar\'e Hopf algebra reads
\begin{align}
	[M_i, P_j] = i\epsilon_{ijk} P_k,
	\quad
	[M_i,P_0] = 0,
	\quad
	[N_i,P_j] = i\delta_{ij} P_0, 
	\quad
	[N_i,P_0] = i P_i,
\end{align}
\begin{align}
	[M_i,M_j] = i\epsilon_{ijk} M_k,
	\quad
	[M_i,N_j] = i\epsilon_{ijk} N_k,
	\quad
	[N_i,N_j] = i\epsilon_{ijk} M_k.
\end{align}
The co-algebra sector is given by
\begin{align}\label{DeltaPi}
	\Delta P_i
	=
	\frac{1}{\kappa} P_i \otimes P_+
	+
	\mathbb{1} \otimes P_i,
\end{align}
\begin{align}\label{DeltaP0}
	\Delta P_0
	=
	\frac{1}{\kappa} P_0 \otimes P_+
	+
	\sum_k
	\frac{P_k}{P_+}
	\otimes
	P_k
	+
	\frac{\kappa}{P_+} \otimes P_0,
\end{align}
\begin{align}\label{DeltaP4}
	\Delta P_4
	=
	\frac{1}{\kappa} P_4 \otimes P_+
	-
	\sum_k
	\frac{P_k}{P_+}
	\otimes
	P_k
	-
	\frac{\kappa}{P_+} \otimes P_0,
\end{align}
\begin{align}\label{DeltaNclassical}
	\Delta N_i
	=
	N_i \otimes \mathbb{1}
	+
	\frac{\kappa}{P_+} \otimes N_i
	+
	\epsilon_{ijk}
	\frac{P_j}{P_+}
	\otimes 
	M_k,
\end{align}
\begin{align}\label{DeltaMclassical}
	\Delta M_i = M_i \otimes \mathbb{1} + \mathbb{1} \otimes M_i,
\end{align}
and the antipodes by
\begin{align}\label{SP}
	S(P_i)
	=
	-\frac{\kappa P_i}{P_+},
	\qquad
	S(P_0)
	=
	-P_0
	+
	\frac{\mathbf{P}^2}{P_+},
	\qquad
	S(P_4)=P_4,
\end{align}
\begin{align}\label{antiMNclassical}
	S(M_i) = -M_i,
	\qquad
	S(N_i)
	=
	-N_i \frac{P_+}{\kappa}
	+
	\frac{1}{\kappa}
	\epsilon_{ijk}
	P_j M_k.
\end{align}
Once again, the co-units $\epsilon(k)$, $\epsilon(N_i)$, $\epsilon(M_j)$ are all zero. Furthermore, notice also that rotations are   non-deformed in both basis. A very interesting property of the classical basis is the fact that the algebra coincides with the non-deformed Poincar\'e algebra, so the totality of the effect of $\kappa$-deformation is found in the co-product and antipode sector. This also shows that $\kappa$-deformation is important at the Hopf algebra level, and not merely at the algebra level. Because of the construction of $\kappa$-Minkowski showed above, the co-products cover a particularly important role since they allow the definition of $\kappa$-Minkowski spacetime. More details can be found in \cite{Arzano:2021scz}, \cite{Kowalski-Glikman:2002eyl}.

\section{From $\kappa$-Minkowski to $\kappa$-Poincar\'e}\label{kMtkP}

The first step in going from $\kappa$-Minkowski to $\kappa$-Poincar\'e is to find an explicit representation of the $\mathfrak{an}(3)$ algebra in eq. \eqref{kminkcomm}. We choose (with a slight abuse of notation, we call the explicit matrices representing the algebra with the same names as the elements of the algebra itself)
\begin{equation}\label{II.1.18}
	\hat x^0 = -\frac{i}{\kappa} \,\left(\begin{array}{ccc}
		0 & \mathbf{0} & 1 \\
		\mathbf{0}^T & \tilde{\mathbf{0}} & \mathbf{0}^T \\
		1 & \mathbf{0} & 0
	\end{array}\right) \quad
	\hat{x}^i = \frac{i}{\kappa} \,\left(\begin{array}{ccc}
			0 & {(\epsilon^i)\,{}^T} &  0\\
			\epsilon^i & \tilde{\mathbf{0}} & \epsilon^i \\
			0 & -(\epsilon^i)\,{}^T & 0
		\end{array}\right).
\end{equation}
The central $\tilde{\mathbf{0}}$ is a $3\times 3$ null matrix, while $\mathbf{0} = (0,0,0)$ and $\epsilon^i$ is a unit column vector, such that $(\epsilon^1)^T = (1,0,0)$, $(\epsilon^2)^T = (0,1,0)$, and $(\epsilon^1)^T = (0,0,1)$. We now compute the group elements of $AN(3)$
\begin{equation}\label{defek}
	\hat{e}_k =e^{ik_i \hat x^i} e^{ik_0 \hat x^0}\,.
\end{equation}
Notice that because of the non-trivial algebra we need to carefully define the exponentiation, i.e. on how we decide to define the group element. Here we choose the so called \textit{time-to-the-right} convention, in which $e^{i k_j \hat{x}^j}$ is to the left of $e^{i k_0 \hat{x}^0}$. Because of this choice, the objects $k_0, \mathbf{k}$ in the exponent (which are dimensionally momenta) coincide with the bicrossproduct basis introduced in the previous section. With a different choice of group element, for example 
\begin{align}\label{normalbasis}
	\hat{e}_q = e^{iq_\mu \hat{x}^\mu}
\end{align}
one would get different coordinates for momentum space (in this particular case, we would be introducing the so called normal basis \cite{Arzano:2021scz}, but we will not address this further). Furthermore, notice that the representation matrices are $5 \times 5$.

We have
\begin{align}
	\begin{pmatrix}
		0 & \bm{0} & 1 \\
		\bm{0} & \bm{0} & \bm{0} \\
		1 & \bm{0} & 0
	\end{pmatrix}^{2n}
	=
	\begin{pmatrix}
		1 & \bm{0} & 0 \\
		\bm{0} & \bm{0} & \bm{0} \\
		0 & \bm{0} & 1
	\end{pmatrix}
	\qquad 
	\begin{pmatrix}
		0 & \bm{0} & 1 \\
		\bm{0} & \bm{0} & \bm{0} \\
		1 & \bm{0} & 0
	\end{pmatrix}^{2n+1}
	=
	\begin{pmatrix}
		0 & \bm{0} & 1 \\
		\bm{0} & \bm{0} & \bm{0} \\
		1 & \bm{0} & 0
	\end{pmatrix}
\end{align}
and therefore
\begin{align}
	e^{ik_0 \hat x^0}
	&=
	\sum_n \frac{(ik_0 x^0)^n}{n!}
	=
	\sum_n \frac{1}{n!} \left(\frac{k_0}{\kappa}\right)^n
	\begin{pmatrix}
		0 & \bm{0} & 1 \\
		\bm{0} & \bm{0} & \bm{0} \\
		1 & \bm{0} & 0
	\end{pmatrix}^n \\
	&=
	\left(\begin{array}{ccc}
		\cosh\frac{k_0}\kappa & \mathbf{0} & \sinh\frac{k_0}\kappa \\&&\\
		\mathbf{0} & \mathbf{1} & \mathbf{0} \\&&\\
		\sinh\frac{k_0}\kappa\; & \mathbf{0}\; & \cosh\frac{k_0}\kappa
	\end{array}\right)
\end{align}
Similarly one can check that 
\begin{align}
	\begin{pmatrix}
		0 & \epsilon^T & 0 \\
		\epsilon & \bm{0} & \epsilon \\
		0 & -\epsilon^T & 0
	\end{pmatrix}^2
	=
	\begin{pmatrix}
		1 & \bm{0} & 1 \\
		\bm{0} & \bm{0} & \bm{0} \\
		-1 & \bm{0} & -1
	\end{pmatrix}
	\qquad
	\begin{pmatrix}
		0 & \epsilon^T & 0 \\
		\epsilon & \bm{0} & \epsilon \\
		0 & -\epsilon^T & 0
	\end{pmatrix}^3
	=
	0
\end{align}
This behaviour is the reason for the `$N$' in the name '$AN(3)$', which stands for `nilpotent'. Therefore 
\begin{align}
	e^{i k_i \hat{x}^i}  
	=
	1  + i k_i \hat{x}^i + \frac{1}{2} (i k_i \hat{x}^i)^2
	=
	\begin{pmatrix}
		1 + \frac{\mathbf{k}^2}{2\kappa^2} & \frac{\mathbf{k}^T}{\kappa} & \frac{\mathbf{k}^2}{2\kappa^2} \\
		\frac{\mathbf{k}}{\kappa} & \bm{1} & \frac{\mathbf{k}}{\kappa} \\
		-\frac{\mathbf{k}^2}{2\kappa^2} & -\frac{\mathbf{k}^T}{\kappa} & 1-\frac{\mathbf{k}^2}{2\kappa^2}.
	\end{pmatrix}
\end{align}
Multiplying these two matrices we get
\begin{align}\label{ek}
	\hat{e}_k
	=
	\begin{pmatrix}
		\cosh  \frac{k_0}{\kappa} + \frac{\mathbf{k}^2}{2\kappa^2}e^{\frac{k_0}{\kappa}} & \frac{\mathbf{k}^T}{\kappa} & \sinh \frac{k_0}{\kappa} + \frac{\mathbf{k}^2}{2\kappa^2} e^{\frac{k_0}{\kappa}} \\
		\frac{\mathbf{k}}{\kappa} e^{\frac{k_0}{\kappa}} & \bm{1} & \frac{\mathbf{k}}{\kappa}e^{\frac{k_0}{\kappa}} \\
		\sinh \frac{k_0}{\kappa} - \frac{\mathbf{k}^2}{2\kappa^2}e^{\frac{k_0}{\kappa}} & -\frac{\mathbf{k}^T}{\kappa} & \cosh \frac{k_0}{\kappa} - \frac{\mathbf{k}^2}{2\kappa^2} e^{\frac{k_0}{\kappa}}
	\end{pmatrix}
\end{align}
where we used the relation $\cosh x + \sinh x = e^x$. Notice now that we can make the following substitutions
\begin{eqnarray}\label{classicalbasis2}
	{P_0}(k_0, \mathbf{k}) &=&\kappa  \sinh
	\frac{k_0}{\kappa} + \frac{\mathbf{k}^2}{2\kappa}\,
	e^{  {k_0}/\kappa}, \\
	P_i(k_0, \mathbf{k}) &=&   k_i \, e^{  {k_0}/\kappa},\label{II.1.20}\\
	{P_4}(k_0, \mathbf{k}) &=& \kappa \cosh
	\frac{k_0}{\kappa} - \frac{\mathbf{k}^2}{2\kappa}\, e^{
		{k_0}/\kappa}
\end{eqnarray}
to bring the above group element $\hat{e}_k$ into the following form
\begin{align}\label{groupclassical}
	\hat{e}_{P(k)}
	=
	\frac{1}{\kappa}
	\begin{pmatrix}
		\tilde{P}_4 & \kappa\mathbf{P}/P_+ & P_0 \\
		\mathbf{P} & \kappa \times \mathbb{1}_{3\times 3}&  \mathbf{P} \\
		\tilde{P}_0 & -\kappa\mathbf{P}/P_+ & P_4
	\end{pmatrix}
\end{align}
where 
\begin{align}\label{P0eP4tilde}
	\tilde{P}_0
	=
	P_0 - \frac{\mathbf{P}^2}{P_+}
	=
	-S(P_0)
	\qquad
	\tilde{P}_4
	=
	P_4 + \frac{\mathbf{P}^2}{P_+}
\end{align}
Incidentally, one can use the group property $\hat{e}_{P(k)}\hat{e}_{Q(l)} = \hat{e}_{P(k)\oplus Q(l)}$ and $\hat{e}^{-1}_{P} = \hat{e}_{S(P)}$ to obtain the deformed antipode and sum in the classical basis.
\begin{align}\label{antipode0}
	S(P_0) = -P_0 + \frac{\mathbf{P}^2}{P_0+P_4} = \frac{\kappa^2}{P_0+P_4}-P_4\,,
\end{align}
\begin{align}\label{antipodei4}
	S(\mathbf{P}) =-\frac{\kappa \mathbf{P} }{P_0+P_4}\,,\quad S(P_4) = P_4.
\end{align}
\begin{align}
	(P\oplus Q)_0 &= \frac1\kappa\, P_0(Q_0+Q_4) + \frac{\mathbf{P}\mathbf{Q}}{P_0+P_4} +\frac{\kappa}{P_0+P_4}\, Q_0\label{defsum0-nostar}\\
	(P\oplus Q)_i &=\frac1\kappa\, P_i(Q_0+Q_4) + Q_i\label{defsumi-nostar}\\
	(P\oplus Q)_4 &= \frac1\kappa\, P_4(Q_0+Q_4) - \frac{\mathbf{P}\mathbf{Q}}{P_0+P_4} -\frac{\kappa}{P_0+P_4}\, Q_0\label{defsum4-nostar}
\end{align}
One can of course do the same thing in the bicrossproduct basis. Notice the relationship between these equations and the co-products in eq. \eqref{DeltaPi}, \eqref{DeltaP0}, \eqref{DeltaP4}. 

Notice that the momenta defined in eq. \eqref{classicalbasis2} are indeed again the classical basis defined in eq. \eqref{classicalbasis1a}, \eqref{classicalbasis1b}. By construction, therefore, they satisfy the constraints in eq. \eqref{classicalconstraints}. Furthermore, notice that the group elements $\hat{e}_{P(k)}$ are uniquely identified once one knows the quantities $P_0, P_i, P_4$, which are the components on the last column of the matrix \eqref{groupclassical}. Indeed, calling $\mathcal{O} = (0,0,0,0,\kappa)^T$, we have
\begin{align}
	\hat{e}_{P(k)} \mathcal{O} = (P_0,\mathbf{P}, P_4)^T
\end{align} 
and the action of the group $AN(3)$ on $\mathcal{O}$ is transitive, meaning that acting with all $g\in AN(3)$ on $\mathcal{O}$ reaches all the points of the manifold associated to the group. We already know that relations \eqref{classicalconstraints} hold, which means that the 4-dimesional manifold associated to $AN(3)$ is half of de Sitter space, namely the half where $P_+>0$. Notice that the curvature of such a space is constant and equal to $1/\kappa^2$. Furthermore, since each point is uniquely described by the coordinates $(P_0, \mathbf{P})$ (after fixing both $P_0$ and $\mathbf{P}$, $P_4$ is not a free parameter, and can be obtained through the constraint), the group manifold $AN(3)$ is therefore our momentum space manifold. This gives a clear picture of how the non-commutative nature of spacetime coordinates is linked to a constant curvature of momentum space. Notice that to obtain coordinates spanning the other half of de Sitter space defined by the first constraint in eq. \eqref{classicalconstraints}, one applies the same group elements to the new origin $\mathcal{O}^*=  (0,0,0,0,-\kappa)^T$. Alternatively, one can define a matrix
\begin{align}
	\mathfrak{z}
	=
	e^{\pi \kappa \hat{x}^0}
\end{align}
which allows us to define 
\begin{align}\label{estar}
	\hat{e}^*_k = \hat{e}_k \mathfrak{z}
\end{align}
and then apply these new objects $\hat{e}^*$ to the same origin $\mathcal{O}=(0,0,0,0,\kappa)$. In both cases, we obtain the coordinates
\begin{eqnarray}\label{classicalbasis2-star}
	{P^*_0}(k_0, \mathbf{k}) &=&-\kappa  \sinh
	\frac{k_0}{\kappa} - \frac{\mathbf{k}^2}{2\kappa}\,
	e^{  {k_0}/\kappa}, \\
	P^*_i(k_0, \mathbf{k}) &=&   -k_i \, e^{  {k_0}/\kappa},\\
	{P^*_4}(k_0, \mathbf{k}) &=& -\kappa \cosh
	\frac{k_0}{\kappa} + \frac{\mathbf{k}^2}{2\kappa}\, e^{	{k_0}/\kappa}.
\end{eqnarray}
These starred coordinates satisfy $P_+^* = P_0^*+P_4^* <0$, and one can show that 
\begin{align}\label{starredcoord}
	S(P^*) = -S(P)
\end{align} 
and that the deformed sum is not modified if one or both of the momenta is starred. In other words, we have
\begin{align}
	(P^{(*)}\oplus Q^{(*)})_0 &= \frac1\kappa\, P^{(*)}_0(Q^{(*)}_0+Q^{(*)}_4) + \frac{\mathbf{P}^{(*)}\mathbf{Q}^{(*)}}{P^{(*)}_0+P^{(*)}_4} +\frac{\kappa}{P^{(*)}_0+P^{(*)}_4}\, Q^{(*)}_0\label{defsum0}\\
	(P^{(*)}\oplus Q^{(*)})_i &=\frac1\kappa\, P^{(*)}_i(Q^{(*)}_0+Q^{(*)}_4) + Q^{(*)}_i\label{defsumi}\\
	(P^{(*)}\oplus Q^{(*)})_4 &= \frac1\kappa\, P^{(*)}_4(Q^{(*)}_0+Q^{(*)}_4) - \frac{\mathbf{P}^{(*)}\mathbf{Q}^{(*)}}{P^{(*)}_0+P^{(*)}_4} -\frac{\kappa}{P^{(*)}_0+P^{(*)}_4}\, Q^{(*)}_0\label{defsum4},
\end{align}
where the ${}^{(*)}$ indicates that either $P$ or $Q$ or both or neither can be starred momenta, and the deformed sum formula works in the same way regardless \cite{Arzano:2020jro}. 

We now address the issue of how one can get the $\kappa$-Poincar\'e structure out of the $AN(3)$ group, and in particular the co-product and antipode sectors. We will not treat in detail all the coproducts and antipodes for all generators, and we will use either the classical or the bicrossproduct basis depending on which is easier to work with. The interested reader is referred back to \cite{Arzano:2022ewc} for more details. 

We start with the coproduct and antipode of momenta. As was already discussed in the previous section, coproducts describe the way the $\kappa$-Poincar\'e algebra act on products, be it products of coordinates like before, or of states like we will consider now. In particular, we have
\begin{align}
	\Delta k_\mu (|p \rangle \otimes |q \rangle)
	=
	(p\oplus q)_\mu|p \rangle \otimes |q \rangle
\end{align}
where $p\oplus q$ is define through the group multiplication
\begin{align}
	\hat{e}_p \hat{e}_q = \hat{e}_{p\oplus q}.
\end{align}
In the same way, the antipode $S(k)_\mu$ is related to inverse group elements by the definition 
\begin{align}
	\hat{e}^{-1}_k = \hat{e}_{S(k)}.
\end{align}
One can therefore multiply two copies of eq. \eqref{ek} and deduce the sum of momenta, and then take the inverse of eq. \eqref{ek} directly to obtain the antipode. Alternatively, in a more illuminating and less computation intensive way, one can work directly with the definition in eq. \eqref{defek}. Indeed, using the Baker-Campbell-Hausdorff formula one can show that
\begin{align}
	[X,Y] = sY
	\qquad\implies\qquad
	e^X e^Y
	=
	\exp 
	\left(
	X + \frac{s}{1-e^{-s}}Y
	\right)
\end{align}
which means that since because of eq. \eqref{kminkcomm} we have
\begin{align}
	[ik_0x^0, i\mathbf{k}\mathbf{x}]
	=
	-k_0\mathbf{k}[x^0, \mathbf{x}]
	=
	-k_0\mathbf{k}
	\frac{i}{\kappa}
	\mathbf{x}
	=
	-\frac{k_0}{\kappa}
	(i\mathbf{k}\mathbf{x}),
\end{align}
then
\begin{align}
	e^{ik_0x^0}
	e^{i\mathbf{k}\mathbf{x}}
	&=
	\exp
	\left(
	ik_0x^0
	+
	\frac{-\frac{k_0}{\kappa}}{1-e^{\frac{k_0}{\kappa}}}
	i\mathbf{k}\mathbf{x}
	\right)
\end{align}
\begin{align}
	e^{i\mathbf{q}\mathbf{x}}
	e^{iq_0x^0}
	&=
	\exp
	\left(
	iq_0x^0
	+ 
	\frac{\frac{k_0}{\kappa}}{1-e^{-\frac{k_0}{\kappa}}}
	i\mathbf{q}\mathbf{x}
	\right) 
	=
	\exp
	\left(
	iq_0x^0
	+ 
	\frac{-\frac{k_0}{\kappa}}{1-e^{\frac{k_0}{\kappa}}}
	ie^{\frac{k_0}{\kappa}}\mathbf{q}\mathbf{x}
	\right)
\end{align}
and therefore sending $\mathbf{q}\rightarrow \mathbf{k}e^{-\frac{k_0}{\kappa}}$ and $q_0\rightarrow k_0$ we prove 
\begin{align}\label{id1}
	e^{ik_0x^0}
	e^{i\mathbf{k}\mathbf{x}}
	=
	e^{ie^{-\frac{k_0}{\kappa}}\mathbf{k}\mathbf{x}}
	e^{ik_0x^0}.
\end{align}
Using this relation, we get
\begin{align}
	\hat{e}_k\hat{e}_q
	&=
	e^{i\mathbf{k}\mathbf{x}}
	e^{ik_0x^0}
	e^{i\mathbf{q}\mathbf{x}}
	e^{iq_0x^0} 
	=
	e^{i\left(\mathbf{k} + e^{-\frac{k_0}{\kappa}}\mathbf{q} \right)\mathbf{x}}
	e^{i(k_0+q_0)x^0}
\end{align}
Furthermore, we can also recover the inverse group element $\hat{e}_k^{-1}$ using a similar procedure. In fact we can use the same steps as before, but at the end we must impose that both the exponent go to zero, so that we get
\begin{align}
	\mathbf{q}
	=
	- e^{\frac{k_0}{\kappa}}
	\mathbf{k}
	\qquad
	q_0 = -k_0
\end{align}
and therefore
\begin{align}
	\hat{e}_{k}^{-1}
	=
	e^{-ie^{\frac{k_0}{\kappa}}
		\mathbf{k}}
	e^{-ik_0x^0}
\end{align}
We now consider the \textit{bicrossproduct coordinate functions} on $AN(3)$ defined as \cite{Arzano:2021scz}
\begin{align}
	f^B_\mu(g\in AN(3))
	=
	f^B_\mu
	\left(
	e^{i\mathbf{k}\mathbf{x}}
	e^{ik_0x^0}
	\right)
	:=
	k_\mu
	=
	(k_0, \mathbf{k})
\end{align}
These are important because functions on a group have a natural structure of Hopf algebra \cite{Arzano:2021scz}. In fact, using the same notation as \cite{Arzano:2021scz}, the algebra $Fun(G)$ of functions on $G$ with multiplication and identity defined as usual
\begin{align}
	(f_1, f_2)(g)=f_1(g)f_2(g)
	\qquad
	I(g)=1
\end{align}
This algebra $Fun(G)$ is of course commutative and associative (because pointwise multiplication of functions with codomain in $\mathbb{C}$ is of course commutative and associative). From this we can naturally define all of the additional structure needed for a Hopf algebra, namely \textit{coproduct}, \textit{co-unit}, and \textit{antipode}
\begin{align}
	\Delta &: Fun(G) \rightarrow Fun(G\times G) 
	\qquad
	\epsilon : Fun(G) \rightarrow \mathbb{C} 
	\qquad
	S : Fun(G) \rightarrow Fun(G)
\end{align}
with the definitions
\begin{align}
	(\Delta f)(g_1, g_2) &= \sum_\alpha f_\alpha^{(1)} \otimes f_\alpha^{(2)}(g_1,g_2)
	= f(g_1g_2) 
\end{align}
\begin{align}
	\epsilon(g) = f(1) 
	\qquad
	(S(f))(g) = f(g^{-1})
\end{align}
Therefore, using these definitions and the bicrossproduct coordinate functions, we immediately find
\begin{align}
	\Delta\left(f^B_0 \right)(\hat{e}_k, \hat{e}_q)
	&=
	f^B_0(\hat{e}_k \hat{e}_q) \\
	&=
	f^B_0 
	\left(
	e^{i\left(\mathbf{k} + e^{-\frac{k_0}{\kappa}}\mathbf{q} \right)\mathbf{x}}
	e^{i(k_0+q_0)x^0}
	\right) \\
	&=
	k_0 + q_0 \\
	&=
	\left(
	f^B_0 \otimes 1 + 1 \otimes f^B_0
	\right)
	(\hat{e}_k, \hat{e}_q)
\end{align}
\begin{align}
	\Delta\left(\mathbf{f}^B_i \right)(\hat{e}_k, \hat{e}_q) 
	&=
	\mathbf{f}^B_i(\hat{e}_k \hat{e}_q) \\
	&=
	\mathbf{f}^B_i 
	\left(
	e^{i\left(\mathbf{k} + e^{-\frac{k_0}{\kappa}}\mathbf{q} \right)\mathbf{x}}
	e^{i(k_0+q_0)x^0}
	\right) \\
	&=
	\mathbf{k} + e^{-\frac{k_0}{\kappa}}\mathbf{q} \\
	&=
	\left(
	\mathbf{f}^B_i \otimes 1 + e^{-\frac{k_0}{\kappa}} \otimes \mathbf{f}^B_i
	\right)
	(\hat{e}_k, \hat{e}_q)
\end{align}
and similarly for the co-unit
\begin{align}
	\epsilon(f^B_0)
	=
	f^B_0(1)
	=
	f^B_0\left(e^{i0\mathbf{x}} e^{i0x^0}\right)
	=
	0
\end{align}
\begin{align}
	\epsilon(\mathbf{f}^B_i)
	=
	\mathbf{f}^B_i(1)
	=
	\mathbf{f}^B_i\left(e^{i0\mathbf{x}} e^{i0x^0}\right)
	=
	0
\end{align}
and the antipode
\begin{align}
	S(f^B_0)(\hat{e}_k)
	=
	f^B_0(\hat{e}_k^{-1})
	=
	f^B_0
	\left(
	e^{-ie^{\frac{k_0}{\kappa}}
		\mathbf{k}}
	e^{-ik_0x^0}
	\right)
	=
	-k_0
\end{align}
\begin{align}
	S(\mathbf{f}^B_i)(\hat{e}_k)
	=
	f_i(\hat{e}_k^{-1})
	=
	f_i
	\left(
	e^{-ie^{\frac{k_0}{\kappa}}
		\mathbf{k}}
	e^{-ik_0x^0}
	\right)
	=
	-e^{\frac{k_0}{\kappa}}
	\mathbf{k}
\end{align}
With a slight abuse of notation, these relations are traditionally written as
\begin{align}
	\Delta k_0 = k_0 \otimes \mathbb{1} + \mathbb{1} \otimes k_0
\qquad
	\Delta \mathbf{k}_i
	=
	\mathbf{k}_i \otimes \mathbb{1} + e^{-\frac{k_0}{\kappa}}\otimes \mathbf{k}_i
\end{align}
\begin{align}
	S(k)_0 = -k_0
	\qquad
	S(\mathbf{k})_i
	=
	-e^{\frac{k_0}{\kappa}}\mathbf{k}_i
\end{align}
so that we recover eq. \eqref{momcopbi} and \eqref{antik}. In the same way, using the classical basis, we recover eq. \eqref{DeltaPi}, \eqref{DeltaP0}, \eqref{SP}.

To get the coproduct of boosts and rotations, we follow a slightly different route. Indeed, knowing that $AN(3)$ is our momentum manifold, we want to find some action of $SO(1,3)$ on it. The trick to find it is to notice that both $AN(3)$ and $SO(1,3)$ are subgroups of $SO(1,4)$. We can then use something called \textit{Iwasawa decomposition}
\begin{align*}
	\mathfrak{so}(1,4) = \mathfrak{so}(1,3) \oplus \mathfrak{a} \oplus \mathfrak{n}
\end{align*}
where $\mathfrak{a}$ is generated by $\hat{x}^0$, $\mathfrak{n}$ is generated by $\hat{x}^i$. In other words, given $g\in SO(1,4)$, then given $K_g\in SO(1,3)$ and $g\in AN(3)$, then there exist two \textit{unique} elements $K'_g \in SO(1,3)$ and $g' \in AN(3)$ such that 
\begin{align*}
	K_g g = g' K'_g
	\qquad
	\Leftrightarrow \qquad
	K_g g (K'_g)^{-1} = g'.
\end{align*}
In this way, we obtain the action of $SO(1,3)$ on a single group element. To get the co-product, we then need to define the action on the product $gh$, where $g,h\in AN(3)$. We do this by the following relation.
\begin{align}
	(gh)'
	=
	K_g gh K'_{gh}
	=
	(K_g g {K'}_g^{-1})
	(K'_g h K'_{gh})
\end{align}
We will see shortly that $K'_{gh}$ is irrelevant for the definition of the coproducts of $SO(1,3)$ generators, so that we only need $K_g$ and $K'_g$. We are only interested in infinitesimal transformations, so 
we assume
\begin{align}
	K_g \approx 1 + i \xi^a \mathfrak{k}_a
	\qquad
	K'_g \approx 1 + i \xi^a h^b_a(g) \mathfrak{k}_b.
\end{align}
Notice that $h^b_a(g)$ is in general momentum dependent, a fact which is responsible for the non-trivial boost co-product. Concentrating on boosts, and in particular choosing a boost in the direction 1, and using the classical basis for simplicity, explicitly we have 
\begin{align}
	&\begin{pmatrix}
		1 & \xi & 0 & 0 & 0 \\
		\xi & 1 & 0 & 0 & 0 \\
		0 & 0 & 1 & 0 & 0 \\
		0 & 0 & 0 & 1 & 0 \\
		0 & 0 & 0 & 0 & 1
	\end{pmatrix}
	\begin{pmatrix}
		\bar{P}_4 & \frac{\kappa \mathbf{P}_i}{P_+} & P_0 \\
		\mathbf{P}_i & \kappa & \mathbf{P}_i \\
		\bar{P}_0 & -\frac{\kappa \mathbf{P}_i}{P_+} & P_4
	\end{pmatrix} \\
	=&
	\begin{pmatrix}
		\bar{P}'_4 & \frac{\kappa \mathbf{P}'_i}{P'_+} & P'_0 \\
		\mathbf{P}'_i & \kappa & \mathbf{P}'_i \\
		\bar{P}'_0 & -\frac{\kappa \mathbf{P}'_i}{P'_+} & P'_4
	\end{pmatrix}
	\begin{pmatrix}
		1 & \bar{\xi}^1 & \bar{\xi}^2 & \bar{\xi}^3 & 0 \\
		\bar{\xi}^1 & 1 & \bar{\rho}^3 & -\bar{\rho}^2 & 0\\
		\bar{\xi}^2 & -\bar{\rho}^3 & 1 & \bar{\rho}^1 & 0\\
		\bar{\xi}^3 & \bar{\rho}^2 & -\bar{\rho}^1 & 1 & 0\\
		0 & 0 & 0 & 0 & 1
	\end{pmatrix}
\end{align}
where the parameters $\bar{\xi}^i$ and $\bar{\rho}^j$ can be uniquely determined by dong the product and comparing both sides of the equivalence \cite{Arzano:2021scz}. After doing the computations, one obtains
\begin{align}
	K_g'
	=
	\begin{pmatrix}
		1 & \frac{\kappa}{P_+}\xi & 0 & 0 & 0 \\
		\frac{\kappa}{P_+}\xi & 1 & \frac{P_2}{P_+}\xi & \frac{P_3}{P_+}\xi & 0 \\
		0 & -\frac{P_2}{P_+}\xi & 1 & 0 & 0 \\
		0 & -\frac{P_3}{P_+}\xi & 0 & 1 & 0 \\
		0 & 0 & 0 & 0 & 1
	\end{pmatrix}
	=
	\mathbb{1} + \frac{\kappa}{P_+} \xi N_1
	+
	\epsilon_{ijk} \frac{P_j}{P_+} M_k.
\end{align}
In other words, we have up to first order
\begin{align}
	(gh)'
	=
	(\mathbb{1}+\xi N_1) g \left(\mathbb{1} + \frac{\kappa}{P_+} \xi N_1
	+
	\epsilon_{ijk} \frac{P_j}{P_+} M_k\right) h K'_{gh}
\end{align}
Now, notice that a multiplication from the right by $K'_{gh}$ cannot influence the momenta in a group element $h$. This is due to the fact that $K'_{gh}$ will have all zero entries in the last column and row, except $(K'_{gh})_{44} = 1$, and a multiplication from the right of such a matrix cannot influence the last row of $h$, which contains the momenta. Therefore, we have up to first order  
\begin{align}\label{firstorderDeltaN}
	(gh)' = gh + 
	\xi
	\left(
	(N_1 g)(h)
	+
	\left(\frac{\kappa}{P_+} g \right)(N_1 h) \right)
	+
	\epsilon_{1jk}
	\left(\frac{P_j}{P_+} g\right)(M_k h).
\end{align}
and generalizing this relation to a boost in a generic direction we get immediately the co-product in eq. \eqref{DeltaNclassical}. Finally, for the antipode, one can just switch $g\mapsto g^{-1}$ and $h\mapsto g$ into eq. \eqref{firstorderDeltaN}, and impose that the result gives $\mathbb{1}$, which means that we get the antipode in eq. \eqref{antiMNclassical}. For the rotation, the whole process is repeated, and one gets that the co-product and the antipode are not deformed, in accordance with eq. \eqref{DeltaMclassical} and \eqref{antiMNclassical}.

\section{Finite boosts for two-particle states}

What we computed above is the co-product of generators of the algebra on tensor product states. To get a finite boost for two particles, one needs to exponentiate this infinitesimal generators to get a finite transformation. However, due to the non-trivial nature of co-products both in the classical and bicrossproduct basis, this is often not practical. Luckily for us, however, the Iwasawa decomposition that was used in section \ref{kMtkP} holds in general, and we only used a first order expansion for the involved matrices because we only needed the generators co-products \cite{Bevilacqua:2023pqz}. Here, we employ once again the same decomposition, but this time we will use the full transformation matrices to obtain a finite boost for a two-particle state. Contrary to the previous sections, the computations and discussions for the rest of the chapter are original results. Because of the importance of the final result for the kinematics of a two-particle state in the $\kappa$-deformed context, we will explicitly perform all the computations in full.

We start therefore from
\begin{align}\label{iwasawa}
	L g = g' L'_g
\end{align}
where the boost on the RHS can in general be momentum dependent.


We need to recover what is $L'_g$. Written explicitly in terms of matrices, and considering without loss of generality $L$ as a boost in the $x^1$ direction, we have
\begin{align}\label{L}
	L
	=
	\begin{pmatrix}
		\gamma & -\beta\gamma & 0 & 0 & 0 \\
		-\beta\gamma & \gamma & 0 & 0 & 0 \\
		0 & 0 & 1 & 0 & 0 \\
		0 & 0 & 0 & 1 & 0 \\
		0 & 0 & 0 & 0 & 1
	\end{pmatrix}
\end{align}
\begin{align}\label{g}
	g
	=
	\begin{pmatrix}
		\tilde{P}_4 & \kappa\mathbf{P}_1/P_+ & \kappa\mathbf{P}_2/P_+ & \kappa\mathbf{P}_3/P_+ & P_0 \\
		\mathbf{P}_1 & \kappa & 0 & 0 & \mathbf{P}_1 \\
		\mathbf{P}_2 & 0 & \kappa & 0 & \mathbf{P}_2 \\
		\mathbf{P}_3 & 0 & 0 & \kappa & \mathbf{P}_3 \\
		\tilde{P}_0 & -\kappa\mathbf{P}_1/P_+ & -\kappa\mathbf{P}_2/P_+ & -\kappa\mathbf{P}_3/P_+ & P_4
	\end{pmatrix}
\end{align}
\begin{align}
	g'
	=
	\begin{pmatrix}
		\tilde{P}'_4 & \kappa\mathbf{P}'_1/P'_+ & \kappa\mathbf{P}'_2/P'_+ & \kappa\mathbf{P}'_3/P'_+ & P'_0 \\
		\mathbf{P}'_1 & \kappa & 0 & 0 & \mathbf{P}'_1 \\
		\mathbf{P}'_2 & 0 & \kappa & 0 & \mathbf{P}'_2 \\
		\mathbf{P}'_3 & 0 & 0 & \kappa & \mathbf{P}'_3 \\
		\tilde{P}'_0 & -\kappa\mathbf{P}'_1/P'_+ & -\kappa\mathbf{P}'_2/P'_+ & -\kappa\mathbf{P}'_3/P'_+ & P'_4
	\end{pmatrix}
\end{align}
where $\tilde{P}_4$ and $\tilde{P}_0$ are defined in eq. \eqref{P0eP4tilde}. For what concerns the matrix $L'_g$, we will use the generic matrix
\begin{align}
	L'_g
	=
	\begin{pmatrix}
		L_{00} & L_{01} & L_{02} & L_{03} & 0 \\
		L_{10} & L_{11} & L_{12} & L_{13} & 0 \\
		L_{20} & L_{21} & L_{22} & L_{23} & 0 \\
		L_{30} & L_{31} & L_{32} & L_{33} & 0 \\
		0 & 0 & 0 & 0 & 1
	\end{pmatrix}
\end{align}

\subsection{Fifth column:}
\begin{align}
	\begin{pmatrix}
		\gamma P_0 - \beta\gamma \mathbf{P}_1 \\
		\gamma \mathbf{P}_1 - \beta\gamma P_0 \\
		\mathbf{P}_2 \\
		\mathbf{P}_3 \\
		P_4
	\end{pmatrix}
	\overset{!}{=}
	\begin{pmatrix}
		P'_0 \\
		\mathbf{P}'_1 \\
		\mathbf{P}'_2 \\
		\mathbf{P}'_3 \\
		P'_4
	\end{pmatrix}
\end{align}
which translates into the equalities 
\begin{align}\label{boost}
	P'_0 &= \gamma P_0 - \beta\gamma \mathbf{P}_1 \\
	\mathbf{P}'_1 &= \gamma \mathbf{P}_1 - \beta\gamma P_0 \\
	\mathbf{P}'_2 &= \mathbf{P}_2 \\
	\mathbf{P}'_3 &= \mathbf{P}_3 \\
	P'_4 &= P_4
\end{align}

\subsection{Fourth column:} 

\begin{align}
	\begin{pmatrix}
		\gamma\kappa \mathbf{P}_3/P_+ \\
		-\beta \gamma \kappa \mathbf{P}_3/P_+ \\
		0 \\
		\kappa \\
		-\kappa \mathbf{P}_3/P_+
	\end{pmatrix}
	\overset{!}{=}
	\begin{pmatrix}
		L_{03} \tilde{P}'_4
		+
		L_{13} \kappa\mathbf{P}'_1/P_+
		+
		L_{23} \kappa\mathbf{P}'_2/P_+
		+
		L_{33} \kappa\mathbf{P}'_3/P_+ \\
		L_{03} \mathbf{P}'_1 + L_{13} \kappa \\
		L_{03} \mathbf{P}'_2 + L_{23} \kappa \\
		L_{03} \mathbf{P}'_3 + L_{33} \kappa \\
		L_{03} \tilde{P}'_0
		-
		L_{13} \kappa\mathbf{P}'_1/P_+
		-
		L_{23} \kappa\mathbf{P}'_2/P_+
		-
		L _{33} \kappa\mathbf{P}'_3/P_+ \\
	\end{pmatrix}
\end{align}
Summing the equations coming from the first and fifth row we get
\begin{align}
	L_{03} (\tilde{P}'_4 + \tilde{P}'_0)
	=
	(\gamma-1) \kappa \mathbf{P}_3/P_+
\end{align}
and using eq. \eqref{P0eP4tilde} and \eqref{boost} we get
\begin{align}
	L_{03} = (\gamma-1)\kappa \frac{\mathbf{P}_3}{P_+P_+'}
\end{align}
and therefore
\begin{align}\label{L03}
	\boxed{
	L_{03} 
	= 
	(\gamma-1) \kappa \frac{\mathbf{P}_3}{P_+(\gamma P_0 -\beta\gamma \mathbf{P}_1 + P_4)}
}
\end{align}
From the third row we get
\begin{align}
	L_{23} = -\frac{\mathbf{P}_2}{\kappa} L_{03}
\end{align}
and therefore
\begin{align}\label{L23}
	\boxed{
	L_{23} 
	=
	(1-\gamma) \frac{\mathbf{P}_2 \mathbf{P}_3}{P_+(\gamma P_0 -\beta\gamma \mathbf{P}_1 + P_4)}
}
\end{align}
From the fourth row we get
\begin{align}
	L_{33} = -\frac{\mathbf{P}_3}{\kappa}L_{03} + 1
\end{align}
and therefore
\begin{align}\label{L33}
	\boxed{
	L_{33} 
	=
	(1-\gamma) \frac{\mathbf{P}_3^2}{P_+(\gamma P_0 -\beta\gamma \mathbf{P}_1 + P_4)} + 1
}
\end{align}
Finally, from the second row we get
\begin{align}
	L_{13}
	=
	-\beta \gamma \frac{\mathbf{P}_3}{ P_+}
	-
	\frac{\mathbf{P}'_1}{\kappa} L_{03}
\end{align}
and therefore
\begin{align}\label{L13}
	\boxed{
	L_{13}
	=
	-\beta \gamma \frac{\mathbf{P}_3}{ P_+}
	+
	(1-\gamma) \frac{(\gamma \mathbf{P}_1 -\beta\gamma P_0) \mathbf{P}_3}{P_+(\gamma P_0 -\beta\gamma \mathbf{P}_1 + P_4)}
}
\end{align}

\subsection{Third column:}

\begin{align}
	\begin{pmatrix}
		\gamma\kappa \mathbf{P}_2/P_+ \\
		-\beta \gamma \kappa \mathbf{P}_2/P_+ \\
		\kappa \\
		0 \\
		-\kappa \mathbf{P}_2/P_+
	\end{pmatrix}
	\overset{!}{=}
	\begin{pmatrix}
		L_{02} \tilde{P}'_4
		+
		L_{12} \kappa\mathbf{P}'_1/P_+
		+
		L_{22} \kappa\mathbf{P}'_2/P_+
		+
		L_{32} \kappa\mathbf{P}'_3/P_+ \\
		L_{02} \mathbf{P}'_1 + L_{12} \kappa \\
		L_{02} \mathbf{P}'_2 + L_{22} \kappa \\
		L_{02} \mathbf{P}'_3 + L_{32} \kappa \\
		L_{02} \tilde{P}'_0
		-
		L_{12} \kappa\mathbf{P}'_1/P_+
		-
		L_{22} \kappa\mathbf{P}'_2/P_+
		-
		L _{32} \kappa\mathbf{P}'_3/P_+ \\
	\end{pmatrix}
\end{align}
Summing the equations coming from the first and fifth row we get
\begin{align}
	L_{02} (\tilde{P}'_4 + \tilde{P}'_0)
	=
	(\gamma-1) \kappa \mathbf{P}_2/P_+
\end{align}
and using eq. \eqref{P0eP4tilde} and \eqref{boost} we get
\begin{align}
	L_{02} = (\gamma-1)\kappa \frac{\mathbf{P}_2}{P_+P_+'}
\end{align}
and therefore
\begin{align}\label{L02}
	\boxed{
		L_{02} 
		= 
		(\gamma-1) \kappa \frac{\mathbf{P}_2}{P_+(\gamma P_0 -\beta\gamma \mathbf{P}_1 + P_4)}
	}
\end{align}
From the fourth row we get
\begin{align}
	L_{32} = -\frac{\mathbf{P}_3}{\kappa} L_{02}
\end{align}
and therefore
\begin{align}\label{L32}
	\boxed{
		L_{32} 
		=
		(1-\gamma) \frac{\mathbf{P}_2 \mathbf{P}_3}{P_+(\gamma P_0 -\beta\gamma \mathbf{P}_1 + P_4)}
	}
\end{align}
From the third row we get
\begin{align}
	L_{22} = -\frac{\mathbf{P}_2}{\kappa}L_{02} + 1
\end{align}
and therefore
\begin{align}\label{L22}
	\boxed{
		L_{22} 
		=
		(1-\gamma) \frac{\mathbf{P}_2^2}{P_+(\gamma P_0 -\beta\gamma \mathbf{P}_1 + P_4)} + 1
	}
\end{align}
Finally, from the second row we get
\begin{align}
	L_{12}
	=
	-\beta \gamma \frac{\mathbf{P}_2}{ P_+}
	-
	\frac{\mathbf{P}'_1}{\kappa} L_{02}
\end{align}
and therefore
\begin{align}\label{L12}
	\boxed{
		L_{12}
		=
		-\beta \gamma \frac{\mathbf{P}_2}{ P_+}
		+
		(1-\gamma) \frac{(\gamma\mathbf{P}_1 -\beta\gamma P_0) \mathbf{P}_2}{P_+(\gamma P_0 -\beta\gamma \mathbf{P}_1 + P_4)}
	}
\end{align}

\subsection{Second column:}

\begin{align}
	\begin{pmatrix}
		\gamma\kappa \mathbf{P}_1/P_+ - \beta\gamma\kappa \\
		-\beta \gamma \kappa \mathbf{P}_1/P_+ + \gamma\kappa \\
		0 \\
		0 \\
		-\kappa \mathbf{P}_1/P_+
	\end{pmatrix}
	\overset{!}{=}
	\begin{pmatrix}
		L_{01} \tilde{P}'_4
		+
		L_{11} \kappa\mathbf{P}'_1/P_+
		+
		L_{21} \kappa\mathbf{P}'_2/P_+
		+
		L_{31} \kappa\mathbf{P}'_3/P_+ \\
		L_{01} \mathbf{P}'_1 + L_{11} \kappa \\
		L_{01} \mathbf{P}'_2 + L_{21} \kappa \\
		L_{01} \mathbf{P}'_3 + L_{31} \kappa \\
		L_{01} \tilde{P}'_0
		-
		L_{11} \kappa\mathbf{P}'_1/P_+
		-
		L_{21} \kappa\mathbf{P}'_2/P_+
		-
		L _{31} \kappa\mathbf{P}'_3/P_+ \\
	\end{pmatrix}
\end{align}
Summing the equations coming from the first and fifth row we get
\begin{align}
	L_{01} (\tilde{P}'_4 + \tilde{P}'_0)
	=
	(\gamma-1) \kappa \mathbf{P}_1/P_+
	-
	\beta\gamma\kappa
\end{align}
and using eq. \eqref{P0eP4tilde} and \eqref{boost} we get
\begin{align}
	L_{01} = (\gamma-1)\kappa \frac{\mathbf{P}_1}{P_+P_+'} - \frac{\beta\gamma\kappa}{P'_+}
\end{align}
and therefore
\begin{align}\label{L01}
	\boxed{
		L_{01} 
		= 
		(\gamma-1) \kappa \frac{\mathbf{P}_1}{P_+(\gamma P_0 -\beta\gamma \mathbf{P}_1 + P_4)}
		-\frac{\beta\gamma\kappa}{\gamma P_0 -\beta\gamma \mathbf{P}_1 + P_4}
	}
\end{align}
From the fourth row we get
\begin{align}
	L_{31} = -\frac{\mathbf{P}_3}{\kappa} L_{01}
\end{align}
and therefore
\begin{align}\label{L31}
	\boxed{
		L_{31} 
		=
		(1-\gamma) \frac{\mathbf{P}_1 \mathbf{P}_3}{P_+(\gamma P_0 -\beta\gamma \mathbf{P}_1 + P_4)}
		+
		\frac{\beta\gamma \mathbf{P}_3}{\gamma P_0 -\beta\gamma \mathbf{P}_1 + P_4}
	}
\end{align}
From the third row we get
\begin{align}
	L_{21} = -\frac{\mathbf{P}_2}{\kappa}L_{01}
\end{align}
and therefore
\begin{align}\label{L21}
	\boxed{
		L_{21} 
		=
		(1-\gamma) \frac{\mathbf{P}_1 \mathbf{P}_2}{P_+(\gamma P_0 -\beta\gamma \mathbf{P}_1 + P_4)}
		+
		\frac{\beta\gamma \mathbf{P}_2}{\gamma P_0 -\beta\gamma \mathbf{P}_1 + P_4}
	}
\end{align}
Finally, from the second row we get
\begin{align}
	L_{11}
	=
	-\beta \gamma \frac{\mathbf{P}_1}{ P_+}
	-
	\frac{\mathbf{P}'_1}{\kappa} L_{01}
	+
	\gamma
\end{align}
and therefore
\begin{align}\label{L11}
	\boxed{
		L_{11}
		=
		-\beta \gamma \frac{\mathbf{P}_1}{ P_+}
		+
		(1-\gamma) \frac{(\gamma\mathbf{P}_1 -\beta\gamma P_0) \mathbf{P}_1}{P_+(\gamma P_0 -\beta\gamma \mathbf{P}_1 + P_4)}
		+
		\frac{\beta\gamma (\gamma\mathbf{P}_1 - \beta\gamma P_0)}{\gamma P_0 -\beta\gamma \mathbf{P}_1 + P_4}
		+
		\gamma
	}
\end{align}

\subsection{First column:}

\begin{align}
	\begin{pmatrix}
		\gamma\tilde{P}_4 - \beta\gamma\mathbf{P}_1 \\
		\gamma \mathbf{P}_1 - \beta\gamma \tilde{P}_4 \\
		\mathbf{P}_2 \\
		\mathbf{P}_3 \\
		\tilde{P}_0
	\end{pmatrix}
	\overset{!}{=}
	\begin{pmatrix}
		L_{00} \tilde{P}'_4
		+
		L_{10} \kappa\mathbf{P}'_1/P_+
		+
		L_{20} \kappa\mathbf{P}'_2/P_+
		+
		L_{30} \kappa\mathbf{P}'_3/P_+ \\
		L_{00} \mathbf{P}'_1 + L_{10} \kappa \\
		L_{00} \mathbf{P}'_2 + L_{20} \kappa \\
		L_{00} \mathbf{P}'_3 + L_{30} \kappa \\
		L_{00} \tilde{P}'_0
		-
		L_{10} \kappa\mathbf{P}'_1/P_+
		-
		L_{20} \kappa\mathbf{P}'_2/P_+
		-
		L _{30} \kappa\mathbf{P}'_3/P_+ \\
	\end{pmatrix}
\end{align}
Summing the equations coming from the first and fifth row we get
\begin{align}
	L_{00} = \gamma \frac{P_4}{P'_+} + \frac{P_0}{P'_+} + (\gamma-1) \frac{\mathbf{P}^2}{P_+P'_+} - \beta\gamma\frac{\mathbf{P}_1}{P'_+}
\end{align}
and therefore
\begin{subequations}\label{L00}
	\begin{empheq}[box=\widefbox]{align}
		L_{00}
		&=
		\frac{\gamma P_4 + P_0}{\gamma P_0 -\beta\gamma\mathbf{P}_1 + P_4}
		+
		(\gamma-1)
		\frac{\mathbf{P}^2}{P_+(\gamma P_0 -\beta\gamma\mathbf{P}_1 + P_4)} \nonumber \\
		&-
		\beta\gamma \frac{\mathbf{P}_1}{\gamma P_0 -\beta\gamma\mathbf{P}_1 + P_4}
	\end{empheq}
\end{subequations}
From the third row we have
\begin{align}
	L_{20}
	=
	\frac{\mathbf{P}_2}{\kappa}(1-L_{00})
\end{align}
and therefore
\begin{subequations}\label{L20}
	\begin{empheq}[box=\widefbox]{align}
		L_{20}
		&=
		\frac{\mathbf{P}_2}{\kappa}
		\Big[1-
		\Big(
		\frac{\gamma P_4 + P_0}{\gamma P_0 -\beta\gamma\mathbf{P}_1 + P_4} \nonumber \\
		&
		+
		(\gamma-1)
		\frac{\mathbf{P}^2}{P_+(\gamma P_0 -\beta\gamma\mathbf{P}_1 + P_4)} 
		-
		\beta\gamma \frac{\mathbf{P}_1}{\gamma P_0 -\beta\gamma\mathbf{P}_1 + P_4}
		\Big)
		\Big]
	\end{empheq}
\end{subequations}
From the fourth row we have
\begin{align}
	L_{30}
	=
	\frac{\mathbf{P}_3}{\kappa}(1-L_{00})
\end{align}
and therefore
\begin{subequations}\label{L30}
	\begin{empheq}[box=\widefbox]{align}
		L_{30}
		&=
		\frac{\mathbf{P}_3}{\kappa}
		\Big[1-
		\Big(
		\frac{\gamma P_4 + P_0}{\gamma P_0 -\beta\gamma\mathbf{P}_1 + P_4}\nonumber \\
		&
		+
		(\gamma-1)
		\frac{\mathbf{P}^2}{P_+(\gamma P_0 -\beta\gamma\mathbf{P}_1 + P_4)} 
		-
		\beta\gamma \frac{\mathbf{P}_1}{\gamma P_0 -\beta\gamma\mathbf{P}_1 + P_4}
		\Big)
		\Big]
	\end{empheq}
\end{subequations}

Finally, from the second row we have
\begin{align}
	L_{10}
	=
	\frac{1}{\kappa}
	\left(\gamma\mathbf{P}_1 - \beta\gamma P_4 - \beta\gamma\frac{\mathbf{P}^2}{P_+}
	-(\gamma\mathbf{P}_1 - \beta\gamma P_0)L_{00}
	\right)
\end{align}
and therefore
\begin{subequations}\label{L10}
	\begin{empheq}[box=\widefbox]{align}
		L_{10}
		&=
		\frac{1}{\kappa}
		\Big[\gamma\mathbf{P}_1 - \beta\gamma P_4 - \beta\gamma\frac{\mathbf{P}^2}{P_+} 
		-(\gamma\mathbf{P}_1 - \beta\gamma P_0)
		\Big(
		\frac{\gamma P_4 + P_0}{\gamma P_0 -\beta\gamma\mathbf{P}_1 + P_4} \nonumber \\
		&+
		(\gamma-1)
		\frac{\mathbf{P}^2}{P_+(\gamma P_0 -\beta\gamma\mathbf{P}_1 + P_4)}
		-
		\beta\gamma \frac{\mathbf{P}_1}{\gamma P_0 -\beta\gamma\mathbf{P}_1 + P_4}
		\Big)
		\Big]
	\end{empheq}
\end{subequations}

\subsection{Expansion of the final matrix up to $1/\kappa$}

In most phenomenological applications, an expansion of the full results up to first order in $\frac{1}{\kappa}$ is in general useful for numerical estimations of the effect of $\kappa$-deformation on canonical, non-deformed quantities. Because of this, we here expand up to first order the matrix elements computed in the previous section. We make use of the following relations.
\begin{align}
	\frac{a}{b + cP_4} \approx \frac{a}{c\kappa}.
\end{align}
More in general, we also have
\begin{align}
	\frac{a + b P_4}{c + d P_4}
	&\approx
	\frac{b}{d}
	+ \frac{b}{d\kappa}
	\left(
	\frac{ad}{b}
	-
	\frac{c}{d}
	\right)
\end{align}
We also have
\begin{align}
	\frac{a\kappa}{P_+(b+P_4)}
	\approx
	\frac{a}{\kappa}
\end{align}
\begin{align}
	\frac{a \kappa}{b + P_4}
	\approx
	a - \frac{ab}{\kappa}
\end{align}
where $a,b,c,d$ do not depend on $\kappa$. We have
\begin{align}
	\eqref{L00} \mapsto
	\boxed{
	L_{00}
	\approx
	\gamma
	+
	\frac{(1-\gamma)}{\kappa}
	\left[
	(1+\gamma)P_0-\beta\gamma \mathbf{P}_1
	\right]
}
\end{align}
\begin{align}
	\eqref{L10} \mapsto
	\boxed{
	L_{10}
	\approx
	-\beta\gamma +  \frac{\gamma}{\kappa}\mathbf{P}_1
	-
	\frac{\gamma}{\kappa}(\gamma\mathbf{P}_1 - \beta\gamma P_0)
}
\end{align}
\begin{align}
	\eqref{L20} \mapsto
	\boxed{
	L_{20}
	\approx
	\frac{(1-\gamma)}{\kappa} \mathbf{P}_2
}
\end{align}
\begin{align}
	\eqref{L30} \mapsto
	\boxed{
	L_{30}
	\approx
	\frac{(1-\gamma)}{\kappa} \mathbf{P}_3
}
\end{align}
\begin{align}
	\eqref{L01} \mapsto
	\boxed{
	L_{01}
	\approx
	-\beta\gamma +  \frac{\gamma}{\kappa}\mathbf{P}_1
	-
	\frac{\gamma}{\kappa}(\gamma\mathbf{P}_1 - \beta\gamma P_0)
	=
	L_{10}
}
\end{align}
\begin{align}
	\eqref{L11} \mapsto
	\boxed{
	L_{11}
	\approx
	\gamma
	+
	\beta\gamma\frac{(\gamma-1)}{\kappa} \mathbf{P}_1
	-
	\frac{\beta^2\gamma^2}{\kappa}P_0
}
\end{align}
\begin{align}
	\eqref{L21} \mapsto
	\boxed{
	L_{21}
	\approx
	\frac{\beta\gamma \mathbf{P}_2}{\kappa}
}
\end{align}
\begin{align}
	\eqref{L31} \mapsto
	\boxed{
	L_{31}
	\approx
	\frac{\beta\gamma \mathbf{P}_3}{\kappa}
}
\end{align}
\begin{align}
	\eqref{L02} \mapsto
	\boxed{
	L_{02} 
	\approx 
	(\gamma-1) \frac{\mathbf{P}_2}{\kappa}
	=
	-L_{20}
}
\end{align}
\begin{align}
	\eqref{L12} \mapsto
	\boxed{
	L_{12}
	\approx
	-\beta \gamma \frac{\mathbf{P}_2}{\kappa}
}
\end{align}
\begin{align}
	\eqref{L22} \mapsto
	\boxed{
	L_{22} 
	\approx
	1
}
\end{align}
\begin{align}
	\eqref{L32} \mapsto
	\boxed{
	L_{32} 
	\approx
	0
}
\end{align}
\begin{align}
	\eqref{L03} \mapsto
	\boxed{
	L_{03} 
	\approx
	(\gamma-1) \frac{\mathbf{P}_3}{\kappa}
}
\end{align}
\begin{align}
	\eqref{L13} \mapsto
	\boxed{
	L_{13}
	\approx
	-\beta \gamma \frac{\mathbf{P}_3}{\kappa}
}
\end{align}
\begin{align}
	\eqref{L23} \mapsto
	\boxed{
	L_{23} 
	\approx 
	0
}
\end{align}
\begin{align}
	\eqref{L33} \mapsto
	\boxed{
	L_{33} 
	\approx
	1
}
\end{align}

We can now finally write the matrix in extended form. We have
\begin{align}
	{\tiny
	L'_g
	=
	\begin{pmatrix}
		\gamma
		+
		\frac{(1-\gamma)}{\kappa}
		\left[
		(1+\gamma) P_0-\beta\gamma \mathbf{P}_1
		\right] 
		& 
		-\beta\gamma +  \frac{\gamma}{\kappa}\mathbf{P}_1
		-
		\frac{\gamma}{\kappa}(\gamma\mathbf{P}_1 - \beta\gamma P_0)
		&
		(\gamma-1) \kappa \frac{\mathbf{P}_2}{\kappa}
		&
		(\gamma-1) \frac{\mathbf{P}_3}{\kappa}
		&
		0 \\
		-\beta\gamma +  \frac{\gamma}{\kappa}\mathbf{P}_1
		-
		\frac{\gamma}{\kappa}(\gamma\mathbf{P}_1 - \beta\gamma P_0)
		&
		\gamma
		+
		\beta\gamma\frac{(\gamma-1)}{\kappa} \mathbf{P}_1
		-
		\frac{\beta^2\gamma^2}{\kappa}P_0 
		&
		-\beta \gamma \frac{\mathbf{P}_2}{\kappa}
		&
		-\beta \gamma \frac{\mathbf{P}_3}{\kappa}
		&
		0 \\
		\frac{(1-\gamma)}{\kappa} \mathbf{P}_2
		&
		\frac{\beta\gamma \mathbf{P}_2}{\kappa}
		&
		1
		&
		0 & 0 \\
		\frac{(1-\gamma)}{\kappa} \mathbf{P}_3
		&
		\frac{\beta\gamma \mathbf{P}_3}{\kappa}
		&
		0 & 1 & 0 \\
		0 & 0 & 0 & 0 & 1
	\end{pmatrix}
}%
\end{align}
This can be rewritten more simply as a contribution with $\kappa\rightarrow\infty$ plus the first order in $\frac{1}{\kappa}$ obtaining
\begin{align}\label{Lpgtot}
	L'_g
	=
	L'_g(\kappa=\infty)
	+
	\frac{1}{\kappa} \tilde{L}'_g
\end{align}
where
\begin{align}\label{Lpginfinity}
	L'_g(\kappa=\infty)
	=
	\begin{pmatrix}
		\gamma & -\beta\gamma &	0 & 0 & 0 \\
		-\beta\gamma & \gamma & 0 & 0 & 0 \\
		0 & 0 & 1 & 0 & 0 \\
		0 & 0 & 0 & 1 & 0 \\
		0 & 0 & 0 & 0 & 1
	\end{pmatrix}
\end{align}
\begin{align}\label{Lpg}
	{\tiny
	\frac{1}{\kappa}
	\tilde{L}'_g
	=
	\frac{1}{\kappa}
	\begin{pmatrix}
		(1-\gamma)
		\left[
		(1+\gamma) P_0-\beta\gamma \mathbf{P}_1
		\right] 
		& 
		\gamma\mathbf{P}_1
		-
		\gamma(\gamma\mathbf{P}_1 - \beta\gamma P_0)
		&
		(\gamma-1) \mathbf{P}_2
		&
		(\gamma-1) \mathbf{P}_3
		&
		0 \\
		\gamma\mathbf{P}_1
		-
		\gamma(\gamma\mathbf{P}_1 - \beta\gamma P_0)
		&
		\beta\gamma(\gamma-1) \mathbf{P}_1
		-
		\beta^2\gamma^2P_0 
		&
		-\beta \gamma \mathbf{P}_2
		&
		-\beta \gamma \mathbf{P}_3
		&
		0 \\
		(1-\gamma) \mathbf{P}_2
		&
		\beta\gamma \mathbf{P}_2
		&
		0
		&
		0 & 0 \\
		(1-\gamma) \mathbf{P}_3
		&
		\beta\gamma \mathbf{P}_3
		&
		0 & 0 & 0 \\
		0 & 0 & 0 & 0 & 0
	\end{pmatrix}
}%
\end{align}

\subsection{Check of the consistency of the first order expansion of $L'_g$}\label{cons}

Notice that if $L'_g$ is a legitimate Lorentz transformation, then by definition one must have
\begin{align}
	L'_g \eta {L'}^T_g = \eta
\end{align}
where $\eta = \text{diag}(+--- \text{const})$. We know that the zeroth order in perturbation satisfies the relation
\begin{align}
	L'_g(\kappa=\infty) \eta {L'}^T_g(\kappa=\infty) = \eta
\end{align}
by definition, which means that we must have (ignoring terms of order $1/\kappa^2$)
\begin{align}
	L'_g(\kappa=\infty) \eta (\tilde{L}')^T_g + \tilde{L}'_g \eta [L'_g(\kappa=\infty)]^T = 0
\end{align}
We have
\begin{align}
	L'_g(\kappa=\infty) \eta (\tilde{L}')^T_g
	=
	\begin{pmatrix}
		A 
		& 
		B
		&
		(1-\gamma) \mathbf{P}_2
		&
		(1-\gamma) \mathbf{P}_3
		&
		0 \\
		C
		&
		D
		&
		\beta \gamma \mathbf{P}_2
		&
		\beta \gamma \mathbf{P}_3
		&
		0 \\
		(1-\gamma) \mathbf{P}_2
		&
		\beta\gamma \mathbf{P}_2
		&
		0
		&
		0 & 0 \\
		(1-\gamma) \mathbf{P}_3
		&
		\beta\gamma \mathbf{P}_3
		&
		0 & 0 & 0 \\
		0 & 0 & 0 & 0 & 0
	\end{pmatrix}
\end{align}
where 
\begin{align}
	A &= 
	\gamma
	(1-\gamma)
	\left[
	(1+\gamma)P_0-\beta\gamma \mathbf{P}_1
	\right]
	+
	\beta\gamma
	[\gamma\mathbf{P}_1
	-
	\gamma(\gamma\mathbf{P}_1 - \beta\gamma P_0)]
	\\
	C &=
	-\beta \gamma
	(1-\gamma)
	\left[
	(1+\gamma) P_0-\beta\gamma \mathbf{P}_1
	\right]
	-
	\gamma
	[\gamma\mathbf{P}_1
	-
	\gamma(\gamma\mathbf{P}_1 - \beta\gamma P_0)]
	\\
	B &=
	\gamma
	[\gamma\mathbf{P}_1
	-
	\gamma(\gamma\mathbf{P}_1 - \beta\gamma P_0)]
	+
	\beta\gamma
	[\beta\gamma(\gamma-1) \mathbf{P}_1
	-
	\beta^2\gamma^2P_0 ]
	\\
	D &=
	-\beta\gamma
	[\gamma\mathbf{P}_1
	-
	\gamma(\gamma\mathbf{P}_1 - \beta\gamma P_0)]
	-
	\gamma
	[\beta\gamma(\gamma-1) \mathbf{P}_1
	-
	\beta^2\gamma^2P_0 ]
\end{align}
Furthermore we have
\begin{align}
	\tilde{L}'_g \eta (L'_g(\kappa=\infty))^T
	=
	\begin{pmatrix}
		\tilde{A} 
		& 
		\tilde{B}
		&
		(\gamma-1) \mathbf{P}_2
		&
		(\gamma-1) \mathbf{P}_3
		&
		0 \\
		\tilde{C}
		&
		\tilde{D}
		&
		-\beta \gamma \mathbf{P}_2
		&
		-\beta \gamma \mathbf{P}_3
		&
		0 \\
		(\gamma-1) \mathbf{P}_2
		&
		-\beta\gamma \mathbf{P}_2
		&
		0
		&
		0 & 0 \\
		(\gamma-1) \mathbf{P}_3
		&
		-\beta\gamma \mathbf{P}_3
		&
		0 & 0 & 0 \\
		0 & 0 & 0 & 0 & 0
	\end{pmatrix}
\end{align}
where 
\begin{align}
	\tilde{A} &= 
	\gamma
	(1-\gamma)
	\left[
	(1+\gamma) P_0-\beta\gamma \mathbf{P}_1
	\right]
	+
	\beta\gamma
	[\gamma\mathbf{P}_1
	-
	\gamma(\gamma\mathbf{P}_1 - \beta\gamma P_0)]
	\\
	\tilde{B} &=
	-\beta \gamma
	(1-\gamma)
	\left[
	(1+\gamma) P_0-\beta\gamma \mathbf{P}_1
	\right]
	-
	\gamma
	[\gamma\mathbf{P}_1
	-
	\gamma(\gamma\mathbf{P}_1 - \beta\gamma P_0)]
	\\
	\tilde{C} &=
	\gamma
	[\gamma\mathbf{P}_1
	-
	\gamma(\gamma\mathbf{P}_1 - \beta\gamma P_0)]
	+
	\beta\gamma
	[\beta\gamma(\gamma-1) \mathbf{P}_1
	-
	\beta^2\gamma^2P_0 ]
	\\
	\tilde{D} &=
	-\beta\gamma
	[\gamma\mathbf{P}_1
	-
	\gamma(\gamma\mathbf{P}_1 - \beta\gamma P_0)]
	-
	\gamma
	[\beta\gamma(\gamma-1) \mathbf{P}_1
	-
	\beta^2\gamma^2P_0 ]
\end{align}
The sum of the two matrices is therefore
\begin{align}
	\begin{pmatrix}
		A + \tilde{A} & B + \tilde{B} & 0 & 0 & 0 \\
		C + \tilde{C} & D + \tilde{D} & 0 & 0 & 0 \\
		0 & 0 & 0 & 0 & 0 \\
		0 & 0 & 0 & 0 & 0 \\
		0 & 0 & 0 & 0 & 0
	\end{pmatrix}
\end{align}
so that we only need to compute the remaining quantities. We do it one by one. First notice that $A = \tilde{A}$, so we need to show that $A=0$. Indeed, we have
\begin{align}
	A
	&=\gamma
	(1-\gamma)
	\left[
	(1+\gamma) P_0-\beta\gamma \mathbf{P}_1
	\right]
	+
	\beta\gamma
	[\gamma\mathbf{P}_1
	-
	\gamma(\gamma\mathbf{P}_1 - \beta\gamma P_0)] \\
	&=
	\mathbf{P}_1
	[-\beta \gamma^2(1-\gamma) + \beta\gamma^2 - \beta \gamma^3] \\
	&+ P_0
	[\gamma(1-\gamma^2) + \beta^2\gamma^3] \\
	&=
	0
\end{align}
where we used the fact that $1-\gamma^2 = (1-\beta^2-1)/(1-\beta^2) = -\beta^2\gamma^2$.
Secondly, notice that also $D = \tilde{D}$, and therefore we must have $D=0$. Indeed we have
\begin{align}
	D
	&=-\beta\gamma
	[\gamma\mathbf{P}_1
	-
	\gamma(\gamma\mathbf{P}_1 - \beta\gamma P_0)]
	-
	\gamma
	[\beta\gamma(\gamma-1) \mathbf{P}_1
	-
	\beta^2\gamma^2P_0 ] \\
	&=
	\mathbf{P}_1
	[-\beta\gamma^2 + \beta\gamma^3 - \beta\gamma^3 + \beta\gamma^2] \\
	&+
	P_0[-\beta^2\gamma^3 + \beta^2\gamma^3]
	\\
	&=0
\end{align}
Then we need to check $B+\tilde{B}$ and $C+\tilde{C}$. Starting with the first one we have
\begin{align}
	B+\tilde{B}
	&=
	\gamma
	[\gamma\mathbf{P}_1
	-
	\gamma(\gamma\mathbf{P}_1 - \beta\gamma P_0)]
	+
	\beta\gamma
	[\beta\gamma(\gamma-1) \mathbf{P}_1
	-
	\beta^2\gamma^2P_0 ] \\
	&+ 
	\left\{
	-\beta \gamma
	(1-\gamma)
	\left[
	(1+\gamma) P_0-\beta\gamma \mathbf{P}_1
	\right]
	-
	\gamma
	[\gamma\mathbf{P}_1
	-
	\gamma(\gamma\mathbf{P}_1 - \beta\gamma P_0)]
	\right\}
	\\ \nonumber \\
	&=
	\beta\gamma
	[\beta\gamma(\gamma-1) \mathbf{P}_1
	-
	\beta^2\gamma^2P_0 ]
	-\beta \gamma
	(1-\gamma)
	\left[
	(1+\gamma) P_0-\beta\gamma \mathbf{P}_1
	\right]
	\\ \nonumber \\
	&=
	\mathbf{P}_1
	[\beta^2\gamma^3 - \beta^2\gamma^2+\beta^2\gamma^2 - \beta^2\gamma^3] + P_0
	[-\beta^3\gamma^3 - \beta\gamma(1-\gamma^2)] \\
	&=
	0
\end{align}
where we used the fact that $1-\gamma^2 = (1-\beta^2-1)/(1-\beta^2) = -\beta^2\gamma^2$. Finally, we have
\begin{align}
	C+\tilde{C}
	&=
	-\beta \gamma
	(1-\gamma)
	\left[
	(1+\gamma) P_0-\beta\gamma \mathbf{P}_1
	\right]
	-
	\gamma
	[\gamma\mathbf{P}_1
	-
	\gamma(\gamma\mathbf{P}_1 - \beta\gamma P_0)] \nonumber  \\
	&+
	\gamma
	[\gamma\mathbf{P}_1
	-
	\gamma(\gamma\mathbf{P}_1 - \beta\gamma P_0)]
	+
	\beta\gamma
	[\beta\gamma(\gamma-1) \mathbf{P}_1
	-
	\beta^2\gamma^2P_0 ]
	\\ 
	&=
	B+\tilde{B} = 0
\end{align}
which concludes the check. Therefore, the matrix 
\begin{align}
	L'_g
	=
	L'_g(\kappa=\infty)
	+
	\frac{1}{\kappa} \tilde{L}'_g
\end{align}
satisfies 
\begin{align}
	L'_g \eta (L'_g)^T = \eta
\end{align}
and it is therefore a valid Lorentz transformation.

\section{Two-particle kinematics}\label{twopartkin}

We now have the finite boost for a two-particle state, and we can therefore turn to understanding the phenomenological consequences of our computations.

As we already discussed in the context of the boost co-product in section \ref{kMtkP}, multiplication of the matrices $L g, L'_g h$ from the right by an arbitrary Lorentz matrix does not change the entries of their last columns so that $|(L g  (L'_g)^{-1})(P)\rangle = |(L g  )(P)\rangle$ and in writing the states we can neglect the Lorentz group element on the right of $g$. Therefore we define the action of the finite Lorentz transformation $\mathcal{L}$ on a two particle state  as \cite{Bevilacqua:2023pqz}
\begin{align}\label{coprodboost}
	\Delta\mathcal{L} \triangleright |g(P) \rangle\otimes |h(Q) \rangle=	|(L g)(P) \rangle |(L'_g h)(Q) \rangle
\end{align}
where $(L g)(P)$ denotes (in the case of the boost along the first axis) the four top components of the last column of the product of the matrices $L$ (which is a canonical boost) and $g$ (eq. \eqref{g}), in components explicitly  
\begin{align}
	Lg(P)
	=
	(\gamma P_0 -\beta \gamma P_1, -\beta\gamma P_0 + \gamma P_1, P_2, P_3)
\end{align}
The components of $L'_g h$ can be similarly computed using eq. \eqref{Lpgtot}, \eqref{Lpginfinity}, and \eqref{Lpg}, as we will discuss in more details shortly.

We now consider the decay of a particle of mass $M$, originally at rest, into two particles of mass $m$. Starting from the first order expansion in $1/\kappa$ of the four-momentum deformed composition rule $(P\oplus Q)_0$ and $(P\oplus Q)_i$ we get 
\begin{align}
	(P\oplus Q)_0
	\approx
	P_0 + Q_0 + \frac{\mathbf{P}\mathbf{Q}}{\kappa} 
	\qquad
	(P\oplus Q)_i
	\approx
	P_i + Q_i + \frac{P_i Q_0}{\kappa}
\end{align}
and we need to impose 
\begin{align}\label{ident}
	P_0 + Q_0 + \frac{\mathbf{P}\mathbf{Q}}{\kappa}
	=
	M
	\qquad
	P_i + Q_i + \frac{P_i Q_0}{\kappa}
	=
	0.
\end{align}
We need to find the spatial momenta such that these two relations are satisfied.
The conservation of spatial momenta tells us that the momenta $\mathbf{P}$ and $\mathbf{Q}$ are parallel, which means that we can align them with one of our axis, and we can write $P_\mu = (P_0,P,0,0)$ and $Q_\mu=(Q_0,Q,0,0)$ without loss of generality. We have therefore
\begin{align}
	P_0 + Q_0 + \frac{PQ}{\kappa}
	=
	M
	\qquad
	P + Q + \frac{P Q_0}{\kappa}
	=
	0.
\end{align}
From the conservation of momenta, one can get the relation
\begin{align}\label{pq}
	P \approx
	-Q + \frac{Q Q_0}{\kappa}
\end{align}
which substituted back into the sum of energies gives (at first order in $1/\kappa$)
\begin{align}\label{Q}
	Q
	\approx
	\sqrt{M^2-4m^2}
	\left(
	\frac{1}{2}
	+
	\frac{3M}{8\kappa}
	\right)
\end{align}
Substituting back into \eqref{pq}, we also get
\begin{align}\label{P}
	P
	\approx
	-\sqrt{M^2-4m^2}
	\left(	
	\frac{1}{2}
	+
	\frac{M}{8\kappa}
	\right).
\end{align}
Notice that the modulus $P$ differs from that of $Q$ by $\sqrt{M^2-4m^2}\, M/4\kappa$. Notice also that another solution would have been to switch the RHS of eq. \eqref{Q} and \eqref{P}, which however amounts to the same solution. Notice that in the formal limit $\kappa\rightarrow\infty$ the spatial momenta are equal and opposite. The deformation effects is here considered only up to first order in $1/\kappa$, but can be understood without needing to approximate. Indeed, since we are imposing $P\oplus Q = 0$, it is clear that one must either have $P = P'$ and $Q = S(P')$ or vice versa. What we obtain in eq. \eqref{Q} and \eqref{P} is just a first order expansion of this all-orders result. We will see in chapter \ref{Chapter2} that this is not only a general kinematical property due to the presence of $\kappa$-deformation, but it is also comes out naturally when dealing with particles and antiparticles at the same time. 

We can now boost the two particles. Since the boost is in the $x^1$ direction but the two particles can go in arbitrary direction in the center of mass frame, we consider spherical coordinates with the polar axis oriented along the $x$ axis, so that the coordinates of the momenta $P_\mu$ and $Q_\mu$ are given by
\begin{align}
	P_\mu &= (P_0, P\cos\theta, P\sin\theta \sin\phi, P\sin\theta \cos\phi) =: (P_0, P_1, P_2, P_3) \\
	Q_\mu &= (Q_0, Q\cos\theta, Q\sin\theta \sin\phi, Q\sin\theta \cos\phi) =: (Q_0, Q_1, Q_2, Q_3).
\end{align}
When computing the boost, we have two possibilities, depending on which particle is the first and which is second in \eqref{coprodboost}, namely $L\triangleright P_\mu$ and $L'_g\triangleright Q_\mu$, or $L\triangleright Q_\mu$ and $L'_g\triangleright P_\mu$. Since the   non-deformed boost acts on momenta as usual, we will only concentrate in the action of $L'_g$. Furthermore, since the second case can be obtained from the first one by simply switching $P\longleftrightarrow Q$, we will only consider the first one. Using equations \eqref{L}, \eqref{Lpgtot}, \eqref{Lpginfinity}, \eqref{Lpg}, \eqref{Q}, \eqref{P}, we can write down the first order expansion in powers of $1/\kappa$ of the momenta boosted using $L'_g$. We have
\begin{align}
	\label{Qcoordinates}
	(L'_g\triangleright Q)_0
	&=
	\gamma(Q_0-\beta  {Q}_1)
	+
	\frac{1}{\kappa}
	\Big[
	(\gamma-1)( {P}_2 {Q}_2 +  {P}_3 {Q}_3 - \gamma  {P}_1 {Q}_1 + \beta\gamma Q_0  {P}_1) \nonumber \\
	&+
	P_0(Q_0 - Q_0\gamma^2 + \beta\gamma^2  {Q}_1)
	\Big]
	\\ 
	(L'_g\triangleright Q)_1
	&=
	\gamma( {Q}_1-\beta Q_0)
	+
	\frac{1}{\kappa}
	\Big[
	-\beta\gamma( {P}_2 {Q}_2) +  {P}_3 {Q}_3 -  {P}_1(Q_0 - \beta {Q}_1)(\gamma-1)\gamma \nonumber \\
	&+
	P_0( {Q}_1- {Q}_1\gamma^2 + Q_0\beta\gamma^2)
	\Big]
	\\ 
	(L'_g\triangleright Q)_2
	&=
	{Q}_2 
	+
	\frac{1}{\kappa}
	[ {P}_2(Q_0-Q_0\gamma +  {Q_1}\beta\gamma)]
	\\ 
	\label{Qcoordinates1}
	(L'_g\triangleright Q)_3
	&=
	{Q}_3 
	+
	\frac{1}{\kappa}
	[ {P}_3(Q_0-Q_0\gamma +  {Q_1}\beta\gamma)]
\end{align}
One can also explicitly write the expressions of $L\triangleright P$ and $L'_g\triangleright Q$ obtained by substituting eq. \eqref{Q}, \eqref{P} in the formulae, which are the following.
\begin{align}
	(L'_g\triangleright Q)_0
	&=
	\frac{1}{2} \gamma \Big(\beta \cos (\theta ) \sqrt{M^2-4 m^2} +\sqrt{5 M^2-4 m^2}\Big) \nonumber \\
	&+
	\frac{1}{40 \kappa }\Big[5 \beta \gamma \cos (\theta ) \sqrt{M^2-4 m^2} \Big(3 M-2 \sqrt{5 M^2-4 m^2}\Big) \nonumber \\ 
	&+
	4 m^2 \Big(5 \gamma ^2-\frac{12 \gamma M}{\sqrt{5 M^2-4 m^2}}-5\Big)+3 M \Big(\gamma \sqrt{5 M^2-4 m^2}-15 \Big(\gamma ^2-1\Big) M\Big) \nonumber \\
	&-
	5 (\gamma -1)^2 \cos (2 \theta ) \Big(4 m^2-M^2\Big)\Big]
	+O\Big(\Big(\frac{1}{\kappa }\Big)^2\Big)
	\\ \nonumber \\
	(L'_g\triangleright Q)_1
	&=
	-\frac{1}{2} \gamma \Big(\beta \sqrt{5 M^2-4 m^2}+\cos (\theta ) \sqrt{M^2-4 m^2}\Big) \nonumber \\
	&+
	\frac{1}{8 \kappa \sqrt{5 M^2-4 m^2}}\Big[\beta \gamma \Big((\gamma -2) \cos (2 \theta ) \Big(4 m^2-M^2\Big) \sqrt{5 M^2-4 m^2} \nonumber \\
	&+
	4 m^2 \Big(3 M-\gamma \sqrt{5 M^2-4 m^2}\Big)-3 M^2 \Big(M-3 \gamma \sqrt{5 M^2-4 m^2}\Big)\Big) \nonumber \\
	&+
	\cos (\theta ) \sqrt{M^2-4 m^2} \Big(M \Big(10 (\gamma -1) M-3 \gamma \sqrt{5 M^2-4 m^2}\Big) \nonumber \\
	&-8 (\gamma -1) m^2\Big)\Big]+O\Big(\Big(\frac{1}{\kappa }\Big)^2\Big)
	\\ \nonumber \\
	(L'_g\triangleright Q)_2
	&=
	\frac{1}{2} \sin (\theta ) \sqrt{M^2-4 m^2} \cos (\phi ) \nonumber \\
	&-\frac{1}{8 \kappa }\Big[\sin (\theta ) \sqrt{M^2-4 m^2} \cos (\phi ) \Big(2 \beta \gamma \cos (\theta ) \sqrt{M^2-4 m^2} \nonumber \\
	&+
	2 (\gamma -1) \sqrt{5 M^2-4 m^2}-3 M\Big)\Big]+O\Big(\Big(\frac{1}{\kappa }\Big)^2\Big)
	\\  \nonumber \\
	(L'_g\triangleright Q)_3
	&=
	\frac{1}{2} \sin (\theta ) \sqrt{M^2-4 m^2} \sin (\phi ) \nonumber \\
	&-
	\frac{1}{8 \kappa }\Big[\sin (\theta ) \sqrt{M^2-4 m^2} \sin (\phi ) \Big(2 \beta \gamma \cos (\theta ) \sqrt{M^2-4 m^2} \nonumber \\
	&+
	2 (\gamma -1) \sqrt{5 M^2-4 m^2}-3 M\Big)\Big]+O\Big(\Big(\frac{1}{\kappa }\Big)^2\Big)
\end{align}
\begin{align}
	(L \triangleright P)_0
	&=
	\frac{1}{2} \gamma \left(\beta \cos (\theta ) \sqrt{M^2-4 m^2}+\sqrt{5 M^2-4 m^2}\right) \nonumber \\
	&+
	\frac{\gamma M \left(\beta \cos (\theta ) \sqrt{M^2-4 m^2} +\frac{M^2-4 m^2}{\sqrt{5 M^2-4 m^2}}\right)}{8 \kappa }+O\Big(\Big(\frac{1}{\kappa }\Big)^2\Big)
	\\ \nonumber \\
	(L \triangleright P)_1
	&=
	-\frac{1}{2} \gamma \left(\beta \sqrt{5 M^2-4 m^2}+\cos (\theta ) \sqrt{M^2-4 m^2}\right) \nonumber \\
	&+\frac{\gamma M \left(\frac{\beta \left(4 m^2-M^2\right)}{\sqrt{5 M^2-4 m^2}}-\cos (\theta ) \sqrt{M^2-4 m^2}\right)}{8 \kappa }+O\Big(\Big(\frac{1}{\kappa }\Big)^2\Big)
	\\ \nonumber \\
	(L \triangleright P)_2
	&=
	-\frac{1}{2} \sin (\theta ) \sqrt{M^2-4 m^2} \cos (\phi ) \nonumber \\
	&-\frac{M \sin (\theta ) \sqrt{M^2-4 m^2} \cos (\phi )}{8 \kappa }+O\Big(\Big(\frac{1}{\kappa }\Big)^2\Big)
	\\ \nonumber \\
	(L \triangleright P)_3
	&=
	-\frac{1}{2} \sin (\theta ) \sqrt{M^2-4 m^2} \sin (\phi ) \nonumber \\
	&-\frac{M \sin (\theta ) \sqrt{M^2-4 m^2} \sin (\phi )}{8 \kappa }+O\Big(\Big(\frac{1}{\kappa}\Big)^2\Big)
\end{align}
Notice that, in absence of deformation, the components of momenta along $x^2$ and $x^3$ (i.e. those components perpendicular to the boost direction) of $L'_g\triangleright Q$ and $L\triangleright P$ are equal and opposite. However, deformation introduces additional terms which make even these components not equal and opposite after boost, which in turn gives a deformed angular distribution. To show this, we plot the angular dependence of the modulus of the deformed boosted momenta to compare it to the   non-deformed case. In figures below all the quantities on the axes are expressed in GeV. 

\begin{figure}[h]
	\centering
	\includegraphics[width=0.6\textwidth]{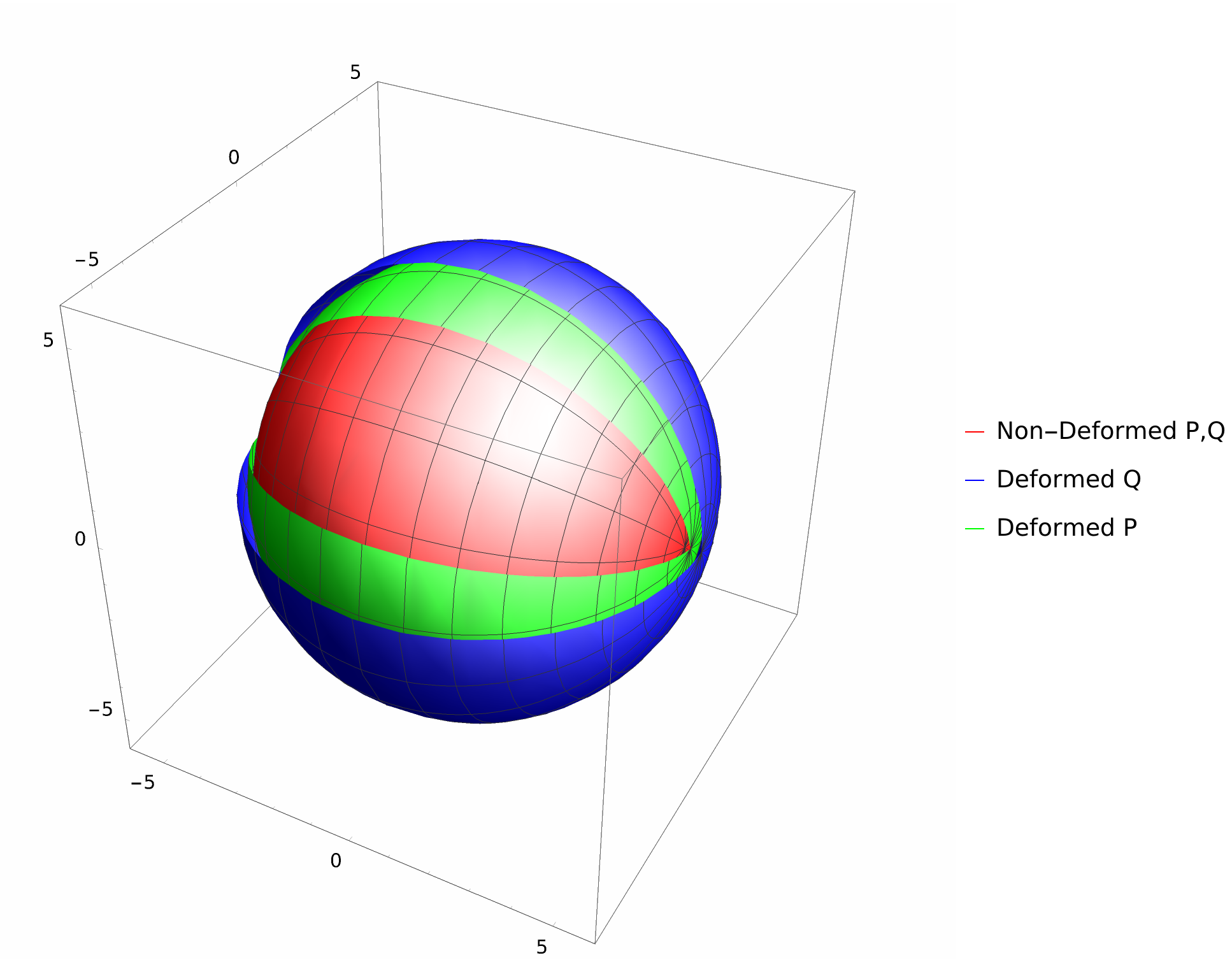}
	\caption{Angular distribution of deformed and   non-deformed momenta in the COM frame, with parameters $M=10 \;\mbox{GeV}$, $m=0.1 \;\mbox{GeV}$, $\kappa=10^2 \;\mbox{GeV}$. Notice the onion-like structure, with the   non-deformed momentum being innermost, followed by the deformed $P$, and finally the deformed $q$. Here, $P$ and $Q$ are antipodal also in the $\kappa$-deformed context (see eq. \eqref{ident}). To give a more concrete numerical estimate, the modulus of the   non-deformed momentum is $5.00$ GeV, the modulus of $P$ is $5.12$ GeV, while the modulus of $Q$ is $5.37$ GeV.}
	\label{fig:PQcomFig}
\end{figure}

\begin{figure}[h]
	\centering
	\includegraphics[width=0.88\textwidth]{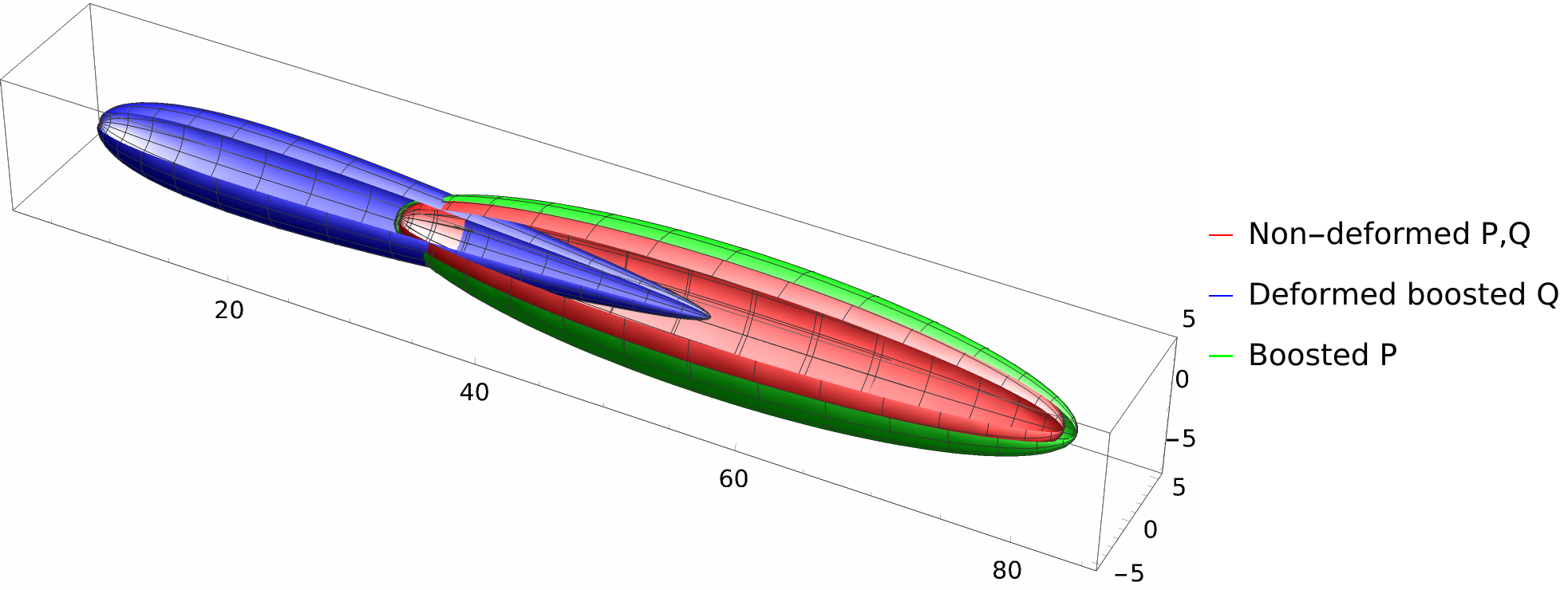}
	\caption{Angular distribution of deformed and   non-deformed momenta in the boosted frame, with parameters $\gamma=5$, $M=10 \;\mbox{GeV}$, $m=0.1 \;\mbox{GeV}$, $\kappa=10^2 \;\mbox{GeV}$. Recall that both $P$ and $Q$ feel the effects of deformation already in the modulus (see eq. \eqref{P} and \eqref{Q}). When boosting, $P$ gets boosted with a canonical non-deformed boost, while $Q$ is instead boosted in a deformed way.}
	\label{fig:PQboostFig}
\end{figure}

In the center of mass (COM) frame, the two momenta are distributed on a sphere. Because of deformation, however, the spatial momenta $P$ and $Q$ are different in modulus (see eq. \eqref{Q}, \eqref{P}). In Figure \ref{fig:PQcomFig} we show the distribution of momenta in the COM frame, together with the distribution in the   non-deformed case. To highlight the qualitative difference, for the next plots we chose hypothetically $M=10 \;\mbox{GeV}$, $m=0.1 \;\mbox{GeV}$, $\kappa=10^2 \;\mbox{GeV}$. Furthermore, we chose the domains $\phi\in \left[-\frac{\pi}{10}, \frac{3\pi}{2}+\frac{\pi}{10}\right]$ and $\phi\in \left[0, \frac{3\pi}{2}\right]$ for the distributions of the deformed $P$ and the deformed $Q$ respectively, in order to clearly show how they are related to the   non-deformed one. Notice that the   non-deformed case contains only a single surface because both spatial momenta have the same modulus. 
We then boost our momenta. In particular, $P_\mu$ will be canonically boosted while $Q_\mu$ will be boosted in a deformed way (recall however that both $P_\mu$ and $Q_\mu$ still have some effects of deformation in their modulus). In this case we chose $\gamma=5$ while the other parameters remain the same. We also restricted the domain in $\phi$ of all distributions to better show the features of the surfaces, and we obtain Figure \ref{fig:PQboostFig}. Notice that the deformed boost not only modifies the amplitude of the boosted momentum, but also its angular distribution. Finally, notice that switching $P\leftrightarrow Q$ in the above formula makes manifest that the $\kappa$-deformed boost contribution is much greater than the center of mass deformation of the moduli.

From a phenomenological point of view, eq. \eqref{Qcoordinates} and \eqref{Qcoordinates1} show that the difference between the two boosted momenta mostly depend on the $\frac{1}{\kappa}$ contribution coming from the deformed boost (large values of boost parameter $\gamma$ highlight these contributions). This in turn is important for the study of two-particle correlations and interference patterns, since these are often expressed as a function of the difference of momenta (or lifetime) of the involved particles. Of particular relevance, the study of favoured mesons from the decays $\Phi^0 (1020)\rightarrow K^0 \bar K^0$, $\Psi(3770)\rightarrow D^0\bar D^0$, $\Upsilon(10580)\rightarrow B^0\bar B^0$ and $\Upsilon(10860)\rightarrow B^0_s\bar B^0_s$.

\chapter{Complex scalar field on $\kappa$-Minkowski spacetime} 

\label{Chapter2}

\section{Introduction}\label{roadmap2}

Given $\kappa$-Minkowski spacetime, one can begin building physical quantities on top of it. In particular, one can introduce fields over $\kappa$-Minkowski spacetime. The whole process is however not trivial, and because of the many computations involved, it will be easy to lose track of the progress or the direction of our investigations. As such, in the remainder of this section we will provide the reader with a roadmap of this chapter. This chapter will be divided in the following main sections:
\begin{itemize}
	\item[$i)$] In order to introduce a field in $\kappa$-Minkowski spacetime, one could directly work in the bicrossproduct basis using the coordinate representation given in eq. \eqref{II.1.18}. It is however much simpler to work with canonical, commuting coordinates, with the caveat that one needs to switch to the classical basis. One can do this through the so called Weyl map (the canonical product of functions is also deformed to a non-commutative $\star$ product under Weyl map). Therefore, the first step will be to introduce the Weyl map and its properties, after which the rest of the discussion will be more transparent. This will be done in section \ref{Wm};
	\item[$ii)$] We are now in the position to  introduce the action of a complex scalar field, and we will do so immediately after having introduced the Weyl map. Because of the properties of the $\star$ product, our total action $S$ will be defined in terms of two possible partial actions $S_1$ and $S_2$. This will be done in section \ref{Introaction};
	\item[$iii)$] The first thing to understand about the action is how to compute physically meaningful quantities starting from it, like for example the equations of motion or the Noether charges. However, because of the non-trivial co-product, the derivatives do not satisfy the Leibniz rule. We will therefore first introduce the action of derivatives on $\star$ products from in section \ref{Actionofderonstar}. After this we can immediately introduce the integration-by-parts relations, from section \ref{Intbyparts1} to \ref{Intbypartsm};
	\item[$iv)$] We now have both an action, and the integration-by-parts relation which means that we can vary the action introduced in section \ref{Introaction} to obtain both the equations of motion (EoM) and the surface terms. We will do this in section \ref{eomandsurf};
	\item[$v)$] We can now build the on-shell field since we know the EoM. Once we have it, we can also discuss its properties under the discrete transformations $C$, $P$, and $T$. We will do this in section \ref{csfCPT};
	\item[$vi)$] Going a bit off-topic, it is useful to introduce at this point the off-shell momentum-space action, in section \ref{offshellactionsect}. This is not needed for the rest of this chapter, but will be needed in chapter \ref{Chapter3} when we will obtain the propagator;
	\item[$vii)$] Since we now have the EoM and the on-shell field, the other important quantities to obtain from the action defined in section \ref{Introaction} are the charges. We will proceed along the following steps:
	\begin{itemize}
		\item[$a)$] To compute the charges, we introduce the symplectic form in section \ref{sympformsect}. The role of this object is twofold. On one hand it allows us to compute the creation/annihilation operators algebra. On the other hand, we can use it to compute the charges using the geometric approach, which we describe in section \ref{chargesfromsympform}. This will lead us to introduce deformed rules for the contraction of vector fields with forms. Using them, we will obtain both the translation (section \ref{chargesfromsympform}), boost (section \ref{boostsect}) and the rotation charges (section \ref{rotsec});
		\item[$b)$]  We will conclude the computation of the charges with some mathematical considerations on the properties of the Poisson brackets of creation/annihilation operators, and the role of the antipode in the interaction between forms and vector fields (section \ref{dsf} and \ref{dercreannalg}).
	\end{itemize}
	\item[$viii)$] We now finally have all the charges, as well as the creation/annihilation operators algebra. The natural next step is to check what is the algebra satisfied by the charges. We check it in section \ref{chargesalgebra};
	\item[$ix)$] We now have all the charges coming from continuous symmetries, and as we discussed in point $v)$ we also have discrete symmetries. We are therefore ready to discuss the interaction between them. The first surprising result is the non-commutativity of the $C$ operator with the boost charge $\mathcal{N}_i$. We show the detailed computations of this fact in section \ref{CNcomm};
	\item[$x)$] Another important and non-trivial result is related to the Greenberg's theorem, which (in canonical axiomatic QFT) shows how invariance $CPT$ transformation is equivalent to Lorentz invariance. Because our result of point $ix)$, we already know that in our model $CPT$ invariance is subtly violated. We investigate the details of Greenberg's theorem and its application to the deformed context in section \ref{green};
	\item[$xi)$] We will conclude with some phenomenological considerations concerning possible signatures of the fact that $[C, \mathcal{N}_i]\neq 0$ in section \ref{def-cpt-pheno}.
\end{itemize}
The Weyl map introduced in point $i)$ of the above list is a quantity which (at least in canonical quantum mechanics) has been known for decades. Here we just review some of the material in the literature which we need to proceed forward, but it does not constitute original work. Similarly, the deformation of the Lorenz rule for derivatives and the integration-by-parts obtained in point $iii)$ have been already discussed in \cite{Freidel:2007hk} in the context of the bicrossproduct basis, and only their translation to the classical basis discussed in this work is original. All other points are original results.

\section{Weyl maps}\label{Wm}

We begin with point $i)$ in the roadmap. In our context, a Weyl map is a function which allows for every object built on top of the group manifold $AN(3)$ to be translated into commutative spacetime with $\star$ product. This includes functions of $x$, integrals, derivatives. Since this is a standard construction, we only pinpoint the most important features and definitions, referring the reader to \cite{Freidel:2007hk}, \cite{Amelino-Camelia:2001rtw}, \cite{Arzano:2018gii}, \cite{Arzano:2020jro} for a more complete discussion. Conceptually speaking, a Weyl map is not strictly needed for any of the computations in the present and following chapters. Indeed, since $AN(3)$ is a Lie group, there are well defined concepts of differentiation and integration, which suffice for our purpose. Nevertheless, being able to switch to a more conventional spacetime, where coordinates are not matrices any more, greatly simplifies the computations, and allows for a more clear interpretation of the results. For clarity, since we will need to talk about the parity transformation $P$, from here on we will indicate the classical basis with $p$ and not $P$.

By definition, we have
\begin{align}
	\mathcal{W}(\hat{e}_k  (\hat{x})) = e_{p  (k)}(x), 
	\quad
	\mathcal{W}^{-1}(e_{p  (k)}(x)) = \hat{e}  _k(\hat{x}),
\end{align}
where
\begin{align}
	e_{p  (k)(x)} = e^{-i(\omega_\mathbf{p}  t - \mathbf{p}  \mathbf{x})},
\end{align}
and where $k$ are momenta in the bicrossproduct basis, and $p$ are the momenta in the classical basis as a function of the $k$ momenta, see eq. \eqref{classicalbasis2}, \eqref{classicalbasis2-star} (recall also the definition of $\hat{e}^*$ in eq. \eqref{estar}). Of course, we need to take care also of the sum of momenta and antipodes, and we have the definitions (we will only write down the transformation in one direction, since the inverse is then obvious, and we also avoid writing the dependence on both $x$ and $\hat{x}$, since they are also obvious)
\begin{align}
	\mathcal{W}(({\hat{e}  }_k)^{-1}) = e_{S(p  (k))}, 
	\qquad
	\mathcal{W}(\hat{e}  _k\hat{e}  _l)
	=
	e_{p  (k)\oplus q  (l)} =: e_{p  } \star e_{q  }
\end{align}
where the second equation defines the $\star$ product. Notice that, since the product in $AN(3)$ is not commutative, then neither is the $\star$ product. Furthermore, notice that since by definition we have
\begin{align}
	e_p^\dag \star e_p = e_p \star e_p^\dag = 1
\end{align}
then we have
\begin{align}
	e_p^\dag = e_{S(p)}.
\end{align}
Furthermore, calling $\hat{\partial}$ the derivatives in the context of the $AN(3)$ group defined by 
\begin{align}
	\hat\partial_\mu\, \hat e_k &= ip_\mu(k)\, \hat e_k\,,\quad \hat\partial_4\, \hat e_k =i(\kappa- p_4(k))\, \hat e_k\nonumber\\
	\hat\partial_\mu\, \hat e^*_k &= ip^*_\mu(k)\, \hat e^*_k\,,\quad \hat\partial_4\, \hat e^*_k =i(\kappa - p^*_4(k))\, \hat e^*_k \label{dergroup}
\end{align}
we define the action of the Weyl map action by
\begin{align}
	\mathcal{W}(\hat{\partial}_\mu \hat{e}  _k)
	=
	\partial_\mu e_{p  }
	\qquad
	\mathcal{W}(\hat{\partial}_4 \hat{e}  _k)
	=
	\partial_4 e_{p  },
	\quad
	\partial_4 e_{p  }
	=
	i(\kappa - p  _4) e_{p  }.
\end{align}
Finally, we have the definition 
\begin{align}
	\widehat{\int} \hat{e}_k
	=
	\int_{\mathbb{R}^4} d^4x 
	\, \mathcal{W}(\hat{e}_k(\hat{x}))
\end{align}
for plane waves. Notice that the integral on the LHS is not an integral over $AN(3)$. This in turn allows us to define fields using the Fourier transform \cite{Freidel:2007hk}, \cite{Amelino-Camelia:2001rtw}, \cite{Arzano:2018gii}, \cite{Arzano:2020jro}, \cite{Oriti:2018bwr}, \cite{Guedes:2013vi}
\begin{align}
	\phi(x) = 
	\int_{AN(3)}
	d\mu(p)
	\tilde{\phi}(p) e_p(x)
	\qquad
	d\mu(p)
	=
	\frac{d^4p}{p_4/\kappa}
\end{align}
restricted by the constraints in eq. \eqref{classicalconstraints}. Of course, one can define the same thing at the group level and not in canonical spacetime by
\begin{align}
	\hat{\phi}(\hat{x}) = 
	\int_{AN(3)}
	d\mu(p)
	\tilde{\phi}(p) \hat{e}_k(\hat{x})
\end{align}
with the definitions
\begin{align}
	\mathcal{W}(\hat{\phi}(\hat{x})) = \phi(x)
	\qquad
	\mathcal{W}(\hat{\phi}(\hat{x})\hat{\psi}(\hat{x}))
	=
	\phi(x) \star \psi(x)
\end{align}
and therefore
\begin{align}
	\widehat{\int}\hat{\phi}(\hat{x})\hat{\psi}(\hat{x})
	=
	\int_{ \mathbb{R}^4} d^4x \, \phi(x) \star \psi(x).
\end{align}

In other words, we can safely translate quantities defined purely in terms of (derivatives and integrals of) group elements in $AN(3)$ into more familiar terms, with canonical Riemannian integrals defined in $\mathbb{R}^4$ and derivatives acting on canonical plane waves, paying attention to the fact that products of different objects need to be defined as $\star$ product. 
Of course, such a set of definitions is not unique, for example here we chose to relate canonical plane waves with group elements defined according to the time-to-the-right convention (recall eq. \eqref{defek}). Instead, we could have chosen a Weyl map with the definition 
\begin{align}
	\mathcal{W}^{-1}(e_p) = e^{ip_\mu \hat{x}^\mu}
\end{align}
which, as we saw in eq. \eqref{normalbasis}, corresponds to a different choice of basis in momentum space. Different Weyl maps can offer different advantages and disadvantages, depending on the problem at hand, for a more in depth discussion see \cite{Freidel:2007hk}, \cite{Amelino-Camelia:2001rtw}, \cite{Arzano:2018gii}, \cite{Arzano:2020jro}. 

In what follows, we will use the Weyl map defined as above, and we will assume that it has already been applied. 

\section{Action for the complex scalar field}\label{Introaction}

We can now go to point $ii)$ in the roadmap in section \ref{roadmap2}. Because of what has been said in section \ref{Wm}, and recalling that we need to get back the non-deformed action in the formal $\kappa\rightarrow \infty$ limit, we have two possible choices for the action, namely
\begin{align}
	S_1 
	&= 
	\int_{\mathbb{R}^4} d^4x \,\, (\partial^\mu \phi)^\dag \star (\partial_\mu \phi) - m^2 \phi^\dag \star \phi \label{S1}\\
	S_2 
	&= 
	\int_{\mathbb{R}^4} d^4x \,\ (\partial_\mu \phi) \star (\partial^\mu \phi)^\dag  - m^2 \phi \star \phi^\dag. \label{S2}
\end{align}
Because of this, and for reasons that will become apparent later, we choose the following action.
\begin{align}
	\boxed{S = \frac{1}{2}
	\int_{\mathbb{R}^4} d^4x \,\, 
	[
	(\partial^\mu \phi)^\dag \star (\partial_\mu \phi) 
	+
	(\partial^\mu \phi) \star (\partial_\mu \phi)^\dag 
	- 
	m^2 (\phi^\dag \star \phi + \phi \star \phi^\dag
	)
	]}.\label{Stot}
\end{align}
We now need to extract the equations of motion (EoM) and the conserved Noether charges associated with continuous symmetries out of this action. Notice that the quantities $\partial^\dag$ are non-trivial objects which will be defined (and whose properties will be investigated) in the next section.

\section{Action of derivatives on $\star$ products}\label{Actionofderonstar}

We now have a star product in a commutative spacetime, and we want to know how derivatives act on the star product of two functions. We are at point $iii)$ of our roadmap in section \ref{roadmap2}. We know that the action cannot be Leibnizian, since otherwise one would get the following contradiction.
\begin{align}\label{contrad1}
	i(p\oplus q)_\mu e_{p\oplus q}
	&=
	\partial_\mu (e_p \star e_q)
	=
	(\partial_\mu e_p) \star e_q
	+
	e_p \star \partial_\mu e_q =
	i(p+q)_\mu e_{p\oplus q}
\end{align}
Indeed, from the first equality in eq. \eqref{contrad1} we already see that the action must be defined in terms of the co-product. Since we are using plane waves whose momenta are in the classical basis, we will need to model the action of derivatives on products over eq. \eqref{DeltaPi}, \eqref{DeltaP0}, \eqref{DeltaP4}. We define therefore the following $\kappa$-deformed Leibniz rules, in analogy with those defined already\footnote{In \cite{Freidel:2007hk} the co-product rules in the bicrossproduct basis are used to define the action of derivatives on products, instead of the classical basis that we use here. Furthermore, the model considered in \cite{Freidel:2007hk} is completely different from the present one, since they consider only one of the two orderings of the action (the equivalent of our eq. \eqref{S1}) and they have a completely different field decomposition. The action of derivatives on products, however, is only a manifestation of the co-product, and it is therefore model-independent.} in \cite{Freidel:2007hk}.
\begin{align}
	&\boxed{\partial_0 (\phi\star \psi) = \frac{1}{\kappa}
		(\partial_0\phi) \star (\Delta_+\psi)
		+ \kappa
		(\Delta_+^{-1}\phi)\star(\partial_0\psi)
		+i
		(\Delta_+^{-1}\partial_i\phi)\star(\partial_i\psi)} \label{4.13bis}\\
	&\qquad \qquad \qquad \boxed{\partial_i (\phi\star \psi) =
		\frac{1}{\kappa}(\partial_i \phi)\star (\Delta_+\psi)
		+
		\phi\star(\partial_i\psi)} \label{4.14bis}\\
	&\qquad \qquad \qquad \qquad
	\boxed{\Delta_+(\phi \star \psi) = \frac{1}{\kappa}(\Delta_+\phi) \star (\Delta_+\psi)}. \label{4.15bis}
\end{align}
Furthermore, we introduce the objects $\partial^\dagger$ which are related to antipodes. More in detail, are defined in such a way that 
\begin{align}
	\partial e_p \propto p e_p
	\qquad
	\Leftrightarrow
	\qquad
	\partial^\dag e_p \propto S(p)e_p.
\end{align}
As such, we expect them to be related to eq. \eqref{SP}. We define 
\begin{align}\label{4.16}
	\boxed{\partial_i^\dagger = \kappa \Delta_+^{-1} \partial_i},
	\qquad
	\boxed{
		\partial_0^\dagger = \partial_0 - i\Delta_+^{-1}\partial^2},
	\qquad
	\boxed{\partial_4^\dagger = -\partial_4},
	\qquad
	\boxed{\Delta_+^\dagger = \kappa^2 \Delta_+^{-1}}
\end{align}
\begin{align}\label{partial+}
	\boxed{\Delta_+ = i\partial_0 + \kappa +i\partial_4}.
\end{align}
Notice that for the definition of $\Delta_+$ we used the conventions highlighted in eq. \eqref{dergroup}. If we define the linear combinations
\begin{equation}\label{Fa}
	\Xi(x) 
	= 
	a_\mathbf{p}\, e^{-i(\omega_pt-\mathbf{p}\mathbf{x})} 
	+ 
	b^\dag_\mathbf{p^*}\, e^{i(S(\omega^*_p)t-S(\mathbf{p}^*)\mathbf{x})} \equiv 
	\Xi_{(+)}(x) + \Xi_{(-)}(x)
\end{equation}
\begin{equation}\label{Fb}
	\Xi^\dag(x) 
	= 
	a^\dag_\mathbf{p}\, e^{-i(S(\omega_p)t-S(\mathbf{p})\mathbf{x})} 
	+ 
	b_\mathbf{p^*}\, e^{i(\omega^*_pt-\mathbf{p}^*\mathbf{x})}
	\equiv 
	\Xi^\dag_{(+)}(x) + \Xi^\dag_{(-)}(x)
\end{equation}
then we have
\begin{align}\label{bbdagEbdbd}
	(\partial_A \Xi)^\dag = \partial_A^\dag \Xi^\dag.
\end{align}
The reason for this particular choice of linear combination $\Xi$ will become clear after we introduce the complex fields in subsequent sections (at this point, $a,a^\dag,b,b^\dag$ are just some complex coefficients). We check the validity of eq. \eqref{bbdagEbdbd} for $A=0,i,4$.

\subsection{$A=0$}\label{mu=0}

In this case 
\begin{align}
	\partial_0 \Xi
	=
	-i\omega_p a_\mathbf{p}\, e^{-i(\omega_pt-\mathbf{p}\mathbf{x})} 
	+ 
	iS(\omega_p^*) b^\dag_\mathbf{p^*}\, e^{i(S(\omega^*_p)t-S(\mathbf{p}^*)\mathbf{x})}
\end{align}
which means that
\begin{align}
	(\partial_0 \Xi)^\dag
	=
	i\omega_p a_\mathbf{p}^\dag\, e^{-i(S(\omega_p)t-S(\mathbf{p})\mathbf{x})} 
	-
	iS(\omega_p^*) b_\mathbf{p^*}\, e^{i(\omega^*_pt-\mathbf{p}^*\mathbf{x})}.
\end{align}
On the other hand
\begin{align}\label{defp4}
	\partial_0^\dag \Xi^\dag 
	&= 
	(\partial_0 - i \Delta_+^{-1}\partial^2)\Xi^\dag \\
	&=
	\underbrace{\left(-i S(\omega_p) + \frac{i}{S(\omega_p) + p_4} S(\mathbf{p})^2\right)}_{=iS(S(\omega_p))}
	a^\dag_\mathbf{p}\, e^{-i(S(\omega_p)t-S(\mathbf{p})\mathbf{x})} \nonumber \\
	&+ 
	\underbrace{\left(i \omega_p^* + \frac{i}{-\omega_p^* - p_4^*}(\mathbf{p}^*)^2\right)}_{=-iS(\omega_p^*)}
	b_\mathbf{p^*}\, e^{i(\omega^*_pt-\mathbf{p}^*\mathbf{x})}
\end{align}
where we also used $S(p_4)=p_4$ which is true both off-shell and on-shell, so that we only need to verify that $S(S(\omega_p))=\omega_p$. We have
\begin{align}
	S(S(\omega_p)) 
	&=
	-S(\omega_p) + \frac{1}{S(\omega_p) + p_4} S(\mathbf{p})^2 \\
	&=
	\omega_p - \frac{\mathbf{p}^2}{\omega_p + p_4} + \frac{\omega_p + p_4}{\kappa^2} \frac{\kappa^2 \mathbf{p}^2}{(\omega_p + p_4)^2} \\
	&=
	\omega_p 
\end{align}
where we used 
\begin{align}
	S(p_0)+p_4  = -p_0 + \frac{\mathbf{p}^2}{p_0+p_4} +p_4=\frac{\kappa^2}{p_0+p_4}.
\end{align}
Notice the important detail that the definition $\hat\partial_4\, \hat e^*_k =i(\kappa - p^*_4(k))\, \hat e^*_k$ in equation \eqref{dergroup} translated through the Weyl map states that
\begin{align}
	\partial_4\, e^{-i(\omega_p^*t - \mathbf{p}^*\mathbf{x})} =i(\kappa - p^*_4(k))\, e^{-i(\omega_p^*t - \mathbf{p}^*\mathbf{x})}
\end{align} 
while in eq. \eqref{defp4}, when we use $\Delta_+^{-1}$ in the second term in the RHS, we are applying $\partial_4$ to $e^{i(\omega_p^*t - \mathbf{p}^*\mathbf{x})} = e^{-i(\omega_pt - \mathbf{p}\mathbf{x})}$, so that in the second term in the RHS of eq. \eqref{defp4} we get the quantity
\begin{align}
	\frac{1}{\omega_p + p_4} = \frac{1}{-\omega_p^* -p_4^*}.
\end{align}

\subsection{$A=i$}\label{mu=i}
In this case 
\begin{align}
	\partial_i \Xi
	=
	i\mathbf{p}_i
	a_\mathbf{p}\, e^{-i(\omega_pt-\mathbf{p}\mathbf{x})} 
	-
	iS(\mathbf{p}^*)_i
	b^\dag_\mathbf{p^*}\, e^{i(S(\omega^*_p)t-S(\mathbf{p}^*)\mathbf{x})}
\end{align}
and therefore
\begin{align}
	(\partial_i \Xi)^\dag
	=
	-i\mathbf{p}_i
	a_\mathbf{p}^\dag\, e^{-i(S(\omega_p)t-S(\mathbf{p})\mathbf{x})} 
	+ 
	iS(\mathbf{p}^*)_i
	b_\mathbf{p^*}\, e^{i(\omega^*_pt-\mathbf{p}^*\mathbf{x})}.
\end{align}
On the other hand we have
\begin{align}
	\partial_i^\dag \Xi^\dag 
	&=
	(\kappa \Delta_+^{-1} \partial_i) \Xi^\dag \\
	&=
	\underbrace{\left(
		\kappa \frac{i}{S(\omega_p) + p_4} S(\mathbf{p})_i
		\right)}_{=-iS(S(\mathbf{p}))_i}
	a^\dag_\mathbf{p}\, e^{-i(S(\omega_p)t-S(\mathbf{p})\mathbf{x})} \nonumber \\
	&+ 
	\underbrace{\left(
		-\kappa \frac{i}{-\omega_p^*-p_4^*} \mathbf{p}^*_i
		\right)}_{=iS(\mathbf{p})_i}
	b_\mathbf{p^*}\, e^{i(\omega^*_pt-\mathbf{p}^*\mathbf{x})}
\end{align}
and also
\begin{align}
	\kappa \frac{i}{S(\omega_p) + p_4} S(\mathbf{p})_i
	=
	-i\kappa
	\frac{\omega_p + p_4}{\kappa^2}
	\frac{\kappa\mathbf{p}_i}{\omega_p + p_4}
	=
	-i\mathbf{p}_i
\end{align}
which concludes the verification.

\subsection{$A = 4$}\label{A=4}
We have
\begin{align}
	\partial_4 \Xi
	=
	i(\kappa-p_4)
	a_\mathbf{p}\, e^{-i(\omega_pt-\mathbf{p}\mathbf{x})} 
	+ 
	i\underbrace{(\kappa - S(p_4))}_{=\kappa - p_4}
	b^\dag_\mathbf{p^*}\, e^{i(S(\omega^*_p)t-S(\mathbf{p}^*)\mathbf{x})} 
\end{align}
where in the first term in the RHS we just used eq. \eqref{dergroup} and in the second term we used the fact that
\begin{align}
	e^{i(S(\omega^*_p)t-S(\mathbf{p}^*)\mathbf{x})} 
	=
	e^{-i(S(\omega_p)t-S(\mathbf{p})\mathbf{x})} 
\end{align}
and then we used again the definition \eqref{dergroup}. Therefore we have
\begin{align}
	(\partial_4 \phi)^\dag
	= 
	-i(\kappa-p_4)
	a_\mathbf{p}^\dag\, e^{-i(S(\omega_p)t-S(\mathbf{p})\mathbf{x})} 
	-i(\kappa - p_4)
	b_\mathbf{p^*}\, e^{i(\omega^*_pt-\mathbf{p}^*\mathbf{x})} 
\end{align}
On the other hand we have
\begin{align}
	\partial_4^\dag \phi^\dag &= -\partial_4 \phi^\dag \\
	&=
	-i(\kappa - S(p_4))
	a^\dag_\mathbf{p}\, e^{-i(S(\omega_p)t-S(\mathbf{p})\mathbf{x})} 
	-i(\kappa - p_4)
	b_\mathbf{p^*}\, e^{i(\omega^*_pt-\mathbf{p}^*\mathbf{x})}
\end{align}
which concludes the check.

\subsection{$A = +$}\label{A=+}
From subsection \eqref{mu=0} and \eqref{A=4} together with the definition \eqref{partial+} we have that $(\Delta_+ \phi)^\dag = \Delta_+^\dag \phi^\dag$. In fact, we know that $(\partial_0 \phi)^\dag = \partial_0^\dag \phi^\dag$ then we also know that $(\alpha \partial_0 \phi)^\dag = \alpha^\dag (\partial_0 \phi)^\dag = \alpha^\dag \partial_0^\dag \phi^\dag$ for any constant $\alpha$, and the same goes for $\partial_4$.

\subsection{$\Delta_+^\dagger = \kappa^2 \Delta_+^{-1}$}
We have
\begin{align}
	\Delta_+^\dag
	&=
	-i\partial_0^\dag +\kappa -i \partial_4^\dag \\
	&=
	-i
	\left(
	\partial_0 -i \Delta_+^{-1} \partial^2
	\right)
	+\kappa +i \partial_4 \\
	&=
	-i\partial_0
	-\frac{\partial^2}{i\partial_0 + \kappa + i\partial_4}
	+\kappa + i\partial_4 \\
\end{align}
Furthermore, recall that by definition we have
\begin{align}
	(\kappa + i\partial_4)^2 = \kappa^2 -\partial_0^2 + \partial^2
\end{align}
and therefore we end up with
\begin{align}
	\Delta_+^\dag = 
	\frac{\partial_0^2 - \partial^2 + \kappa^2 -\partial_0^2 + \partial^2}{\Delta_+} = \frac{\kappa^2}{\Delta_+}
\end{align}
which proves the formula.

\section{Integration-by-parts relation for $\partial_i$ going $\leftarrow$}\label{Intbyparts1}
We now need explicitly what are the integration by parts relations that one can get using eq. \eqref{4.13bis}, \eqref{4.14bis}, \eqref{4.15bis}, \eqref{4.16}, \eqref{partial+}. This is a preparatory step, and we will need it after the introduction of the action of the complex scalar field in order to compute the continuous charges. 

For the spatial integration by parts we have
\begin{align}\label{ibp-pdagp-i}
	\boxed{(\partial_i\phi)^\dag \star (\partial_i\psi)
		=
		\partial_i\left[
		(\partial_i\phi)^\dagger \star \psi
		\right]
		-
		\frac{\Delta_+}{\kappa}
		\left[(\partial^2\phi)^\dag \star \psi \right]}.
\end{align}
which can be obtained using the following steps
\begin{align}
	(\partial_i\phi)^\dag \star (\partial_i\psi)
	&\overset{\eqref{4.14bis}}{=}
	\partial_i\left[
	(\partial_i\phi)^\dagger \star \psi
	\right]
	-
	\frac{1}{\kappa}[\partial_i(\partial_i\phi)^\dag] \star (\Delta_+\psi) \\
	&\overset{\eqref{4.15bis}}{=}
	\partial_i\left[
	(\partial_i\phi)^\dagger \star \psi
	\right]
	-
	\Delta_+
	\left[(\Delta_+^{-1}\partial_i(\partial_i\phi)^\dag) \star \psi \right] \\
	&=
	\partial_i\left[
	(\partial_i\phi)^\dagger \star \psi
	\right]
	-
	\frac{\Delta_+}{\kappa}
	\left[(\kappa \Delta_+^{-1}\partial_i(\partial_i\phi)^\dag) \star \psi \right] \\
	&\overset{\eqref{4.16}}{=}
	\partial_i\left[
	(\partial_i\phi)^\dagger \star \psi
	\right]
	-
	\frac{\Delta_+}{\kappa}
	\left[(\partial_i^\dag(\partial_i\phi)^\dag) \star \psi \right] \label{nextolastibis}\\
	&=
	\partial_i\left[
	(\partial_i\phi)^\dagger \star \psi
	\right]
	-
	\frac{\Delta_+}{\kappa}
	\left[(\partial^2\phi)^\dag \star \psi \right]\label{resibis}.
\end{align}

\section{Integration-by-parts relation for $\partial_0$ going $\leftarrow$}
We do the same thing for the temporal part. We have
\begin{align}\label{ibp-pdagp-0}
	\boxed{(\partial_0\phi)^\dag \star (\partial_0 \psi)
		=
		\frac{\partial_0}{\kappa}\left[
		(\Delta_+(\partial_0\phi)^\dag) \star \psi
		\right]
		- 
		i\partial_i[(\Delta_+^{-1}\partial_i{\partial_0\phi})^\dag\star \psi]
		-
		\frac{\Delta_+}{\kappa}\left[
		(\partial_0^2\phi)^\dag \star \psi
		\right]}
\end{align}
which can be proven using the following steps. 
\begin{align}
	&\kappa \, (\partial_0\phi)^\dag \star (\partial_0 \psi) \\
	&\overset{\eqref{4.13bis}}{=}
	\partial_0\left[
	(\Delta_+(\partial_0\phi)^\dag) \star \psi
	\right]
	-
	\frac{1}{\kappa}
	(\Delta_+\partial_0(\partial_0\phi)^\dag)\star (\Delta_+\psi)
	-i
	(\partial_i(\partial_0\phi)^\dag) \star (\partial_i\psi) \\
	&\overset{\eqref{4.15bis}}{=}
	\partial_0\left[
	(\Delta_+(\partial_0\phi)^\dag) \star \psi
	\right]
	-
	\Delta_+
	[(\partial_0(\partial_0\phi)^\dag)\star \psi]
	-i
	(\partial_i(\partial_0\phi)^\dag) \star (\partial_i\psi)  \\
	&\overset{\eqref{4.14bis}}{=}
	\partial_0\left[
	(\Delta_+(\partial_0\phi)^\dag) \star \psi
	\right]
	-
	\Delta_+
	[(\partial_0(\partial_0\phi)^\dag)\star \psi]
	-i
	\partial_i[(\partial_i({\partial_0\phi})^\dag)\star \psi] \nonumber \\ 
	&+
	\frac{i}{\kappa}  
	(\partial^2({\partial_0\phi})^\dag)\star (\Delta_+ \psi) \\
	&\overset{\eqref{4.15bis}}{=}
	\partial_0\left[
	(\Delta_+(\partial_0\phi)^\dag) \star \psi
	\right]
	-
	\Delta_+
	[(\partial_0(\partial_0\phi)^\dag)\star \psi]
	-i
	\partial_i[(\partial_i({\partial_0\phi})^\dag)\star \psi] \nonumber \\
	&+
	i\Delta_+  
	[(\Delta_+^{-1}\partial^2({\partial_0\phi})^\dag)\star \psi] \\
	&=
	\partial_0\left[
	(\Delta_+(\partial_0\phi)^\dag) \star \psi
	\right]
	-
	i\partial_i[(\partial_i({\partial_0\phi})^\dag)\star \psi] \nonumber \\
	&-
	\Delta_+\left\{
	[(\partial_0 -i \Delta_+^{-1}\partial^2)(\partial_0\phi)^\dag] \star \psi
	\right\} \\
	&\overset{\eqref{4.16}}{=}
	\partial_0\left[
	(\Delta_+(\partial_0\phi)^\dag) \star \psi
	\right]
	-
	i\partial_i[\Delta_+ \kappa^{-1}(\partial_i{\partial_0\phi})^\dag\star \psi] \nonumber \\
	&-
	\Delta_+\left\{
	[(\partial_0 -i \Delta_+^{-1}\partial^2)(\partial_0\phi)^\dag] \star \psi
	\right\}  \\
	&\overset{\eqref{4.16}}{=}
	\partial_0\left[
	(\Delta_+(\partial_0\phi)^\dag) \star \psi
	\right]
	-
	i\kappa\partial_i[(\Delta_+^{-1} \partial_i{\partial_0\phi})^\dag\star \psi] \nonumber \\
	&-
	\Delta_+\left\{
	[(\partial_0 -i \Delta_+^{-1}\partial^2)(\partial_0\phi)^\dag] \star \psi
	\right\}  \\
	&\overset{\eqref{4.16}}{=}
	\partial_0\left[
	(\Delta_+(\partial_0\phi)^\dag) \star \psi
	\right]
	-i \kappa
	\partial_i[(\Delta_+^{-1}\partial_i{\partial_0\phi})^\dag\star \psi]
	-
	\Delta_+\left\{
	[\partial_0^\dag(\partial_0\phi)^\dag] \star \psi
	\right\} \label{nextolast0bis}\\
	&=
	\partial_0\left[
	(\Delta_+(\partial_0\phi)^\dag) \star \psi
	\right]
	-i \kappa
	\partial_i[(\Delta_+^{-1}\partial_i{\partial_0\phi})^\dag\star \psi]
	-
	\Delta_+\left[
	(\partial_0^2\phi)^\dag \star \psi
	\right].\label{res0bis}
\end{align}

\section{Integration-by-parts relation for $\partial_i$ going $\rightarrow$}
Working in the same way, we obtain
\begin{align}\label{ibp-ppdag-i}
	\boxed{(\partial_i \psi) \star (\partial_i \phi)^\dag
	=
	\kappa \partial_i (\psi \star [\Delta_+^{-1}(\partial_i \phi)^\dag])
	-
	\psi \star (\partial^2 \phi)^\dag}.
\end{align}
Indeed, we have
\begin{align}
	\frac{1}{\kappa}(\partial_i \psi) \star (\partial_i \phi)^\dag
	\overset{\eqref{4.14bis}}{=}
	\partial_i (\psi \star [\Delta_+^{-1}(\partial_i \phi)^\dag])
	-
	\psi \star [\Delta_+^{-1} \partial_i (\partial_i \phi)^\dag]
\end{align}
and therefore
\begin{align}
	(\partial_i \psi) \star (\partial_i \phi)^\dag
	&\overset{\eqref{4.14bis}}{=}
	\kappa \partial_i (\psi \star [\Delta_+^{-1}(\partial_i \phi)^\dag])
	-
	\psi \star [\kappa \Delta_+^{-1} \partial_i (\partial_i \phi)^\dag] \\
	&\overset{\eqref{4.16}}{=}
	\kappa \partial_i (\psi \star [\Delta_+^{-1}(\partial_i \phi)^\dag])
	-
	\psi \star (\partial^2 \phi)^\dag.
\end{align}

\section{Integration-by-parts relation for $\partial_0$ going $\rightarrow$}

One can show that 
\begin{subequations}\label{ibp-ppdag-0}
	\begin{empheq}[box=\widefbox]{align}
	(\partial_0 \psi) \star (\partial_0 \phi)^\dag
	&=
	\partial_0 (\psi \star [\kappa\Delta_+^{-1}(\partial_0\phi)^\dag])
	-
	i
	\partial_i (\psi \star [\Delta_+^{-1}\partial_i(\partial_0 \phi)^\dag])\nonumber \\
	&
	-
	[\psi \star (\partial_0^2 \phi)^\dag] 
	+
	\left(\frac{i}{\kappa}\partial_0 + i\frac{\partial_4}{\kappa}\right)
	[\psi \star (\partial_0^2 \phi)^\dag]
	\end{empheq}
\end{subequations}
Indeed we have
\begin{align}
	\frac{1}{\kappa}(\partial_0 \psi) \star (\partial_0 \phi)^\dag
	&\overset{\eqref{4.13bis}}{=}
	\partial_0 (\psi \star [\Delta_+^{-1}(\partial_0\phi)^\dag])
	-
	\kappa(\Delta_+^{-1}\psi) \star (\Delta_+^{-1}\partial_0(\partial_0\phi)^\dag)
	\nonumber \\
	&-
	i(\Delta_+^{-1}\partial_i \psi) \star (\Delta_+^{-1}\partial_i (\partial_0\phi)^\dag) \\ \nonumber \\
	&\overset{\eqref{4.15bis}}{=}
	\partial_0 (\psi \star [\Delta_+^{-1}(\partial_0\phi)^\dag])
	-
	\Delta_+^{-1}[\psi \star (\partial_0(\partial_0\phi)^\dag)] \nonumber \\
	&-
	i\kappa^{-1}\Delta_+^{-1}[(\partial_i \psi) \star (\partial_i (\partial_0\phi)^\dag)] \\ \nonumber \\
	&\overset{\eqref{4.14bis}}{=}
	\partial_0 (\psi \star [\Delta_+^{-1}(\partial_0\phi)^\dag])
	-
	\Delta_+^{-1}[\psi \star (\partial_0(\partial_0\phi)^\dag)] \nonumber \\
	&-
	i\Delta_+^{-1}
	\Big[
	\partial_i (\psi \star [\Delta_+^{-1}\partial_i(\partial_0 \phi)^\dag])
	-
	\psi \star [\Delta_+^{-1} \partial^2 (\partial_0 \phi)^\dag]
	\Big] \\ \nonumber \\
	&=
	\partial_0 (\psi \star [\Delta_+^{-1}(\partial_0\phi)^\dag])
	-
	i\Delta_+^{-1}
	\partial_i (\psi \star [\Delta_+^{-1}\partial_i(\partial_0 \phi)^\dag]) \nonumber \\
	&-
	\Delta_+^{-1}
	[\psi \star (\partial_0 - i \Delta_+^{-1}\partial^2)(\partial_0\phi)^\dag] \\ \nonumber \\
	&\overset{\eqref{4.16}}{=}
	\partial_0 (\psi \star [\Delta_+^{-1}(\partial_0\phi)^\dag])
	-
	i\Delta_+^{-1}
	\partial_i (\psi \star [\Delta_+^{-1}\partial_i(\partial_0 \phi)^\dag]) \nonumber \\
	&-
	\Delta_+^{-1}
	[\psi \star (\partial_0^2 \phi)^\dag]
\end{align}
We now expand $\Delta_+^{-1}$ (recall that $\Delta_+ = i\partial_0 + \kappa +i\partial_4$). By expanding we mean writing the differential operator $\kappa\Delta_+^{-1}$ in terms of a sum of derivatives in the numerator. In other words, after applying $\kappa\Delta_+^{-1}$ to a star product, we get some function of momenta at the denominator, and we expand such a function around a point in momentum space which is convenient for us. At this point, we convert this function back to action of derivatives on the original star product. The net effective result is that we have expanded out original differential operator as if it was already a function of momenta. We have
\begin{align}
	\kappa\Delta_+^{-1} = \frac{1}{i\frac{\partial_0}{\kappa} + 1 +i\frac{\partial_4}{\kappa}}
\end{align}
Notice that we cannot expand around $(\partial_0, \partial_i)=(0,0)$ (here we abuse notation a bit by addressing derivatives as if they are already momenta, but one should always keep in mind our previous discussion) because in this case $m = 0$ on-shell, and we are assuming $m\neq 0$ instead. Furthermore, notice that if we expand around $\partial_0 = 0$, then the expansion up to first order would be
\begin{align}
	\kappa \Delta_+^{-1}
	=
	\kappa \Delta_+^{-1}(\partial_0 = 0) + \frac{\delta \kappa\Delta_+^{-1}}{\delta \partial_0}\Bigg|_{\partial_0=0} \partial_0
	+
	O(\partial_0^2)
\end{align}
where we used eq. \eqref{dergroup}, and in particular the zeroth order would become
\begin{align}
	\kappa \Delta_+^{-1}(\partial_0 = 0) = \frac{1}{1-\sqrt{1+\frac{m^2}{\kappa^2}}} \neq 1
\end{align}
which is however problematic, because one would not recover the correct equation of motion (since the temporal component of the kinetic term will have a different rescaling than the spatial part). We need to find an expansion point $p$ such that $\kappa\Delta_+^{-1}(p) = 1$. Therefore, we expand around $\partial_0 = -\partial_4 = \partial_4^\dag$. We have
\begin{align}
	\kappa \Delta_+^{-1}
	&=
	\kappa \Delta_+^{-1}(\partial_0 = -\partial_4) + \frac{\delta \kappa\Delta_+^{-1}}{\delta \partial_0}\Bigg|_{\partial_0=-\partial_4} (\partial_0 - (-\partial_4))
	+
	O((\partial_0 + \partial_4)^2) \\
	&=
	1 - \frac{\frac{i}{\kappa}}{\left( i\frac{\partial_0}{\kappa} + 1 +i\frac{\partial_4}{\kappa}\right)^2}\Bigg|_{\partial_0 = -\partial_4} (\partial_0 + \partial_4)
	=
	1 - \frac{i}{\kappa}\partial_0 
	- i\frac{\partial_4}{\kappa}
\end{align}
Coming back to our integration-by-parts relation, without loss of generality we can write
\begin{align}
	(\partial_0 \psi) \star (\partial_0 \phi)^\dag
	&=
	\partial_0 (\psi \star [\kappa\Delta_+^{-1}(\partial_0\phi)^\dag])
	-
	i
	\partial_i (\psi \star [\Delta_+^{-1}\partial_i(\partial_0 \phi)^\dag])
	-
	[\psi \star (\partial_0^2 \phi)^\dag] \nonumber\\
	&
	+
	i
	\partial_i 
	\left(\frac{i}{\kappa}\partial_0 + i\frac{\partial_4}{\kappa} \right)
	(\psi \star [\Delta_+^{-1}\partial_i(\partial_0 \phi)^\dag]) \nonumber \\
	&+
	\left(\frac{i}{\kappa}\partial_0 + i\frac{\partial_4}{\kappa}\right)
	[\psi \star (\partial_0^2 \phi)^\dag]
\end{align}
However, the term $i
\partial_i 
\left(\frac{i}{\kappa}\partial_0 + i\frac{\partial_4}{\kappa} \right)
(\psi \star [\Delta_+^{-1}\partial_i(\partial_0 \phi)^\dag])$ can be ignored because it contain a double total derivative (we are considering $\partial_4$ as an independent derivative with respect to the others, because indeed $p_4$ can have a value independent from the value of $p_0$ or $p_i$). This proves eq. \eqref{ibp-ppdag-0}.

Notice that, since we do not have any prefactor in front of the kinetic part of the on-shell operator, we also don't need to modify the mass term to obtain the correct on-shell relation, like it was needed in the previous ordering. Furthermore, notice also that we only expanded the $\Delta_+^{-1}$ which acted globally on each star product, while there would be no meaning in expanding the factors $\Delta_+^{-1}$ inside the star product. Furthermore, this expansion does not mean that our results are approximate. Indeed, higher order term in the expansion of $\kappa\Delta_+^{-1}$ would contribute factors proportional to $\partial^2$ or higher, and all these contributions vanish in any case because of the integral.

\section{Mass terms integration by parts}\label{Intbypartsm}
Recalling that $\Delta_+ = i\partial_0 + \kappa +i\partial_4$ for the first ordering we have
\begin{align}\label{ibp-pdagp-m}
	\boxed{m^2 \phi^\dag \star \psi
	=
	-
	\frac{i\partial_0}{\kappa}
	(m^2 \phi^\dag \star \psi)
	-
	\frac{i\partial_4}{\kappa}
	(m^2 \phi^\dag \star \psi)
	+
	\frac{\Delta_+}{\kappa}
	(m^2 \phi^\dag \star \psi)}
\end{align}
which can be proven trivially from
\begin{align}
	m^2 \phi^\dag \star \psi
	&=
	-\left(\frac{\Delta_+}{\kappa} - 1\right)(m^2 \phi^\dag \star \psi )
	+
	\frac{\Delta_+}{\kappa}
	(m^2 \phi^\dag \star \psi).
\end{align}
For the second ordering we don't have a contribution coming from this reasoning.

\section{EoM and surface terms $\Pi^0$, $\Pi^i$, and $\Pi^4$ for both orderings of the action}\label{eomandsurf}

We now have all the necessary tools, i.e. eq. \eqref{ibp-pdagp-i}, \eqref{ibp-pdagp-0}, \eqref{ibp-ppdag-i}, \eqref{ibp-ppdag-0}, \eqref{ibp-pdagp-m}, to integrate by parts both orderings of the action, i.e. eq. \eqref{S1} and \eqref{S2}, which we rewrite here for simplicity
\begin{align}
	\eqref{S1} \rightarrow \qquad
	S_1 
	&= 
	\int_{\mathbb{R}^4} d^4x \,\, (\partial^\mu \phi)^\dag \star (\partial_\mu \phi) - m^2 \phi^\dag \star \phi 
\end{align}
\begin{align}
	\eqref{S2} \rightarrow \qquad
	S_2 
	&= 
	\int_{\mathbb{R}^4} d^4x \,\ (\partial_\mu \phi) \star (\partial^\mu \phi)^\dag  - m^2 \phi \star \phi^\dag .
\end{align}
We are at point $iv)$ of our roadmap in section \ref{roadmap2}. We can now compute the variation of the action in eq. \eqref{Stot}
\begin{align}
	\delta S_1 = \frac12\int_{ \mathbb{R}^4}d^4x\,
	(\partial_\mu \delta \phi)^\dag\star\partial^\mu \phi
	+
	(\partial_\mu \phi)^\dag\star\partial^\mu \delta \phi
	-
	m^2 \delta \phi^\dag\star \phi
	-
	m^2 \phi^\dag\star \delta \phi \label{4.35}
\end{align}
which can be immediately rewritten using integration by parts \eqref{ibp-pdagp-i}, \eqref{ibp-pdagp-0}, \eqref{ibp-pdagp-m} as\footnote{One can integrate by parts one of the two terms, and then obtain the other by taking the Hermitian conjugate. Of course, the same relation is obtained if one starts with the co-product rules for the derivatives $\partial^\dag$, and then proceeds from there.}
\begin{align}
	\delta S_1 = \frac12\int_{ \mathbb{R}^4}d^4x\,
	\Bigg\{
	-&\frac{\Delta_+}{\kappa}
	\left[
	(\partial_\mu^\dag (\partial^\mu)^\dag - m^2) \phi^\dag
	\star \delta\phi
	\right]
	+ \partial_A \left( \Pi_1^A \star \delta \phi \right) \nonumber \\
	-&
	\frac{\kappa}{\Delta_+}
	\left[
	\delta\phi^\dag
	\star
	(\partial_\mu\partial^\mu - m^2) \phi
	\right]
	+
	\partial_A^\dag
	\left(
	\delta\phi^\dag \star \left(\Pi_1^A\right)^\dag
	\right)
	\Bigg\} \label{4.36}
\end{align}
where
\begin{align}
	\Pi^0_1 = (\Pi_0)_1 &= \frac{1}{\kappa} (\Delta_+ \partial_0^\dag +i m^2) \phi^\dag \label{4.37} \\
	\Pi^i_1 = -(\Pi_i)_1 &= (-\partial_i (1 + i\Delta_+^{-1}\partial_0))\phi^\dag \label{4.38} \\
	\Pi^4_1=(\Pi_4)_1 &= -i\frac{m^2\phi^\dag}{\kappa}. \label{4.39}
\end{align}
For the other ordering we proceed in a similar manner, with the only difference that we use the integration-by-parts relations \eqref{ibp-ppdag-i}, \eqref{ibp-ppdag-0}. We have
\begin{align}
	\delta S_2 = \frac12\int_{ \mathbb{R}^4}d^4x\,
	\partial^\mu \phi \star (\partial_\mu \delta \phi)^\dag
	+
	\partial^\mu \delta \phi \star (\partial_\mu \phi)^\dag
	-
	m^2  \phi \star \delta\phi^\dag
	-
	m^2 \delta \phi \star \phi^\dag \label{deltaS2}
\end{align}
which can be rewritten as
\begin{align}
	\delta S_2 = \frac12\int_{ \mathbb{R}^4}d^4x\,
	\Bigg\{
	-&
	\left[
	\delta\phi \star
	(\partial_\mu^\dag (\partial^\mu)^\dag - m^2) \phi^\dag
	\right]
	+ \partial_A \left( \delta \phi \star \Pi^A  \right) \nonumber \\
	-&
	\left[
	(\partial_\mu\partial^\mu - m^2) \phi
	\star
	\delta\phi^\dag
	\right]
	+
	\partial_A^\dag
	\left(
	\left(\Pi^A\right)^\dag
	\star 
	\delta\phi^\dag
	\right)
	\Bigg\} \label{4.36bis}
\end{align}
where
\begin{align}
	\Pi^0_2 = (\Pi_0)_2 &= \left(\frac{\kappa}{\Delta_+} \partial_0^\dag +\frac{i}{\kappa}(\partial_0^\dag)^2
	\right)\phi^\dag \label{4.37bis} \\
	\Pi^i_2 = -(\Pi_i)_2 &= - \frac{\kappa}{\Delta_+} (\partial_i^\dag + i\partial_i\partial_0^\dag)\phi^\dag \label{4.38bis} \\
	\Pi^4_2=(\Pi_4)_2 &= +i\frac{(\partial_0^\dag)^2}{\kappa} \phi^\dag . \label{4.39bis}
\end{align}
We see that the equations of motion are the canonical Klein-gordon equation
\begin{align}\label{EoM-KG}
	(\partial_\mu\partial^\mu - m^2) \phi = 0 
	\qquad
	(\partial_\mu^\dag(\partial^\mu)^\dag - m^2) \phi^\dag = 0.
\end{align}
Notice that in momentum space, $\partial_\mu^\dag (\partial^\mu)^\dag \leftrightarrow S(p)_\mu S(p)^\mu = p_\mu p^\mu \leftrightarrow \partial_\mu\partial^\mu$), where one can check explicitly that 
\begin{align}
	S(p)_\mu S(p)^\mu 
	&=
	S(\omega_p)^2 - S(\mathbf{p})^2 \\
	&=
	\left(-\omega_p + \frac{\mathbf{p}^2}{\omega_p + p_4}\right)^2
	-
	\left(
	-\frac{\kappa\mathbf{p}}{\omega_p + p_4}
	\right)^2 \\
	&=
	\omega_p^2 -2\frac{\omega_p \mathbf{p}^2}{\omega_p + p_4}
	+
	\frac{\mathbf{p}^4}{(\omega_p + p_4)^2}
	-
	\frac{\kappa^2\mathbf{p}^2}{(\omega_p+p_4)^2} \\
	&=
	\omega_p^2 + \mathbf{p}^2 \frac{-(\omega_p + p_4)^2}{(\omega_p + p_4)^2} \\
	&=
	p_\mu p^\mu
\end{align}
and therefore
\begin{align}\label{on-shell-antipode}
	S(p)_\mu S(p)^\mu
	=
	p_\mu p^\mu
\end{align}
The fields satisfying such equations are usually written as linear combinations of plane waves and their Hermitian conjugate. Here, because of the curved nature of momentum space, it is not so automatic. Nevertheless, we will see in the next section that fields satisfying these equations, in the context of $\kappa$-deformation, can be described in a particularly clear way.

\section{Complex scalar field and its properties}\label{csfCPT}

We are now in point $v)$ of our roadmap. 
As it is clear from eq. \eqref{classicalconstraints}, the $AN(3)$ group manifold is a 4-dimensional submanifold of a 5-dimensional space. Furthermore, we have plane waves, i.e. group elements $g\in AN(3)$. To define a field on it which satisfies the EoM in eq. \eqref{EoM-KG} we can use both these facts and define \cite{Arzano:2020jro} 
\begin{align}
	\phi(x)
	=
	\int d^5p
	\,\,
	2\kappa
	\delta(p_A p^A + \kappa^2)\theta(p_+) \theta(p_4)
	\delta(p_\mu p^\mu - m^2)
	\tilde{\phi}(p) 
	e^{-ipx}
\end{align}
where $\phi(x)$ is a function in commutative spacetime. Notice that the quantity 
\begin{align}
	d^5p
	\,\,
	2\kappa
	\delta(p_A p^A + \kappa^2)
\end{align}
is the metric in $\mathbb{R}^5$ restricted to the group manifold with the constraint in eq. \eqref{classicalconstraints} (the $\kappa$ is only present for dimensional reasons). In the same way, the Heaviside $\theta$ functions restrict ourselves to the correct part of the manifold, namely where $P_+>0$ and $P_4>0$. Finally, the remaining Dirac delta imposes the on-shell condition. In order to write this field in a more familiar way, one can first split the constraints as follows.
\begin{align}
	\delta(p_\mu p^\mu - m^2)
	=
	\delta(p_\mu p^\mu - m^2)
	\theta(p_0-m)
	+
	\delta(p_\mu p^\mu - m^2)
	\theta(-p_0-m)
\end{align}
We use the notation
\begin{align}
	\phi_+(x)
	=
	\int d^5p
	\,\,
	2\kappa
	\delta(p_A p^A + \kappa^2)\theta(p_+) \theta(p_4)
	\delta(p_\mu p^\mu - m^2)
	\theta(p_0-m)
	\tilde{\phi}(p) 
	e^{-ipx}
\end{align}
\begin{align}
	\phi_-(x)
	=
	\int d^5p
	\,\,
	2\kappa
	\delta(p_A p^A + \kappa^2)\theta(p_+) \theta(p_4)
	\delta(p_\mu p^\mu - m^2)
	\theta(-p_0-m)
	\tilde{\phi}(p) 
	e^{-ipx}
\end{align}
for simplicity. We can then change variable in $\phi_-$ by sending $p\mapsto S(p)$. Because of eq. \eqref{on-shell-antipode} and \eqref{SP}, we know that the arguments of the Dirac delta are invariant. One can also show starting from eq. \eqref{SP} that
\begin{align}
	S(p_+) = -p_0 + \frac{\mathbf{p}^2}{p_+} + S(p)_4 =
	\frac{\kappa^2}{p_+}
\end{align}
so that also the Heaiside $\theta(p_+)$ and $\theta(p_4)$ are left invariant. One can then change variable $p\mapsto -p=p^*$ and pick up some additional factors coming from the determinant of the Jacobian obtained in the coordinate change $p\mapsto S(p)$. Furthermore, one can apply the Dirac deltas, reducing the integration from $d^5p$ to $d^3p$, with other factors added to the integrand (more details can be found in \cite{Arzano:2020jro}). The result is that the field can be written as 
\begin{align}\label{fa0}
	\phi(x) &= \int \frac{d^3p}{\sqrt{2\omega_\mathbf{p}}}\, \,\zeta(p)\, a_{\mathbf{p}}\, e^{-i(\omega_\mathbf{p}t-\mathbf{p}\mathbf{x})}  + \int \frac{d^3p^*}{\sqrt{2|\omega_{\mathbf{p}^*}|}}\, \,\zeta(p) b^\dag_{\mathbf{p}^*}\, e^{i(S(\omega_{\mathbf{p}^*})t-S(\mathbf{p}^*)\mathbf{x})}
\end{align}
\begin{align}\label{fb0}
	\phi^\dag(x) &= \int \frac{d^3p}{\sqrt{2\omega_\mathbf{p}}}\, \,\zeta(p)\, a^\dag_{\mathbf{p}}\, e^{-i(S(\omega_\mathbf{p})t-S(\mathbf{p})\mathbf{x})}  + \int \frac{d^3p^*}{\sqrt{2|\omega_{\mathbf{p}^*}|}}\, \,\zeta(p) b_{\mathbf{p}^*}\, e^{i(\omega_{\mathbf{p}^*}t-\mathbf{p}^*)\mathbf{x}} 
\end{align}
Notice that if $p>0$, then $S(p^*)>0$, so that both $\phi(x)$ and $\phi^\dag(x)$ are a linear combination of positive and negative energy states. Notice that there is an additional function $\zeta(p)$ which will be important when talking about the action. Furthermore, notice also that the ${}^*$ can be eliminated by all the expressions by simply change variable $p^*\mapsto -p$. Therefore, it may be convenient to use instead the fields
\begin{align}\label{fa}
	\phi(x) &= \int \frac{d^3p}{\sqrt{2\omega_\mathbf{p}}}\, \,\zeta(p)\, a_{\mathbf{p}}\, e^{-i(\omega_\mathbf{p}t-\mathbf{p}\mathbf{x})}  + \int \frac{d^3p}{\sqrt{2\omega_{\mathbf{p}}}}\, \,\zeta(p) b^\dag_{\mathbf{p}}\, e^{-i(S(\omega_{\mathbf{p}})t-S(\mathbf{p})\mathbf{x})}
\end{align}
\begin{align}\label{fb}
	\phi^\dag(x) &= \int \frac{d^3p}{\sqrt{2\omega_\mathbf{p}}}\, \,\zeta(p)\, a^\dag_{\mathbf{p}}\, e^{-i(S(\omega_\mathbf{p})t-S(\mathbf{p})\mathbf{x})}  + \int \frac{d^3p}{\sqrt{2\omega_{\mathbf{p}}}}\, \,\zeta(p) b_{\mathbf{p}}\, e^{-i(\omega_{\mathbf{p}}t-\mathbf{p})\mathbf{x}} 
\end{align}
for simplicity. The reason for using the variables $p^*$ instead of $-p$ is to make clear that the minus sign in $-p$ needs to also be present in front of $p_4$, and not only $p_0, \mathbf{p}$. This is important because we need the relation defined in \eqref{starredcoord} to hold. Having clarified this point, we can safely switch from $p^*$ to $-p$ without issues.

\subsection{$C,P,T$ in the non-deformed context: a short review}

We provide a short review of the definition of $C,P,T$ transformations in the non-deformed case, which will be the basis for our treatment of the deformed case. The non-deformed fields are the same as eq. \eqref{fa}, \eqref{fb} but in the limit $\kappa\rightarrow\infty$, so we have
\begin{align}\label{fa-u}
	\phi_U(x) &= \int \frac{d^3p}{\sqrt{2\omega_\mathbf{p}}}\, \, a_{\mathbf{p}}\, e^{-i(\omega_\mathbf{p}t-\mathbf{p}\mathbf{x})}  + \int \frac{d^3p}{\sqrt{2\omega_{\mathbf{p}}|}}\, \, b^\dag_{\mathbf{p}}\, e^{i(\omega_{\mathbf{p}}t-\mathbf{p})\mathbf{x})}
\end{align}
\begin{align}\label{fb-u}
	\phi_U^\dag(x) &= \int \frac{d^3p}{\sqrt{2\omega_\mathbf{p}}}\, \, a^\dag_{\mathbf{p}}\, e^{i(\omega_\mathbf{p}t-\mathbf{p}\mathbf{x})}  + \int \frac{d^3p}{\sqrt{2\omega_{\mathbf{p}}}}\, \, b_{\mathbf{p}}\, e^{-i(\omega_{\mathbf{p}}t-\mathbf{p})\mathbf{x}}.
\end{align}
\begin{itemize}
	\item[$(P)$] By definition, a $P$ transformation is defined by
	 $(t,\mathbf{x}) \mapsto (t, -\mathbf{x})$ and therefore $\mathcal{P}\phi(t,\mathbf{x})\mathcal{P}^{-1} = \phi(t,-\mathbf{x})$. 
	 \begin{align}
	 	\mathcal{P}\phi_U(t,\mathbf{x})\mathcal{P}^{-1}
	 	&=
	 	\int\frac{d^3p}{\sqrt{2\omega_p}} 
	 	\mathcal{P}a_\mathbf{p} \mathcal{P}^{-1}
	 	e^{-i(\omega_p t - \mathbf{p}\mathbf{x})} 
	 	+
	 	\mathcal{P}b_\mathbf{p}^\dag \mathcal{P}^{-1}
	 	e^{i(\omega_p t - \mathbf{p}\mathbf{x})}  \\
	 	&\overset{!}{=}
	 	\int\frac{d^3p}{\sqrt{2\omega_p}} 
	 	a_\mathbf{p}
	 	e^{-i(\omega_p t + \mathbf{p}\mathbf{x})} 
	 	+
	 	b_\mathbf{p}^\dag
	 	e^{i(\omega_p t + \mathbf{p}\mathbf{x})} 
	 \end{align}
	 which means that
	 \begin{align}
	 	\mathcal{P}a_{\mathbf{p}}\mathcal{P}^{-1} = a_{-\mathbf{p}}
	 	\qquad
	 	\mathcal{P}b_{\mathbf{p}}\mathcal{P}^{-1} = b_{-\mathbf{p}}
	 \end{align} 
	 and the same for $a^\dag, b^\dag$.

	 \item[$(T)$] By definition, a $T$ transformation is defined by $(t,\mathbf{x}) \mapsto (-t, \mathbf{x})$ and therefore $\mathcal{T}\phi(t,\mathbf{x})\mathcal{T}^{-1} = \phi(-t,\mathbf{x})$. Recall that this is an anti-unitary operator. One can show easily this because $\mathcal{T}x\mathcal{T}^{-1}=x$, $\mathcal{T}\mathbf{p}\mathcal{T}^{-1}=-\mathbf{p}$ imply
	  \begin{align}
	  	[x,p]=i \xrightarrow{\mathcal{T}} -[x,p]=\mathcal{T}i\mathcal{T}^{-1} \overset{!}{=} -i.
	  \end{align}
	  We have
	  \begin{align}
	  	\mathcal{T}\phi(t,\mathbf{x})\mathcal{T}^{-1}
	  	&=
	  	\int\frac{d^3p}{\sqrt{2\omega_p}} 
	  	\mathcal{T}a_\mathbf{p} \mathcal{T}^{-1}
	  	e^{+ i(\omega_p t - \mathbf{p}\mathbf{x})} 
	  	+
	  	\mathcal{T}b_\mathbf{p}^\dag \mathcal{T}^{-1}
	  	e^{- i(\omega_p t - \mathbf{p}\mathbf{x})}  \\
	  	&\overset{!}{=}
	  	\int\frac{d^3p}{\sqrt{2\omega_p}} 
	  	a_\mathbf{p}
	  	e^{-i(-\omega_p t - \mathbf{p}\mathbf{x})} 
	  	+
	  	b_\mathbf{p}^\dag
	  	e^{i(-\omega_p t + \mathbf{p}\mathbf{x})} 
	  \end{align}
	  which means that
	  \begin{align}
	  	\mathcal{T}a_{\mathbf{p}}\mathcal{T}^{-1} = a_{-\mathbf{p}}
	  	\qquad
	  	\mathcal{T}b_{\mathbf{p}}\mathcal{T}^{-1} = b_{-\mathbf{p}}
	  \end{align} 
	  and the same for $a^\dag, b^\dag$.

	  \item[$(C)$]
	  Contrary to the previous two transformation, this is not related to spacetime coordinates. Its action is defined by the relation $\mathcal{C}\phi \mathcal{C}^{-1} = \phi^\dag$. We have	  
	  \begin{align}
	  	\mathcal{C}\phi(t,\mathbf{x})\mathcal{C}^{-1}
	  	&=
	  	\int\frac{d^3p}{\sqrt{2\omega_p}} 
	  	\mathcal{C}a_\mathbf{p} \mathcal{C}^{-1}
	  	e^{-i(\omega_p t - \mathbf{p}\mathbf{x})} 
	  	+
	  	\mathcal{C}b_\mathbf{p}^\dag \mathcal{C}^{-1}
	  	e^{+ i(\omega_p t - \mathbf{p}\mathbf{x})}  \\
	  	&\overset{!}{=}
	  	\int\frac{d^3p}{\sqrt{2\omega_p}} 
	  	a_\mathbf{p}^\dag
	  	e^{+i(\omega_p t - \mathbf{p}\mathbf{x})} 
	  	+
	  	b_\mathbf{p}
	  	e^{-i(\omega_p t + \mathbf{p}\mathbf{x})}  
	  \end{align}
	  which means that
	  \begin{align}
	  	\mathcal{C}a_{\mathbf{p}}\mathcal{C}^{-1} = b_{\mathbf{p}}
	  	\qquad
	  	\mathcal{C}b_{\mathbf{p}}\mathcal{C}^{-1} = a_{\mathbf{p}}
	  \end{align} 
	  and the same for $a^\dag, b^\dag$.
\end{itemize}

\subsection{$C,P,T$ in the $\kappa$-deformed context}

The $P$ and $T$ transformations can consistently be defined as acting like in the non-deformed case (they leave $[\hat{x}^0,\hat{x}^i]=\frac{i}{\kappa}\hat{x}^i$ invariant). Therefore, they behave as in the non-deformed case analysed before
\begin{align}
	\mathcal{T}\phi(t,\mathbf{x})\mathcal{T}^{-1}
	&=
	\phi(-t,\mathbf{x})
	\quad \implies \quad 
	\mathcal{T}a_{\mathbf{p}}\mathcal{T}^{-1}
	=
	a_{-\mathbf{p}} \\
	\mathcal{P}\phi(t,\mathbf{x})\mathcal{P}^{-1}
	&=
	\phi(t,-\mathbf{x})
	\quad \implies \quad 
	\mathcal{P}a_{\mathbf{p}}\mathcal{P}^{-1}
	=
	a_{-\mathbf{p}}
\end{align}
The only non-trivial transformation is therefore the $C$ transformation. However, because of the presence of the antipode $S(\cdot)$ in the fields in eq. \eqref{fa}, \eqref{fb}, also $C$ can be shown to behave like in the non-deformed case (in its action on $a, a^\dag, b, b^\dag$) when acting on fields.  

Indeed, take the $\phi_+(x)$ component first
\begin{align}
	\mathcal{C} \phi_+(x) \mathcal{C}^{-1}
	=
	\int\frac{d^3p}{\sqrt{2\omega_p}} \zeta(p) \,
	\mathcal{C}a_\mathbf{p} \mathcal{C}^{-1}
	e^{-i(\omega_p t - \mathbf{p}\mathbf{x})}.
\end{align}
On the other hand we have
\begin{align}
	\phi^\dag_-(x)
	=
	\int \frac{d^3p}{\sqrt{2\omega_{\mathbf{p}}}}\, \,\zeta(p) b_{\mathbf{p}}\, e^{-i(\omega_{\mathbf{p}}t-\mathbf{p})\mathbf{x}}
\end{align}
which means that
\begin{align}
	\mathcal{C}a_{\mathbf{p}} \mathcal{C}^{-1}
	=
	b_{\mathbf{p}}.
\end{align}
At the same time, we also have
\begin{align}
	\mathcal{C} \phi_-(x) \mathcal{C}^{-1}
	=
	\int \frac{d^3p}{\sqrt{2\omega_{\mathbf{p}}}}\, \,\zeta(p) \mathcal{C}b^\dag_{\mathbf{p}}\mathcal{C}^{-1}\, e^{-i(S(\omega_{\mathbf{p}})t-S(\mathbf{p})\mathbf{x})}
\end{align}
and
\begin{align}
	\phi^\dag_+(x)
	=
	\int \frac{d^3p}{\sqrt{2\omega_\mathbf{p}}}\, \,\zeta(p)\, a^\dag_{\mathbf{p}}\, e^{-i(S(\omega_\mathbf{p})t-S(\mathbf{p})\mathbf{x})}
\end{align}
which means that
\begin{align}
	\mathcal{C}b^\dag_{\mathbf{p}} \mathcal{C}^{-1}
	=
	a^\dag_{\mathbf{p}}.
\end{align}
Notice that the definition of the $C$ transformation is trivial also in the deformed case, and we stress once again that this is due to the presence of the antipode in the field definition. This is in contrast with other approaches to charge conjugation, which do not give rise to such simple transformation rules.

\section{Off-shell action in momentum space}\label{offshellactionsect}

It is instructive (and we will need it in chapter \ref{Chapter3}) to compute the off-shell action in momentum space (we are in point $vi)$ of our roadmap in section \ref{roadmap2}). This gives a unique insight on the on-shell relation in momentum space, as well as a clear picture of the invariance of the action under both discrete $C,P,T$ transformation, and continuous $\kappa$-Poincar\'e transformations. 

To do so we will need the off-shell field. We cannot use the same construction as before because it includes the Dirac delta $\delta(p_\mu p^\mu - m^2)$. Therefore, we define 
\begin{align}\label{fa-off-simp}
	\phi_{off}(x) &= \int_{\mathcal{I}^+} \frac{d^4p}{p_4/\kappa}\, \,\zeta(p)\,
	\left\{ a_{\mathbf{p}}\, e^{-i(\omega_\mathbf{p}t-\mathbf{p}\mathbf{x})}  
	+
	b^\dag_{\mathbf{p}}\, e^{-i(S(\omega_{\mathbf{p}})t-S(\mathbf{p})\mathbf{x})}
	\right\}
\end{align}
\begin{align}\label{fb-off-simp}
	\phi_{off}^\dag(x) &= \int_{\mathcal{I}^+} \frac{d^4p}{p_4/\kappa}\, \,\zeta(p)\,
	\left\{ a^\dag_{\mathbf{p}}\, e^{-i(S(\omega_\mathbf{p})t-S(\mathbf{p})\mathbf{x})}  
	+
	b_{\mathbf{p}}\, e^{-i(\omega_{\mathbf{p}}t-\mathbf{p}\mathbf{x})}
	\right\}.
\end{align}
Since we have two ordering of the action, namely eq. \eqref{S1} and \eqref{S2}, we will compute them separately.

\subsection{First ordering of the action}\label{firstorderaction}

Using the fact that $(\partial_A \phi)^\dag \equiv \partial_A^\dag \phi^\dag$ and substituting eq. \eqref{fa-off-simp}, \eqref{fb-off-simp} into \eqref{S1} we get
\begin{align}
	S = 
	\int d^4x &
	\int
	\frac{d^4 p}{p_4/\kappa}
	\frac{d^4 q}{q_4/\kappa}
	\Bigg\{
	(p_\mu q^\mu -m^2)
	a_\mathbf{p}^\dag a_\mathbf{q}
	e^{-i(S(\omega_p) \oplus \omega_q)t}
	e^{i (S(\mathbf{p}) \oplus \mathbf{q})\mathbf{x}} \\
	& - 
	(p_\mu S(q)^\mu -m^2)
	a_\mathbf{p}^\dag b_{\mathbf{q}}^\dag 
	e^{-i(S(\omega_p) \oplus S(\omega_q))t}
	e^{i (S(\mathbf{p}) \oplus S(\mathbf{q}))\mathbf{x}} \label{elim1} \\
	& -
	(S(p)_\mu q^\mu - m^2)
	b_{\mathbf{p}} a_\mathbf{q}
	e^{-i(\omega_p \oplus \omega_q)t} 
	e^{i (\mathbf{p} \oplus \mathbf{q})\mathbf{x}} \label{elim2} \\
	& +
	(S(p)_\mu S(q)^\mu -m^2)
	b_{\mathbf{p}} b_{\mathbf{q}}^\dag
	e^{-i(\omega_p \oplus S(\omega_q))t} 
	e^{-i(\mathbf{p} \oplus S(\mathbf{q}))\mathbf{x}}
	\Bigg\}\zeta(p)\zeta(q)
\end{align}
For the moment we will ignore the terms \eqref{elim1} and \eqref{elim2}. Their treatment is quite non-trivial, and we will defer their treatment to future studies.
Therefore, integrating in $d^4x$,
\begin{align}
	S 
	=&
	\int
	\frac{d^4 p}{p_4/\kappa}
	\frac{d^4 q}{q_4/\kappa}
	\Bigg\{
	(p_\mu q^\mu -m^2)
		a_\mathbf{p}^\dag a_\mathbf{q}
		\delta(S(\omega_p) \oplus \omega_q)
		\delta(S(\mathbf{p}) \oplus \mathbf{q}) \\
	&
	+
	(S(p)_\mu S(q)^\mu -m^2)
		b_{\mathbf{p}} b_{\mathbf{q}}^\dag
		\delta(\omega_p \oplus S(\omega_q)) 
		\delta(\mathbf{p} \oplus S(\mathbf{q}))
	\Bigg\}\zeta(p)\zeta(q) \\
	=&
	\int
	\frac{d^4 p}{p_4/\kappa}
	\frac{d^4 q}{q_4/\kappa}
	\Bigg\{
	(p_\mu q^\mu -m^2)
		a_\mathbf{p}^\dag a_\mathbf{q}
		\delta(S(p)\oplus q) \\
	&
	+
	(S(p)_\mu S(q)^\mu -m^2)
		b_{\mathbf{p}} b_{\mathbf{q}}^\dag
		\delta(p\oplus S(q))
	\Bigg\}\zeta(p)\zeta(q)
\end{align}
where we used the notation
\begin{align}
	\delta(S(\omega_p) \oplus \omega_q)
	\delta(S(\mathbf{p}) \oplus \mathbf{q})
	&=
	\delta(S(p)\oplus q) \\
	\qquad
	\delta(\omega_p \oplus S(\omega_q)) 
	\delta(\mathbf{p} \oplus S(\mathbf{q}))
	&=
	\delta(p\oplus S(q)).
\end{align}
We now use a known result in the literature \cite{Arzano:2018gii} which states that 
\begin{align}
	\delta(p\oplus S(q)) = \frac{|p_+|^3}{\kappa^3}\delta(S(p)\oplus q)
	\qquad
	\delta(S(p)\oplus q) = \frac{p_4}{\kappa} \delta(p-q)
\end{align}
The additional $\frac{p_4}{\kappa}$ factor eliminates one of the factors in one of the two metrics, so that the final expression for the firs off-shell action in momentum space is 
\begin{align}\label{S1-momentumspace}
	S
	=
	\int
	\frac{d^4 p}{p_4/\kappa}
	\zeta^2(p)
	\left[
	(p_\mu p^\mu -m^2)
	a_{\mathbf{p}}^\dag a_{\mathbf{p}}
	+
	(S(p)_\mu S(p)^\mu -m^2)
	b_{\mathbf{p}} b_{\mathbf{p}}^\dag
	\frac{|p_+|^3}{\kappa^3}
	\right]
\end{align}
where the red term is the factor coming from the delta.

\subsection{Second ordering of the action}\label{secondorderaction}
Using the switched convention, i.e. substituting eq. \eqref{fa-off-simp} and \eqref{fb-off-simp} into eq. \eqref{S2}, we obtain
\begin{align}
	S = 
	\int d^4 x&
	\int
	\frac{d^4 p}{p_4/\kappa}
	\frac{d^4 q}{q_4/\kappa}
	\Bigg\{
	(p_\mu q^\mu -m^2)
	a_\mathbf{q}a_\mathbf{p}^\dag
	e^{-i(\omega_qt-\mathbf{q}\mathbf{x})} 
	\star 
	e^{-i(S(\omega_p)t-S(\mathbf{p})\mathbf{x})}\\
	& +
	(S(p)_\mu S(q)^\mu -m^2)
	b_{\mathbf{q}}^\dag b_{\mathbf{p}}  
	e^{i(S(\omega_q)t-S(\mathbf{q})\mathbf{x})}
	\star 
	e^{i(\omega_pt-\mathbf{p}\mathbf{x})} 
	\Bigg\}\zeta(p)\zeta(q) \\
	=
	\int d^4x&
	\int
	\frac{d^4 p}{p_4/\kappa}
	\frac{d^4 q}{q_4/\kappa}
	\Bigg\{
	(p_\mu q^\mu -m^2)
	a_\mathbf{q}a_\mathbf{p}^\dag 
	e^{-i(\omega_q \oplus S(\omega_p))t}
	e^{i (\mathbf{q} \oplus S(\mathbf{p}))\mathbf{x}} \\
	& +
	(S(p)_\mu S(q)^\mu -m^2)
	b_{\mathbf{q}}^\dag b_{\mathbf{p}} 
	e^{-i(S(\omega_q) \oplus \omega_p)t} 
	e^{-i(S(\mathbf{q})\oplus \mathbf{p})\mathbf{x}}
	\Bigg\}\zeta(p)\zeta(q)
\end{align}
The two missing terms have been eliminated for the same reason as before, and we will study them in detail in forthcoming publications.

These are exactly the same terms as in the previous calculations with only two exceptions:
\begin{itemize}
	\item[1)] In the previous action we had the products $a_\mathbf{p}^\dag a_\mathbf{q}$ and $b_{\mathbf{p}} b_{\mathbf{q}}^\dag$, while with this switched convention we have the other order, i.e. $a_\mathbf{q}a_\mathbf{p}^\dag$ and $b_{\mathbf{q}}^\dag b_{\mathbf{p}}$;
	\item[2)] The Dirac deltas are switched with respect to the computations with the previous ordering. More explicitly, the exponential multiplying the term $a_\mathbf{q}a_\mathbf{p}^\dag$ here is the same that was multiplying the term $b_{\mathbf{p}} b_{\mathbf{q}}^\dag$, and vice versa.
\end{itemize}
Therefore, all the computations proceed in the same way but the extra coefficient is now found with the $a$, $a^\dag$ operators, i.e. we have
\begin{align}\label{S2-momentumspace}
	S
	=
	\int
	\frac{d^4 p}{p_4/\kappa}
	\zeta^2(p)
	\left[
	\frac{|p_+|^3}{\kappa^3}
	(p_\mu p^\mu -m^2)
	a_{\mathbf{p}}^\dag a_{\mathbf{p}}
	+
	(S(p)_\mu S(p)^\mu -m^2)
	b_{\mathbf{p}} b_{\mathbf{p}}^\dag
	\right]
\end{align}
Notice that the on-shell conditions are not swapped, because their formulation only depends on the property $(\partial_A \phi)^\dag = \partial_A^\dag \phi^\dag$, and not on the ordering of the star products.

\subsection{Sum of the two actions}\label{actionsum}

Using the definition of the total action given by eq. \eqref{Stot}, and substituting eq. \eqref{S1-momentumspace} and \eqref{S2-momentumspace}, we obtain (recall that we are still in the classical case, but upon quantization one should remember that normal-ordering should be applied)
\begin{align}\label{Stot-momentumspace}
	S =
	\int
	\frac{d^4 p}{p_4/\kappa}
	\left[1+\frac{|p_+|^3}{\kappa^3}\right]
	\zeta^2(p)
	\left\{
	(p_\mu p^\mu -m^2)
	a_{\mathbf{p}}^\dag a_{\mathbf{p}}
	+
	(S(p)_\mu S(p)^\mu -m^2)
	b_{\mathbf{p}}^\dag b_{\mathbf{p}}
	\right\}.
\end{align}
If we decided to go on-shell, the computations would proceed as usual, but with an additional Dirac delta imposing the shell condition, which means that the action would be exactly zero.

\section{Symplectic form for the two actions}\label{sympformsect}

We are now at point $vii)a)$ of our roadmap in section \ref{roadmap2}. 

\subsection{Example in the  non-deformed case}

Let us consider a complex scalar field with the Lagrangian
\begin{equation}\label{Symp1}
	L = - \left(\partial^\mu \phi^\dag\partial_\mu\phi + m^2 \phi^\dag\phi\right)
\end{equation}
The variation of this Lagrangian is
\begin{equation}\label{Symp2}
	\delta L = - \left(\partial^\mu \delta\phi^\dag\partial_\mu\phi +\partial^\mu \phi^\dag\partial_\mu\delta\phi + m^2 \delta\phi^\dag\phi+ m^2\phi^\dag\delta\phi\right)
\end{equation}
This variation contains terms proportional to the field equations and the total derivative term
\begin{equation}\label{Symp3}
	\delta L = {\rm EoM}  - \partial^\mu\left( \delta\phi^\dag\partial_\mu\phi\right) -\partial_\mu\left(\partial^\mu \phi^\dag\delta\phi \right)
\end{equation}
Therefore the variation of the action is
\begin{equation}\label{Symp4}
	\delta S = \int_M{\rm EoM}  -\int_{\partial M} d\sigma_\mu\left( \delta\phi^\dag\partial^\mu\phi+\partial^\mu \phi^\dag\delta\phi \right)
\end{equation}
The second, boundary term in this expression is the so-called presymplectic structure or Liouville form $\theta$, if we choose the boundary $\partial M$ to be a Cauchy surface, for example the surface of constant time $t=0$. Then we can find the symplectic form
\begin{equation}\label{Symp5}
	\omega = \delta \theta =\frac12 \int_{t=0} d^3x \, \left( \delta\phi^\dag\wedge \frac{\partial}{\partial t}\delta\phi-\frac{\partial}{\partial t}\delta \phi^\dag\wedge\delta\phi \right)_{t=0}
\end{equation}
Now we take the field decomposition
\begin{equation}\label{Symp6}
	\phi(t,\mathbf{x}) = \int \frac{d^3p}{\sqrt{2E}}\, a_\mathbf{p}\, e^{-i(Et - {\mathbf{p}}{\mathbf{x}})}+ b^\dag_\mathbf{p}\, e^{i(Et -{\mathbf{p}}{\mathbf{x}})}
\end{equation}
so that
\begin{equation}\label{Symp7}
	\delta \phi(t,\mathbf{x}) = \int \frac{d^3p}{\sqrt{2E}}\,\delta a_\mathbf{p}\, e^{-i(Et - {\mathbf{p}}{\mathbf{x}})}+ \delta b^\dag_\mathbf{p}\, e^{i(Et -{\mathbf{p}}{\mathbf{x}})}
\end{equation}
and plug it to \eqref{Symp5} to obtain the symplectic form, whose inverse defines for us the Poisson brackets, and (upon quantization) the commutators.
\begin{align}
	\omega &= \frac i2\int d^3x\int  d^3p\,d^3q \left(\delta a^\dag_\mathbf{p}\, e^{-i {\mathbf{p}}{\mathbf{x}}}+ \delta b_\mathbf{p}\, e^{i{\mathbf{p}}{\mathbf{x}}}\right)\wedge\left(-\delta a_\mathbf{q}\, e^{i {\mathbf{q}}{\mathbf{x}}}+ \delta b^\dag_\mathbf{q}\, e^{-i{\mathbf{q}}{\mathbf{x}}}\right)\nonumber\\
	&-\frac i2\int d^3x\int d^3q\,d^3p \left(\delta a^\dag_\mathbf{p}\, e^{-i {\mathbf{p}}{\mathbf{x}}}- \delta b_\mathbf{p}\, e^{i{\mathbf{p}}{\mathbf{x}}}\right)\wedge\left(\delta a_\mathbf{q}\, e^{i {\mathbf{q}}{\mathbf{x}}}+ \delta b^\dag_\mathbf{q}\, e^{-i{\mathbf{q}}{\mathbf{x}}}\right)\nonumber
\end{align}
Defining
\begin{equation}\label{Symp8a}
	\Omega = \int {d^3p}d^3q \,\, \omega(p,q)
\end{equation}
we find
\begin{equation}\label{Symp8}
	\Omega(p,q) \sim -\delta a^\dag_\mathbf{p}\wedge\delta a_\mathbf{q} - \delta b^\dag_\mathbf{p}\wedge\delta b_\mathbf{q}
\end{equation}
from which the commutational relations
\begin{equation}\label{Symp9}
	[ a_\mathbf{p},a^\dag_\mathbf{q}]\sim [ b_\mathbf{p}, b^\dag_\mathbf{q}]\sim\delta(p-q)
\end{equation}

\subsection{The $\kappa$-deformed case}

The extension of this reasoning to the case of fields in the context of $\kappa$-Poincar\'e algebra is straightforward. In particular we have already computed the surface term in previous sections, and we are only interested in those terms whose boundary is spacelike, i.e. we only need to consider the following presymplectic forms.
\begin{align}\label{Symp10}
	& \qquad \theta_1 = -\int_{\mathbb{R}^3} d^3x
	\left(
	\Pi^0_1 \star \delta\phi
	+
	\delta\phi^\dag \star \left(\Pi^0_1\right)^\dag
	\right) \nonumber \\
	=
	-\int_{\mathbb{R}^3} d^3x&
	\left\{
	\frac{1}{\kappa} \left(\Delta_+ \partial_0^\dag +i m^2\right)	\phi^\dag \star \delta\phi
	+
	\delta \phi^\dagger \star
	\left[
	\frac{1}{\kappa} \left(\Delta_+ \partial_0^\dag +i m^2\right)	\phi^\dag
	\right]^\dag
	\right\}.
\end{align}
\begin{align}\label{Symp11}
	& \qquad \qquad \theta_2 = -\int_{\mathbb{R}^3} d^3x
	\left(
	\delta\phi \star \Pi^0_2 
	+
	\left(\Pi^0_2\right)^\dag \star \delta\phi^\dag
	\right) \nonumber \\
	=
	-\int_{\mathbb{R}^3} d^3x&
	\left\{
	\delta \phi \star
	\left(\frac{\kappa}{\Delta_+} \partial_0^\dag
	+\frac{i}{\kappa}(\partial_0^\dag)^2
	\right)\phi^\dag 
	+
	\left[
	\left(
	\frac{\kappa}{\Delta_+} \partial_0^\dag	
	+\frac{i}{\kappa}(\partial_0^\dag)^2
	\right)\phi^\dag
	\right]^\dag
	\star \delta \phi^\dagger
	\right\}.
\end{align}
Notice that the sign in the RHS of these two presymplectic forms is such that we get the correct symplectic form in the canonical limit $\kappa \rightarrow \infty$ (and therefore the correct canonical commutator between $a$ and $a^\dag$,a nd same for $b$ and $b^\dag$). Furthermore, we are not writing down explicitly the wedge product for the moment because at the moment it is not present, it will be when we will consider $\delta \theta_1$ and $\delta \theta_2$. To simplify the computations, notice that 
\begin{align}
	\Pi_1^0
	&=
	\left(
	\frac{\Delta_+}{\kappa} \partial_0^\dag
	+
	\frac{im^2}{\kappa}
	\right)\phi^\dag \\
	&=	
	\left(
	\Delta_+
	(\partial_0 -i\Delta_+^{-1}\partial^2)
	+
	im^2
	\right)
	\frac{\phi^\dag}{\kappa} \\
	&=
	\left(
	\frac{\Delta_+}{\kappa} \partial_0
	+ \frac{i}{\kappa}(m^2-\partial^2)
	\right)\phi^\dagger \\
	&=
	\left(
	\frac{\Delta_+}{\kappa} \partial_0
	- \frac{i\partial_0^2}{\kappa}
	\right)\phi^\dagger  \label{SPi2}\\
	&=
	\left(
	\frac{i\partial_0 + \kappa +i\partial_4}{\kappa} \partial_0
	- \frac{i\partial_0^2}{\kappa}
	\right)\phi^\dagger \\
	&=
	\left[
	\frac{\kappa + i\partial_4}{\kappa} \partial_0 
	\right] \phi^\dag.
\end{align}
and in the same way one gets
\begin{align}
	\left[
	\frac{\kappa + i\partial_4}{\kappa} \partial_0 
	\right]
	=
	\left(
	\frac{\Delta_+}{\kappa} \partial_0^\dag
	+
	\frac{im^2}{\kappa}
	\right)
	=
	\left(
	\frac{\kappa}{\Delta_+} \partial_0^\dag
	+ \frac{i(\partial_0^\dag)^2}{\kappa}
	\right)^\dag
\end{align}
and therefore
\begin{align}
	\left(
	\frac{\kappa}{\Delta_+} \partial_0^\dag
	+ \frac{i(\partial_0^\dag)^2}{\kappa}
	\right)
	=
	\left[
	\frac{\kappa + i\partial_4}{\kappa} \partial_0 
	\right]^\dag. 
\end{align} 
Notice that the quantity $\partial_4$ acting on any plane wave brings down $i\kappa - ip_4$, and since $p_4=S(p_4)$, then $\kappa + i\partial_4 \rightarrow \kappa + i(i\kappa - ip_4) = p_4$ (recall that $S(p_4)=p_4$). The presymplectic forms can therefore be simplified to 
\begin{align}\label{pre-Symp10}
	\theta_1 = -\int_{\mathbb{R}^3} d^3x
	\frac{p_4}{\kappa}
	\left(
	\partial_0\phi^\dag \star \delta\phi
	+
	\delta\phi^\dag \star \partial_0^\dag \phi 
	\right)
\end{align}
\begin{align}\label{pre-Symp11}
	\theta_2 = -\int_{\mathbb{R}^3} d^3x
	\frac{p_4}{\kappa}
	\left(
	\delta\phi \star \partial_0^\dag\phi^\dag
	+
	\partial_0\phi \star \delta\phi^\dag
	\right).
\end{align}
Taking the delta, we get
\begin{align}\label{Symp10-simp}
	\delta\theta_1 = -\int_{\mathbb{R}^3} d^3x
	\frac{p_4}{\kappa}
	\left(
	\delta\partial_0\phi^\dag \overset{\wedge}{\star} \delta\phi
	-
	\delta\phi^\dag \overset{\wedge}{\star} \delta\partial_0^\dag \phi 
	\right)
\end{align}
\begin{align}\label{Symp11-simp}
	\delta\theta_2 = -\int_{\mathbb{R}^3} d^3x
	\frac{p_4}{\kappa}
	\left(
	-\delta\phi \overset{\wedge}{\star} \partial_0^\dag \delta\phi^\dag
	+
	\partial_0\delta \phi \overset{\wedge}{\star} \delta\phi^\dag
	\right).
\end{align}
The notation $\overset{\wedge}{\star}$ means that we are still taking the $\star$ product among plane waves, and also the wedge product between $\delta a, \delta a^\dag, \delta b, \delta b^\dag.$ We can now explicitly compute both symplectic forms using the fields in eq. \eqref{fa}, \eqref{fb}. Starting from the first, we have
\begin{align}
	\delta\theta_1 
	&=
	-\int_{\mathbb{R}^3}
	\frac{p_4}{\kappa}
	\left\{
	\delta \partial_0
	\phi^\dag \overset{\wedge}{\star} \delta\phi
	-
	\delta\phi^\dagger \overset{\wedge}{\star}
	\delta\partial_0^\dag
	\phi
	\right\}  \\ \nonumber \\
	&=
	-i\int_{\mathbb{R}^3}
	\int \frac{d^3p}{\sqrt{2\omega_\mathbf{p}}}
	\frac{d^3q}{\sqrt{2\omega_\mathbf{q}}}
	\,\zeta(q)\,\zeta(p)\
	\frac{p_4}{\kappa} \times \nonumber \\
	&\Big\{
	(-S(\omega_\mathbf{q})-S(\omega_\mathbf{p}))
	a^\dag_\mathbf{q} \wedge  a_\mathbf{p}
	e^{-i(S(\omega_\mathbf{q})\oplus \omega_\mathbf{p})t}
	e^{i(S(\mathbf{q})\oplus\mathbf{p})\mathbf{x}} \\
	&
	+
	(-\omega_\mathbf{q}-\omega_\mathbf{p})
	b_\mathbf{q} \wedge b^\dag_\mathbf{p}
	e^{-i(\omega_\mathbf{q}\oplus S(\omega_\mathbf{p}))t}
	e^{i(\mathbf{q}\oplus S(\mathbf{p}))\mathbf{x}} \\
	&-
	(S(\omega_\mathbf{q})+\omega_\mathbf{p})
	a^\dag_\mathbf{q} \wedge  b^\dag_\mathbf{p}
	e^{-i(S(\omega_\mathbf{q})\oplus S(\omega_\mathbf{p}))t}
	e^{i(S(\mathbf{q})\oplus S(\mathbf{p}))\mathbf{x}} \\
	&-
	(\omega_\mathbf{q} + S(\omega_\mathbf{p}))
	a_\mathbf{q} \wedge b_\mathbf{p}
	e^{-i(\omega_\mathbf{q}\oplus \omega_\mathbf{p})t}
	e^{i(\mathbf{q}\oplus\mathbf{p})\mathbf{x}}
	\Big\}
\end{align}
We now use 
\begin{align}\label{DD1}
	\int_{\mathbb{R}^3}
	e^{-i(S(\omega_q)t-S(\mathbf{q})\mathbf{x})}
	\star
	e^{-i(\omega_pt-\mathbf{p}\mathbf{x})}
	=
	\delta
	\left(
	\mathbf{p}
	-
	\mathbf{q}
	\right)
\end{align} 
\begin{align}\label{DD2}
	\int_{\mathbb{R}^3}
	e^{-i(\omega_pt-\mathbf{p}\mathbf{x})}
	\star 
	e^{-i(S(\omega_q)t-S(\mathbf{q})\mathbf{x})}
	=
	\frac{|q_+|^3}{\kappa^3}
	\delta
	\left(\mathbf{p} - \mathbf{q}\right)
\end{align} 
so that the first two terms give 
\begin{align}\label{Symp12}
	\boxed{\delta\theta_1
		=
		i\int \frac{d^3p}{2\omega_\mathbf{p}}
		\zeta^2(p)
		\frac{p_4}{\kappa}
		\left(
		2S(\omega_\mathbf{p}) 
		a_\mathbf{p}^\dag \wedge a_\mathbf{p}
		+
		\frac{p_+^3}{\kappa^3}
		2\omega_\mathbf{p}
		b_\mathbf{p} \wedge b_\mathbf{p}^\dag
		\right)}
\end{align}
and the other two go away because the Dirac delta impose on-shell that $\omega_\mathbf{q} = -S(\omega_\mathbf{p})$. Indeed, one can check for example that
\begin{align}
	\int_{\mathbb{R}^3}
	e^{-i\omega_p t + i\mathbf{p} \mathbf{x}}
	\star 
	e^{-i\omega_q t + i\mathbf{q} \mathbf{x}} 
	=
	e^{-i(\omega_p \oplus \omega_q)t}
	\delta(\mathbf{p} \oplus \mathbf{q}) 
	=	
	e^{-i(\omega_p \oplus \omega_q)t}
	\delta\left( \frac{q_+}{\kappa} \mathbf{p} + \mathbf{q} \right)
\end{align}
If we first integrate in $d^3p$, the quantity $|q_+|^3/\kappa^3$ can be brought outside the delta obtaining $\frac{\kappa^3}{|q_+|^3} \delta(\mathbf{p} + \mathbf{q} \frac{\kappa}{q_+})$. Therefore we have
\begin{align}
	e^{-i(\omega_p \oplus \omega_q)t}
	\delta\left( \frac{q_+}{\kappa} \mathbf{p} + \mathbf{q} \right)
	=
	\frac{\kappa^3}{|q_+|^3}
	e^{-i(\omega_p \oplus \omega_q)t}
	\delta\left( \mathbf{p} + \frac{\kappa}{q_+}\mathbf{q} \right)
\end{align}
The Dirac deltas imply also a modification of the energy $\omega_p$. In fact, we have
\begin{align}\label{1.77}
	\omega_q^2 - \mathbf{q}^2 = m^2 = \omega_p^2 -\mathbf{p}^2 = \omega_p^2 - \frac{\kappa^2}{q_+^2} \mathbf{q}^2
\end{align}
and therefore
\begin{align}\label{1.78}
	\omega_p^2 = m^2 + \frac{\kappa^2}{q_+^2}\mathbf{q}^2
	=
	S(\omega_q)^2.
\end{align}
In other words, the Dirac delta imposes
\begin{align}\label{firstoffdiagdel}
	\omega_p = -S(\omega_q).
\end{align}
Notice that the LHS contains a minus sign because on-shell we have $\omega_p>0$ and $S(\omega_q)<0$, so taking the square root we need to take this into account. The other exponential is treated in the same way, since one can change variables $p\mapsto S(p)$, obtaining the same result, so that the time-dependent parts go away.

For the second one, we proceed in a similar manner, and we have
\begin{align}
	\delta\theta_2 
	&=
	\int_{\mathbb{R}^3}
	\frac{p_4}{\kappa}
	\left\{
	\delta \phi \overset{\wedge}{\star} 
	\partial_0^\dag
	\delta\phi^\dag
	-
	\delta\partial_0 
	\phi
	\overset{\wedge}{\star} \delta\phi^\dag
	\right\} \\ \nonumber \\
	&=
	i\int_{\mathbb{R}^3}
	\int \frac{d^3p}{\sqrt{2\omega_\mathbf{p}}}
	\frac{d^3q}{\sqrt{2\omega_\mathbf{q}}}
	\,\zeta(q)\,\zeta(p)\
	\frac{p_4}{\kappa} \times \nonumber \\
	&\Big\{
	(\omega_\mathbf{q}+\omega_\mathbf{p})
	\delta a_\mathbf{p} \wedge \delta a^\dag_\mathbf{q} 
	e^{-i( \omega_\mathbf{p} \oplus S(\omega_\mathbf{q}))t}
	e^{i(\mathbf{p} \oplus S(\mathbf{q}))\mathbf{x}} \\
	&+
	(S(\omega_\mathbf{p}) + S(\omega_\mathbf{q}))
	\delta b^\dag_\mathbf{p} \delta \delta b_\mathbf{q} 
	e^{-i(S(\omega_\mathbf{p}) \oplus \omega_\mathbf{q})t}
	e^{i(S(\mathbf{p}) \oplus \mathbf{q})\mathbf{x}} \\
	&+
	(S(\omega_\mathbf{p})+\omega_\mathbf{q})
	\delta b^\dag_\mathbf{p} \delta \delta a^\dag_\mathbf{q} 
	e^{-i(S(\omega_\mathbf{p}) \oplus S(\omega_\mathbf{q}))t}
	e^{i(S(\mathbf{p}) \oplus S(\mathbf{q}))\mathbf{x}} \\
	&+
	(\omega_\mathbf{p} + S(\omega_\mathbf{q}))
	\delta b_\mathbf{p} \delta \delta a_\mathbf{q} 
	e^{-i(\omega_\mathbf{p}\oplus \omega_\mathbf{q})t}
	e^{i(\mathbf{p}\oplus\mathbf{q})\mathbf{x}}
	\Big\}
\end{align}
so that, proceeding in the same way as before, we have
\begin{align}\label{Symp13}
	\boxed{\delta\theta_2
		=
		\int \frac{d^3p}{2\omega_\mathbf{p}}
		\zeta^2(p)
		\frac{p_4}{\kappa}
		\left(
		\frac{p_+^3}{\kappa^3}
		2\omega_\mathbf{p}
		\delta a_\mathbf{p} \wedge \delta a_\mathbf{p}^\dag
		+
		2S(\omega_\mathbf{p})
		\delta b_\mathbf{p}^\dag \wedge \delta b_\mathbf{p}
		\right)}
\end{align}
Because of the definition of the action, the total symplectic form will be
\begin{align}
	\delta\theta_{T} = \frac{\delta\theta_1+\delta\theta_2}{2}
\end{align}
and therefore we have
\begin{align}
	\boxed{\delta\theta_T
	=
	i\int \frac{d^3p}{2\omega_p}
	\zeta(p)^2
	\left[
	\frac{|p_+|^3}{\kappa^3}
	\omega_p - S(\omega_p)
	\right]	\frac{p_4}{\kappa}
	\left\{
	a_\mathbf{p}\wedge
	a_\mathbf{p}^\dag\,
	-
	b_{\mathbf{p}}^\dag\wedge
	b_{\mathbf{p}}\,
	\right\}}.
\end{align}
This result can be simplified using the definition
\begin{align}\label{alpha}
	\boxed{\alpha(p)
		=
		\frac{\zeta(p)^2}{2\omega_p}
		\left[
		\frac{|p_+|^3}{\kappa^3}\omega_p - S(\omega_p)
		\right]\frac{p_4}{\kappa}}
\end{align}
and we get
\begin{align}\label{Symp14}
	\boxed{\delta\theta_T
		=
		i\int d^3p \,\,
		\alpha(p)
		\left\{
		a_\mathbf{p}\wedge
		a_\mathbf{p}^\dag\,
		-
		b_{\mathbf{p}}^\dag\wedge
		b_{\mathbf{p}}\,
		\right\}}.
\end{align}
This result would imply the Poisson brackets
\begin{align}
	\left\{
	a_\mathbf{p}, a_\mathbf{q}^\dag
	\right\}
	&= - i 
	\frac{1}{\xi(p)^2} \frac{\kappa}{p_4} \frac{2\omega_p}{\frac{|p_+|^3}{\kappa^3}\omega_p - S(\omega_p)} 
	\delta(\mathbf{p}-\mathbf{q}) \label{Symp15}\\
	\left\{
	b_{\mathbf{p}}, b_{\mathbf{q}}^\dag
	\right\}
	&= - i 
	\frac{1}{\xi(p)^2}\frac{\kappa}{p_4} \frac{2\omega_p}{\frac{|p_+|^3}{\kappa^3}\omega_p - S(\omega_p)} 
	\delta(\mathbf{p}-\mathbf{q}) . \label{Symp16}
\end{align}
which can also be simplified using eq. \eqref{alpha} as follows
\begin{align}
	\left\{
	a_\mathbf{p}, a_\mathbf{q}^\dag
	\right\}
	&= -\frac{i}{\alpha(p)}
	\delta(\mathbf{p}-\mathbf{q}) \label{Symp15-simp}\\
	\left\{
	b_{\mathbf{p}}, b_{\mathbf{q}}^\dag
	\right\}
	&= -\frac{i}{\alpha(p)} 
	\delta(\mathbf{p}-\mathbf{q}) . \label{Symp16-simp}
\end{align}
Notice that, if one uses the canonical quantization map $[A,B] = i\hbar \{A,B\}$ and assume $\hbar=1$, then we would get the commutators
\begin{align}
	\left[
	a_\mathbf{p}, a_\mathbf{q}^\dag
	\right]
	&= 
	\frac{1}{\alpha(p)}
	\delta(\mathbf{p}-\mathbf{q}) \label{Symp15-t}\\
	\left[
	b_{\mathbf{p}}, b_{\mathbf{q}}^\dag
	\right]
	&= 
	\frac{1}{\alpha(p)}
	\delta(\mathbf{p}-\mathbf{q}) . \label{Symp16-t}
\end{align}
where the LHS is positive and the coefficient goes to 1 for $\kappa\rightarrow \infty$, as it should. We will comment on how to technically get these commutations relations in section \ref{dercreannalg} below.

\section{Conserved charges: geometric approach}\label{chargesfromsympform}

The starting point is the symplectic form in eq. \eqref{Symp14}. For simplicity, we will call $\Omega = \delta\theta_T = \frac{\delta\theta_1 + \delta\theta_2}{2}$. In order to get the charges from the symplectic form we need to do the following.

Assuming that the charges come from a symmetry described by some continuous vector field $\xi$ in spacetime, then we compute \cite{Harlow:2019yfa} (this approach is also used quite often in GR, for example see \cite{Wald:1993nt}, \cite{Iyer:1994ys})
\begin{align}\label{chargesdef}
	-\delta_\xi \lrcorner \, \Omega  \overset{!}{=} \delta Q_\xi
\end{align}
where $\delta$ is the exterior derivative in phase space (i.e. the set of solutions of the EoM) and $Q_\xi$ is the charge associated to the vector $\xi$. The object $\delta_\xi$ is a vector field in phase space generated by the killing vector field $\xi$. In other words, $\delta_\xi A$ measures the infinitesimal variation of the object $A$ in phase space due to the symmetry due to the action along $\xi$ in spacetime. The symbol $\lrcorner$ is used to indicete contraction of the vector field with forms. Notice that each charge obtained in this way is a symmetry of the symplectic form. Indeed one can easily show that 
\begin{align}
	\pounds_{\delta_\xi} \Omega
	=
	\delta_{\xi} \lrcorner \, \delta\Omega
	+
	\delta(\delta_\xi \lrcorner \, \Omega)
	=
	0
	+
	\delta^2 Q_\xi
	=
	0
\end{align}
where we used the fact that $\delta^2 = 0$ since $\delta$ is the exterior derivative in phase space, and also $\delta\Omega=0$ because $\Omega$ is an exact 2-form. The charges built in this way are independent of the dynamics because the symplectic form $\Omega$ and the vector field $\delta_\xi$ are both time independent and non-dynamical by construction, and they are defined on-shell.

To see how the procedure of getting the charges using this formalism, we first tackle the problem in the canonical non-deformed QFT context.

\subsection{Non-deformed case}

In this case, the symplectic form is given by the $\kappa \rightarrow \infty$ limit of eq. \eqref{Symp14} above, and it reads
\begin{align}\label{Symp14-undefboost}
	\Omega^{\kappa \rightarrow \infty}
	&=
	i\int d^3p \,
	(
	\delta a_\mathbf{p}\wedge
	\delta a_\mathbf{p}^\dag\,
	-
	\delta b_{\mathbf{p}}^\dag\wedge
	\delta b_{\mathbf{p}}).
\end{align}

\subsubsection{Translation charges}

In this case, the translation can be described by the vectors $\partial_0$ or $\partial_i$, depending on whether we are talking about time translation or spatial translations. 

Starting with the time translations, we first need to understand what $\delta_{\partial_0}a_\mathbf{p}$ is (it will be analogous for $b$). We know that after a translation we have
\begin{align}\label{atransf}
	a_\mathbf{p} \mapsto e^{i\epsilon \omega_p}a_\mathbf{p}
	=
	a_\mathbf{p}
	+
	i\epsilon \omega_p a_\mathbf{p}
\end{align}
and therefore
\begin{align}
	\delta_{\partial_0} a_\mathbf{p} = i\epsilon \omega_p a_\mathbf{p}
	\quad
	\Leftrightarrow
	\quad
	\delta_{\partial_0} a_\mathbf{p}^\dag
	=
	-
	i\epsilon \omega_p a_\mathbf{p}^\dag
\end{align}
where the second expression can be obtained from the first one by simply taking the hermitian conjugate. Therefore, we have (the wedge product goes away after the contraction because the creation/annihilation operators are not themselves forms)
\begin{align}
	-\delta_{\partial_0} \lrcorner \, \Omega^{\kappa \rightarrow \infty}
	&=
	-i\int d^3p \,
	(
	\delta_{\partial_0} a_\mathbf{p}
	\delta a_\mathbf{p}^\dag\,
	-
	\delta a_\mathbf{p}
	\delta_{\partial_0} a_\mathbf{p}^\dag
	-
	\delta_{\partial_0} b_{\mathbf{p}}^\dag
	\delta b_{\mathbf{p}}
	+
	\delta b_{\mathbf{p}}^\dag
	\delta_{\partial_0} b_{\mathbf{p}}) \\
	&=
	-i\int d^3p \,
	[
	i\epsilon \omega_p a_\mathbf{p}
	\delta a_\mathbf{p}^\dag\,
	-
	\delta a_\mathbf{p}
	(-i\epsilon \omega_p a_\mathbf{p}^\dag) \nonumber \\
	&-
	(-i\epsilon \omega_p b_{\mathbf{p}}^\dag)
	\delta b_{\mathbf{p}}
	+
	\delta b_{\mathbf{p}}^\dag
	i\epsilon \omega_p b_{\mathbf{p}}] \\
	&=
	\epsilon 
	\int d^3p \,
	[
	\omega_p a_\mathbf{p}
	\delta a_\mathbf{p}^\dag\,
	+
	\delta a_\mathbf{p}
	\omega_p a_\mathbf{p}^\dag
	+
	\omega_p b_{\mathbf{p}}^\dag
	\delta b_{\mathbf{p}}
	+
	\delta b_{\mathbf{p}}^\dag
	\omega_p b_{\mathbf{p}}] \\
	&=
	\epsilon 
	\delta
	\int d^3p \,
	\omega_p 
	(
	a_\mathbf{p}
	a_\mathbf{p}^\dag\,
	+
	b_{\mathbf{p}}^\dag
	b_{\mathbf{p}}
	)
\end{align}
and therefore we get the canonical charge apart from the irrelevant prefactor $\epsilon$. 
\begin{align}
	\mathcal{P}^{\kappa\rightarrow\infty}_0
		=
		\int d^3p \,\,
		\omega_p
		\left\{
		a_\mathbf{p}^\dag\,
		a_\mathbf{p}\,
		+
		b_{\mathbf{p}}\,
		b_{\mathbf{p}}^\dag\,
		\right\}
\end{align}
We can also check that eq. \eqref{atransf} gives the correct field transformation.
\begin{align}
	\delta_{\partial_\mu} \phi
	&=
	\int \frac{d^3p}{\sqrt{2\omega_\mathbf{p}}}\, 
	\left[
	i \epsilon^\mu p_\mu\, a_{\mathbf{p}}\, e^{-i(\omega_\mathbf{p}t-\mathbf{p}\mathbf{x})}  -   i \epsilon^\mu p_\mu b^\dag_{\mathbf{p}}\, e^{i(\omega_\mathbf{p}t-\mathbf{p}\mathbf{x})}
	\right]
	\nonumber \\
	&=
	\epsilon^\mu\partial_\mu\, \phi(x).
\end{align}
Of course, once we quantize these charges, the operators will need to be put in normal order. The same exact reasoning will work for the spatial translation charges, it is sufficient to substitute $\omega_p$ with $\mathbf{p}_i$ everywhere.

\subsubsection{Boost and rotation charges}

In this case, the procedure is again the same, but this time we need to use as a vector field $\xi$ the generator of boost transformation, and then we need to understand how creation/annihilation operator change under this transformation, i.e. we need to understand what is the object $\delta_\Lambda a_\mathbf{p}$.

In the case of boosts the field Fourier components transform as follows
\begin{align}\label{assumedboost}
	\delta^B a_\mathbf{p} =i\omega_{\mathbf p}\, \lambda^i
	\frac{\partial a_\mathbf{p}}{\partial \mathbf{p}^i}\,
	+
	i\, a_\mathbf{p}
	\lambda^i
	\frac{\mathbf{p}_i}{2\omega_\mathbf{p}}
	,\quad \nonumber \\ 
	\delta^B a^\dag_\mathbf{p} =i\omega_{\mathbf p}\, \lambda^i
	\frac{\partial a^\dag_\mathbf{p}}{\partial \mathbf{p}^i}
	+
	i\, a_\mathbf{p}^\dag
	\lambda^i
	\frac{\mathbf{p}_i}{2\omega_\mathbf{p}}
\end{align}
and analogous expressions for $b_\mathbf{p}$, $b^\dag_\mathbf{p}$.
We have therefore
\begin{align}\label{boostcomp}
	-	\delta^B \lrcorner \, \Omega^{\kappa \rightarrow \infty}
	&=
	i	\int d^3p \,
	(
	\delta^B a^\dag_\mathbf{p}
	\delta a_\mathbf{p}\,
	-
	\delta a_\mathbf{p}^\dag
	\delta^B a_\mathbf{p}
	-
	\delta^B b_{\mathbf{p}}
	\delta b^\dag_{\mathbf{p}}
	+
	\delta b_{\mathbf{p}}
	\delta^B b^\dag_{\mathbf{p}}) \nonumber\\
	&=
	{\lambda}^i \int d^3p \,
	\omega_{\mathbf p}
	\left(
	\frac{\partial a_\mathbf{p}}{\partial \mathbf{p}^i}
	\delta a_\mathbf{p}^\dag\,
	-
	\delta a_\mathbf{p}
	\frac{\partial a_\mathbf{p}^\dag}{\partial \mathbf{p}^i}
	-
	\frac{\partial b_{\mathbf{p}}^\dag}{\partial \mathbf{p}^i}
	\delta b_{\mathbf{p}}
	+
	\delta b_{\mathbf{p}}^\dag
	\frac{\partial b_{\mathbf{p}}}{\partial \mathbf{p}^i}
	\right) \nonumber \\
	&+
	{\lambda}^i \int d^3p \,
	\frac{\mathbf{p}_i}{2\omega_\mathbf{p}}
	\left(
	a_\mathbf{p}
	\delta a_\mathbf{p}^\dag\,
	-
	\delta a_\mathbf{p}
	a_\mathbf{p}^\dag
	+
	\delta b_\mathbf{p}^\dag\,
	b_\mathbf{p}
	-
	b_\mathbf{p}^\dag
	\delta b_\mathbf{p}
	\right) \nonumber \\
	&=
	\frac{1}{2}
	{\lambda}^i \delta \int d^3p \,
	\omega_{\mathbf p}
	\left(
	\frac{\partial a_\mathbf{p}}{\partial \mathbf{p}^i}
	a_\mathbf{p}^\dag\,
	-
	a_\mathbf{p}
	\frac{\partial a_\mathbf{p}^\dag}{\partial \mathbf{p}^i}
	-
	\frac{\partial b_{\mathbf{p}}^\dag}{\partial \mathbf{p}^i}
	b_{\mathbf{p}}
	+
	b_{\mathbf{p}}^\dag
	\frac{\partial b_{\mathbf{p}}}{\partial \mathbf{p}^i}
	\right)
\end{align}
which gives the non-deformed boost charge
\begin{align}\label{canonicalboost}
	{\cal N}_i^{\kappa \rightarrow \infty} =
	\frac{1}{2}
	\int d^3p \,
	\omega_{\mathbf p}
	\left(
	\frac{\partial a_\mathbf{p}}{\partial \mathbf{p}^i}
	a_\mathbf{p}^\dag\,
	-
	a_\mathbf{p}
	\frac{\partial a_\mathbf{p}^\dag}{\partial \mathbf{p}^i}
	-
	\frac{\partial b_{\mathbf{p}}^\dag}{\partial \mathbf{p}^i}
	b_{\mathbf{p}}
	+
	b_{\mathbf{p}}^\dag
	\frac{\partial b_{\mathbf{p}}}{\partial \mathbf{p}^i}
	\right).
\end{align}
The case of rotations $R$ is analogous, and starting from the transformations
\begin{align}\label{dlorentzu}
	\delta^R a_\mathbf{p}
	&=
	i \epsilon^{ijk} \rho_i
	\mathbf{p}_{j} \frac{\partial}{\partial \mathbf{p}^{k}}
	a_\mathbf{p} 
	\qquad
	\delta^R a_\mathbf{p}^\dag
	=
	i \epsilon^{ijk} \rho_i
	\mathbf{p}_{j} \frac{\partial}{\partial \mathbf{p}^{k}}
	a_\mathbf{p}^\dag 
\end{align}
and analogously for $b_\mathbf{p}, b^\dag_\mathbf{p}$, one gets the charge
\begin{align}
	{\cal M}_i =
	\frac{1}{8}
	\int d^3p \,\epsilon_{ijk}
	\mathbf{p}^j
	\left(
	\frac{\partial a_\mathbf{p}}{\partial \mathbf{p}_k}
	a_\mathbf{p}^\dag\,
	-
	a_\mathbf{p}
	\frac{\partial a_\mathbf{p}^\dag}{\partial \mathbf{p}_k}
	-
	\frac{\partial b_{\mathbf{p}}^\dag}{\partial \mathbf{p}_k}
	b_{\mathbf{p}}
	+
	b_{\mathbf{p}}^\dag
	\frac{\partial b_{\mathbf{p}}}{\partial \mathbf{p}_k}
	\right).
\end{align}
Once again, we can obtain the field transformations from eq. \eqref{assumedboost} and \eqref{dlorentzu}, and one respectively gets
\begin{align}
	\delta^B\phi(x) 
	& =
	i\lambda_i\, 
	\int \frac{d^3p}{\sqrt{2\omega_\mathbf{p}}}\, \,
	\left[
	\omega_{\mathbf p}\, 
	\frac{\partial a_\mathbf{p}}{\partial \mathbf{p}^i}\,
	+
	a_\mathbf{p}
	\frac{\mathbf{p}_i}{2\omega_\mathbf{p}}
	\right]
	e^{-i(\omega_\mathbf{p}t-\mathbf{p}\mathbf{x})}  
	+
	\left[
	\omega_{\mathbf p}\, 
	\frac{\partial b^\dag_\mathbf{p}}{\partial \mathbf{p}^i}
	+
	b_\mathbf{p}^\dag
	\frac{\mathbf{p}_i}{2\omega_\mathbf{p}}
	\right]
	e^{i(\omega_\mathbf{p}t-\mathbf{p}\mathbf{x})} \nonumber \\
	&=
	-i\lambda_i\, 
	\int \frac{d^3p}{\sqrt{2\omega_\mathbf{p}}}\, 
	\left[
	a_{\mathbf{p}}\,
	\omega_\mathbf{p} \frac{\partial}{\partial \mathbf{p}_i}
	e^{-i(\omega_\mathbf{p}t-\mathbf{p}\mathbf{x})}  
	+
	b^\dag_{\mathbf{p}}\,
	\omega_\mathbf{p} \frac{\partial}{\partial \mathbf{p}_i}\, 
	e^{i(\omega_\mathbf{p}t-\mathbf{p}\mathbf{x})}
	\right] \nonumber \\
	&=i \lambda_i\, x^i\frac{\partial }{\partial t}\, \phi(x)
\end{align}
and 
\begin{align}
	\delta^R \phi(x) 
	&=
	i\rho^i\, \epsilon_{ik}{}^j\, 
	\int \frac{d^3p}{\sqrt{2\omega_\mathbf{p}}}\, 
	\left[
	a_{\mathbf{p}}\,
	\mathbf{p}_k \frac{\partial}{\partial \mathbf{p}_j}
	e^{-i(\omega_\mathbf{p}t-\mathbf{p}\mathbf{x})}  
	+
	b^\dag_{\mathbf{p}}\,
	\mathbf{p}_k \frac{\partial}{\partial \mathbf{p}_j}\, 
	e^{i(\omega_\mathbf{p}t-\mathbf{p}\mathbf{x})}
	\right] \nonumber\\
	&=
	i\rho^i\, \epsilon_{ik}{}^j\,  x^j\frac\partial{\partial x^k}\, \phi(x)
\end{align}
which are the standard spacetime boost and rotation transformations.

Notice that for the boosts we actually have antisymmetrization when writing them in spacetime. In fact, we only wrote down the contribution from $\omega_\mathbf{p} \frac{\partial}{\partial \mathbf{p}_i}$, but we should also include the contribution of $\mathbf{p}_i \frac{\partial}{\partial \omega_\mathbf{p}}$. However, since we are only integrating in $d^3p$, we first need to transform the derivative in $\omega_\mathbf{p}$ into a derivative in $\mathbf{p}_i$ (which will allow us to integrate by parts). Moreover, since we are on-shell (which means that $\omega_\mathbf{p} = \sqrt{\mathbf{p}^2+m^2}$), one can show by chain rule that 
\begin{align}
	\frac{\partial }{\partial \mathbf{p}_i}
	=
	\frac{\partial \omega_\mathbf{p}}{\partial \mathbf{p}_i}
	\frac{\partial}{\partial \omega_\mathbf{p}}
	=
	\frac{\mathbf{p}^i}{\omega_\mathbf{p}}
	\frac{\partial}{\partial \omega_\mathbf{p}}
	=
	-\frac{\mathbf{p}_i}{\omega_\mathbf{p}}
	\frac{\partial}{\partial \omega_\mathbf{p}}
\end{align}
and we get the same contribution as before.

Given the importance of the boost for the following discussion, as a double check of the correctness of eq. \eqref{canonicalboost}, we also sketch the computation of the boost charge from scratch using Noether theorem in non-deformed QFT. The action is
\begin{align}
	\mathcal{L} = (\partial_\mu \phi)^\dag (\partial^\mu \phi) - m^2 \phi^\dag \phi
\end{align}
and the on-shell fields for such an action are given by
\begin{align}\label{canfield}
	\phi = \int
	\frac{d^3 p}{\sqrt{2\omega_p}}
	a_{\mathbf{p}}
	e^{-i(\omega_p t - \mathbf{p}\mathbf{x})}
	+
	b_{\mathbf{p}}^\dag
	e^{i(\omega_p t - \mathbf{p}\mathbf{x})}
\end{align}
\begin{align}\label{canfieldag}
	\phi^\dag = \int
	\frac{d^3 p}{\sqrt{2\omega_p}}
	a_{\mathbf{p}}^\dag
	e^{i(\omega_p t - \mathbf{p}\mathbf{x})}
	+
	b_{\mathbf{p}}
	e^{-i(\omega_p t - \mathbf{p}\mathbf{x})}.
\end{align}
The energy-momentum tensor is given by
\begin{align}\label{tmunu}
	T_\mu^\nu
	&=
	\frac{\partial \mathcal{L}}{\partial (\partial_\nu \phi)} \partial_\mu \phi 
	+
	\frac{\partial \mathcal{L}}{\partial (\partial_\nu \phi^\dag)} \partial_\mu \phi^\dag 
	-
	\delta^\nu_\mu \mathcal{L} \\
	&=
	\frac{1}{2}
	(\partial_\mu \phi)^\dag (\partial^\nu \phi)
	+
	\frac{1}{2}
	(\partial_\mu \phi) (\partial^\nu \phi)^\dag
	-
	\frac{1}{2} (\partial_\alpha \phi)^\dag (\partial^\alpha \phi) 
	+ 
	\frac{1}{2}
	m^2 \phi^\dag \phi.
\end{align}
By Noether theorem, the charge is given by
\begin{align}
	\mathcal{L}_{0i} 
	= 
	i\int_{\mathbb{R}^3} (x_{0}T^0_{i} 
	-
	x_{i} T^0_{0} )
\end{align}
(the additional $i$ factor is added by hand, and it does not modify the discussion) and using eq. \eqref{tmunu} and eq. \eqref{canfield}, \eqref{canfieldag} one can do the explicit computations, which we will not do explicitly here. Notice however the following subtlety. In the term $x_i T^0_0$, the $x_i$ factor can be described as a derivative in momentum space acting on the spatial exponent, but since this exponent will contain $\mathbf{p}-\mathbf{q}$ (the two momenta are necessary because $T_{\mu\nu}$ contains product of fields, each of which is an integral in momentum space, so we need two integration variables) we have two ways of going so. This will result in two possible ways to compute the coefficient $x_i T^0_0$, namely 
\begin{align}
	-\int_{\mathbb{R}^3}
	x_i T^0_0
	&=
	-
	\int
	\frac{d^3 p}{\sqrt{2\omega_p}}
	\frac{d^3 q}{\sqrt{2\omega_q}}
	\Bigg\{
	(\omega_p\omega_q + \mathbf{p}\mathbf{q} + m^2) \times \nonumber \\
	&\times \left(
	-ia_{\mathbf{p}} a_{\mathbf{q}}^\dag
	e^{-i(\omega_p - \omega_q) t} 
	+
	i b_{\mathbf{p}}^\dag b_{\mathbf{q}}
	e^{i(\omega_p - \omega_q) t} 
	\right)
	\frac{\partial}{\partial \mathbf{p}^i}
	\delta(\mathbf{p}-\mathbf{q})
	\Bigg\} 
\end{align}
\begin{align}
	-\int_{\mathbb{R}^3}
	x_i T^0_0
	&=
	-
	\int
	\frac{d^3 p}{\sqrt{2\omega_p}}
	\frac{d^3 q}{\sqrt{2\omega_q}}
	\Bigg\{
	(\omega_p\omega_q + \mathbf{p}\mathbf{q} + m^2) \times \nonumber \\
	& \times \left(
	ia_{\mathbf{p}} a_{\mathbf{q}}^\dag
	e^{-i(\omega_p - \omega_q) t} 
	-
	i b_{\mathbf{p}}^\dag b_{\mathbf{q}}
	e^{i(\omega_p - \omega_q) t} 
	\right)
	\frac{\partial}{\partial \mathbf{q}^i}
	\delta(\mathbf{p}-\mathbf{q}) 
	\Bigg\}
\end{align}
After integration by parts, in both cases one term will eliminate the coefficient $x_0 T^0_i$, and the final charges will be 
\begin{align}\label{boostpartialres1}
	N_i^{(p)}
	=
	-\int
	d^3p \,
	\Bigg\{
	\omega_p
	\left(
	-\frac{\partial a_{\mathbf{p}}}{\partial \mathbf{p}^i} a_{\mathbf{p}}^\dag
	+
	\frac{\partial b_{\mathbf{p}}^\dag}{\partial \mathbf{p}^i} b_{\mathbf{p}}
	\right)
	\Bigg\}.
\end{align}
\begin{align}\label{boostpartialres2}
	N_i^{(q)}
	=
	\int
	d^3p \,
	\Bigg\{
	\omega_p
	\left(
	- a_{\mathbf{p}}\frac{\partial a_{\mathbf{p}}^\dag}{\partial \mathbf{p}^i} 
	+
	b_{\mathbf{p}}^\dag \frac{\partial b_{\mathbf{p}}}{\partial \mathbf{p}^i} 
	\right)
	\Bigg\}.
\end{align}
However, this means that we can write without loss of generality that the boost charge $N_i$ is given by
\begin{align}
	N_i = \frac{1}{2} (N_i^{(p)} + N_i^{(q)})
\end{align}
which is exactly eq. \eqref{canonicalboost}.

\paragraph{Crucial step}\label{crucialstep1}
Notice that in the last passage of Eq. \eqref{boostcomp} we used the fact that (we consider only the $a, a^\dag$ case since the same reasoning applies to $b, b^\dag$)
\begin{align}
	\frac{\partial a_\mathbf{p}}{\partial \mathbf{p}^i}
	\delta a_\mathbf{p}^\dag\,
	-
	\delta a_\mathbf{p}
	\frac{\partial a_\mathbf{p}^\dag}{\partial \mathbf{p}^i}
	=
	\frac{1}{2}
	\delta
	\left(
	\frac{\partial a_\mathbf{p}}{\partial \mathbf{p}^i}
	a_\mathbf{p}^\dag\,
	-
	a_\mathbf{p}
	\frac{\partial a_\mathbf{p}^\dag}{\partial \mathbf{p}^i}
	\right) 
	+
	\frac{1}{2}
	\frac{\partial }{\partial \mathbf{p}^i}
	\left(
	a_\mathbf{p}
	\delta a_\mathbf{p}^\dag\,
	-
	\delta a_\mathbf{p}
	a_\mathbf{p}^\dag
	\right).
\end{align}
The second term on the RHS of the last equation is still not a total derivative because of the presence of $\omega_\mathbf{p}$ inside the integrand.
After integration by parts, this term contributes a factor 
\begin{align}
	-\frac{\mathbf{p}_i}{2\omega_\mathbf{p}}
	\left(
	a_\mathbf{p}
	\delta a_\mathbf{p}^\dag\,
	-
	\delta a_\mathbf{p}
	a_\mathbf{p}^\dag
	\right)
\end{align}
to the integrand, which cancels the second term in the second passage of \eqref{boostcomp}.

\subsection{$\kappa$-deformed case}

Here we work with the full symplectic form in eq. \eqref{Symp14}. In obtaining this equation we used the following property of the wedge product of two $1$-forms $e$ and $f$.
\begin{align}
	e \wedge f = - f \wedge e.
\end{align}
In particular, this property was used when summing the contribution to the symplectic form coming from the two different orderings of the action. This is a non-trivial assumption in the presence of deformation, and we will discuss it more in section \ref{dsf}.

\subsubsection{Translation charges}

The idea is basically the same, with the only exception that now we need to consider the presence of antipodes. Because of this, the first thing to do is to understand how the creation/annihilation operator transform under translation.

Judging by the definition of the fields in eq. \eqref{fa} and \eqref{fb}, we assume the following behaviour
\begin{align}
	a_\mathbf{p} \mapsto e^{i \epsilon \omega_p} a_\mathbf{p}
	\qquad
	a_\mathbf{p}^\dag \mapsto e^{i \epsilon S(\omega_p)} a_\mathbf{p}^\dag
	\qquad
	b_{\mathbf{p}}^\dag \mapsto e^{i \epsilon S(\omega_p)} b_{\mathbf{p}}^\dag
	\qquad
	b_{\mathbf{p}} \mapsto e^{i \epsilon \omega_p} b_{\mathbf{p}}
\end{align}
and the same for spatial momenta, or in other words
\begin{align}\label{defodeltat}
	\delta_{\partial_0} a_\mathbf{p} &= i \epsilon \omega_p a_\mathbf{p}
	\qquad
	\delta_{\partial_0} a_\mathbf{p}^\dag = i \epsilon S(\omega_p) a_\mathbf{p}^\dag \nonumber 
\\
	\delta_{\partial_0} b_{\mathbf{p}}^\dag &= i \epsilon S(\omega_p) b_{\mathbf{p}}^\dag
	\qquad
	\delta_{\partial_0} b_{\mathbf{p}} = i \epsilon \omega_p b_{\mathbf{p}}
\end{align}
These transformations are correspond to the field transformation
\begin{align}\label{defactionofgenP}
	\delta_{\partial_\mu} \phi
	&=
	\int \frac{d^3p}{\sqrt{2\omega_\mathbf{p}}}\, \zeta(p) 
	\left[
	i \epsilon^\mu p_\mu\, a_{\mathbf{p}}\, e^{-i(\omega_\mathbf{p}t-\mathbf{p}\mathbf{x})}  +   i \epsilon^\mu S(p)_\mu b^\dag_{\mathbf{p}}\, e^{-i(S(\omega_\mathbf{p})t-S(\mathbf{p})\mathbf{x})}
	\right]
	\nonumber \\
	&=
	-\epsilon^\mu\partial_\mu\, \phi(x)
\end{align}
for the index $\mu=0$. However, for these transformations to be well defined, then in order to ensure the invariance of the products $a a^\dag$ and $bb^\dag$ we need to interpret them as a star products. This star product is defined in the phase space and not in canonical spacetime, otherwise the steps leading to the computation of the charges are faulted by the fact that we cannot write $a e^{(\cdot)} \star b e^{[\cdot]} = ab e^{(\cdot)}\star e^{[\cdot]}$. However, things are easily solved by defining
	\begin{align}
		a e^{(\cdot)} \star b e^{[\cdot]} = (a \star_{PS} b) \, e^{(\cdot)}\star e^{[\cdot]}.
	\end{align}
In fact, we have
\begin{align}\label{PSstar}
	a_\mathbf{p} a_\mathbf{p}^\dag
	:=
	a_\mathbf{p} \star_{PS} a_\mathbf{p}^\dag
	\overset{def.}{\mapsto}
	(a_\mathbf{p}
	\star_{PS}
	a_\mathbf{p}^\dag)
	\cdot
	\left(
	e^{i \epsilon \omega_p} \star e^{i \epsilon S(\omega_p)}
	\right)
	=
	a_\mathbf{p} \star_{PS} a_\mathbf{p}^\dag
\end{align}
This definition of $aa^\dag$ as $a \star_{PS} a^\dag$ does not influence the creation/annihilation operators algebra or the definition of normal ordering, so everything remains well defined. 

However, there is now the issue of how to treat the contraction of a vector field $\delta_\xi$ with the form $\delta a_\mathbf{p}\wedge
\delta a_\mathbf{p}^\dag$. In fact, the objects inside the wedge product are now to be treated in the deformed case.

\paragraph{Non-deformed contraction:} While we are in the context of forms, the product between the forms $\delta a$ and $\delta a^\dag$ is treated using the canonical wedge product with its properties. However, when we contract with $\delta_\xi$, one has canonically that
\begin{align}\label{oneofthedaxioms}
	\delta_\xi \lrcorner \, (\delta a_\mathbf{p}\wedge
	\delta a_\mathbf{p}^\dag)
	=
	\delta_\xi a_\mathbf{p}
	\delta a_\mathbf{p}^\dag
	-
	\delta a_\mathbf{p}
	\delta_\xi a_\mathbf{p}^\dag
\end{align}
and the wedge product is not present. Notice that the change of sign in the second term can be understood in terms of the following steps. 
\begin{itemize}
	\item The vector field always acts on the first objects it encounters in the wedge product. Because of this, to compute its action on the second term of the wedge product, we first need to use the formula
	\begin{align}
		v \wedge w = (-)^{p\cdot q} w \wedge v
	\end{align} 
	where $v$ and $w$ are respectively a $p$-form and a $q$-form. In our case, we have the wedge product of two $1$-forms, and therefore we have
	\begin{align}
		\delta a_\mathbf{p}\wedge
		\delta a_\mathbf{p}^\dag
		=
		-
		\delta a_\mathbf{p}^\dag \wedge
		\delta a_\mathbf{p}.
	\end{align}
	\item Now the vector field acts on the first component, giving
	\begin{align}
		- (\delta_\xi  a_\mathbf{p}^\dag) 
		\delta a_\mathbf{p}.
	\end{align}
	\item Now the object $\delta_\xi  a_\mathbf{p}^\dag$ is just some function, i.e. a $0$-form, and therefore it commutes with $\delta a$, so its placement is arbitrary.
\end{itemize}

\paragraph{Deformed contraction:} In this case, we define:\footnote{We write them down for $a$, but the same goes for $b$}
\begin{align}\label{deformedcontraction}
	\delta_\xi \lrcorner \, (\delta a_\mathbf{p}\wedge
	\delta a_\mathbf{p}^\dag)
	=
	(\delta_\xi a_\mathbf{p})
	\delta a_\mathbf{p}^\dag
	+
	\delta a_\mathbf{p} [S
	(
	\delta_\xi 
	)
	a_\mathbf{p}^\dag ].
\end{align}
The main difference between this axiom and the one in eq. \eqref{oneofthedaxioms} is the presence of the antipode instead of the normal minus sign in the second term. Notice that, since $S(AB)=S(A)S(B)$, and since $S(A)S(B) \xrightarrow{\kappa\rightarrow\infty}(-A)(-B) = AB$, the object $S(\delta_\xi)$ in eq. \eqref{deformedcontraction} must contain an additional minus sign when necessary, for consistency with the  non-deformed axiom. For example, if $\delta_\xi = \epsilon \mathbf{p}_i \frac{\partial}{\partial \mathbf{p}_i}$ where $\epsilon$ is just some constant, then $S(\delta_\xi) = -\epsilon S(\mathbf{p})_i \frac{\partial}{\partial S(\mathbf{p})_i}$, which has the correct $\kappa\rightarrow\infty$ limit.

Coming back to the charges, we first notice that we can rewrite
\begin{align}
	\Omega
	&=
	i\int d^3p \,
	\alpha \, (
	\delta a_\mathbf{p}\wedge
	\delta a_\mathbf{p}^\dag\,
	-
	\delta b_{\mathbf{p}}^\dag\wedge
	\delta b_{\mathbf{p}})\label{sign1} \\
	&=
	-i\int d^3p \,
	\alpha \, (
	\delta a_\mathbf{p}^\dag\wedge
	\delta a_\mathbf{p}\,
	-
	\delta b_{\mathbf{p}}\wedge
	\delta b_{\mathbf{p}}^\dag). \label{sign2}
\end{align}
Then, contracting with the vector field $\delta_{\partial_\mu}$, using the assumption in eq. \eqref{deformedcontraction}, and using eq. \eqref{defodeltat} we finally get
\begin{align}\label{ttranschargefromsymp}
	-\delta_{\partial_0} \lrcorner \, \Omega
	&=
	i\int d^3p \,\alpha \,
	(
	\delta_{\partial_0} a_\mathbf{p}^\dag
	\delta a_\mathbf{p}\,
	+
	\delta a_\mathbf{p}^\dag
	S(\delta_{\partial_0}) a_\mathbf{p}
	-
	\delta_{\partial_0} b_{\mathbf{p}}
	\delta b_{\mathbf{p}}^\dag
	-
	\delta b_{\mathbf{p}}
	S(\delta_{\partial_0}) b_{\mathbf{p}}^\dag) \\
	&=
	i\int d^3p \,\alpha \,
	[
	i\epsilon S(\omega_p) a_\mathbf{p}^\dag
	\delta a_\mathbf{p}\,
	+
	\delta a_\mathbf{p}^\dag
	(i\epsilon S(\omega_p) a_\mathbf{p}) \nonumber \\
	&-
	(i\epsilon \omega_p b_{\mathbf{p}}^\dag)
	\delta b_{\mathbf{p}}
	-
	\delta b_{\mathbf{p}}^\dag
	i\epsilon \omega_p b_{\mathbf{p}}] \\
	&=
	-\epsilon \int d^3p \,\alpha \,
	[
	S(\omega_p) a_\mathbf{p}^\dag
	\delta a_\mathbf{p}\,
	+
	\delta a_\mathbf{p}^\dag
	S(\omega_p) a_\mathbf{p}
	-
	\omega_p b_{\mathbf{p}}^\dag
	\delta b_{\mathbf{p}}
	-
	\delta b_{\mathbf{p}}^\dag
	\omega_p b_{\mathbf{p}}] \\
	&=
	\epsilon \delta
	\int d^3p \,\alpha \,
	[
	-S(\omega_p) a_\mathbf{p}
	a_\mathbf{p}^\dag\,
	+
	\omega_p b_{\mathbf{p}}^\dag
	b_{\mathbf{p}}
	]
\end{align}
which, apart from the irrelevant $\epsilon$ factor, gives the correct time translation charge (of course, in order to compare the final expression for the charges, normal ordering is always assumed. As stated before, the normal ordering is not affected by whether we consider the product of creation/annihilation operators as $aa^\dag$ or $a \star a^\dag$.). Notice also that we have used the fact that $\delta(AB) = (\delta A)B + A\delta B$. Of course, the same reasoning gives the correct space translation charges, since the procedure works in exactly the same way. We get therefore the following translation charges
\begin{align}\label{timetrans}
	\boxed{\mathcal{P}_0
		=
		\int d^3p \,\,
		\alpha(p)
		\left\{
		-S(\omega_p)
		a_\mathbf{p}^\dag\,
		a_\mathbf{p}\,
		+
		\omega_p
		b_{\mathbf{p}}\,
		b_{\mathbf{p}}^\dag\,
		\right\}}
\end{align}
\begin{align}\label{spatialtrans}
	\boxed{\mathcal{P}_i
		=
		\int d^3p \,\,
		\alpha(p)
		\left\{
		-S(\mathbf{p})_i
		a_\mathbf{p}^\dag\,
		a_\mathbf{p}\,
		+
		\mathbf{p}_i
		b_{\mathbf{p}}\,
		b_{\mathbf{p}}^\dag
		\right\}}
\end{align}
Notice that by construction we have the formal limit ($\zeta$ goes to $1$ for $\kappa\rightarrow\infty$ by definition)
\begin{align}
	\lim_{\kappa\rightarrow\infty} \alpha(p) = 1.
\end{align}
Furthermore, since $\omega_p>0$ and $-S(\omega_p)>0$, the energy has the correct sign, and in the limit $\kappa\rightarrow\infty$ we get back the correct translation charges of non-deformed complex scalar field. As a final comment, in \cite{Arzano:2020jro} the same charges were calculated using a direct approach based on the Noether theorem. However, it became clear while writing this thesis that the direct approach presented in \cite{Arzano:2020jro} only reproduces some of the terms of the full charges computed here (the charges in \cite{Arzano:2020jro} are however still correct, in light of the geometric approach presented here). The source of this incongruence is an interesting topic, deserving of further investigation, and it will be tackled in future publications.

\subsubsection{Notes on the above (overall) procedure}

There are several things to notice about the above procedure. 
\begin{itemize}
	\item[1)] The most obvious thing to address is that there seems to be no reason as to why we chose to write 
	\begin{align}\label{omega1}
		\Omega
		&=
		-i\int d^3p \,
		\alpha \, (
		\delta a_\mathbf{p}^\dag\wedge
		\delta a_\mathbf{p}\,
		-
		\delta b_{\mathbf{p}}\wedge
		\delta b_{\mathbf{p}}^\dag)
	\end{align}
	and not for example 
	\begin{align}\label{omega2}
		\Omega
		=
		-i\int d^3p \,
		\alpha \, (
		\delta a_\mathbf{p}^\dag\wedge
		\delta a_\mathbf{p}\,
		+
		\delta b_{\mathbf{p}}^\dag\wedge
		\delta b_{\mathbf{p}})
	\end{align}
	which seems at least as good. 
	The only sensible answer to this is to look at the single symplectic forms $\delta \theta_1$ and $\delta \theta_2$, we have
	\begin{align}
		\delta \theta_1 \propto \delta \phi^\dag \wedge \delta\phi
		\qquad
		\delta \theta_2 \propto \delta\phi \wedge \delta \phi^\dag
	\end{align}
	In both cases, the off diagonal terms go away and we are only left with the wedge products between $a, a^\dag$ and between $b, b^\dag$. However, both in $\delta \theta_1$ and $\delta \theta_2$, since $\phi$ contains $a$ and $b^\dag$ (and of course $\phi^\dag$ contains $a^\dag$ and $b$), each time that we have $\delta a \wedge \delta a^\dag$ we also have $\delta b^\dag \wedge \delta b$. Therefore, the correct final symplectic charge must be of the kind in eq. \eqref{omega1} and not the one in eq. \eqref{omega2}. 
	
	Of course, in non-deformed QFT, either \eqref{omega1} or \eqref{omega2} are equally fine, but in that case there is no antipode in the formula for the contraction of the symplectic form and an external vector field $\delta_\xi$, while there is one in the deformed case in \eqref{deformedcontraction}. 
	
	Therefore, because of the presence of the antipode in \eqref{deformedcontraction}, it seems that the structure of the symplectic form in the deformed case is more rigid than in the non-deformed case, because although we are free to chose the overall sign, we cannot arbitrarily change the order of only a part of the symplectic form before contracting with $\delta_\xi$.
	
	\item[2)] Having tackled the first important point, then there is the issue of why we chose the overall sign as in eq. \eqref{sign2} and not the one in \eqref{sign1}. The answer is that one can go from \eqref{sign1} to \eqref{sign2} by means of the following exchange
	\begin{align}
		a \leftrightarrow b
		\qquad
		a^\dag \leftrightarrow b^\dag
	\end{align}
	so that, performing the same exact computations, instead of the charge 
	\begin{align}
		Q_1
		=
		\int d^3p \,\alpha \,
		[
		-S(\omega_p) a_\mathbf{p}
		a_\mathbf{p}^\dag\,
		+
		\omega_p b_{\mathbf{p}}^\dag
		b_{\mathbf{p}}
		]
	\end{align}
	we would end up with the charge
	\begin{align}
		Q_2
		=
		\int d^3p \,\alpha \,
		[
		-S(\omega_p) b_{\mathbf{p}}^\dag
		b_{\mathbf{p}}
		+
		\omega_p 
		a_\mathbf{p}
		a_\mathbf{p}^\dag\,
		].
	\end{align}
	However, for all practical purposes, the charge $Q_2$ is the same as $Q_1$ because both $a, a^\dag$ and $b, b^\dag$ live on the same manifold. The important thing is that one of them has energy $\omega_p$ and the other $-S(\omega_p)$ while living on the same manifold, the name of the operators are not important in this context. 
	
	\item[3)] One last thing to notice is that the assumption \eqref{deformedcontraction} is not in contradiction with the properties of the differential operator $d_F = \epsilon^A \partial_A + \omega_{\mu\nu}L_{\mu\nu}$ that we used in the computation of the charges (at leat at first sight). In fact, while $d_F$ is the exterior derivative in spacetime, we are dealing with (the properties of) the exterior derivative $\delta$ in phase space, which is however the space of solutions to the EoM. Therefore, the behaviour of $\delta$ can be different than the one of $d_F$.
\end{itemize}

\subsubsection{Boosts}\label{boostsect}

Analogously to what has been done before, we can now concentrate on the boost charges. Once again, we need to understand what is the action of $\delta_\Lambda$ on $a, a^\dag, b, b^\dag$.

One approach would be to start from the assumed Lorentz transformations
\begin{align}\label{boostphi}
	\delta^B\phi(x)  &=i \lambda_i\, x^i\frac{\partial }{\partial t}\, \phi(x)\nonumber\\
	& =
	-i\lambda_i\, 
	\int \frac{d^3p}{\sqrt{2\omega_\mathbf{p}}}\, \,\zeta(p)\,
	a_{\mathbf{p}}\,
	\omega_\mathbf{p} \frac{\partial}{\partial \mathbf{p}_i}
	e^{-i(\omega_\mathbf{p}t-\mathbf{p}\mathbf{x})}  \nonumber \\
	&-
	i\lambda_i\, 
	\int \frac{d^3p}{\sqrt{2\omega_\mathbf{p}}}\, \,\zeta(p)\, 
	b^\dag_{\mathbf{p}}\,
	S(\omega_\mathbf{p}) \frac{\partial}{\partial S(\mathbf{p})_i}\, 
	e^{-i(S(\omega_\mathbf{p})t-S(\mathbf{p})\mathbf{x})}.
\end{align}
and integrating by parts obtaining
\begin{align*}
	&\delta^B\phi(x) 
	 =
	i\lambda_i\, 
	\int \frac{d^3p}{\sqrt{2\omega_\mathbf{p}}}\, \,
	\omega_\mathbf{p}
	\,\zeta(p)\,
	\left[
	\frac{\partial a_{\mathbf{p}}}{\partial \mathbf{p}_i}
	+
	a_{\mathbf{p}}
	\frac{\sqrt{2\omega_\mathbf{p}}}{\zeta(p) \omega_p}
	\frac{\partial }{\partial \mathbf{p}_i}
	\left(
	\frac{\omega_\mathbf{p}}{\sqrt{2 \omega_\mathbf{p}}}
	\zeta(p)
	\right)
	\right]
	e^{-i(\omega_\mathbf{p}t-\mathbf{p}\mathbf{x})}  \nonumber \\
	&+
	i\lambda_i\, 
	\int \frac{d^3p}{\sqrt{2\omega_\mathbf{p}}}\, \,
	S(\omega_\mathbf{p})
	\,\zeta(p)\,
	\left[
	\frac{\partial b_{\mathbf{p}}^\dag}{\partial S(\mathbf{p})_i}
	+
	b_{\mathbf{p}}^\dag
	\frac{\sqrt{2\omega_\mathbf{p}}}{\zeta(p)S(\omega_p)}
	\frac{\partial }{\partial S(\mathbf{p})_i}
	\left(
	\frac{S(\omega_\mathbf{p})}{\sqrt{2 \omega_\mathbf{p}}}
	\zeta(p)
	\right)
	\right]
	e^{-i(S(\omega_\mathbf{p})t-S(\mathbf{p})\mathbf{x})}.
\end{align*}
\begin{align*}
	&\delta^B\phi(x) 
	=
	i\lambda_i\, 
	\int \frac{d^3p}{\sqrt{2\omega_\mathbf{p}}}\, \,
	\,\zeta(p)\,
	\Bigg\{
	\omega_\mathbf{p}
	\frac{\partial a_{\mathbf{p}}}{\partial \mathbf{p}_i}
	+
	a_{\mathbf{p}}
	\frac{\sqrt{2\omega_\mathbf{p}}}{\zeta(p) }
	\frac{\partial }{\partial \mathbf{p}_i}
	\left(
	\frac{\omega_\mathbf{p}}{\sqrt{2 \omega_\mathbf{p}}}
	\zeta(p)
	\right) 
	\Bigg\}
	e^{-i(\omega_\mathbf{p}t-\mathbf{p}\mathbf{x})}  \nonumber \\
	&+
	i\lambda_i\, 
	\int \frac{d^3p}{\sqrt{2\omega_\mathbf{p}}}\, \,
	\,\zeta(p)\,
	\Bigg\{
	S(\omega_\mathbf{p})
	\frac{\partial b_{\mathbf{p}}^\dag}{\partial S(\mathbf{p})_i}
	+
	b_{\mathbf{p}}^\dag
	\frac{\sqrt{2\omega_\mathbf{p}}}{\zeta(p)}
	\frac{\partial }{\partial S(\mathbf{p})_i}
	\left(
	\frac{S(\omega_\mathbf{p})}{\sqrt{2 \omega_\mathbf{p}}}
	\zeta(p)
	\right)
	\Bigg\}
	e^{-i(S(\omega_\mathbf{p})t-S(\mathbf{p})\mathbf{x})}.
\end{align*}

Therefore, limiting ourselves to $\phi$ since $\phi^\dag$ is analogous, we would have the following relations 
\begin{align}
	\delta^B a_\mathbf{p} 
	&=
	-i \lambda^i \, 
	\omega_p 
	\left[
	\frac{\partial}{\partial \mathbf{p}^i}
	+
	\frac{1}{\zeta(p) \sqrt{\omega_\mathbf{p}}}
	\frac{\partial }{\partial \mathbf{p}^i}
	(\zeta(p)\sqrt{\omega_\mathbf{p}})
	\right]a_\mathbf{p} \label{dboost-f1} \\
	\delta^B a_\mathbf{p}^\dag
	&=
	-i \lambda^i \,
	S(\omega_p) 
	\left[
	\frac{\partial }{\partial S(\mathbf{p})^i}
	+
	\frac{\sqrt{\omega_\mathbf{p}}}{\zeta(p) S(\omega_p)}
	\frac{\partial }{\partial S(\mathbf{p})^i}
	\left(
	\frac{S(\omega_\mathbf{p})}{\sqrt{ \omega_\mathbf{p}}}
	\zeta(p)
	\right)
	\right]a_{\mathbf{p}}^\dag \label{dboost-f2} \\
	\delta^B b_\mathbf{p} 
	&=
	-i \lambda^i \, 
	\omega_p 
	\left[
	\frac{\partial}{\partial \mathbf{p}^i}
	+
	\frac{1}{\zeta(p) \sqrt{\omega_\mathbf{p}}}
	\frac{\partial }{\partial \mathbf{p}^i}
	(\zeta(p)\sqrt{\omega_\mathbf{p}})
	\right]b_\mathbf{p} \label{dboost-f3} \\
	\delta^B b_\mathbf{p}^\dag
	&=
	-i \lambda^i \,
	S(\omega_p) 
	\left[
	\frac{\partial }{\partial S(\mathbf{p})^i}
	+
	\frac{\sqrt{\omega_\mathbf{p}}}{\zeta(p) S(\omega_p)}
	\frac{\partial }{\partial S(\mathbf{p})^i}
	\left(
	\frac{S(\omega_\mathbf{p})}{\sqrt{ \omega_\mathbf{p}}}
	\zeta(p)
	\right)
	\right]b_\mathbf{p}^\dag \label{dboost-f4}
\end{align}
We will however now see that the above transformations need to be modified into
\begin{align}
	\delta^B a_\mathbf{p} 
	&=
	-i \lambda^i \, 
	\omega_\mathbf{p}
	\left[
	\frac{\partial}{\partial \mathbf{p}^i}
	+
	\frac{1}{2}
	\frac{1}{\omega_\mathbf{p}}
	\frac{\partial [\omega_\mathbf{p} S(\alpha(p))]}{\partial \mathbf{p}^i}
	\right]a_\mathbf{p} \label{dboosta1} \\
	\delta^B a_\mathbf{p}^\dag
	&=
	-i \lambda^i \,
	S(\omega_\mathbf{p}) 
	\left[
	\frac{\partial }{\partial S(\mathbf{p})^i}
	+
	\frac{1}{2}
	\frac{1}{S(\omega_\mathbf{p})}
	\frac{\partial [S(\omega_\mathbf{p}) \alpha(p)]}{\partial S(\mathbf{p})^i}
	\right]a_{\mathbf{p}}^\dag \label{dboosta2} \\
	\delta^B b_\mathbf{p} 
	&=
	-i \lambda^i \, 
	\omega_\mathbf{p}
	\left[
	\frac{\partial}{\partial \mathbf{p}^i}
	+
	\frac{1}{2}
	\frac{1}{\omega_\mathbf{p}}
	\frac{\partial [\omega_\mathbf{p} \alpha(p)]}{\partial \mathbf{p}^i}
	\right]b_\mathbf{p} \label{dboosta3} \\
	\delta^B b_\mathbf{p}^\dag
	&=
	-i \lambda^i \,
	S(\omega_\mathbf{p}) 
	\left[
	\frac{\partial }{\partial S(\mathbf{p})^i}
	+
	\frac{1}{2}
	\frac{1}{S(\omega_\mathbf{p})}
	\frac{\partial [S(\omega_\mathbf{p}) S(\alpha(p))]}{\partial S(\mathbf{p})^i}
	\right]b_\mathbf{p}^\dag \label{dboosta4}
\end{align}
in order to obtain a conserved charge. The reasoning here is as follows. We know that the action is invariant under $\kappa$-Poincar\'e transformations, which means that by Noether theorem there are conserved charges. More in detail, there is a one-to-one correspondence between symmetries and conserved charges. One usually goes in one direction, obtaining the conserved charges as a result of exploiting the symmetries of the action. Here we go in the other direction, obtaining the charges, and then defining the symmetry of the system by showing that charges are conserved. 

Notice that in this whole process, there is no way to know a priori the transformation laws of the fields under a symmetry of the action. A symmetry of the action does not necessarily translate to a symmetry of the field inside the action; quite the opposite, the fields change under some transformation, and the action is such combination of fields and their derivatives which is invariant. Eq. \eqref{boostphi} represents a canonical field transformation under non-deformed boost; it is therefore not surprising that it does not give rise to a charge since it is not a symmetry of our system. 

We now first show that eq. \eqref{dboosta1}, \eqref{dboosta2}, \eqref{dboosta3}, \eqref{dboosta4} give rise to a conserved charge, and then we will show to what field transformation they correspond to. We have
\begin{align}
	-&\delta^B \lrcorner \, \Omega
	=
	-	i\int d^3p \,\alpha(p)
	\left(
	\delta^B a_\mathbf{p}^\dag
	\delta a_\mathbf{p}\,
	+
	\delta a_\mathbf{p}^\dag
	S(\delta^B) a_\mathbf{p}
	-
	\delta^B b_{\mathbf{p}}
	\delta b_{\mathbf{p}}^\dag
	-
	\delta b_{\mathbf{p}}
	S(\delta^B) b_{\mathbf{p}}^\dag\right)  \nonumber\\
	&=
	-\lambda^i\,\int d^3p \,\alpha(p)
	\left(
	S(\omega_p) \frac{\partial a_\mathbf{p}^\dag}{\partial S(\mathbf{p})^i}
	\delta a_\mathbf{p}\,
	-
	\delta a_\mathbf{p}^\dag \,
	S(\omega_p) \frac{\partial a_\mathbf{p}}{\partial S(\mathbf{p})^i}
	-
	\omega_p\frac{\partial b_{\mathbf{p}}}{\partial \mathbf{p}^i}
	\delta b_{\mathbf{p}}^\dag
	+
	\delta b_{\mathbf{p}}
	\omega_p \frac{\partial b_{\mathbf{p}}^\dag}{\partial \mathbf{p}^i}
	\right)\nonumber\\
	&
	+
	\lambda^i\,\int d^3p  \frac{1}{2}
	\frac{\partial [S(\omega_\mathbf{p}) \alpha(p)]}{\partial S(\mathbf{p})^i}
	\left(
	a_\mathbf{p}
	\delta a_\mathbf{p}^\dag\,
	-
	\delta a_\mathbf{p}
	a_\mathbf{p}^\dag
	\right)
	-
	\frac{1}{2}
	\frac{\partial [\omega_\mathbf{p} \alpha(p)]}{\partial \mathbf{p}^i}
	\left(
	b_{\mathbf{p}}
	\delta b_{\mathbf{p}}^\dag
	-
	\delta b_{\mathbf{p}} b_{\mathbf{p}}^\dag
	\right)  \\
	&=
	-\frac{1}{2}{\lambda}^i \delta \int d^3p \,\alpha(p)
	\left\{
	S(\omega_p) \left[
	\frac{\partial a_\mathbf{p}^\dag}{\partial S(\mathbf{p})^i}
	a_\mathbf{p}\,
	-
	a_\mathbf{p}^\dag \, \frac{\partial a_\mathbf{p}}{\partial S(\mathbf{p})^i}
	\right]
	+
	\omega_p
	\left[
	b_{\mathbf{p}}
	\frac{\partial b_{\mathbf{p}}^\dag}{\partial \mathbf{p}^i}
	-
	\frac{\partial b_{\mathbf{p}}}{\partial \mathbf{p}^i}
	b_{\mathbf{p}}^\dag
	\right]
	\right\}
\end{align}
which gives the boost charge
\begin{align}\label{defboostfromsymp}
	\boxed{{\cal   N }_i
	=
	- \frac{1}{2}
	\int d^3p \,\alpha(p)
	\left\{
	S(\omega_p) \left[
	\frac{\partial a_\mathbf{p}^\dag}{\partial S(\mathbf{p})^i}
	a_\mathbf{p}\,
	-
	a_\mathbf{p}^\dag \, \frac{\partial a_\mathbf{p}}{\partial S(\mathbf{p})^i}
	\right]
	+
	\omega_p
	\left[
	b_{\mathbf{p}}
	\frac{\partial b_{\mathbf{p}}^\dag}{\partial \mathbf{p}^i}
	-
	\frac{\partial b_{\mathbf{p}}}{\partial \mathbf{p}^i}
	b_{\mathbf{p}}^\dag
	\right]
	\right\}}.
\end{align}

\paragraph{Crucial step}\label{defboostdisc}

Once again, in order to be able to take the exterior derivative outside, we used the fact that 
\begin{align}\label{fund1def}
	\frac{\partial a_\mathbf{p}}{\partial S(\mathbf{p})^i}
	\delta a_\mathbf{p}^\dag\,
	-
	\delta a_\mathbf{p}
	\frac{\partial a_\mathbf{p}^\dag}{\partial S(\mathbf{p})^i}
	&=
	\delta
	\left(
	\frac{\partial a_\mathbf{p}}{\partial S(\mathbf{p})^i}
	a_\mathbf{p}^\dag\,
	-
	a_\mathbf{p}
	\frac{\partial a_\mathbf{p}^\dag}{\partial S(\mathbf{p})^i}
	\right) \nonumber \\
	&-
	\frac{\partial \delta a_\mathbf{p}}{\partial S(\mathbf{p})^i}
	a_\mathbf{p}^\dag\,
	+
	a_\mathbf{p}
	\frac{\partial \delta a_\mathbf{p}^\dag}{\partial S(\mathbf{p})^i}
\end{align}
and 
\begin{align}\label{fund2def}
	\frac{\partial a_\mathbf{p}}{\partial S(\mathbf{p})^i}
	\delta a_\mathbf{p}^\dag\,
	-
	\delta a_\mathbf{p}
	\frac{\partial a_\mathbf{p}^\dag}{\partial S(\mathbf{p})^i}
	&=
	\frac{\partial }{\partial S(\mathbf{p})^i}
	\left(
	a_\mathbf{p}
	\delta a_\mathbf{p}^\dag\,
	-
	\delta a_\mathbf{p}
	a_\mathbf{p}^\dag
	\right) \nonumber \\
	&+
	\frac{\partial \delta a_\mathbf{p}}{\partial S(\mathbf{p})^i}
	a_\mathbf{p}^\dag\,
	-
	a_\mathbf{p}
	\frac{\partial \delta a_\mathbf{p}^\dag}{\partial S(\mathbf{p})^i}
\end{align}
and therefore, adding \eqref{fund1def} and \eqref{fund2def} we finally get
\begin{align}
	\frac{\partial a_\mathbf{p}}{\partial S(\mathbf{p})^i}
	\delta a_\mathbf{p}^\dag\,
	-
	\delta a_\mathbf{p}
	\frac{\partial a_\mathbf{p}^\dag}{\partial S(\mathbf{p})^i}
	&=
	\frac{1}{2}
	\delta
	\left(
	\frac{\partial a_\mathbf{p}}{\partial S(\mathbf{p})^i}
	a_\mathbf{p}^\dag\,
	-
	a_\mathbf{p}
	\frac{\partial a_\mathbf{p}^\dag}{\partial S(\mathbf{p})^i}
	\right) \nonumber \\ 
	&+
	\frac{1}{2}
	\frac{\partial }{\partial S(\mathbf{p})^i}
	\left(
	a_\mathbf{p}
	\delta a_\mathbf{p}^\dag\,
	-
	\delta a_\mathbf{p}
	a_\mathbf{p}^\dag
	\right).
\end{align}
Of course, the same exact reasoning works for the $b, b^\dag$ operators. Here, notice that after integration by parts the second term produces the following contribution to $-\delta_{\partial_\Lambda} \, \lrcorner \, \Omega$
\begin{align}\label{rel1}
	- \frac{1}{2}
	\frac{\partial [S(\omega_\mathbf{p}) \alpha(p)]}{\partial S(\mathbf{p})^i}
	\left(
	a_\mathbf{p}
	\delta a_\mathbf{p}^\dag\,
	-
	\delta a_\mathbf{p}
	a_\mathbf{p}^\dag
	\right)
	+
	\frac{1}{2}
	\frac{\partial [\omega_\mathbf{p} \alpha(p)]}{\partial \mathbf{p}^i}
	\left(
	b_{\mathbf{p}}
	\delta b_{\mathbf{p}}^\dag
	-
	\delta b_{\mathbf{p}} b_{\mathbf{p}}^\dag
	\right),
\end{align}
and once again it eliminates the relevant term in the computation of the boost charge.

\bigskip

We can now turn to the transformation of the field due to eq. \eqref{dboosta1}, \eqref{dboosta2}, \eqref{dboosta3}, \eqref{dboosta4}. We have
\begin{align}
	&\delta^B\phi(x)  
	=
	i \lambda_i\, x^i\frac{\partial }{\partial t}\, \phi(x) \nonumber \\
	&-
	i\lambda_i\, 
	\int \frac{d^3p}{\sqrt{2\omega_\mathbf{p}}}\, \,
	\,\zeta\,
	\Bigg\{
	\omega_\mathbf{p}
	\left[
	\frac{1}{2}
	\frac{1}{\omega_\mathbf{p}}
	\frac{\partial [\omega_\mathbf{p} S(\alpha)]}{\partial \mathbf{p}^i}
	-
	\frac{\sqrt{\omega_\mathbf{p}}}{\zeta \omega_p}
	\frac{\partial }{\partial \mathbf{p}^i}
	\left(
	\frac{\omega_\mathbf{p}}{\sqrt{ \omega_\mathbf{p}}}
	\zeta
	\right)
	\right]a_{\mathbf{p}}
	e^{-i(\omega_\mathbf{p}t-\mathbf{p}\mathbf{x})}  \nonumber \\
	&-
	S(\omega_\mathbf{p})
	\left[
	\frac{1}{2}
	\frac{1}{S(\omega_\mathbf{p})}
	\frac{\partial [S(\omega_\mathbf{p}) S(\alpha)]}{\partial S(\mathbf{p})^i}
	-
	\frac{\sqrt{\omega_\mathbf{p}}}{\zeta S(\omega_p)}
	\frac{\partial }{\partial S(\mathbf{p})^i}
	\left(
	\frac{S(\omega_\mathbf{p})}{\sqrt{ \omega_\mathbf{p}}}
	\zeta
	\right)
	\right]b_{\mathbf{p}}^\dag
	e^{-i(S(\omega_\mathbf{p})t-S(\mathbf{p})\mathbf{x})}
	\Bigg\} \label{active-defboost1}
\end{align}
\begin{align}
	&\delta^B\phi^\dag(x)  
	=
	-i \lambda_i\, x^i\frac{\partial }{\partial t}\, \phi^\dag(x) \nonumber \\
	&-
	i\lambda_i\, 
	\int \frac{d^3p}{\sqrt{2\omega_\mathbf{p}}}\, \,
	\,\zeta\,
	\Bigg\{
	\omega_\mathbf{p}
	\left[
	\frac{1}{2}
	\frac{1}{\omega_\mathbf{p}}
	\frac{\partial [\omega_\mathbf{p} \alpha]}{\partial \mathbf{p}^i}
	-
	\frac{\sqrt{\omega_\mathbf{p}}}{\zeta \omega_p}
	\frac{\partial }{\partial \mathbf{p}^i}
	\left(
	\frac{\omega_\mathbf{p}}{\sqrt{ \omega_\mathbf{p}}}
	\zeta
	\right)
	\right]b_{\mathbf{p}}
	e^{-i(\omega_\mathbf{p}t-\mathbf{p}\mathbf{x})}  \nonumber \\
	&-
	S(\omega_\mathbf{p})
	\left[
	\frac{1}{2}
	\frac{1}{S(\omega_\mathbf{p})}
	\frac{\partial [S(\omega_\mathbf{p}) \alpha]}{\partial S(\mathbf{p})^i}
	-
	\frac{\sqrt{\omega_\mathbf{p}}}{\zeta S(\omega_p)}
	\frac{\partial }{\partial S(\mathbf{p})^i}
	\left(
	\frac{S(\omega_\mathbf{p})}{\sqrt{ \omega_\mathbf{p}}}
	\zeta
	\right)
	\right]a_{\mathbf{p}}^\dag
	e^{-i(S(\omega_\mathbf{p})t-S(\mathbf{p})\mathbf{x})}
	\Bigg\}\label{active-defboost2}
\end{align}
To understand these additional terms, let us use eq. \eqref{alpha}, and let us assume that we want to impose
\begin{align}\label{convention0}
	\xi(p) \,\, : \,\, \alpha(p) = 1.
\end{align}
This is a natural choice, since it eliminates this factor from the charges. In this case, we have 
\begin{align}\label{convention}
	\zeta(p)^{-2} = \frac{1}{2\omega_{\mathbf{p}}}\, \frac{p_4}{\kappa}\, \left[
	\frac{|p_+|^3}{\kappa^3}
	\omega_{\mathbf{p}} - S(\omega_{\mathbf{p}})
	\right]	
\end{align}
and one can show that, to the leading order in $\frac{1}{\kappa}$, the boost transformation becomes
\begin{align}\label{LOphibo}
	\delta^B\phi(x)  
	&=
	i \lambda_i\, x^i\frac{\partial }{\partial t}\, \phi(x) 
	+
	i\lambda_i\, 
	\int \frac{d^3p}{\sqrt{2\omega_\mathbf{p}}}\, \,
	\Bigg\{
	\frac{\mathbf{p}_i}{\kappa}
	\left(
	\frac{m^2}{\omega_\mathbf{p}^2}
	-
	2
	\right)
	a_{\mathbf{p}}
	e^{-i(\omega_\mathbf{p}t-\mathbf{p}\mathbf{x})}  \nonumber \\
	&+
	\frac{\mathbf{p}_i}{\kappa}
	\left(
	\frac{5}{2}
	-
	\frac{m^2 }{2\omega_\mathbf{p}^2}
	\right)
	b_{\mathbf{p}}^\dag
	e^{-i(S(\omega_\mathbf{p})t-S(\mathbf{p})\mathbf{x})}
	\Bigg\}
\end{align}
\begin{align}\label{LOphidagbo}
	\delta^B\phi^\dag(x)
	&=
	-i \lambda_i\, x^i\frac{\partial }{\partial t}\, \phi(x)^\dag 
	+
	i\lambda^i\, 
	\int \frac{d^3p}{\sqrt{2\omega_\mathbf{p}}}\, \,
	\Bigg\{
	\frac{\mathbf{p}_i}{\kappa}
	\left(
	\frac{m^2}{\omega_\mathbf{p}^2}
	-
	2
	\right)
	b_{\mathbf{p}}
	e^{-i(\omega_\mathbf{p}t-\mathbf{p}\mathbf{x})}
	\nonumber \\
	&+
	\frac{\mathbf{p}_i}{\kappa}
	\left(
	\frac{5}{2}
	-
	\frac{m^2 }{2\omega_\mathbf{p}^2}
	\right)
	a_{\mathbf{p}}^\dag
	e^{-i(S(\omega_\mathbf{p})t-S(\mathbf{p})\mathbf{x})} 
	\Bigg\}
\end{align}
We see that particles and antiparticles, upon boost, get an additional translation. Furthermore, notice also that using the convention in eq. \eqref{convention}, this additional translation is the same for both $a^\dag$ and $b^\dag$. This is already apparent looking at eq. \eqref{dboosta2}, \eqref{dboosta4}, since if $\alpha=1$ the quantities $\alpha$ and $S(\alpha)$ disappear from the formulae\footnote{The antipode does not act on objects independent of momenta, so that $S(1)=1$.}. If we used instead the convention
\begin{align}\label{convention1}
	\zeta(p) \,\, : \,\, \zeta^2(p) \left[1 + \frac{|p_+|^3}{\kappa}\right] \frac{\kappa}{p_4} = 1
\end{align}
then one gets different translations for $a^\dag$ and $b^\dag$, as is again apparent from eq. \eqref{dboosta2}, \eqref{dboosta4}. In this case, in fact, we would have $S(\alpha)\neq \alpha$. 


Repeating the same calculations as before, in this case one gets the following boost transformation for the fields. 
\begin{align}\label{LOphibo-conv2}
	\delta^B\phi(x)  
	&=
	i \lambda_i\, x^i\frac{\partial }{\partial t}\, \phi(x) 
	+
	i\lambda_i\, 
	\int \frac{d^3p}{\sqrt{2\omega_\mathbf{p}}}\, \,
	\Bigg\{
	\frac{\mathbf{p}_i}{\kappa}
	\left(
	-\frac{19}{16\sqrt{2}}
	\right)
	a_{\mathbf{p}}
	e^{-i(\omega_\mathbf{p}t-\mathbf{p}\mathbf{x})}  \nonumber \\
	&+
	\frac{\mathbf{p}_i}{\kappa}
	\left[
	\frac{3}{16\sqrt{2}}
	\left(
	9+\frac{4m^2}{\omega_\mathbf{p}^2}
	\right)
	\right]
	b_{\mathbf{p}}^\dag
	e^{-i(S(\omega_\mathbf{p})t-S(\mathbf{p})\mathbf{x})}
	\Bigg\}
\end{align}
\begin{align}\label{LOphidagbo-conv2}
	\delta^B\phi^\dag(x)
	&=
	-i \lambda_i\, x^i\frac{\partial }{\partial t}\, \phi(x)^\dag 
	+
	i\lambda^i\, 
	\int \frac{d^3p}{\sqrt{2\omega_\mathbf{p}}}\, \,
	\Bigg\{
	\frac{\mathbf{p}_i}{\kappa}
	\left(
	-\frac{11}{16\sqrt{2}}
	\right)
	b_{\mathbf{p}}
	e^{-i(\omega_\mathbf{p}t-\mathbf{p}\mathbf{x})}
	\nonumber \\
	&+
	\frac{\mathbf{p}_i}{\kappa}
	\left(
	\frac{19}{16\sqrt{2}}
	+
	\frac{3m^2}{4\sqrt{2} \omega_\mathbf{p}^2}
	\right)
	a_{\mathbf{p}}^\dag
	e^{-i(S(\omega_\mathbf{p})t-S(\mathbf{p})\mathbf{x})}
	\Bigg\}.
\end{align}
In this case, the additional translations for $a^\dag$ and $b^\dag$ are different.

\paragraph{The (very) peculiar features of boosts}

We already saw in eq. \eqref{L} that a single particle state gets boosted canonically. This is not just an assumption, since it can be proven by using the algebra of the charges. Anticipating a little the results of section \ref{chargesalgebra}, we will show that all the charges that we compute in this section satisfy the canonical, non-deformed Poincar\'e algebra. This means in particular that $[\mathcal{N}_i, \mathcal{P}_j] = -i \eta_{ij} \mathcal{P}_0$. We now define single-particle states as usual, namely given the vacuum state $|0\rangle$ we have
\begin{align}
	a^\dag_\mathbf{p} |0\rangle
	=
	|\mathbf{p} \rangle_a 
	\qquad
	b^\dag_\mathbf{p} |0\rangle
	=
	|\mathbf{p} \rangle_b.
\end{align}
Since we already computed the translation charges (see eq. \eqref{timetrans}, \eqref{spatialtrans}), we can use them to obtain what is the eigenvalue of the above momentum eigenstates. For example, recalling eq. \eqref{Symp15-t}, \eqref{Symp16-t}, we have
\begin{align}
	\mathcal{P}_0 |\mathbf{q}\rangle_a
	&=
	\int \, d^3p \, \alpha [-S(\omega_\mathbf{p}) a^\dag_\mathbf{p} a_\mathbf{p}] a^\dag_\mathbf{q}|0\rangle \nonumber \\
	&=
	\int \, d^3p \, \alpha [-S(\omega_\mathbf{p})\frac{1}{\alpha} \delta(\mathbf{p}-\mathbf{q})] a^\dag_\mathbf{p} |0\rangle \nonumber \\
	&=
	-S(\omega_\mathbf{p}) |\mathbf{p}\rangle_a
\end{align}
and in the same way one obtains
\begin{align}
	\mathcal{P}_i |\mathbf{p}\rangle_a
	=
	-S(\mathbf{p})_i |\mathbf{p}\rangle_a
	\qquad
	\mathcal{P}_i |\mathbf{p}\rangle_b
	=
	\mathbf{p}_i |\mathbf{p}\rangle_b \\
	\mathcal{P}_0 |\mathbf{p}\rangle_a
	=
	-S(\omega_\mathbf{p}) |\mathbf{p}\rangle_a
	\qquad
	\mathcal{P}_0 |\mathbf{p}\rangle_b
	=
	\omega_\mathbf{p} |\mathbf{p}\rangle_b.
\end{align}
We can now use the commutation relation $[\mathcal{N}_i, \mathcal{P}_j] = -i \eta_{ij} \mathcal{P}_0$ to show that 
\begin{align}
	-i\lambda^j	\mathcal{N}_j\,	\mathcal{P}_i |\mathbf{p}\rangle_a &=iS(p)_i \lambda^j	\mathcal{N}_j	 |\mathbf{p}\rangle_a =-i\mathcal{P}_i \,\lambda^j	\mathcal{N}_j	 |\mathbf{p}\rangle_a + \lambda_{i} S(\omega_\mathbf{p})|\mathbf{p}\rangle_a\\
	-i\lambda^j	\mathcal{N}_j\,	\mathcal{P}_0 |\mathbf{p}\rangle_a &=iS(\omega_\mathbf{p}) \lambda^j	\mathcal{N}_j	 |\mathbf{p}\rangle_a =-i\mathcal{P}_0 \,\lambda^j	\mathcal{N}_j	 |\mathbf{p}\rangle_a + \lambda^{i} S(p)_i |\mathbf{p}\rangle_a
\end{align}
which immediately imply 
\begin{equation}\label{Naction}
	-i  \lambda^j	\mathcal{N}_j	 |\mathbf{p}\rangle_a = |\mathbf{p}+ \bm{\lambda}\omega_\mathbf{p} \rangle_a
\end{equation}
which is the canonical action of an infinitesimal boosts. One can then obtain the finite boost, which is therefore the canonical one.

If we now define a wave packet using the fields in eq. \eqref{fa}, \eqref{fb} (we can concentrate on particles created by $a^\dag$, since the same considerations will be valid about those created by $b^\dag$), we have
\begin{align}
	\phi^\dag|0\rangle
	&=
	\int \frac{d^3p}{\sqrt{2\omega_\mathbf{p}}}
	\, \zeta(p) \,
	a^\dag_\mathbf{p} e^{-i(S(\omega_\mathbf{p}) - S(\mathbf{p})\mathbf{x})} |0\rangle \nonumber \\
	&=
	\int \frac{d^3p}{\sqrt{2\omega_\mathbf{p}}}
	\,\zeta(p) e^{-i(S(\omega_\mathbf{p}) - S(\mathbf{p})\mathbf{x})} |\mathbf{p}\rangle \nonumber \\
	&=
	|\varphi(x)\rangle
\end{align}
and acting now with a boost, since single particle states behave canonically, we will get a canonically boosted single particle wave packet. The $\varphi$ in this equation describes the packet obtained by an appropriate definition of $\zeta(p)$. On the other hand, if we first boost the field (obtaining eq. \eqref{LOphidagbo} or \eqref{LOphidagbo-conv2}, or any other expression based on different choices of $\zeta(p)$), and only then apply the boosted field to the vacuum state, we get a wave packet with a different distribution due to the presence of the additional factors in eq. \eqref{LOphidagbo}, \eqref{LOphidagbo-conv2}.

This is a puzzling property of boosts which needs to be more deeply understood. One of the possible solutions would be to understand whether there are differences between active and passive boosts. Indeed, one can consider the boost acting on the state $|\varphi(x)\rangle$ as a passive boost (the wave packet has already been created, and only later a boost is applied, which changes the coordinates describing the distribution). On the other hand, eq. \eqref{LOphidagbo}, \eqref{LOphidagbo-conv2} may be considered as active boosts (the field is boosted before any particle is created). In a passive boost, the particle does not need to exchange energy with the environment, since we (the observers) are moving at a different speed, describing what we see from a different point of view. On the other hand, to physically create an active boost on a field one needs to interact with it in some way, for example an electron can be actively boosted by switching on an electric field in order to accelerate it, and then switching it off, obtaining an electron in uniform rectilinear motion. In canonical, non-deformed QFT, the net result is the same and indeed active and passive transformations are completely equivalent. It is however possible that in the deformed context there may be different contributions in the two cases, leading to the difference just described. However in this context we are only considering free particles, so that such considerations would be premature. We will leave the investigation of this puzzle to forthcoming publications.




\paragraph{The role of $\zeta(p)$}\label{role-of-zeta}

Notice that in order to get a better idea for the boost of a field in the deformed context we needed to assume a specific value for $\zeta(p)$. In the above discussions, we used two natural definitions, namely eq. \eqref{convention} and \eqref{convention1}. At first sight, therefore, it would seem that the physical results of our model depend on our choice of the arbitrary factor $\zeta(p)$, which is of course absurd. 

This is however not the case. Notice, in fact, that the factor $\zeta(p)$ can be found inside $\alpha$ in eq. \eqref{alpha}, which is in turn present in the charges (up to now we only have the translation charges in eq. \eqref{timetrans}, \eqref{spatialtrans}, and the boost charge in eq. \eqref{defboostfromsymp}, but we will see that it also appears in the rotation charge), symplectic form (see eq. \eqref{Symp14}), and creation/annihilation operators algebra (eq. \eqref{Symp15-t}, \eqref{Symp16-t}). However, it also appears in the off-shell action in momentum space in eq. \eqref{Stot-momentumspace}. This means that if we sue the convention for $\zeta(p)$ in eq. \eqref{convention}, then the generators of the algebra are much simpler, but of course the off-shell action now has a global prefactor in the integrand given by 
\begin{align}
	\left(1+\frac{|p_+|^3}{\kappa^3}\right)
	\left\{
	\frac{1}{2\omega_p}
	\left[
	\frac{|p_+|^3}{\kappa^3}\omega_p - S(\omega_p)
	\right]\frac{p_4}{\kappa}
	\right\}^{-1}.
\end{align}
On the other hand, with convention \eqref{convention1}, the global prefactor in the momentum-space off-shell action in eq. \eqref{Stot-momentumspace} is 1, but the charges now contain a global prefactor in the integrand
\begin{align}
	\left(1+\frac{|p_+|^3}{\kappa^3}\right)^{-1}
	\frac{1}{2\omega_p}
	\left[
	\frac{|p_+|^3}{\kappa^3}\omega_p - S(\omega_p)
	\right]\frac{p_4}{\kappa}.
\end{align}
Therefore, there is a do ut des scenario between the algebra generators and the off-shell momentum-space action, which is important because, as we will see in chapter \ref{Chapter3}, it is used in the definition of the propagator by using the path integral in the $\kappa$-deformed context. 

In other words, if we are in the situation described by eq. \eqref{convention}, then after a boost particles and antiparticles receive the same additional translation, but their propagator is modified by some non-trivial prefactor. On the other hand, we can choose to work with convention given by eq. \eqref{convention1}, in which case after a boost particles and antiparticles receive a different additional translation, but the propagator will be the traditional one. One may very well choose another convention, in which case the behaviour of particles and antiparticles under boosts and their propagator have mixed features, in between the two extremal case discussed above.

\subsubsection{Rotations}\label{rotsec}

For rotations, we can proceed in the same manner as for boost, and we only need to understand how to write $\delta_R a$ and $\delta_R a^\dag$ (the case of $b$ and $b^\dag$ will be analogous). 
We start from
\begin{align}\label{dlorentz}
	\delta^R a_\mathbf{p}
	&=
	i \epsilon^{ijk} \rho_i
	\mathbf{p}_{j} \frac{\partial}{\partial \mathbf{p}^{k}}
	a_\mathbf{p} 
	\qquad
	\delta^R a_\mathbf{p}^\dag
	=
	i \epsilon^{ijk} \rho_i
	S(\mathbf{p})_{j} \frac{\partial}{\partial S(\mathbf{p})^{k}}
	a_\mathbf{p}^\dag \nonumber \\
	\delta^R b_\mathbf{p}
	&=
	i \epsilon^{ijk} \rho_i
	\mathbf{p}_{j} \frac{\partial}{\partial \mathbf{p}^{k}}
	b_\mathbf{p} 
	\qquad
	\delta^R b_\mathbf{p}^\dag
	=
	i \epsilon^{ijk} \rho_i
	S(\mathbf{p})_{j} \frac{\partial}{\partial S(\mathbf{p})^{k}}
	b_\mathbf{p}^\dag
\end{align} 
and we have
\begin{align}
	-\delta_R \lrcorner \, \Omega
	&=
	i\int d^3p \,\alpha \,
	(
	\delta_{\partial_R} a_\mathbf{p}^\dag
	\delta a_\mathbf{p}\,
	+
	\delta a_\mathbf{p}^\dag
	S(\delta_{\partial_R}) a_\mathbf{p} \nonumber \\
	&-
	\delta_{\partial_R} b_{\mathbf{p}}
	\delta b_{\mathbf{p}}^\dag
	-
	\delta b_{\mathbf{p}}
	S(\delta_{\partial_R}) b_{\mathbf{p}}^\dag)  \\
	&=
	i \epsilon^{ij} \int d^3p \,\alpha \,
	\Big(
	S(\mathbf{p})_{[i} \frac{\partial a_\mathbf{p}^\dag}{\partial S(\mathbf{p})^{j]}} 
	\delta a_\mathbf{p}\,
	-
	\delta a_\mathbf{p}^\dag \,
	S(\mathbf{p})_{[i} \frac{\partial a_\mathbf{p}}{\partial S(\mathbf{p})^{j]}}  \nonumber \\
	&-
	\mathbf{p}_{[i} \frac{\partial b_{\mathbf{p}}}{\partial \mathbf{p}^{j]}} 
	\delta b_{\mathbf{p}}^\dag
	+
	\delta b_{\mathbf{p}}
	\mathbf{p}_{[i} \frac{\partial b_{\mathbf{p}}^\dag}{\partial \mathbf{p}^{j]}}  
	\Big) \\
	&=
	i \epsilon^{ij} \delta \int d^3p \,\alpha \,
	\Big(
	S(\mathbf{p})_{[i} \frac{\partial a_\mathbf{p}^\dag}{\partial S(\mathbf{p})^{j]}} 
	a_\mathbf{p}\,
	-
	a_\mathbf{p}^\dag \,
	S(\mathbf{p})_{[i} \frac{\partial a_\mathbf{p}}{\partial S(\mathbf{p})^{j]}} \nonumber \\ 
	&-
	\mathbf{p}_{[i} \frac{\partial b_{\mathbf{p}}}{\partial \mathbf{p}^{j]}} 
	b_{\mathbf{p}}^\dag
	+
	b_{\mathbf{p}}
	\mathbf{p}_{[i} \frac{\partial b_{\mathbf{p}}^\dag}{\partial \mathbf{p}^{j]}}  
	\Big)
\end{align}
which give the rotation charge
\begin{align}\label{rotcharge}
	{\cal	M}_{i}=
	-\epsilon_i{}^{jk}
	\frac{1}{8}
	\int  d^3q \,
	\left(
	S(\mathbf{q})_{j}
	\frac{\partial a_{\mathbf{q}}^\dag}{\partial S(\mathbf{q})^{k}}
	a_{\mathbf{q}}
	-
	a_{\mathbf{q}}^\dag
	S(\mathbf{q})_{j}
	\frac{\partial a_{\mathbf{q}}}{\partial S(\mathbf{q})^{k}}
	+
	b_{\mathbf{q}}
	\mathbf{q}_{j}
	\frac{\partial b_{\mathbf{q}}^\dag}{\partial \mathbf{q}^{k}}
	-
	\mathbf{q}_{j}
	\frac{\partial b_{\mathbf{q}}}{\partial \mathbf{q}^{k}}
	b_{\mathbf{q}}^\dag
	\right).
\end{align}
The considerations in this case are the same as for the boost charges.

\subsection{Is it $\delta A \wedge \delta B = - \delta B \wedge \delta A$, or $\delta A \wedge \delta B = S(\delta B \wedge \delta A)$?}\label{dsf}

We are at point $vii)b)$ of our roadmap in section \ref{roadmap2}. 
We have introduced the antipode in the contraction rule in eq. \eqref{deformedcontraction}, but so far we have not discussed whether this inclusion is consistent with other rules concerning forms, vector fields, and their interaction. 

The first thing to tackle is the one reported in the title of this subsection. Indeed, in obtaining the symplectic form in eq. \eqref{Symp14}, we used the fact that $\delta A \wedge \delta B = - \delta B \wedge \delta A$ (here $A, B$ represent the quantities $a, a^\dag, b, b^\dag$). Furthermore, we used it in two ways. The first one was when we were taking the exterior derivative of an object like $(\delta B)A$. Indeed canonically we have
\begin{align}\label{non-def-wedge}
	\delta[(\delta B)A] = -\delta B \wedge \delta A
\end{align}
but we did not address whether this should instead be substituted by
\begin{align}\label{def-wedge}
	\delta[(\delta B)A] = S(\delta B \wedge \delta A).
\end{align}
The second time was when we added the two symplectic forms $\delta\theta_1$ and $\delta\theta_2$ in eq. \eqref{Symp12} and \eqref{Symp13} coming from the two action orderings, since one of the two had a global minus sign to compensate a global ordering change in the wedge products. In other words, to go from eq. \eqref{Symp12} and \eqref{Symp13} to eq. \eqref{Symp14} we used the fact that
\begin{align}\label{antisymm-wedge}
	\delta A \wedge \delta B
	=
	-\delta B \wedge \delta A.
\end{align}
Notice however that these two relations are not uncorrelated. Indeed eq. \eqref{non-def-wedge} can be considered as a consequence of eq. \eqref{antisymm-wedge} because the exterior derivative only acts on the first object, and then we switch places using eq. \eqref{antisymm-wedge}. In other words, we have
\begin{align}
	\delta (\delta B A)
	=
	\delta (A \delta B)
	=
	\delta A \wedge \delta B
	=
	-
	\delta B \wedge \delta A.
\end{align}

In what follows, we will consider what happens when we try to obtain a symplectic form using the assumption \eqref{def-wedge}. We will show that the only way to obtain an explicitly time-independent symplectic form is by imposing
\begin{align}\label{S-on-wed}
	S(\delta A \wedge \delta B) = -\delta A \wedge \delta B.
\end{align}
In other words, the antipode acting on wedge products (without any contraction) behaves like a canonical minus sign. We stress that the time independence of the symplectic form is a necessary requirement if one wants time-independent commutators between creation/annihilation operators, as well as time independent charges.

Indeed, using eq. \eqref{def-wedge}, and keeping in mind eq. \eqref{pre-Symp10} and \eqref{pre-Symp11}, one can easily see that the time dependent coefficients in both $\delta\theta_1$ and $\delta\theta_2$ are of the type\footnote{Here we discuss about only one of these time-dependent terms, but for all the others the same considerations hold.}
\begin{align}
	\left[\omega_\mathbf{q} \delta b_\mathbf{q} \wedge \delta a_\mathbf{p}
	-
	S(\omega_\mathbf{p})
	S(\delta b_\mathbf{q} \wedge \delta a_\mathbf{p})\right]
	e^{i(\mathbf{q}\oplus \mathbf{p})\mathbf{x}}
	e^{-i(\omega_\mathbf{q}\oplus\omega_\mathbf{p})t}.
\end{align}
We know from eq. \eqref{firstoffdiagdel} that the Dirac delta which comes out of the exponential imposes $\omega_\mathbf{q}=-S(\omega_\mathbf{p})$, so that the time-dependent term becomes 
\begin{align}
	-
	S(\omega_\mathbf{p})
	\left[\delta b_{S(\mathbf{p})} \wedge \delta a_\mathbf{p}
	+
	S(\delta b_{S(\mathbf{p})} \wedge \delta a_\mathbf{p})\right]
	e^{-i([-S(\omega_\mathbf{p})]\oplus\omega_\mathbf{p})t}.
\end{align}
Notice that, without the assumption in eq. \eqref{def-wedge}, this term would be identically zero. If instead we use the assumption in eq. \eqref{def-wedge}, then this time-dependent term\footnote{Crucially $[-S(\omega_\mathbf{p})]\oplus\omega_\mathbf{p} \neq 0$, this can be easily realized by noticing that the $\kappa\rightarrow\infty$ of this sum is $\omega_\mathbf{p}+\omega_\mathbf{p} \neq 0$.} would remain, unless
\begin{align}\label{possiblesolution}
	\boxed{
	S(\delta a_\mathbf{q} \wedge \delta b_{S(\mathbf{q})}\,)
	=
	-
	\delta a_\mathbf{q} \wedge \delta b_{S(\mathbf{q})}
}
\end{align}
i.e. if the antipode acting on a wedge product is exactly the same as the canonical minus acting on it, which is what we wanted to show.

\subsection{And what if we add contractions with vector fields?} \label{dercreannalg}

In the previous subsections we have seen that 
\begin{align}\label{fundassum}
	\delta B \wedge \delta A 
	= 
	S(\delta A \wedge \delta B)
	=
	-
	\delta A \wedge \delta B.
\end{align}
Eq. \eqref{fundassum} however, only deals with wedge products in the absence of contraction with vector fields. In case we need to contract with some vector fields, because of the antipode in eq. \eqref{deformedcontraction}, one needs to take care of whether to use the antipode or the canonical minus sign when changing the order of the wedge product involved in a contraction. 

Given any two functions $f,g$ in phase space, their Poisson bracket is defined as \cite{sympbook}
\begin{align}\label{defPoisson}
	\{f,g\} = \omega(X_f, X_g)
\end{align}
where $X_f$ and $X_g$ are the vector fields generated by $f$ and $g$ respectively, i.e.
\begin{align}
	X_f \, \lrcorner \, \omega = -\delta f
	\qquad
	X_g \, \lrcorner \, \omega = -\delta g
\end{align}
and where $\omega$ is the symplectic 2-form (compare the above equations with eq. \eqref{chargesdef}). In our case, we will use the symplectic form \eqref{Symp14}. In this case the functions on momentum space would be the creation/annihilation operators themselves. 

For simplicity, we first concentrate on the $a, a^\dag$ part. In this case, recalling the definition \eqref{deformedcontraction} it is easy to see that, assuming the symplectic form in eq. \eqref{Symp14}, we have
\begin{align}
	X_{a_\mathbf{q}} = i\frac{1}{S(\alpha(q))} \frac{\partial}{\partial a_\mathbf{q}^\dag}
	\qquad
	X_{a^\dag_\mathbf{q}} = i \frac{1}{\alpha(q)} \frac{\partial}{\partial a_\mathbf{q}}
\end{align}
In fact we have
\begin{align}
	X_{a_\mathbf{q}} \, \lrcorner \, \Omega
	&=
	i \int d^3p \, \alpha(p) \,
	\left[
	i\frac{1}{S(\alpha(q))} \frac{\partial}{\partial a_\mathbf{q}^\dag} 
	\,\lrcorner\,
	(\delta a_\mathbf{p}\wedge
	\delta a_\mathbf{p}^\dag)
	\right] \\
	&=
	i \int d^3p \, \alpha(p) \,
	\left[
	0
	+
	\delta a_\mathbf{p}
	S\left(
		i\frac{1}{S(\alpha(q))} \frac{\partial }{\partial a_\mathbf{q}^\dag}
	\right)
	a_\mathbf{p}^\dag
	\right] \\
	&=
	- \int d^3p \, \alpha(p) \,
	\delta a_\mathbf{p}
	\frac{1}{\alpha(q)} \underbrace{\frac{\partial a_\mathbf{p}^\dag}{\partial a_\mathbf{q}^\dag}}_{=\delta(\mathbf{p}-\mathbf{q})} \\
	&=
	-\delta a_\mathbf{q}
\end{align}
\begin{align}\label{Xadagdef}
	X_{a_\mathbf{q}^\dag} \, \lrcorner \, \Omega
	&=
	i \int d^3p \, \alpha(p) \,
	\left[
	i\frac{1}{\alpha(q)} \frac{\partial}{\partial a_\mathbf{q}} 
	\,\lrcorner\,
	(\delta a_\mathbf{p}\wedge
	\delta a_\mathbf{p}^\dag)
	\right] \\
	&=
	i \int d^3p \, \alpha(p) \,
	\left[
	i\frac{1}{\alpha(q)} \underbrace{\frac{\partial a_\mathbf{p}}{\partial a_\mathbf{q}}}_{=\delta(\mathbf{p}-\mathbf{q})}
	\delta a_\mathbf{p}^\dag
	+
	0
	\right] \\
	&=
	-\delta a_\mathbf{q}^\dag
\end{align}
Notice that here we used the assumption $S(a)=a$ and $S(a^\dag)=a^\dag$, so that the antipode does not act on the single creation/annihilation operator but only on their wedge product. There are some subtleties that need to be sorted out. 

First of all, notice that the vector field $X_{a_\mathbf{q}}$ is defined in such a way that it requires the application of the antipode to `work as intended'. For this reason, in order to get the correct value for the commutator $\{a, a^\dag\}$ we must first contract $\Omega$ with $X_{a_\mathbf{q}}$ and only after this we can contract with $X_{a_\mathbf{q}^\dag}$, and not vice versa. In this way, we immediately get 
\begin{align}\label{firstPoissonbrack}
	\{a_\mathbf{p}, a_\mathbf{q}^\dag\}
	=
	X_{a_\mathbf{q}^\dag}
	\,\lrcorner\,
	(X_{a_\mathbf{p}}
	\,\lrcorner\,
	\Omega
	)
	=
	-\frac{i}{\alpha}
	\delta(\mathbf{p}-\mathbf{q})
\end{align}
which is in accordance with the canonical result (see for example eq. \eqref{Symp15-simp}). It is important to notice that at first sight one has 
\begin{align}
	X_{a_\mathbf{q}^\dag}
	\,\lrcorner\,
	(X_{a_\mathbf{p}}
	\,\lrcorner\,
	\Omega
	)
	\neq
	-
	X_{a_\mathbf{p}}
	\,\lrcorner\,
	(X_{a_\mathbf{q}^\dag}
	\,\lrcorner\,
	\Omega
	)
\end{align}
since, as can be seen from eq. \eqref{Xadagdef}, we have
\begin{align}
	X_{a_\mathbf{p}}
	\,\lrcorner\,
	(X_{a_\mathbf{q}^\dag}
	\,\lrcorner\,
	\Omega
	)
	=
	-\frac{i}{S(\alpha)}
	\delta(\mathbf{p}-\mathbf{q}).
\end{align}
However, as explained above, intuitively speaking one can say that in this case the vector field $X_{a_\mathbf{q}}$ is not `working in the way it was built to work', since it requires the application of an antipode. More technically, keeping in mind eq. \eqref{deformedcontraction}, notice that when we are contracting with vector fields, we are not free in general to switch the order of the wedge products unless we use the antipode. An example of this fact is given by the computation of the charges using the symplectic form. Take for example the time translation charge (the computations of the charges are done in section \ref{chargesfromsympform}, here we just use an example for the argument's sake). The symplectic form is given by \eqref{Symp14}, and the creation/annihilation operator transformations are given by \eqref{defodeltat}, and at the end we get the charge in eq. \eqref{timetrans}. 

However, if one wants to change the order of the wedge products involved in a contraction with an external vector field (this point is crucial), then one gets exactly the same result if one uses eq. \eqref{fundassum} (with the antipode). Indeed, using the symplectic form (consider just the $a, a^\dag$ case and ignoring the factor $\alpha$ for simplicity)
\begin{align}
	\Omega
	=
	-i
	\int d^3p \,\alpha \,
	S(\delta a_\mathbf{p}\wedge \delta a_\mathbf{p}^\dag)
\end{align}
then the charge becomes
\begin{align}
	-\delta_{\partial_0} \lrcorner \, \Omega
	&=
	i \int d^3p \,\,\,\alpha \,
	S(\delta_{\partial_0}) a_\mathbf{p}
	\delta a_\mathbf{p}\,
	+
	\delta a_\mathbf{p}
	S(S(\delta_{\partial_0})) a_\mathbf{p}^\dag \\
	&=
	i \int d^3p \,\,\,\alpha \,
	S(\delta_{\partial_0}) a_\mathbf{p}
	\delta a_\mathbf{p}\,
	+
	\delta a_\mathbf{p}
	\delta_{\partial_0} a_\mathbf{p}^\dag \\
	&=
	-\epsilon
	\int d^3p \,\,\,\alpha \,
	S(\omega_p) a_\mathbf{p}
	\delta a_\mathbf{p}\,
	+
	\delta a_\mathbf{p}
	S(\omega_p) a_\mathbf{p}^\dag \\
	&=
	\epsilon \delta
	\int d^3p \,\,\,\alpha \,
	[-S(\omega_p) a_\mathbf{p} a_\mathbf{p}^\dag]
\end{align}
which gives exactly the same result. Notice that, in the presence of a contraction with a vector field, the prescription in eq. \eqref{fundassum} needs means that the antipode is applied \textit{last}, meaning that 
\begin{align}\label{fundassmucontr}
	X \, \lrcorner \, (\delta A \wedge \delta B)
	=
	S(X \, \lrcorner \, (\delta B \wedge \delta A)).
\end{align}
In other words, given the contraction
\begin{align}\label{fundassmucontr-2}
	X \, \lrcorner \, (\delta A \wedge \delta B)
	=
	(XA) \delta B + \delta A(S(X)B)
\end{align}
we define 
\begin{align}\label{fundassmucontr-3}
	\boxed{S(X \, \lrcorner \, (\delta B \wedge \delta A))
	=
	(S(X)A)\delta B + \delta A (XB)}.
\end{align}
Notice that, under the assumption that $S$ does not act on $A, \delta A, B, \delta B$ individually (in this section they represent $a, b, a^\dag, b^\dag$), eq. \eqref{fundassmucontr} explains the above computations regarding charges. Alternatively, one could drop the assumption that $S(\cdot)$ does not act on $A, \delta A, B, \delta B$ individually, and simply define 
\begin{align}
	\boxed{X\, \lrcorner \, (\delta A \wedge \delta B) = S(X) \, \lrcorner \, (\delta B \wedge \delta A)}.
\end{align}

In light of these considerations, and also keeping in mind that eq. \eqref{fundassum} implies that 
\begin{align}\label{fundassumod}
	\delta A \wedge \delta B
	=
	-S
	(\delta A \wedge \delta B)
\end{align} 
and, if we apply the same point of view that antipodes should be applied last as before, we see that there are several ways to define the contraction of a 2-form with two vector fields. In particular, since $X_{a_\mathbf{q}}$ has been designed to account for the presence of an antipode, and since as shown in eq. \eqref{fundassmucontr} the antipode is applied last, 
notice that we have
\begin{align}
	-S(X_{a_\mathbf{p}}
	\,\lrcorner\,
	(X_{a_\mathbf{q}^\dag}
	\,\lrcorner\,
	\Omega
	))
	=
	\frac{i}{S(S(\alpha))}
	\delta(\mathbf{p}-\mathbf{q})
	=
	\frac{i}{\alpha}
	\delta(\mathbf{p}-\mathbf{q})
\end{align}
which gives the correct value for the Poisson bracket defined as
\begin{align}\label{inverseorder}
	\{a_\mathbf{q}^\dag, a_\mathbf{p}\}
	=
	-S(X_{a_\mathbf{p}}
	\,\lrcorner\,
	(X_{a_\mathbf{q}^\dag}
	\,\lrcorner\,
	\Omega
	))
	=
	\frac{i}{\alpha}
	\delta(\mathbf{p}-\mathbf{q}).
\end{align}
Therefore, once a prescription for the contraction of two vector fields with a 2-form is given, when switching the order of contractions one needs to impose an additional global $-S$ to the computations. Notice that we indeed recover the property
\begin{align}
	\{a_\mathbf{q}^\dag, a_\mathbf{p}\}
	=
	-
	\{a_\mathbf{p}, a_\mathbf{q}^\dag\}
\end{align}

Of course, if instead of starting from the symplectic form in eq. \eqref{Symp14} we started from the one in eq. \eqref{sign1}, then the vector fields $X_{a_\mathbf{q}}, X_{a_\mathbf{q}^\dag}$ would now be given by
\begin{align}
	X_{a_\mathbf{q}} = -i\frac{1}{\alpha(q)} \frac{\partial}{\partial a_\mathbf{q}^\dag}
	\qquad
	X_{a^\dag_\mathbf{q}} = -i \frac{1}{S(\alpha(q))} \frac{\partial}{\partial a_\mathbf{q}}
\end{align}
In fact we would have
\begin{align}
	X_{a_\mathbf{q}} \, \lrcorner \, \Omega
	&=
	-i \int d^3p \, \alpha(p) \,
	\left[
	-i\frac{1}{\alpha(q)} \frac{\partial}{\partial a_\mathbf{q}^\dag} 
	\,\lrcorner\,
	(\delta a_\mathbf{p}^\dag\wedge
	\delta a_\mathbf{p})
	\right] \\
	&=
	- \int d^3p \, \alpha(p) \,
	\delta a_\mathbf{p}
	\frac{1}{\alpha(q)} \underbrace{\frac{\partial a_\mathbf{p}^\dag}{\partial a_\mathbf{q}^\dag}}_{=\delta(\mathbf{p}-\mathbf{q})} \\
	&=
	-\delta a_\mathbf{q}
\end{align}
\begin{align}
	X_{a_\mathbf{q}^\dag} \, \lrcorner \, \Omega
	&=
	-i \int d^3p \, \alpha(p) \,
	\left[
	-i\frac{1}{S(\alpha(q))} \frac{\partial}{\partial a_\mathbf{q}} 
	\,\lrcorner\,
	(\delta a_\mathbf{p}^\dag\wedge
	\delta a_\mathbf{p})
	\right] \\
	&=
	-i \int d^3p \, \alpha(p) \,
	\left[
	0
	+
	\delta a_\mathbf{p}^\dag
	S\left(
	-i\frac{1}{S(\alpha(q))} \frac{\partial}{\partial a_\mathbf{q}} 
	\right)
	\delta a_\mathbf{p}
	\right] \\
	&=
	-i \int d^3p \, \alpha(p) \,
	\left[
	0
	-
	i\frac{1}{\alpha(q)} \underbrace{\frac{\partial a_\mathbf{p}}{\partial a_\mathbf{q}}}_{=\delta(\mathbf{p}-\mathbf{q})}
	\delta a_\mathbf{p}^\dag
	\right] \\
	&=
	-\delta a_\mathbf{q}^\dag
\end{align}
Once again, following the fact that this time $X_{a_\mathbf{q}^\dag}$ has been defined to accommodate for the antipode, it must be contracted first, so that we end up with 
\begin{align}
	\{a^\dag_\mathbf{p}, a_\mathbf{q}\}
	=
	X_{a_\mathbf{q}}
	\,\lrcorner\,
	(X_{a_\mathbf{p}^\dag}
	\,\lrcorner\,
	\Omega
	)
	=
	\frac{i}{\alpha}
	\delta(\mathbf{p}-\mathbf{q})
\end{align}
which is in perfect agreement with eq. \eqref{inverseorder}. Once again, if we switch the order of contraction, we need to add a global $-S$, and indeed one has
\begin{align}
	\{a_\mathbf{p}, a_\mathbf{q}^\dag\}
	=
	-
	S(
	X_{a_\mathbf{q}^\dag}
	\,\lrcorner\,
	(X_{a_\mathbf{p}}
	\,\lrcorner\,
	\Omega
	))
	=
	-\frac{i}{\alpha}
	\delta(\mathbf{p}-\mathbf{q})
\end{align}
which is in accordance with eq. \eqref{firstPoissonbrack}. Of course, the same reasoning gives the correct propagators for $b, b^\dag$, and all the cross-commutators involving mixing of $a, a^\dag$ with $b, b^\dag$ are zero. Furthermore, one also gets $\{a,a\} = \{b,b\} = 0$.

To clarify and summarize the above discussion, we write the main steps below:
\begin{itemize}
	\item[1)] We start from some symplectic form $\omega$ and from two quantities $f,g$ of which we want to compute the Poisson brackets. For the moment, we restrict our attention to the case when $f,g$ are some creation/annihilation operators which are part of the phase space of the theory (and therefore $\delta f, \delta g$ appear somewhere in the definition of $\omega$);
	\item[2)] Use the definitions $X_f \, \lrcorner \, \omega = -\delta f$ and $X_g \, \lrcorner \, \omega = -\delta g$ to obtain the vector fields $X_g, X_f$;
	\item[3)] The definition of the vector fields themselves will imply that there is one correct order of contraction of $X_g, X_f$ with $\omega$. Using this contraction, one defines the relevant Poisson bracket, say for example $\{f,g\} = X_g \,\lrcorner\, (X_f \,\lrcorner\, \Omega)$;
	\item[4)] We now want to relate the brackets $\{f,g\}$ and $\{g,f\}$. Notice that, because of the previous point, only $\{f,g\}$ contains the appropriate contractions of $X_f, X_g$ with $\omega$. In order to correctly define $\{g,f\}$, therefore, we make use of a global $-S$ to be added to the reversed contraction, defining $\{g,f\} = -S(X_f \,\lrcorner\, (X_g \,\lrcorner\, \Omega))$, i.e. 
	\begin{align}
		\boxed{\{f,g\} = X_g \,\lrcorner\, (X_f \,\lrcorner\, \Omega)
		\qquad
		\Leftrightarrow
		\qquad
		\{g,f\} = -S(X_f \,\lrcorner\, (X_g \,\lrcorner\, \Omega))}.
	\end{align}
	Notice of course that this gives the correct $\kappa\rightarrow \infty$ limit. 
	\item[5)] When $f,g$ are creation/annihilation operators, the above procedure ensures that one gets the crucial property $\{f,g\} = -\{g,f\}$, as well as the correct value for the Poisson brackets. 
\end{itemize}

\section{The algebra of the $\kappa$-deformed charges}\label{chargesalgebra}

We are now at point $viii)$ of our roadmap in section \ref{roadmap2}. We now verify that the translation, rotation, and boost charges, which are respectively eq. \eqref{timetrans}, \eqref{spatialtrans}, \eqref{rotcharge}, \eqref{defboostfromsymp}, satisfy the canonical, non-deformed Poincar\'e algebra. 
\begin{align}
	[\mathcal{M}_i, \mathcal{P}_j] = i \epsilon_{ijk} \mathcal{P}_k
	\qquad 
	[\mathcal{M}_i, \mathcal{P}_0] = 0
	\qquad 
	[\mathcal{N}_i, \mathcal{P}_j] = -i \eta_{ij} \mathcal{P}_0
	\qquad
	[\mathcal{N}_i, \mathcal{P}_0] = - i \mathcal{P}_i
\end{align}
\begin{align}
	[\mathcal{M}_i, \mathcal{M}_j] = i \epsilon_{ijk} \mathcal{M}_k
	\qquad
	[\mathcal{M}_i, \mathcal{N}_j] = i \epsilon_{ijk} \mathcal{N}_k
	\qquad
	[\mathcal{N}_i, \mathcal{N}_j] = -i \epsilon_{ijk} \mathcal{M}_k
\end{align}
Recall that we are using the convention $\eta = diag(+---)$.

Since the computations are long but straightforward, we will only explicitly verify two of these commutators, but the rest of them can be verified using the same exact procedures.

\subsubsection{$[\mathcal{N}_i, \mathcal{P}_j] = - i \eta_{ij} \mathcal{P}_0$}

The commutator $[\mathcal{N}_i, \mathcal{P}_j]$ contains two components, i.e. 
\begin{align}
	-i\alpha(p)\alpha(q) S(\omega_p) S(\mathbf{q})_j
	\Bigg\{
	\left[
	\frac{\partial a_{\mathbf{p}}^\dag}{\partial S(\mathbf{p})^i} a_{\mathbf{p}},
	a_{\mathbf{q}}^\dag a_{\mathbf{q}}
	\right]
	-
	\left[
	a_{\mathbf{p}}^\dag
	\frac{\partial a_{\mathbf{p}}}{\partial S(\mathbf{p})^i} ,
	a_{\mathbf{q}}^\dag a_{\mathbf{q}} 
	\right]
	\Bigg\}
\end{align}
\begin{align}
	i\, \alpha(p)\alpha(q) \omega_p \mathbf{q}_j
	\Bigg\{
	\left[
	b_{\mathbf{p}}
	\frac{\partial b_{\mathbf{p}}^\dag}{\partial \mathbf{p}^i}  ,
	b_{\mathbf{q}}b_{\mathbf{q}}^\dag 
	\right]
	-
	\left[
	\frac{\partial  b_{\mathbf{p}}}{\partial \mathbf{p}^i}
	b_{\mathbf{p}}^\dag
	,
	b_{\mathbf{q}}b_{\mathbf{q}}^\dag 
	\right]
	\Bigg\}
\end{align}
Starting from the first one, we have
\begin{align}
	&-i\alpha(p)\alpha(q) S(\omega_p) S(\mathbf{q})_j
	\left[
	\frac{\partial a_{\mathbf{p}}^\dag}{\partial S(\mathbf{p})^i} a_{\mathbf{p}},
	a_{\mathbf{q}}^\dag a_{\mathbf{q}} 
	\right] \\
	&=
	-i\alpha(p)\alpha(q) S(\omega_p) S(\mathbf{q})_j
	\Bigg\{
	\frac{\partial a_{\mathbf{p}}^\dag}{\partial S(\mathbf{p})^i} \underbrace{[a_{\mathbf{p}},
		a_{\mathbf{q}}^\dag]}_{=\frac{1}{\alpha}\delta(\mathbf{p}-\mathbf{q})}
	a_{\mathbf{q}}
	+
	a_{\mathbf{q}}^\dag
	\underbrace{\left[
		\frac{\partial a_{\mathbf{p}}^\dag}{\partial S(\mathbf{p})^i}
		,
		a_{\mathbf{q}}
		\right]}_{= -\frac{1}{\alpha} \frac{\partial}{\partial S(\mathbf{p})^i} \delta(\mathbf{q}-\mathbf{p})}
	a_{\mathbf{p}}
	\Bigg\} \\
	&=
	-i\alpha(p) S(\omega_p) S(\mathbf{q})_j
	\frac{\partial a_{\mathbf{p}}^\dag}{\partial S(\mathbf{p})^i} \delta(\mathbf{p}-\mathbf{q})
	a_{\mathbf{q}}\nonumber \\
	&+
	i\alpha(p) S(\omega_p) S(\mathbf{q})_j
	a_{\mathbf{q}}^\dag
	\left(
	\frac{\partial}{\partial S(\mathbf{p})^i} \delta(\mathbf{q}-\mathbf{p})
	\right)
	a_{\mathbf{p}}\label{interm}
\end{align}
Integrating by parts the last term in the last passage above we get
\begin{align}
	-
	i\alpha(p) S(\omega_p) S(\mathbf{q})_j
	a_{\mathbf{q}}^\dag
	\delta(\mathbf{q}-\mathbf{p})
	\frac{\partial}{\partial S(\mathbf{p})^i}
	a_{\mathbf{p}} 
	-
	i
	\frac{\partial (\alpha(p) S(\omega_p))}{\partial S(\mathbf{p})^i}
	S(\mathbf{q})_j
	a_{\mathbf{q}}^\dag
	\delta(\mathbf{q}-\mathbf{p})
	a_{\mathbf{p}} 
\end{align}
and substituting back in eq. \eqref{interm} we get (we can now safely apply the Dirac delta since no derivative acts on it)
\begin{align}
&
	\left(
	-i\alpha(p) S(\omega_p) S(\mathbf{p})_j
	\frac{\partial }{\partial S(\mathbf{p})^i}
	(a_{\mathbf{p}}^\dag 
	a_{\mathbf{p}})
	-
	i
	\frac{\partial (\alpha(p) S(\omega_p))}{\partial S(\mathbf{p})^i}
	S(\mathbf{p})_j
	a_{\mathbf{p}}^\dag
	a_{\mathbf{p}} 
	\right)
	\delta(\mathbf{q}-\mathbf{p}) \\
	&=
	\Bigg(
	-i
	\frac{\partial }{\partial S(\mathbf{p})^i}
	(\alpha(p) S(\omega_p) S(\mathbf{p})_j a_{\mathbf{p}}^\dag 
	a_{\mathbf{p}})
	+
	i
	S(\mathbf{p})_j 
	a_{\mathbf{p}}^\dag 
	a_{\mathbf{p}}
	\frac{\partial }{\partial S(\mathbf{p})^i}
	(\alpha(p) S(\omega_p) ) \nonumber \\
	&
	+
	i
	\alpha(p) S(\omega_p)
	\eta_{ij}
	a_{\mathbf{p}}^\dag 
	a_{\mathbf{p}}
	-
	i
	\frac{\partial (\alpha(p) S(\omega_p))}{\partial S(\mathbf{p})^i}
	S(\mathbf{p})_j
	a_{\mathbf{p}}^\dag
	a_{\mathbf{p}} 
	\Bigg)
	\delta(\mathbf{q}-\mathbf{p}) \\
	&=
	i
	\alpha(p) S(\omega_p)
	\eta_{ij}
	a_{\mathbf{p}}^\dag 
	a_{\mathbf{p}}
	\delta(\mathbf{q}-\mathbf{p})
\end{align}
where in the last passage we ignored $\frac{\partial }{\partial S(\mathbf{p})^i}
(\alpha(p) S(\omega_p) S(\mathbf{p})_j a_{\mathbf{p}}^\dag 
a_{\mathbf{p}})$ since it is a surface term after the application of the Dirac delta. 

We can analogously proceed for the second term, as follows 
\begin{align}
	&i\, \alpha(p)\alpha(q) \omega_p \mathbf{q}_j
	\left[
	b_{\mathbf{p}}
	\frac{\partial b_{\mathbf{p}}^\dag}{\partial \mathbf{p}^i}  ,
	b_{\mathbf{q}}b_{\mathbf{q}}^\dag 
	\right] \\
	&=
	i\, \alpha(p)\alpha(q) \omega_p \mathbf{q}_j
	\Bigg\{
	b_{\mathbf{p}}
	\underbrace{\left[
		\frac{\partial b_{\mathbf{p}}^\dag}{\partial \mathbf{p}^i}  ,
		b_{\mathbf{q}}
		\right]}_{=-\frac{1}{\alpha} \frac{\partial}{\partial \mathbf{p}^i}\delta(\mathbf{q}-\mathbf{p})}
	b_{\mathbf{q}}^\dag 
	+
	0
	+
	0
	+
	b_{\mathbf{q}}
	\underbrace{[b_{\mathbf{p}},
		b_{\mathbf{q}}^\dag]}_{=\frac{1}{\alpha}\delta(\mathbf{p}-\mathbf{q})}
	\frac{\partial b_{\mathbf{p}}^\dag}{\partial \mathbf{p}^i}
	\Bigg\} \\
	&=
	+
	i\, \alpha(p) \omega_p \mathbf{q}_j
	b_{\mathbf{q}}
	\delta(\mathbf{p}-\mathbf{q})
	\frac{\partial b_{\mathbf{p}}^\dag}{\partial \mathbf{p}^i}
	+
	i\, \alpha(p) \omega_p \mathbf{q}_j
	\frac{\partial b_{\mathbf{p}}}{\partial \mathbf{p}^i}
	\delta(\mathbf{q}-\mathbf{p})
	b_{\mathbf{q}}^\dag \nonumber \\
	&+
	i\,  \mathbf{q}_j
	\frac{\partial (\alpha(p) \omega_p)}{\partial \mathbf{p}^i}
	b_{\mathbf{p}}
	\delta(\mathbf{q}-\mathbf{p})
	b_{\mathbf{q}}^\dag \\
	&=
	\left(
	i\,  \mathbf{p}_j
	\frac{\partial (\alpha(p) \omega_p)}{\partial \mathbf{p}^i}
	b_{\mathbf{p}}
	b_{\mathbf{p}}^\dag 
	+
	i\, \alpha(p) \omega_p \mathbf{q}_j
	\frac{\partial }{\partial \mathbf{p}^i}
	(b_{\mathbf{p}}
	b_{\mathbf{p}}^\dag )
	\right)
	\delta(\mathbf{p}-\mathbf{q}) \\
	&=
	\Big(
	i\,  
	\frac{\partial }{\partial \mathbf{p}^i}
	(\mathbf{p}_j \alpha(p) \omega_p b_{\mathbf{p}}
	b_{\mathbf{p}}^\dag)
	-
	i\,  \mathbf{p}_j \alpha(p) \omega_p 
	\frac{\partial }{\partial \mathbf{p}^i}
	(b_{\mathbf{p}}
	b_{\mathbf{p}}^\dag)
	-
	i\, \eta_{ij}  
	\alpha(p) \omega_p b_{\mathbf{p}}
	b_{\mathbf{p}}^\dag \nonumber \\
	&+
	i\, \alpha(p) \omega_p \mathbf{q}_j
	\frac{\partial }{\partial \mathbf{p}^i}
	(b_{\mathbf{p}}
	b_{\mathbf{p}}^\dag )
	\Big)
	\delta(\mathbf{p}-\mathbf{q}) \\
	&=
	-
	i\, \eta_{ij}  
	\alpha(p) \omega_p b_{\mathbf{p}}
	b_{\mathbf{p}}^\dag
	\delta(\mathbf{p}-\mathbf{q}).
\end{align}
In summary, we have shown that 
\begin{align}
	[N_i, \mathcal{P}_j]
	&=
	\int
	d^3p \, d^3q \,
	\left(
	i
	\alpha(p) S(\omega_p)
	\eta_{ij}
	a_{\mathbf{p}}^\dag 
	a_{\mathbf{p}}
	-
	i\, \eta_{ij}  
	\alpha(p) \omega_p b_{\mathbf{p}}
	b_{\mathbf{p}}^\dag
	\right)
	\delta(\mathbf{q}-\mathbf{p}) \\
	&=
	-i \eta_{ij} \mathcal{P}_0
\end{align}
which is what we wanted to show.

\subsubsection{$[\mathcal{N}_i, \mathcal{N}_j] = -i\epsilon_{ijk} \mathcal{M}_k$}

The commutator $[\mathcal{N}_i, \mathcal{N}_j]$ contains two parts, the $a, a^\dag$ part and the $b, b^\dag$ part. Starting with the one concerning the $a$'s we have
\begin{align}
	&-\alpha(p)\alpha(q)S(\omega_p)S(\omega_q)
	\Bigg[
	\Bigg\{
	\frac{\partial a_{\mathbf{p}}^\dag}{\partial S(\mathbf{p})^i}
	a_{\mathbf{p}}
	-
	a_{\mathbf{p}}^\dag
	\frac{\partial a_{\mathbf{p}}}{\partial S(\mathbf{p})^i}
	\Bigg\}
	,
	\Bigg\{
	\frac{\partial a_{\mathbf{q}}^\dag}{\partial S(\mathbf{q})^j}
	a_{\mathbf{q}}
	-
	a_{\mathbf{q}}^\dag
	\frac{\partial a_{\mathbf{q}}}{\partial S(\mathbf{q})^j}
	\Bigg\}
	\Bigg] \\
	=
	&-\alpha(p)\alpha(q)S(\omega_p)S(\omega_q)
	\Bigg\{
	\Bigg[
	\frac{\partial a_{\mathbf{p}}^\dag}{\partial S(\mathbf{p})^i}
	a_{\mathbf{p}}
	,
	\frac{\partial a_{\mathbf{q}}^\dag}{\partial S(\mathbf{q})^j}
	a_{\mathbf{q}}
	\Bigg]
	+
	\Bigg[
	\frac{\partial a_{\mathbf{p}}^\dag}{\partial S(\mathbf{p})^i}
	a_{\mathbf{p}}
	,
	-
	a_{\mathbf{q}}^\dag
	\frac{\partial a_{\mathbf{q}}}{\partial S(\mathbf{q})^j}
	\Bigg] \nonumber \\
	&+
	\Bigg[
	-
	a_{\mathbf{p}}^\dag
	\frac{\partial a_{\mathbf{p}}}{\partial S(\mathbf{p})^i}
	,
	\frac{\partial a_{\mathbf{q}}^\dag}{\partial S(\mathbf{q})^j}
	a_{\mathbf{q}}
	\Bigg]
	+
	\Bigg[
	-
	a_{\mathbf{p}}^\dag
	\frac{\partial a_{\mathbf{p}}}{\partial S(\mathbf{p})^i}
	,
	-
	a_{\mathbf{q}}^\dag
	\frac{\partial a_{\mathbf{q}}}{\partial S(\mathbf{q})^j}
	\Bigg]
	\Bigg\}
\end{align}
We will compute these commutators one by one starting from the first. Once again, we use 
\begin{align}
	[AB, CD]
	=
	A[B,C]D + [A,C]BD + CA[B,D] + C[A,D]B
\end{align}
and we get
\begin{align}
	&-\alpha(p)\alpha(q)S(\omega_p)S(\omega_q)
	\Bigg[
	\frac{\partial a_{\mathbf{p}}^\dag}{\partial S(\mathbf{p})^i}
	a_{\mathbf{p}}
	,
	\frac{\partial a_{\mathbf{q}}^\dag}{\partial S(\mathbf{q})^j}
	a_{\mathbf{q}}
	\Bigg] \\
	&=
	-\alpha(p)\alpha(q)S(\omega_p)S(\omega_q)
	\Bigg\{
	\frac{\partial a_{\mathbf{p}}^\dag}{\partial S(\mathbf{p})^i}
	\left[
	a_{\mathbf{p}}
	,
	\frac{\partial a_{\mathbf{q}}^\dag}{\partial S(\mathbf{q})^j}
	\right]
	a_{\mathbf{q}}
	+
	\frac{\partial a_{\mathbf{q}}^\dag}{\partial S(\mathbf{q})^j}
	\left[
	\frac{\partial a_{\mathbf{p}}^\dag}{\partial S(\mathbf{p})^i}
	,
	a_{\mathbf{q}}
	\right]
	a_{\mathbf{p}}
	\Bigg\} \\
	&=
	-\alpha(p)\alpha(q)S(\omega_p)S(\omega_q)
	\Bigg\{
	\frac{\partial a_{\mathbf{p}}^\dag}{\partial S(\mathbf{p})^i}
	\left[
	a_{\mathbf{p}}
	,
	\frac{\partial a_{\mathbf{q}}^\dag}{\partial S(\mathbf{q})^j}
	\right]
	a_{\mathbf{q}}
	-
	\frac{\partial a_{\mathbf{q}}^\dag}{\partial S(\mathbf{q})^j}
	\left[
	a_{\mathbf{q}}
	,
	\frac{\partial a_{\mathbf{p}}^\dag}{\partial S(\mathbf{p})^i}
	\right]
	a_{\mathbf{p}}
	\Bigg\}
\end{align}
Now we notice that the second term is just the first one with $\mathbf{p} \leftrightarrow \mathbf{q}$ and $i\leftrightarrow j$, but we can just rename the integration variables in such a way that the second one and the first are the same except with $i \leftrightarrow j$, and therefore the above integral becomes
\begin{align}\label{intermediatestep}
	-\alpha(p)\alpha(q)S(\omega_p)S(\omega_q)
	\frac{\partial a_{\mathbf{p}}^\dag}{\partial S(\mathbf{p})^{[i}}
	\left[
	a_{\mathbf{p}}
	,
	\frac{\partial a_{\mathbf{q}}^\dag}{\partial S(\mathbf{q})^{j]}}
	\right]
	a_{\mathbf{q}}
\end{align}
where we just used the antisymmetrized index to shorten the notation. Now we have a small subtlety related to the presence of the $\frac{1}{\alpha}$ in front of the delta coming from the commutator. In fact, recall from eq. \eqref{Symp15-t} that  
\begin{align}
	\left[
	a_\mathbf{p}, a_\mathbf{q}^\dag
	\right]
	&=
	\frac{1}{\alpha} 
	\delta(\mathbf{p}-\mathbf{q}).
\end{align}
However we have two choices, one where the $\frac{1}{\alpha}$ on the RHS depends on $p$ and the other where it depends on $q$. We are in the following situation
\begin{align}
	f(p,q) \alpha(p)\alpha(q) 
	\left[
	a_\mathbf{p},
	\frac{\partial a_\mathbf{q}^\dag}{\partial S(\mathbf{q})}
	\right]
	=
	f(p,q)
	\alpha(p)\alpha(q)
	\frac{\partial}{\partial S(\mathbf{q})}
	\left(
	\frac{1}{\alpha}
	\delta(\mathbf{p}-\mathbf{q})
	\right)
\end{align}
Now, if the $\alpha$ in the round brackets depends on $p$, then the above formula reduces to 
\begin{align}
	f(p,q)
	\alpha(q)
	\frac{\partial}{\partial S(\mathbf{q})}
	\delta(\mathbf{p}-\mathbf{q})
	\xrightarrow{\text{int. by parts and appl.} \delta}
	-
	\left(
	\frac{\partial}{\partial S(\mathbf{q})}
	f(q) \alpha(q)
	\right)
	\delta(\mathbf{p}-\mathbf{q})
\end{align}
However, if the $\alpha$ in the round brackets depends on $q$, then we have
\begin{align}
	&f(p,q)
	\alpha(p)\alpha(q)
	\frac{\partial}{\partial S(\mathbf{q})}
	\left(
	\frac{1}{\alpha}
	\delta(\mathbf{p}-\mathbf{q})
	\right) \\
	&=
	-
	f(p,q)
	\alpha(p)\alpha(q)
	\frac{1}{\alpha^2(q)}
	\frac{\partial \alpha(q)}{\partial S(\mathbf{q})}
	\delta(\mathbf{p}-\mathbf{q})
	+
	f(p,q)
	\alpha(p)
	\frac{\partial}{\partial S(\mathbf{q})}
	\delta(\mathbf{p}-\mathbf{q}) \\
	&=
	-
	f(q)
	\frac{\partial \alpha(q)}{\partial S(\mathbf{q})}
	\delta(\mathbf{p}-\mathbf{q})
	+
	f(p,q)
	\alpha(p)
	\frac{\partial}{\partial S(\mathbf{q})}
	\delta(\mathbf{p}-\mathbf{q}) \label{needed1}\\
	&=
	-
	f(q)
	\frac{\partial \alpha(q)}{\partial S(\mathbf{q})}
	\delta(\mathbf{p}-\mathbf{q})
	-
	\left(
	\frac{\partial}{\partial S(\mathbf{q})}
	f(p,q)
	\right)
	\alpha(p)
	\delta(\mathbf{p}-\mathbf{q}) \label{needed2}\\
	&=
	-
	\left(
	\frac{\partial}{\partial S(\mathbf{q})}
	f(q) \alpha(q)
	\right)
	\delta(\mathbf{p}-\mathbf{q})
\end{align}
where in eq. \eqref{needed1} we applied the Dirac delta in the first term (because it is not acted upon by any derivatives) and in eq. \eqref{needed2} we integrated by parts the second term on the RHS. One sees that both approaches lead to the same final result. Therefore for simplicity we will always consider the $\frac{1}{\alpha}$ in the definition of $[a, a^\dag]$ to depend on the variable that is not acted upon by the derivation inside the commutator.

Coming back to eq. \eqref{intermediatestep}, we have
\begin{align}
	&-\alpha(p)\alpha(q)S(\omega_p)S(\omega_q)
	\frac{\partial a_{\mathbf{p}}^\dag}{\partial S(\mathbf{p})^{[i}}
	\left[
	a_{\mathbf{p}}
	,
	\frac{\partial a_{\mathbf{q}}^\dag}{\partial S(\mathbf{q})^{j]}}
	\right]
	a_{\mathbf{q}} \\
	&=
	-\alpha(q)S(\omega_p)S(\omega_q)
	\frac{\partial a_{\mathbf{p}}^\dag}{\partial S(\mathbf{p})^{[i}}
	a_{\mathbf{q}}
	\frac{\partial }{\partial S(\mathbf{q})^{j]}}
	\delta(\mathbf{p}-\mathbf{q}) \\
	&=
	\frac{\partial \alpha(q)}{\partial S(\mathbf{q})^{[j}}
	S(\omega_q)^2
	\frac{\partial a_{\mathbf{q}}^\dag}{\partial S(\mathbf{q})^{i]}}
	a_{\mathbf{q}}
	+
	\alpha(q)
	S(\mathbf{q})_{[j}
	\frac{\partial a_{\mathbf{q}}^\dag}{\partial S(\mathbf{q})^{i]}}
	a_{\mathbf{q}}
	+
	\alpha(q)
	S(\omega_q)^2
	\frac{\partial a_{\mathbf{q}}^\dag}{\partial S(\mathbf{q})^{[i}}
	\frac{\partial a_{\mathbf{q}}}{\partial S(\mathbf{q})^{j]}} \label{NLstep}
\end{align}
We can now integrate by parts the first term of the last step, which (ignoring surface terms) will give us
\begin{align}
	&\frac{\partial \alpha(q)}{\partial S(\mathbf{q})^{[j}}
	S(\omega_q)^2
	\frac{\partial a_{\mathbf{q}}^\dag}{\partial S(\mathbf{q})^{i]}}
	a_{\mathbf{q}} \\
	&=
	-
	\alpha(q)
	2
	\frac{\partial S(\omega_q)}{\partial S(\mathbf{q})^{[j}}
	S(\omega_q)
	\frac{\partial a_{\mathbf{q}}^\dag}{\partial S(\mathbf{q})^{i]}}
	a_{\mathbf{q}}
	-
	\alpha(q)
	S(\omega_q)^2
	\frac{\partial a_{\mathbf{q}}^\dag}{\partial S(\mathbf{q})^{[i}}
	\frac{\partial a_{\mathbf{q}}}{\partial S(\mathbf{q})^{j]}}	\\
	&=
	-
	2
	\alpha(q)
	S(\mathbf{q})_{[j}
	\frac{\partial a_{\mathbf{q}}^\dag}{\partial S(\mathbf{q})^{i]}}
	a_{\mathbf{q}}
	-
	\alpha(q)
	S(\omega_q)^2
	\frac{\partial a_{\mathbf{q}}^\dag}{\partial S(\mathbf{q})^{[i}}
	\frac{\partial a_{\mathbf{q}}}{\partial S(\mathbf{q})^{j]}}. 
\end{align}
Notice that there is no term with wo derivatives because the antisymmetrization sends it to zero.
Substituting this back into eq. \eqref{NLstep} we finally obtain that all the terms go away with the exception of one of them, and we get
\begin{align}
	-\alpha(p)\alpha(q)S(\omega_p)S(\omega_q)
	\Bigg[
	\frac{\partial a_{\mathbf{p}}^\dag}{\partial S(\mathbf{p})^i}
	a_{\mathbf{p}}
	,
	\frac{\partial a_{\mathbf{q}}^\dag}{\partial S(\mathbf{q})^j}
	a_{\mathbf{q}}
	\Bigg]
	&=
	-
	\alpha(q)
	S(\mathbf{q})_{[j}
	\frac{\partial a_{\mathbf{q}}^\dag}{\partial S(\mathbf{q})^{i]}}
	a_{\mathbf{q}}\\
	&=
	\alpha(q)
	S(\mathbf{q})_{[i}
	\frac{\partial a_{\mathbf{q}}^\dag}{\partial S(\mathbf{q})^{j]}}
	a_{\mathbf{q}}
\end{align}
which is the expected result.

We now pass to the second and third commutator. We treat them together because it is easy to show that their sum is zero. In fact, we have
\begin{align}
	\Bigg\{
	\Bigg[
	\frac{\partial a_{\mathbf{p}}^\dag}{\partial S(\mathbf{p})^i}
	a_{\mathbf{p}}
	,
	-
	a_{\mathbf{q}}^\dag
	\frac{\partial a_{\mathbf{q}}}{\partial S(\mathbf{q})^j}
	\Bigg] 
	+
	\Bigg[
	-
	a_{\mathbf{p}}^\dag
	\frac{\partial a_{\mathbf{p}}}{\partial S(\mathbf{p})^i}
	,
	\frac{\partial a_{\mathbf{q}}^\dag}{\partial S(\mathbf{q})^j}
	a_{\mathbf{q}}
	\Bigg]
	\Bigg\} \\
	=
	\Bigg\{
	\Bigg[
	\frac{\partial a_{\mathbf{p}}^\dag}{\partial S(\mathbf{p})^i}
	a_{\mathbf{p}}
	,
	-
	a_{\mathbf{q}}^\dag
	\frac{\partial a_{\mathbf{q}}}{\partial S(\mathbf{q})^j}
	\Bigg] 
	-
	\Bigg[
	\frac{\partial a_{\mathbf{q}}^\dag}{\partial S(\mathbf{q})^j}
	a_{\mathbf{q}}
	,
	-
	a_{\mathbf{p}}^\dag
	\frac{\partial a_{\mathbf{p}}}{\partial S(\mathbf{p})^i}
	\Bigg]
	\Bigg\}
\end{align}
and we notice that the second commutator is exactly the same as the first one but with $\mathbf{p} \leftrightarrow \mathbf{q}$. However, as was done previously, since we are integrating in $p$ and $q$, we can just rename them and therefore switch them. Therefore, the sum of the second and third term will be
\begin{align}
	\Bigg\{
	\Bigg[
	\frac{\partial a_{\mathbf{p}}^\dag}{\partial S(\mathbf{p})^i}
	a_{\mathbf{p}}
	,
	-
	a_{\mathbf{q}}^\dag
	\frac{\partial a_{\mathbf{q}}}{\partial S(\mathbf{q})^j}
	\Bigg] 
	-
	\Bigg[
	\frac{\partial a_{\mathbf{p}}^\dag}{\partial S(\mathbf{p})^j}
	a_{\mathbf{p}}
	,
	-
	a_{\mathbf{q}}^\dag
	\frac{\partial a_{\mathbf{q}}}{\partial S(\mathbf{q})^i}
	\Bigg]
	\Bigg\}
	=
	0
\end{align}
which is what we wanted to show.

We can finally deal with the last term. We have
\begin{align}
	&-\alpha(p)\alpha(q)S(\omega_p)S(\omega_q)
	\Bigg[
	a_{\mathbf{p}}^\dag
	\frac{\partial a_{\mathbf{p}}}{\partial S(\mathbf{p})^i}
	,
	a_{\mathbf{q}}^\dag
	\frac{\partial a_{\mathbf{q}}}{\partial S(\mathbf{q})^j}
	\Bigg] \\
	&=
	-\alpha(p)\alpha(q)S(\omega_p)S(\omega_q)
	\Bigg\{
	a_{\mathbf{p}}^\dag
	\left[
	\frac{\partial a_{\mathbf{p}}}{\partial S(\mathbf{p})^i}
	,
	a_{\mathbf{q}}^\dag
	\right]
	\frac{\partial a_{\mathbf{q}}}{\partial S(\mathbf{q})^j}
	+
	a_{\mathbf{q}}^\dag
	\left[
	a_{\mathbf{p}}^\dag
	,
	\frac{\partial a_{\mathbf{q}}}{\partial S(\mathbf{q})^j}
	\right]
	\frac{\partial a_{\mathbf{p}}}{\partial S(\mathbf{p})^i}
	\Bigg\} \\
	&-\alpha(p)\alpha(q)S(\omega_p)S(\omega_q)
	\Bigg\{
	a_{\mathbf{p}}^\dag
	\left[
	\frac{\partial a_{\mathbf{p}}}{\partial S(\mathbf{p})^i}
	,
	a_{\mathbf{q}}^\dag
	\right]
	\frac{\partial a_{\mathbf{q}}}{\partial S(\mathbf{q})^j}
	-
	a_{\mathbf{q}}^\dag
	\left[
	\frac{\partial a_{\mathbf{q}}}{\partial S(\mathbf{q})^j}
	,
	a_{\mathbf{p}}^\dag
	\right]
	\frac{\partial a_{\mathbf{p}}}{\partial S(\mathbf{p})^i}
	\Bigg\}
\end{align}
Once again, we can without problem switch $\mathbf{p} \leftrightarrow \mathbf{q}$ on the second term, obtaining finally
\begin{align}
	&-\alpha(p)\alpha(q)S(\omega_p)S(\omega_q)
	a_{\mathbf{p}}^\dag
	\left[
	\frac{\partial a_{\mathbf{p}}}{\partial S(\mathbf{p})^{[i}}
	,
	a_{\mathbf{q}}^\dag
	\right]
	\frac{\partial a_{\mathbf{q}}}{\partial S(\mathbf{q})^{j]}} 
\end{align}
Proceeding as before, we have
\begin{align}
	&-\alpha(p)\alpha(q)S(\omega_p)S(\omega_q)
	a_{\mathbf{p}}^\dag
	\left[
	\frac{\partial a_{\mathbf{p}}}{\partial S(\mathbf{p})^{[i}}
	,
	a_{\mathbf{q}}^\dag
	\right]
	\frac{\partial a_{\mathbf{q}}}{\partial S(\mathbf{q})^{j]}} \\
	&=
	-\alpha(p)S(\omega_p)S(\omega_q)
	a_{\mathbf{p}}^\dag
	\frac{\partial a_{\mathbf{q}}}{\partial S(\mathbf{q})^{[j}}
	\frac{\partial }{\partial S(\mathbf{p})^{i]}}
	\delta(\mathbf{p}-\mathbf{q}) \\
	&=
	\frac{\partial \alpha(q)}{\partial S(\mathbf{q})^{[i}}
	S(\omega_q)^2
	a_{\mathbf{q}}^\dag
	\frac{\partial a_{\mathbf{q}}}{\partial S(\mathbf{q})^{j]}}
	+
	\alpha(q)
	S(\mathbf{q})_{[i}
	a_{\mathbf{q}}^\dag
	\frac{\partial a_{\mathbf{q}}}{\partial S(\mathbf{q})^{j]}}
	+
	\alpha(q)
	S(\omega_q)^2
	\frac{\partial a_{\mathbf{q}}^\dag}{\partial S(\mathbf{q})^{[i}}
	\frac{\partial a_{\mathbf{q}}}{\partial S(\mathbf{q})^{j]}}
\end{align}
Once again, we integrate by parts the first term, obtaining
\begin{align}
	\frac{\partial \alpha(q)}{\partial S(\mathbf{q})^{[i}}
	S(\omega_q)^2
	a_{\mathbf{q}}^\dag
	\frac{\partial a_{\mathbf{q}}}{\partial S(\mathbf{q})^{j]}}
	=
	-
	2
	\alpha(q)
	S(\mathbf{q})_{[i}
	a_{\mathbf{q}}^\dag
	\frac{\partial a_{\mathbf{q}}}{\partial S(\mathbf{q})^{j]}}
	-
	\alpha(q)
	S(\omega_q)^2
	\frac{\partial a_{\mathbf{q}}^\dag}{\partial S(\mathbf{q})^{[i}}
	\frac{\partial a_{\mathbf{q}}}{\partial S(\mathbf{q})^{j]}}
\end{align}
and substituting back we get the final result
\begin{align}
	-\alpha(p)\alpha(q)S(\omega_p)S(\omega_q)
	a_{\mathbf{p}}^\dag
	\left[
	\frac{\partial a_{\mathbf{p}}}{\partial S(\mathbf{p})^{[i}}
	,
	a_{\mathbf{q}}^\dag
	\right]
	\frac{\partial a_{\mathbf{q}}}{\partial S(\mathbf{q})^{j]}} 
	=
	-
	\alpha(q)
	S(\mathbf{q})_{[i}
	a_{\mathbf{q}}^\dag
	\frac{\partial a_{\mathbf{q}}}{\partial S(\mathbf{q})^{j]}}
\end{align}
Finally, the result of the commutator coming from the $a, a^\dag$ part is given by
\begin{align}
	\frac{1}{4}\int d^3p \, d^3q \,
	\alpha(q)
	S(\mathbf{q})_{[i}
	\frac{\partial a_{\mathbf{q}}^\dag}{\partial S(\mathbf{q})^{j]}}
	a_{\mathbf{q}}
	-
	\alpha(q)
	S(\mathbf{q})_{[i}
	a_{\mathbf{q}}^\dag
	\frac{\partial a_{\mathbf{q}}}{\partial S(\mathbf{q})^{j]}}
\end{align}
which is what was expected.

For the commutators coming from the $b, b^\dag$ part, we have
\begin{align}
	-\alpha(p)\alpha(q)\omega_p\omega_q
	\Bigg\{
	&\left[
	b_{\mathbf{p}} 
	\frac{\partial b_{\mathbf{p}}^\dag}{\partial \mathbf{p}^i}
	,
	b_{\mathbf{q}} 
	\frac{\partial b_{\mathbf{q}}^\dag}{\partial \mathbf{q}^j}
	\right]
	+
	\left[
	b_{\mathbf{p}} 
	\frac{\partial b_{\mathbf{p}}^\dag}{\partial \mathbf{p}^i}
	,
	-
	\frac{\partial b_{\mathbf{q}}}{\partial \mathbf{q}^j} b_{\mathbf{q}}^\dag
	\right] \\
	+&
	\left[
	-
	\frac{\partial b_{\mathbf{p}}}{\partial \mathbf{p}^i} b_{\mathbf{p}}^\dag
	,
	b_{\mathbf{q}} 
	\frac{\partial b_{\mathbf{q}}^\dag}{\partial \mathbf{q}^j}
	\right]
	+
	\left[
	-
	\frac{\partial b_{\mathbf{p}}}{\partial \mathbf{p}^i} b_{\mathbf{p}}^\dag
	,
	-
	\frac{\partial b_{\mathbf{q}}}{\partial \mathbf{q}^j} b_{\mathbf{q}}^\dag
	\right]
	\Bigg\}
\end{align}
The second and third term go away for the same reasons as before, and only the first and last term remain.

Notice that if one (formally) switches $b\rightarrow \tilde{a}^\dag$ and $\tilde{b}^\dag \rightarrow a$, one gets exactly the same commutators as the $a, a^\dag$ case\footnote{The fact that the derivatives and the prefactors do not have the antipode is irrelevant in this context.} but with an additional minus sign. This minus sign comes from the fact that sending $b\rightarrow \tilde{a}^\dag$ and $\tilde{b}^\dag \rightarrow a$ also sends $\delta = [b, b^\dag] \rightarrow [\tilde{a}^\dag, \tilde{a}] = -\delta$. Therefore, one can do the exact same computations as for $a$, with the additional minus sign coming from the deltas. Therefore the final result will be 
\begin{align}
	-\alpha(q)
	\mathbf{q}_{[i}
	\frac{\partial \tilde{a}_{\mathbf{q}}^\dag}{\partial \mathbf{q}^{j]}}
	\tilde{a}_{\mathbf{q}}
	+
	\alpha(q)
	\mathbf{q}_{[i}
	\tilde{a}_{\mathbf{q}}^\dag
	\frac{\partial \tilde{a}_{\mathbf{q}}}{\partial \mathbf{q}^{j]}}
\end{align}
or, in other words, 
\begin{align}
	-\alpha(q)
	\mathbf{q}_{[i}
	\frac{\partial b_{\mathbf{q}}}{\partial \mathbf{q}^{j]}}
	b_{\mathbf{q}}^\dag
	+
	\alpha(q)
	\mathbf{q}_{[i}
	b_{\mathbf{q}}
	\frac{\partial b_{\mathbf{q}}^\dag}{\partial \mathbf{q}^{j]}}
\end{align}
One can also double check the above result by direct computations. Therefore, the $b, b^\dag$ contribution to the commutator is given by
\begin{align}
	\frac{1}{4}
	\int d^3p \, d^3q \,
	\alpha
	\left(
	b_{\mathbf{q}}
	\mathbf{q}_{[i}
	\frac{\partial b_{\mathbf{q}}^\dag}{\partial \mathbf{q}^{j]}}
	-
	\mathbf{q}_{[i}
	\frac{\partial b_{\mathbf{q}}}{\partial \mathbf{q}^{j]}}
	b_{\mathbf{q}}^\dag
	\right)
\end{align}
Finally, from these results, one can immediately say that
\begin{align}
	[\mathcal{N}_i, \mathcal{N}_j]=
		\frac{1}{4}
		\int  d^3q \,
		\alpha(q)
		\left(
		S(\mathbf{q})_{[i}
		\frac{\partial a_{\mathbf{q}}^\dag}{\partial S(\mathbf{q})^{j]}}
		a_{\mathbf{q}} 
		-
		a_{\mathbf{q}}^\dag
		S(\mathbf{q})_{[i}
		\frac{\partial a_{\mathbf{q}}}{\partial S(\mathbf{q})^{j]}}
		+
		b_{\mathbf{q}}
		\mathbf{q}_{[i}
		\frac{\partial b_{\mathbf{q}}^\dag}{\partial \mathbf{q}^{j]}}
		-
		\mathbf{q}_{[i}
		\frac{\partial b_{\mathbf{q}}}{\partial \mathbf{q}^{j]}}
		b_{\mathbf{q}}^\dag
		\right)
\end{align}
Calling the RHS $\mathcal{M}_{ij}$ and noticing that  $M_k = \frac{1}{2}\epsilon_{kij}M_{ij}$ we get the result.

\section{$C$ and its commutator with the boosts}\label{CNcomm}

We are now at point $ix)$ of our roadmap in section \ref{roadmap2}. The $C$ operator in our context is given by
\begin{align}\label{C}
	C = 
	\int \, d^3p
	\, 
	\alpha (
	b^\dag_{\mathbf{p}}
	a_\mathbf{p}
	+
	a^\dag_{\mathbf{p}}
	b_{\mathbf{p}}
	)
\end{align}
In fact, it is easy to show that this operator satisfies the canonical $C$ property 
\begin{align}
	C a^\dag |0\rangle = b^\dag |0\rangle
	\qquad
	C b^\dag |0\rangle = a^\dag |0\rangle
\end{align}
where we used the fact that $C^{-1}|0\rangle =|0\rangle$. In fact, using eq. \eqref{Symp15-t} and \eqref{Symp16-t} we have
\begin{align}
	C a^\dag_{\mathbf{p}} |0\rangle 
	&=
	\int \, d^3q
	\, 
	\alpha (
	b^\dag_{\mathbf{q}}
	a_\mathbf{q}
	+
	a^\dag_{\mathbf{q}}
	b_{\mathbf{q}}
	)a^\dag_{\mathbf{p}}|0\rangle \\
	&=
	\int \, d^3q
	\, 
	\alpha 
	b^\dag_{\mathbf{q}}
	a_\mathbf{q}
	a^\dag_{\mathbf{p}}|0\rangle
	+
	0 \\
	&=
	\int \, d^3q
	\, 
	\alpha 
	b^\dag_{\mathbf{q}}
	\left(
	\frac{1}{\alpha} 
	\delta(\mathbf{p}-\mathbf{q})
	+
	a^\dag_{\mathbf{p}}
	a_\mathbf{q}
	\right)|0\rangle \\
	&=
	b^\dag_{\mathbf{p}} |0\rangle
\end{align}
and the computation for the verification of the relation $C b^\dag |0\rangle = a^\dag |0\rangle$ proceed in the same way. Notice that $C=C^\dag$ and $C^2 = 1$. In fact we have 
\begin{align}
	C^2 a^\dag_\mathbf{k}|0\rangle
	&=
	\int d^3p \, d^3q \,
	\alpha(p)\alpha(q)
	(
	b^\dag_{\mathbf{p}}
	a_\mathbf{p}
	+
	a^\dag_{\mathbf{p}}
	b_{\mathbf{p}}
	)
	(
	b^\dag_{\mathbf{q}}
	a_\mathbf{q}
	+
	a^\dag_{\mathbf{q}}
	b_{\mathbf{q}}
	)a^\dag_\mathbf{k}|0\rangle \\
	&=
	\int d^3p \,
	\alpha(p)
	(
	b^\dag_{\mathbf{p}}
	a_\mathbf{p}
	+
	a^\dag_{\mathbf{p}}
	b_{\mathbf{p}}
	)
	b^\dag_{\mathbf{k}}|0\rangle \\
	&=
	a^\dag_\mathbf{k}|0\rangle
\end{align}
and the same if we started from $b^\dag_{\mathbf{q}}|0\rangle$. Finally, notice that the $C$ operator in eq. \eqref{C} is a symmetry of the off-shell action, since acting by conjugation we only exchange $a^{(\dag)}\leftrightarrow b^{(\dag)}$.

We now compute the commutator between the charge conjugation operator and the boost operator. Using the boost charge in eq. \eqref{defboostfromsymp} we have
\begin{align}
	[N_i, C]
	&=
	\frac{i}{2}
	\int \, d^3p \, d^3q \, 
	\alpha(p)\alpha(q)
	\Bigg\{
	S(\omega_p)
	\left[
	\frac{\partial a_{\mathbf{p}}^\dag}{\partial S(\mathbf{p})^i}
	a_{\mathbf{p}}
	,
	b^\dag_{\mathbf{q}}
	a_\mathbf{q}
	\right]
	+
	S(\omega_p)
	\left[
	\frac{\partial a_{\mathbf{p}}^\dag}{\partial S(\mathbf{p})^i}
	a_{\mathbf{p}}
	,
	a^\dag_{\mathbf{q}}
	b_{\mathbf{q}}
	\right] \\
	&+
	S(\omega_p)
	\left[
	-
	a_{\mathbf{p}}^\dag
	\frac{\partial a_{\mathbf{p}}}{\partial S(\mathbf{p})^i}
	,
	b^\dag_{\mathbf{q}}
	a_\mathbf{q}
	\right]
	+
	S(\omega_p)
	\left[
	-
	a_{\mathbf{p}}^\dag
	\frac{\partial a_{\mathbf{p}}}{\partial S(\mathbf{p})^i}
	,
	a^\dag_{\mathbf{q}}
	b_{\mathbf{q}}
	\right] \\
	&+
	\omega_p
	\left[
	b_{\mathbf{p}} 
	\frac{\partial b_{\mathbf{p}}^\dag}{\partial \mathbf{p}^i}
	,
	b^\dag_{\mathbf{q}}
	a_\mathbf{q}
	\right]
	+
	\omega_p
	\left[
	b_{\mathbf{p}} 
	\frac{\partial b_{\mathbf{p}}^\dag}{\partial \mathbf{p}^i}
	,
	a^\dag_{\mathbf{q}}
	b_{\mathbf{q}}
	\right] \\
	&+
	\omega_p
	\left[
	-
	\frac{\partial b_{\mathbf{p}}}{\partial \mathbf{p}^i} b_{\mathbf{p}}^\dag
	,
	b^\dag_{\mathbf{q}}
	a_\mathbf{q}
	\right]
	+
	\omega_p
	\left[
	-
	\frac{\partial b_{\mathbf{p}}}{\partial \mathbf{p}^i} b_{\mathbf{p}}^\dag
	,
	a^\dag_{\mathbf{q}}
	b_{\mathbf{q}}
	\right]
	\Bigg\}
\end{align}
We compute these commutators one by one using the commutator property
\begin{align}
	[AB,C] = A[B,C] + [A,C]B
\end{align}
We have the following.
\begin{itemize}
	\item \textbf{First}
	\begin{align}
		&\alpha(p)\alpha(q)
		S(\omega_p)
		\left[
		\frac{\partial a_{\mathbf{p}}^\dag}{\partial S(\mathbf{p})^i}
		a_{\mathbf{p}}
		,
		b^\dag_{\mathbf{q} }
		a_\mathbf{q}
		\right] 
		\\
		&=
		0
		+
		\alpha(p)\alpha(q)
		S(\omega_p)
		\left[
		\frac{\partial a_{\mathbf{p}}^\dag}{\partial S(\mathbf{p})^i}
		,
		a_\mathbf{q}
		\right]
		a_{\mathbf{p}}
		b^\dag_{\mathbf{q} } \\
		&=
		-\alpha(p)
		S(\omega_p)
		a_{\mathbf{p}}
		b^\dag_{\mathbf{q} }
		\frac{\partial }{\partial S(\mathbf{p})^i}
		\delta(\mathbf{p}-\mathbf{q}) \\
		&=
		\Bigg\{
		\left(
		\frac{\partial \alpha}{\partial S(\mathbf{p})^i}
		S(\omega_p)
		+
		\alpha
		\frac{S(\mathbf{p})_i}{S(\omega_p)}
		\right)
		a_{\mathbf{p}}
		b^\dag_{\mathbf{q} }
		+
		\alpha(p)
		S(\omega_p)
		\frac{\partial a_{\mathbf{p}}}{\partial S(\mathbf{p})^i}
		b^\dag_{\mathbf{q} } 
		\Bigg\}
		\delta(\mathbf{p}-\mathbf{q}) \\
		&=
		-
		\alpha(p)
		S(\omega_p) 
		a_{\mathbf{p}}
		\frac{\partial b^\dag_{\mathbf{p} } }{\partial S(\mathbf{p})^i}
	\end{align}
where in the last passage we integrated once again by parts.
\item \textbf{Second}
\begin{align}
	&\alpha(p)\alpha(q)
	S(\omega_p)
	\left[
	\frac{\partial a_{\mathbf{p}}^\dag}{\partial S(\mathbf{p})^i}
	a_{\mathbf{p}}
	,
	a_\mathbf{q}^\dag
	b_{\mathbf{q} }
	\right] \\
	&=
	\alpha(p)\alpha(q)
	S(\omega_p)
	\frac{\partial a_{\mathbf{p}}^\dag}{\partial S(\mathbf{p})^i}
	\left[
	a_{\mathbf{p}}
	,
	a_\mathbf{q}^\dag
	\right] 
	b_{\mathbf{q} }
	+
	0 \\
	&=
	\alpha(p)
	S(\omega_p)
	\frac{\partial a_{\mathbf{p}}^\dag}{\partial S(\mathbf{p})^i}
	b_{\mathbf{p} }
\end{align}
\item \textbf{Third}
\begin{align}
	&\alpha(p)\alpha(q)
	S(\omega_p)
	\left[
	-
	a_{\mathbf{p}}^\dag
	\frac{\partial a_{\mathbf{p}}}{\partial S(\mathbf{p})^i}
	,
	b^\dag_{\mathbf{q} }
	a_\mathbf{q}
	\right] \\
	&=
	0
	+
	\alpha(p)\alpha(q)
	S(\omega_p)
	\left[
	-
	a_{\mathbf{p}}^\dag
	,
	a_\mathbf{q}
	\right]
	\frac{\partial a_{\mathbf{p}}}{\partial S(\mathbf{p})^i}
	b^\dag_{\mathbf{q} } \\
	&=
	\alpha(p)
	S(\omega_p)
	\frac{\partial a_{\mathbf{p}}}{\partial S(\mathbf{p})^i}
	b^\dag_{\mathbf{p} } 
\end{align}
\item \textbf{Fourth}
\begin{align}
	&\alpha(p)\alpha(q)
	S(\omega_p)
	\left[
	-
	a_{\mathbf{p}}^\dag
	\frac{\partial a_{\mathbf{p}}}{\partial S(\mathbf{p})^i}
	,
	a_\mathbf{q}^\dag
	b_{\mathbf{q} }
	\right] \\
	&=
	-
	\alpha(p)\alpha(q)
	S(\omega_p)
	a_{\mathbf{p}}^\dag
	\left[
	\frac{\partial a_{\mathbf{p}}}{\partial S(\mathbf{p})^i}
	,\
	a_\mathbf{q}^\dag
	\right]
	b_{\mathbf{q} }
	+
	0 \\
	&=
	-
	\alpha(p)
	S(\omega_p)
	a_{\mathbf{p}}^\dag
	b_{\mathbf{q} }
	\frac{\partial }{\partial S(\mathbf{p})^i}
	\delta(\mathbf{p}-\mathbf{q}) \\
	&=
	\Bigg\{
	\left(
	\frac{\partial \alpha}{\partial S(\mathbf{p})^i}
	S(\omega_p)
	+
	\alpha
	\frac{S(\mathbf{p})_i}{S(\omega_p)}
	\right)
	a_{\mathbf{p}}^\dag
	b_{\mathbf{q} }
	+
	\alpha(p)
	S(\omega_p)
	\frac{\partial a_{\mathbf{p}^\dag}}{\partial S(\mathbf{p})^i}
	b_{\mathbf{q} } 
	\Bigg\}
	\delta(\mathbf{p}-\mathbf{q}) \\
	&=
	-
	\alpha(p)
	S(\omega_p) a_{\mathbf{p}}^\dag
	\frac{\partial b _{\mathbf{p} }}{\partial S(\mathbf{p})^i}
\end{align}
where in the last passage we integrated once again by parts.
\item \textbf{Fifth}
\begin{align}
	\alpha(p)\alpha(q)
	\omega_p
	\left[
	b_{\mathbf{p}} 
	\frac{\partial b_{\mathbf{p}}^\dag}{\partial \mathbf{p}^i}
	,
	b^\dag_{\mathbf{q} }
	a_\mathbf{q}
	\right]
	&=
	0
	+
	\alpha(p)\alpha(q)
	\omega_p
	\left[
	b_{\mathbf{p}} 
	,
	b^\dag_{\mathbf{q} }
	\right]
	\frac{\partial b_{\mathbf{p}}^\dag}{\partial \mathbf{p}^i}
	a_\mathbf{q} \\
	&=
	\alpha(p)
	\omega_p
	\frac{\partial b_{\mathbf{p}}^\dag}{\partial \mathbf{p}^i}
	a_\mathbf{p}
\end{align}
\item \textbf{Sixth}
\begin{align}
	&\alpha(p)\alpha(q)
	\omega_p
	\left[
	b_{\mathbf{p}} 
	\frac{\partial b_{\mathbf{p}}^\dag}{\partial \mathbf{p}^i}
	,
	a_\mathbf{q}^\dag
	b_{\mathbf{q} }
	\right] \\
	&=
	\alpha(p)\alpha(q)
	\omega_p
	b_{\mathbf{p}} 
	\left[
	\frac{\partial b_{\mathbf{p}}^\dag}{\partial \mathbf{p}^i}
	,
	b_{\mathbf{q} }
	\right]
	a_\mathbf{q}^\dag
	+
	0 \\
	&=
	-\alpha(p)
	\omega_p
	b_{\mathbf{p}} 
	a_\mathbf{q}^\dag
	\frac{\partial}{\partial \mathbf{p}^i}
	\delta(\mathbf{p}-\mathbf{q}) \\
	&=
	\Bigg\{
	\left(
	\frac{\partial \alpha}{\partial \mathbf{p}^i}
	\omega_p
	+
	\alpha\frac{\mathbf{p}_i}{\omega_p}
	\right)
	b_{\mathbf{p}} 
	a_\mathbf{q}^\dag
	+
	\alpha(p)\omega_p
	\frac{\partial b_{\mathbf{p}}}{\partial \mathbf{p}^i}
	a_\mathbf{q}^\dag
	\Bigg\}
	\delta(\mathbf{p}-\mathbf{q}) \\
	&=
	-
	\alpha(p)\omega_p b_{\mathbf{p}}
	\frac{\partial a_\mathbf{p}^\dag}{\partial \mathbf{p}^i}
\end{align} 
where in the last passage we integrated once again by parts.
\item \textbf{Seventh}
\begin{align}
	&\alpha(p)\alpha(q)
	\omega_p
	\left[
	-
	\frac{\partial b_{\mathbf{p}}}{\partial \mathbf{p}^i} b_{\mathbf{p}}^\dag
	,
	b^\dag_{\mathbf{q} }
	a_\mathbf{q}
	\right] \\
	&=
	0
	+
	\alpha(p)\alpha(q)
	\omega_p
	\left[
	-
	\frac{\partial b_{\mathbf{p}}}{\partial \mathbf{p}^i} 
	,
	b^\dag_{\mathbf{q} }
	\right]
	b_{\mathbf{p}}^\dag
	a_\mathbf{q} \\
	&=
	-
	\alpha(p)
	\omega_p
	b_{\mathbf{p}}^\dag
	a_\mathbf{q}
	\frac{\partial }{\partial \mathbf{p}^i} 
	\delta(\mathbf{p}-\mathbf{q}) \\
	&=
	\Bigg\{
	\left(
	\frac{\partial \alpha}{\partial \mathbf{p}^i}
	\omega_p
	+
	\alpha\frac{\mathbf{p}_i}{\omega_p}
	\right)
	b_{\mathbf{p}}^\dag
	a_\mathbf{q}
	+
	\alpha(p)\omega_p
	\frac{\partial b_{\mathbf{p}}^\dag}{\partial \mathbf{p}^i}
	a_\mathbf{q}
	\Bigg\}
	\delta(\mathbf{p}-\mathbf{q}) \\
	&=
	-
	\alpha(p)\omega_p b_{\mathbf{p}}^\dag
	\frac{\partial a_\mathbf{p}}{\partial \mathbf{p}^i}
\end{align}
where in the last passage we integrated once again by parts.
\item \textbf{Eighth}
\begin{align}
	\alpha(p)\alpha(q)
	\omega_p
	\left[
	-
	\frac{\partial b_{\mathbf{p}}}{\partial \mathbf{p}^i} b_{\mathbf{p}}^\dag
	,
	a^\dag_{\mathbf{q}}
	b_{\mathbf{q} }
	\right] 
	&=
	-
	\alpha(p)\alpha(q)
	\omega_p
	\frac{\partial b_{\mathbf{p}}}{\partial \mathbf{p}^i}
	\left[
	b_{\mathbf{p}}^\dag
	,
	b_{\mathbf{q} }
	\right] 
	a^\dag_{\mathbf{q}} \\
	&=
	\alpha(p)
	\omega_p
	\frac{\partial b_{\mathbf{p}}}{\partial \mathbf{p}^i}
	a^\dag_{\mathbf{p}} 
\end{align}
\end{itemize} 
Putting all these computations together, we are left with 
\begin{align}
	[\mathcal{N}_i, C]
	&=
	\frac{i}{2}
	\int \, d^3p \, 
	\alpha(p)
	\Bigg\{
	S(\omega_p)
	\left[
	\frac{\partial a_{\mathbf{p}}}{\partial S(\mathbf{p})^i}
	b^\dag_{\mathbf{p} } 
	-
	a_{\mathbf{p}}
	\frac{\partial b^\dag_{\mathbf{p} } }{\partial S(\mathbf{p})^i}
	+
	\frac{\partial a_{\mathbf{p}}^\dag}{\partial S(\mathbf{p})^i}
	b_{\mathbf{p} }
	-
	a_{\mathbf{p}}^\dag
	\frac{\partial b _{\mathbf{p} }}{\partial S(\mathbf{p})^i}
	\right] \nonumber \\
	&+
	\omega_p
	\left[
	\frac{\partial b_{\mathbf{p}}^\dag}{\partial \mathbf{p}^i}
	a_\mathbf{p}
	-
	b_{\mathbf{p}}^\dag
	\frac{\partial a_\mathbf{p}}{\partial \mathbf{p}^i}
	+
	\frac{\partial b_{\mathbf{p}}}{\partial \mathbf{p}^i}
	a^\dag_{\mathbf{p}} 
	-
	b_{\mathbf{p}}
	\frac{\partial a_\mathbf{p}^\dag}{\partial \mathbf{p}^i}
	\right]
	\Bigg\} 
\end{align}
which (as expected) is different than zero.


Finally, notice that in the  non-deformed case, all the computations will proceed in exactly the same way but \textit{without} antipodes (recall that $\alpha(p)$ is a global prefactor such that $\alpha\rightarrow 1$ when $\kappa \rightarrow \infty$). Furthermore, notice also that
\begin{align}
	S(A) \frac{\partial}{\partial S(B)}
	\xrightarrow{\kappa\rightarrow\infty} -A \frac{\partial}{\partial(-B)}= A \frac{\partial}{\partial B}.
\end{align}
Therefore, the above result in the  non-deformed case reduces to 
\begin{align}
	[N_i, C]
	&=
	\frac{i}{2}
	\int \, d^3p \, 
	\Bigg\{
	\omega_p
	\left[
	\frac{\partial a_{\mathbf{p}}}{\partial \mathbf{p}^i}
	b^\dag_{\mathbf{p} } 
	-
	a_{\mathbf{p}}
	\frac{\partial b^\dag_{\mathbf{p} } }{\partial \mathbf{p}^i}
	+
	\frac{\partial a_{\mathbf{p}}^\dag}{\partial \mathbf{p}^i}
	b_{\mathbf{p} }
	-
	a_{\mathbf{p}}^\dag
	\frac{\partial b _{\mathbf{p} }}{\partial \mathbf{p}^i}
	\right] \nonumber \\
	&+
	\omega_p
	\left[
	\frac{\partial b_{\mathbf{p}}^\dag}{\partial \mathbf{p}^i}
	a_\mathbf{p}
	-
	b_{\mathbf{p}}^\dag
	\frac{\partial a_\mathbf{p}}{\partial \mathbf{p}^i}
	+
	\frac{\partial b_{\mathbf{p}}}{\partial \mathbf{p}^i}
	a^\dag_{\mathbf{p}} 
	-
	b_{\mathbf{p}}
	\frac{\partial a_\mathbf{p}^\dag}{\partial \mathbf{p}^i}
	\right]
	\Bigg\}  \\
	&=0
\end{align}
where in the last passage we used the fact that, because of the limit $\kappa \rightarrow \infty$, the two square brackets are equal and opposite.

\section{$\cancel{CPT}$ $\Leftrightarrow$ \cancel{Lorentz}?}\label{green}

In \cite{Greenberg:2002uu} it was shown that $CPT$ violation implies Lorentz violation and vice versa. The theorem is obtained in the context of axiomatic QFT, and in particular on the properties of Wightman functions. The relation between $CPT$ invariance and Lorentz invariance is still not clear in the non-deformed context \cite{Duetsch:2012sd}, \cite{Chaichian:2011fc}. In our context, $\kappa$-deformation introduces non-trivialities already at the one-particle level even in absence of interaction, a fact which can be intuitively understood by noticing that $-S(p)\neq p$. Furthermore, as discussed in the previous section, although the action is manifestly $CPT$ and invariant under $\kappa$-Poincar\'e, the Noether charges obtained from continuous symmetries of the action (and in particular the boost charge) have very non-trivial relations with the discrete symmetries (in this particular case $C$). A theorem like \cite{Greenberg:2002uu} is therefore not automatically satisfied in the $\kappa$-deformed context. In what follows, we reproduce the reasoning in the non-deformed case, explicitly performing all the steps and adding some figures, and at the end we write a discussion on possible departures from the conclusions of the theorem. We are at point $x)$ of our roadmap in section \ref{roadmap2}.

\subsection{Greenberg's argument}

The proof of Greenberg's theorem is based on a theorem by Jost in \cite{Jost:1957zz} (a scan of the original paper in german can be found in \cite{origJosttheorem}, 
see also \cite{Wightman:1963deu} for a description of this argument. Greenberg has also described the theorem and its steps in \cite{Greenberg:2003nv}). First we write down what Jost theorem says without proof, then we write down the setup for Greenberg's argument, and finally the argument itself. Since Jost theorem is fundamental for Greenberg's proof, we will describe it in the next section. The description of this theorem provided in this section is solely based on the provided references, and we will use the same notation and setup.

Jost theorem states the following:

\bigskip

\noindent\fbox{
\parbox{\textwidth}{
\textbf{Jost theorem:} Given the (Wightman) functions
\begin{align}
	W^{(n)}(x_1, x_2, \dots, x_n) 
	:= 
	\langle 0 | \phi_1(x_1)\phi_2(x_2) \dots \phi_n(x_n) |0\rangle
\end{align}
the validity of the \textit{weak local commutativity} condition at Jost points
\begin{align}
	W^{(n)}(x_1, x_2, \dots, x_n)
	=
	W^{(n)}(x_n, x_{n-1}, \dots, x_1)
\end{align}
is equivalent to the validity of the $CPT$ symmetry. 
The Jost points $\{x_i\}$ are a set of points such that the linear combination
\begin{align}
	\sum_i c_i (x_i - x_{i-1})
\end{align}
is always spacelike, for any choice of $c_i \geq 0$ with $\sum_i c_i >0$.
}
}
\bigskip

\textit{Setup}:
We consider only passive transformations. A quantum field theory is considered Lorentz covariant if it is both in-cone covariant and out-of-cone covariant. A quantum field theory is said to be in-cone covariant if the Wightman functions are covariant (notice that the Wightman functions are matrix elements of unordered fields). On the other hand, a quantum field theory is said to be out-of-cone covariant if the matrix elements of time-ordered fields $\tau$ are covariant. In the notation used in \cite{Greenberg:2002uu}, the $\tau$ functions are written as 
\begin{align}\label{tau}
	\tau 
	= 
	\sum_\sigma \theta(x_{\sigma_1}^0, x_{\sigma_2}^0, \dots, x_{\sigma_n}^0)
	W^{(n)}(x_{\sigma_1}, x_{\sigma_2}, \dots, x_{\sigma_n}) 
\end{align}
where the function $\theta(x_{\sigma_1}, x_{\sigma_2}, \dots, x_{\sigma_n})$ ensures that $x_{\sigma_1}^0 \geq x_{\sigma_2}^0 \geq \dots \geq x_{\sigma_n}^0$. The sum goes over all the permutations of $n$ elements.

\bigskip

\textit{Argument}:
The idea is to show that violation of $CPT$ (in the form of violation of the weak local commutativity condition) implies a violation of the out-of-cone covariance condition, and since a theory is Lorentz covariant if it is covariant \textit{both} in-cone and out-of-cone, we get a violation of the Lorentz covariance of the theory.

Since we want to use Jost theorem, consider points $x_1, \dots, x_n$ separated by spacelike intervals. For these points, there can be Lorentz transformations that reverse their time ordering. Therefore, in order for $\tau$ to be invariant under Lorentz transformation, the Wightman functions must not depend of the order of the fields in its argument (when these fields are evaluated at points with spacelike separation). In fact, say that there exist some Lorentz transformation such that one goes from $x_{\sigma_1}^0 \geq x_{\sigma_2}^0$ to $x_{\sigma_2}^0 \geq x_{\sigma_1}^0$. Then the function \eqref{tau} gets sent to 
\begin{align}
	\tau \overset{!}{=} \tilde{\tau}
	&=
	\sum_\sigma \theta(x_{\sigma_2}^0, x_{\sigma_1}^0, \dots, x_{\sigma_n}^0)
	W^{(n)}(x_{\sigma_1}, x_{\sigma_2}, \dots, x_{\sigma_n}) \\
	&=
	\sum_\sigma \theta(x_{\sigma_1}^0, x_{\sigma_2}^0, \dots, x_{\sigma_n}^0)
	W^{(n)}(x_{\sigma_2}, x_{\sigma_1}, \dots, x_{\sigma_n})
\end{align}
and therefore we must have
\begin{align}
	W^{(n)}(x_{\sigma_1}, x_{\sigma_2}, \dots, x_{\sigma_n})
	=
	W^{(n)}(x_{\sigma_2}, x_{\sigma_1}, \dots, x_{\sigma_n}).
\end{align}
Since the same can be repeated for any choice of points, then indeed for spacelike-separated points the $W^{(n)}$ functions do not depend on the order of the fields.

Now we can chose the points separated by spacelike intervals to be Jost sets, and we can chose them in such a way that their successive time differences are all positive. Graphically speaking, Jost points of this kind can be represented for example as follows (in red)
\begin{center}
	\begin{tikzpicture}
		\draw[color=black, -{Stealth[length=2mm]}] (-0.5,0) -- (3,0);
		\draw[color=black, -{Stealth[length=2mm]}] (0,-0.5) -- (0,3);
		\draw[color=blue] (0,0) -- (2.5,2.5);
		
		\filldraw[color=red] (0.5,0.2) circle (0.4mm);
		\filldraw[color=red] (1,0.4) circle (0.4mm);
		\filldraw[color=red] (1.5,0.6) circle (0.4mm);
		\filldraw[color=red] (2,0.8) circle (0.4mm);
	\end{tikzpicture}
\end{center}
Then there exist some Lorentz transformation such that all the successive time differences are all negative, i.e. we are in the following situation.
\begin{center}
	\begin{tikzpicture}
		\draw[color=black, -{Stealth[length=2mm]}, rotate around={30:(0,0)}] (-0.5,0) -- (3,0);
		\draw[color=black, -{Stealth[length=2mm]}, rotate around={-30:(0,0)}] (0,-0.5) -- (0,3);
		\draw[color=blue] (0,0) -- (2.5,2.5);
		
		\filldraw[color=red] (0.5,0.2) circle (0.4mm);
		\filldraw[color=red] (1,0.4) circle (0.4mm);
		\filldraw[color=red] (1.5,0.6) circle (0.4mm);
		\filldraw[color=red] (2,0.8) circle (0.4mm);
	\end{tikzpicture}
\end{center}
However, because of the invariance of $\tau$ under such a transformation, then for these Jost points we must have
\begin{align}
	W^{(n)}(x_1, x_2, \dots, x_n)
	=
	W^{(n)}(x_n, x_{n-1}, \dots, x_1)
\end{align}
which is the weak local commutativity of Jost theorem. Therefore, Lorentz symmetry implies weak local commutativity, which by Jost theorem is equivalent to CPT symmetry, and vice versa if CPT is violated, then the same is true for Lorentz symmetry.

\subsection{Jost theorem}

We use as a reference Greenberg's paper \cite{Greenberg:2003nv} where he describes in detail Jost proof. In this section we will just write down the main steps of the proof, using the same notation used in \cite{Greenberg:2003nv}.

Jost proof follows the following steps:
\begin{itemize}
	\item[1)] The key idea is to use the group $SL(2, \mathbb{C})$, which is the double covering of the proper orthochronous Lorentz group $L_+^{\uparrow}$. One uses $SL(2, \mathbb{C})$ instead of $L_+^{\uparrow}$ because spacetime inversion $PT = -1$ is connected to the identity in $SL(2, \mathbb{C})$, while it is not connected in $L_+^{\uparrow}$. 
	
	\item[2)] To use this extended Lorentz group, one needs to use the analytic continuation of the Wightman functions, i.e. of the vacuum expectation values of products of fields. However, instead of using the canonical $W^{(n)}(x_1, \dots, x_n)$, because of translation invariance one can use the equivalent function $\tilde{W}(x_1-x_2, \dots, x_{n-1}-x_n) := \tilde{W}(\xi_1, \dots, \xi_{n-1})$ where $\xi_j = x_j - x_{j+1}$. In order to obtain an analytic continuation of $\tilde{W}$, we promote each difference $\xi_j$ to a complex variable, and in particular
	\begin{align}
		z_j= \xi_j -i\eta_j 
	\end{align}
	The introduction of the $-i\eta_j$ transforms the distribution $\tilde{W}(\xi_1, \dots, \xi_{n-1})$ into an analytic function $\tilde{W}(z_1, \dots, z_{n-1})$.
	
	Notice that the use of the differences $\xi_j$ is a necessary step for the proof. In fact, the whole argument rests on the existence of real points in the analyticity domain of the analytic function $\tilde{W}(z_1, \dots, z_{n-1})$. These real points $\{\xi_j\}$ are the Jost points discussed above, and are defined by the condition that the sum
	\begin{align}
		s:=\sum_j c_j \xi_j
	\end{align}
	is spacelike, where $c_j\geq 0$ and $\sum_j c_j >0$. Of course, since $s$ must be spacelike for any observers, it must be defined in terms of $\xi_j$ and not $x_j$. 
	
	\item[3)] The newly defined analytic functions  $\tilde{W}(z_1, \dots, z_{n-1})$ are still defined only in a domain which does \textit{not} include real points. However, because of Lorentz invariance with respect to the double covering group $SL(2,\mathbb{C})$, we have
	\begin{align}\label{complexlorentz}
		\tilde{W}(\Lambda z_1, \dots, \Lambda z_{n-1})
		=
		\tilde{W}(z_1, \dots, z_{n-1})
	\end{align}
	This allows us to enlarge the domain, and the new domain does contain the Jost points. Furthermore, because of the fact that in $SL(2,\mathbb{C})$ there is spacetime inversion, there is some $\Lambda$ such that 
	\begin{align}
		\tilde{W}(\Lambda z_1, \dots, \Lambda z_{n-1})
		=
		(-1)^L\tilde{W}(-z_1, \dots, -z_{n-1})
	\end{align}
	for some coefficient $L$ derived in \cite{Greenberg:2003nv}. Writing the above relation at the Jost points, using the vacuum expectation values instead of the Wightman functions and using eq. \eqref{complexlorentz} we get\footnote{The fields $\phi_i$ in this expression can be any kind of field, boson or fermion.}
	\begin{align}\label{identity1}
		\langle 0 | \phi_1(x_1) \dots \phi_n(x_n) |0\rangle
		=
		(-1)^L 
		\langle 0 | \phi_1(-x_1) \dots \phi_n(-x_n) |0\rangle.
	\end{align}

	\item[4)]
	There is now a small but important subtlety in the above equality. In fact, the equality merely states that the value of the function on the LHS is the same as the one on the RHS.	However (essentially because of the antisymmetric nature of the $T$ operator) the function $(-1)^L 
	\langle 0 | \phi_1(-x_1) \dots \phi_n(-x_n) |0\rangle$
	has a different domain of definition than the function
	$\langle 0 |  \phi_1(x_1) \dots \phi_n(x_n) |0\rangle$. The way to solve this is to permute the fields inside the expectation value on the RHS of eq. \eqref{identity1}. Of course, in doing so one needs to use the fact that boson fields commute when evaluated at different spacetime points and fermion fields anticommute, i.e.
	\begin{align}
		[\phi(x),\phi(y)]=[A(x),A(y)] = \{\psi(x),\psi(y)\}=0
		\qquad
		x\neq y
	\end{align} 
	Upon using these relations, one can write that 
	\begin{align}\label{weakcommcond}
		\langle 0 |  \phi_1(x_1) \dots \phi_n(x_n) |0\rangle
		=
		(i)^F
		\langle 0 |  \phi_n(x_n) \dots \phi_n(x_1) |0\rangle
	\end{align} 
	where $F$ is the number of fermions in the expectation value. The identity \eqref{weakcommcond} is called `weak local commutativity', and for the sake of the theorem it only needs to hold at Jost points.
	Notice that weak local commutativity is not needed for the theorem if, instead of considering just Wightman functions, one uses the time ordered products of Wightman functions $\tau$ analogous to the one in eq. \eqref{tau}.

	\item[5)] Using the weak local commutativity, eq. \eqref{identity1} becomes
	\begin{align}\label{jostpointsvev}
		\langle 0 | \phi_1(x_1) \dots \phi_n(x_n) |0\rangle
		=
		(-1)^L 
		(i)^F
		\langle 0 | \phi_n(-x_n) \dots \phi_1(-x_1) |0\rangle.
	\end{align}
	Now both functions in both sides have the same domain of definition, and since they are analytic and coincide in some open subset of their domain (i.e. the neighbour(s) of the Jost points) the above equality implies the same equality of Wightman functions across the whole domain. Therefore we have
	\begin{align}
		\tilde{W}(z_1, \dots, z_n)
		=
		(-1)^L 
		(i)^F
		\tilde{W}(-z_n, \dots, -z_1). 
	\end{align}
	At this point, we can also take the limit $\eta_j \rightarrow 0$ and the analytic functions become once again distributions, and we have
	\begin{align}\label{finalwightman}
		\tilde{W}(\xi_1, \dots, \xi_n)
		=
		(-1)^L 
		(i)^F
		\tilde{W}(-\xi_n, \dots, -\xi_1). 
	\end{align}
	Notice that we couldn't take this limit before using the weak local commutativity because the domains of definitions of the functions were different. Therefore, we would have been stuck to the complex case, without the possibility of taking the limit and coming back to the real (physical) case. 

	\item[6)] We can write eq. \eqref{finalwightman} in terms of vacuum expectation value as
	\begin{align}
		\langle 0 | \phi_1(x_1) \dots \phi_n(x_n) |0\rangle
		=
		(-1)^L 
		(i)^F
		\langle 0 | \phi_n(-x_n) \dots \phi_1(-x_1) |0\rangle
	\end{align}
	Notice that this equation is formally equivalent to eq. \eqref{jostpointsvev}, but while eq. \eqref{jostpointsvev} only made sens in a neighbour of Jost points (i.e. the associated Wightman function was an analytic function) here everything is real (i.e. the Wightman functions are again distributions). One can restore the order of the fields through hermitian conjugation on the RHS obtaining
	\begin{align}\label{CPTinv}
		\langle 0 | \phi_1(x_1) \dots \phi_n(x_n) |0\rangle
		=
		(-1)^L 
		(i)^F
		\langle 0 | \phi_1^\dag(-x_1) \dots \phi_n^\dag(-x_n) |0\rangle^*
	\end{align}
	and from this one can read off the CPT transformation which reads\footnote{The indices $l$ and $f$ refer to the number of dotted and undotted indices in the field representation used in \cite{Greenberg:2003nv}, but it is not important for the proof.}
	\begin{align}
		\theta \phi(x) \theta^\dag
		=
		(-1)^l i^f \phi^\dag(-x). 
	\end{align}

	\item[7)] With the points $1)$ to $6)$ one has proved that (assuming Lorentz symmetry) the validity of the weak local commutativity at Jost points \eqref{weakcommcond} implies CPT symmetry. To show the other direction one can just use the same steps in reverse order, and therefore weak local commutativity in the neighbour of Jost points is equivalent to CPT symmetry.

\end{itemize}



\subsection{Comments on Jost and Greenberg theorems}

There are already some doubts about the validity of Greenberg's argument for canonical quantum field theory \cite{Duetsch:2012sd}, \cite{Chaichian:2011fc}. Apart from these arguments, one can also say the following. 

\begin{itemize}

\item The first problematic step in the whole argument is point 4) of the proof of Jost theorem. In fact, in the  non-deformed context the (anti)commutator of fields reduces to the (anti)commutators between creation/annihilation operators because plane waves are just functions and commute with everything (themselves included). Therefore, the statement about field is just a rewriting of the same statement about creation/annihilation operators. However, in the deformed context, plane waves do not commute and their product is described in terms of star operator and antipode. This fact, coupled with the deformed commutators in \eqref{Symp15-t}, \eqref{Symp16-t}, immediately implies that (anti)commutation relations for (fermion) fields are at best not obvious, and its imposition could very well lead to the destruction of some $\kappa$-deformed effects.

\item On a more pragmatical note, the definition of something like $\langle 0 | \phi_1(x_1) \star \phi_2(x_2) |0\rangle$ requires the knowledge of the commutator $[x_1^0, \mathbf{x}_2]$. Several proposals have been given for this commutator, each with its own pros and cons, highlighting the fact that an objective, natural definition of such commutator is lacking. As such, although any of the definition could be taken as assumption, different choices would result in different mathematical formulations of $\langle 0 | \phi_1(x_1) \star \phi_2(x_2) |0\rangle$, with different physical properties. This issue could be sidestepped by going to momentum space, where a natural definition of two-point function could be given using the path integral approach (see e.g. chapter \ref{Chapter3}). However, if we have more than two fields in the VEV, one has the issue that the Fock space is not well defined, see for example \cite{Arzano:2022vmh}. 

\item As stated in \cite{Greenberg:2003nv}, one does not need to use any assumption related to the (anti)commutators between fields if, instead of using just the Wightman function, one uses the time-ordered product $\tau$ of such functions as in eq. \eqref{tau}. This of course is true also in the deformed context, but in light of the discussion on the previous point it is not clear whether the $\tau$ in \eqref{tau} does indeed correspond to the correct physical $n$-point function. For example, in \cite{Arzano:2018gii} it was shown that the time-ordered vacuum expectation value (VEV) of two fields gives the correct propagator only in some subset of phase space (for trans-Planckian momenta the propagator is not the time-ordered VEV). The model considered in \cite{Arzano:2018gii} is different from the one discussed in this work, nevertheless it highlights the fact that general conclusions about the behaviour of time-ordered VEV are in general groundless, and a lot depends on the considered model. In other words, one could question whether the correct physical quantities to be considered are time ordered products of Wightman functions in the first place. 

	

\item The weak local commutativity in the proof of Jost theorem needs to hold at Jost points which are defined in terms of difference of coordinates. However, how to sum or subtract coordinate vectors is not a trivial matter in the deformed context. 

\end{itemize}

As a concrete proof of the above considerations, in what follows we will explicitly show that the two-point function $\langle 0 | \phi^\dag (x_1) \star \phi(x_2) |0\rangle$ satisfies the CPT invariance relation, in the form of eq. \eqref{CPTinv}, but it is not Lorentz invariant, meaning that eq. \eqref{complexlorentz} (in the real limit) is not satisfied, and therefore both Jost and Greenberg theorem are not valid.

\subsubsection{$CPT$ properties of $\langle 0 | \phi_1^\dag(x_1) \star \phi_2(x_2) |0\rangle$}

We want to show that 
\begin{align}\label{CPTJostdef}
	\langle 0 | \phi^\dag (x_1) \star \phi(x_2) |0\rangle
	=
	\langle 0 | \phi(-x_1) \star \phi^\dag(-x_2) |0\rangle^*
\end{align}
In order to do so, we use the fields in eq. \eqref{fa} and \eqref{fb}, and we have
\begin{align}
	&\langle 0 | \phi^\dag (x_1) \star \phi(x_2) |0\rangle
 \nonumber \\
	&=
	\int
	\frac{d^4q}{q_4/\kappa}\, \frac{d^4p}{p_4/\kappa}\, 
	\theta(p_0)\,
	\zeta(p)\zeta(q) \,
	\langle 0 | b_{\mathbf{q} }b_{\mathbf{p} }^\dag |0\rangle
	e^{-i(\omega_qt_1-\mathbf{q}\mathbf{x}_1)}
	\star 
	e^{-i(S(\omega_p)t_2-S(\mathbf{p})\mathbf{x}_2)} 
\end{align}
\begin{align}
	&\langle 0 | \phi (-x_1) \star \phi^\dag(-x_2) |0\rangle^*
	 \\
	&=
	\int
	\frac{d^4p}{p_4/\kappa}\, \frac{d^4q}{q_4/\kappa}\,  
	\theta(p_0)\,
	\zeta(p)\zeta(q) \,
	\langle 0 | a_{\mathbf{p}} a_{\mathbf{q}}^\dag |0\rangle
	e^{i(\omega_pt_2-\mathbf{p}\mathbf{x}_2)} 
	\star
	e^{i(S(\omega_q)t_1-S(\mathbf{q})\mathbf{x}_1)}
\end{align}
Recalling eq. \eqref{Symp15-t} and \eqref{Symp16-t} we immediately have
\begin{align}
	\langle 0 | a_{\mathbf{p}} a_{\mathbf{q}}^\dag |0\rangle
	=
	\langle 0 | [a_{\mathbf{p}}, a_{\mathbf{q}}^\dag] |0\rangle
	=
	\langle 0 | [b_{\mathbf{q} }, b_{\mathbf{p} }^\dag] |0\rangle
	=
	\langle 0 | b_{\mathbf{q} }b_{\mathbf{p} }^\dag |0\rangle
\end{align}
so we only need to consider the plane waves. Notice that we are working in different spacetime points (but with $p=q$ because of the Dirac delta and we are on-shell). We now use the commutation relation
\begin{align}\label{kappamink-differentpoints}
	[x_1^0, x_2^j] = \frac{i}{\kappa} x_2^j.
\end{align}
We compute the two star products in the bicrossproduct basis for simplicity.
We have (using the time-to-the-right ordering)
\begin{align}
	\hat{e}_k(x) \hat{e}_l(x)
	=
	\hat{e}_{k\oplus l}(x)
	=
	e^{i(\mathbf{k} \mathbf{x} + e^{-\frac{k_0}{\kappa}}\mathbf{l}\mathbf{x})}
	e^{i(k_0x^0 + l_0x^0)}
\end{align}
\begin{align}
	\hat{e}_{S(k)}(x)
	=
	e^{-i e^{\frac{k_0}{\kappa}}\mathbf{k}\mathbf{x}}
	e^{-ik_0x^0}
\end{align}
Before proceeding recall that the inverse Weyl map sends \cite{Arzano:2020jro}
\begin{align}
	\mathcal{W}(\hat{e}_k(x))
	=
	e^{-i(\omega_k t - \mathbf{k}\mathbf{x})}
	\qquad
	\Leftrightarrow
	\qquad
	\mathcal{W}^{-1}
	\left(
	e^{-i(\omega_k t - \mathbf{k}\mathbf{x})}
	\right)
	=
	\hat{e}_k(x)
\end{align}
and therefore we have
\begin{align}
	\mathcal{W}^{-1}
	\left(
	e^{i(\omega_k t - \mathbf{k}\mathbf{x})}
	\right)
	=
	\mathcal{W}^{-1}
	\left(
	e^{-i(\omega_k (-t) - \mathbf{k}(-\mathbf{x}))}
	\right)
	=
	\hat{e}_k(-x)
\end{align}
Using this fact we have
\begin{align}
	e^{i(\omega_pt_2-\mathbf{p}\mathbf{x}_2)} 
	\star
	e^{i(S(\omega_p)t_1-S(\mathbf{p})\mathbf{x}_1)}
	&\mapsto
	\hat{e}_p(-x_1)\hat{e}_{S(p)}(-x_2) 
\end{align}	
and computing the RHS explicitly we have
\begin{align}
	e^{i(\mathbf{p} (-\mathbf{x}_2) + e^{-\frac{p_0}{\kappa}}(- e^{\frac{p_0}{\kappa}}\mathbf{p})(-\mathbf{x}_1))}
	e^{i(p_0(-x^0_2) - p_0(-x^0_1))} 
	=
	e^{i\mathbf{p}(\mathbf{x}_1 - \mathbf{x}_2)}
	e^{ip_0(x^0_1 - x^0_2)}
\end{align}
and furthermore
\begin{align}
	e^{-i(\omega_pt_1-\mathbf{p}\mathbf{x}_1)}
	\star 
	e^{-i(S(\omega_p)t_2-S(\mathbf{p})\mathbf{x}_2)} 
	\mapsto
	\hat{e}_p(x_1)\hat{e}_{S(p)}(x_2)
\end{align}
and the RHS computed explicitly gives
\begin{align}
	e^{i(\mathbf{p} \mathbf{x}_1 + e^{-\frac{p_0}{\kappa}}(- e^{\frac{p_0}{\kappa}}\mathbf{p})\mathbf{x}_2)}
	e^{i(p_0x^0_1 - p_0x^0_2)} 
	&=
	e^{i\mathbf{p}(\mathbf{x}_1 - \mathbf{x}_2)}
	e^{ip_0(x^0_1 - x^0_2)}
\end{align}
Therefore the two integrands are equal and we have shown that eq. \eqref{CPTJostdef} holds. Notice that eq. \eqref{CPTJostdef} is a particular case of the expression
\begin{align}\label{CPTJostdefvera}
	\langle 0 | \phi_1 (x_1) \star \phi_2(x_2) |0\rangle
	=
	\langle 0 | \phi_1^\dag(-x_1) \star \phi_2^\dag(-x_2) |0\rangle^*
\end{align}
with the choice $\phi_1(x_1)=\phi^\dag(x_1)$, $\phi_2(x_2)=\phi(x_2)$. Any other choice leads to zero in our case, because in each case there would be an operator $a$ or $b$ acting on the vacuum (recall that $[a,b]=[a,b^\dag]=0$), or terms of the kind $\langle 0 | a^\dag a^\dag |0\rangle =0$ because $a^\dag a^\dag |0\rangle \neq |0\rangle$ and $a^\dag a^\dag |0\rangle \perp |0\rangle$. The only other choice that does give a non-zero contribution is $\langle 0 | \phi^\dag (x_1) \star \phi(x_2) |0\rangle$, however in this case the computations go in exactly the same way. Therefore, we have proved that eq. \eqref{CPTJostdefvera} holds for any combination of two scalar fields in the deformed context. 

Notice the important observation that, even if we used the commutation relations
\begin{align}\label{ultraloc}
	[x_1^0, x_2^i]
	=
	\begin{cases}
		0, & \text{if}\ x_1\neq x_2 \\
		\frac{i}{\kappa} x_2^i, & \text{if}\ x_1=x_2
	\end{cases}
\end{align}
we would have gotten the same result. Indeed, we compute these expectation values at different points, and if we use the fact that fields at different points behave as if they were non-deformed, we reach exactly the same result as the canonical case.

\subsubsection{Lorentz properties of $\langle 0 | \phi_1^\dag(x_1) \star \phi_2(x_2) |0\rangle$}

Since we are considering products of fields at different spacetime points, in this section we work in the context of the commutation relations in eq. \eqref{ultraloc}. In this way, we get that the boosts act with a trivial co-product on star products. From the first order boost transformations in eq. \eqref{LOphibo} and \eqref{LOphidagbo} we immediately have (up to first order in the parameter $\lambda_i$)
\begin{align}
	&\langle 0 | \Lambda \phi^\dag (x_1) \star \Lambda \phi(x_2) |0\rangle \nonumber  \\
	&\approx
	\langle 0 | \phi^\dag (x_1) \star \phi(x_2) |0\rangle
	\nonumber \\
	&+
	\langle 0 | \left(-i \lambda_i\, x^i_1\frac{\partial }{\partial x^0_1}\,\phi^\dag (x_1)\right) \star \phi(x_2) |0\rangle 
	+
	\langle 0 | \phi^\dag (x_1) \star \left(i \lambda_i\, x^i_2\frac{\partial }{\partial x^0_2}\,\phi(x_2) \right)|0\rangle \nonumber \\ 
	&+
	\int \, \frac{d^3p}{\sqrt{2\omega_\mathbf{p}}} \, \frac{d^3q}{\sqrt{2\omega_\mathbf{q}}}
	\zeta(p)\zeta(q) \,
	\langle 0 | b_{\mathbf{q} }b_{\mathbf{p} }^\dag |0\rangle
	e^{-i(\omega_qt_1-\mathbf{q}\mathbf{x}_1)}
	\star 
	e^{-i(S(\omega_p)t_2-S(\mathbf{p})\mathbf{x}_2)} \times \nonumber \\
	&\times
	\left[
	\zeta(p)
	i\lambda^i
	\frac{\mathbf{q}_i}{\kappa}
	\left(
	\frac{m^2}{\omega_\mathbf{q}^2}
	-
	2
	\right)
	-
	\zeta(q)
	i\lambda^i
	\frac{\mathbf{p}_i}{\kappa}
	\left(
	\frac{5}{2}
	-\frac{m^2 }{2\omega_\mathbf{p}^2}
	\right)
	\right]
	\nonumber \\ \nonumber \\
	&=
	\langle 0 | \phi^\dag (x_1) \star \phi(x_2) |0\rangle
	\nonumber \\
	&
	-i \lambda_i\, x^i_1\frac{\partial }{\partial x^0_1}\,
	\langle 0 | \phi^\dag (x_1) \star \phi(x_2) |0\rangle 
	+
	i \lambda_i\, x^i_2\frac{\partial }{\partial x^0_2}\,
	\langle 0 | \phi^\dag (x_1) \star \phi(x_2) |0\rangle \nonumber \\ 
	&+
	\int \, \frac{d^3p}{2\omega_\mathbf{p}} \,
	e^{i\mathbf{p}(\mathbf{x}_1 - \mathbf{x}_2)}
	e^{ip_0(x^0_1 - x^0_2)} 
	\left[
	i\lambda^i
	\frac{\mathbf{p}_i}{\kappa}
	\left(
	\frac{3m^2}{2\omega_\mathbf{p}^2}
	-
	\frac{9}{2}
	\right)
	\right]
\end{align}
were we used the fact that $[a, a^\dag] = [b, b^\dag]=1$
and the fact that
\begin{align}
	\zeta(p) 
	\approx 
	1
	-
	\frac{1}{\kappa}
	\frac{2\omega_\mathbf{p}^2 + m^2}{4\omega_\mathbf{p}}.
\end{align}
We first consider the terms
\begin{align}\label{remaining}
	-i \lambda_i\, x^i_1\frac{\partial }{\partial x^0_1}\,
	\langle 0 | \phi^\dag (x_1) \star \phi(x_2) |0\rangle 
	+
	i \lambda_i\, x^i_2\frac{\partial }{\partial x^0_2}\,
	\langle 0 | \phi^\dag (x_1) \star \phi(x_2) |0\rangle
\end{align}
Using the CPT property \eqref{CPTJostdef} of the two-point function we can rewrite these two terms as
\begin{align}
	-i \lambda_i\, x^i_1\frac{\partial }{\partial x^0_1}\,
	\langle 0 | \phi^\dag (x_1) \star \phi(x_2) |0\rangle 
	+
	i \lambda_i\, x^i_2\frac{\partial }{\partial x^0_2}\,
	\langle 0 | \phi(-x_1) \star \phi^\dag(-x_2) |0\rangle
\end{align}
Using the fields in eq. \eqref{fa} and \eqref{fb} one an now show that (keeping only the relevant part for the two-point function)
\begin{align}
	x^i\frac{\partial }{\partial x^0}\,\phi^\dag(x) 
	= 
	&\int \frac{d^4q}{q_4/\kappa}\, \theta(q_0)\, \zeta(q)\, 
	a^\dag_\mathbf{q}\,
	\left(
	-iS(\omega_q)
	\frac{\partial}{\partial i S(\mathbf{q})_i}
	\right)
	e^{-i(S(\omega_q)x^0-S(\mathbf{q})\mathbf{x})} 
	\nonumber \\
	+ &\int \frac{d^4q}{q_4/\kappa}\, \theta(q_0)\, 
	\zeta(q)\, 
	b_\mathbf{q }\,
	\left(
	-i\omega_q
	\frac{\partial}{\partial i\mathbf{q}_i}
	\right)
	e^{-i(\omega_qx^0-\mathbf{q}\mathbf{x})}.
\end{align}
\begin{align}
	x^i\frac{\partial }{\partial x^0}\,\phi^\dag(-x) 
	= 
	&\int \frac{d^4q}{q_4/\kappa}\, \theta(q_0)\, \zeta(q)\, 
	a^\dag_\mathbf{q}\,
	\left(
	iS(\omega_q)
	\frac{\partial}{\partial (-i S(\mathbf{q})_i)}
	\right)
	e^{i(S(\omega_q)x^0-S(\mathbf{q})\mathbf{x})} 
	\nonumber \\
	+ &\int \frac{d^4q}{q_4/\kappa}\, \theta(q_0)\, 
	\zeta(q)\, 
	b_\mathbf{q }\,
	\left(
	i\omega_q
	\frac{\partial}{\partial (-i\mathbf{q}_i)}
	\right)
	e^{i(\omega_qx^0-\mathbf{q}\mathbf{x})}.
\end{align}
Hence we have
\begin{align}
	x^i_1\frac{\partial }{\partial x^0_1} \phi^\dag(x_1)
	=
	x^i_2\frac{\partial }{\partial x^0_2} \phi^\dag(-x_2)
\end{align}
and therefore
\begin{align}
	x^i_1\frac{\partial }{\partial x^0_1}\,
	\langle 0 | \phi^\dag (x_1) \star \phi(x_2) |0\rangle 
	=
	x^i_2\frac{\partial }{\partial x^0_2}\,
	\langle 0 | \phi^\dag (x_1) \star \phi(x_2) |0\rangle
\end{align}
Because of this relation, eq. \eqref{remaining} is therefore zero
\begin{align}
	-i \lambda_i\, x^i_1\frac{\partial }{\partial x^0_1}\,
	\langle 0 | \phi^\dag (x_1) \star \phi(x_2) |0\rangle 
	+
	i \lambda_i\, x^i_2\frac{\partial }{\partial x^0_2}\,
	\langle 0 | \phi^\dag (x_1) \star \phi(x_2) |0\rangle
	=0
\end{align}
and the boost transformation property for the two point function becomes 
\begin{align}\label{nnlstep}
	\langle 0 | \Lambda \phi^\dag (x_1) \star \Lambda \phi(x_2) |0\rangle
	&\approx
	\langle 0 | \phi^\dag (x_1) \star \phi(x_2) |0\rangle \\
	&+
	\int \, \frac{d^3p}{2\omega_\mathbf{p}} \,
	e^{i\mathbf{p}(\mathbf{x}_1 - \mathbf{x}_2)}
	e^{ip_0(x^0_1 - x^0_2)} 
	\left[
	i\lambda_i
	\frac{\mathbf{p}_i}{\kappa}
	\left(
	\frac{3m^2}{2\omega_\mathbf{p}^2}
	-
	\frac{9}{2}
	\right)
	\right].
\end{align}
We now treat the second term on the RHS. First of all, recall that we are performing all the computations at equal times, so that we only need to consider 
\begin{align}
	\int \, \frac{d^3p}{2\omega_\mathbf{p}} \,
	e^{i\mathbf{p}(\mathbf{x}_1 - \mathbf{x}_2)}
	\left[
	i\lambda_i
	\frac{\mathbf{p}_i}{\kappa}
	\left(
	\frac{3m^2}{2\omega_\mathbf{p}^2}
	-
	\frac{9}{2}
	\right)
	\right]
\end{align}
Calling $\Delta \mathbf{x} = \mathbf{x}_1 - \mathbf{x}_2$ and assuming without loss of generality that $\Delta \mathbf{x} \parallel \mathbf{z}$ we have
\begin{align}
	-2\pi 
	\frac{\partial}{\partial i\Delta \mathbf{x}^i} 
	\int \, \frac{d|\mathbf{p}| \mathbf{p}^2 d \cos\theta}{2\omega_\mathbf{p}} \,
	e^{i|\mathbf{p}||\Delta \mathbf{x}| \cos\theta}
	\left[
	i\lambda_i
	\frac{1}{\kappa}
	\left(
	\frac{3m^2}{2\omega_\mathbf{p}^2}
	-
	\frac{9}{2}
	\right)
	\right]
\end{align}
Performing the angular integral one gets
\begin{align}
	-2\pi 
	\frac{\partial}{\partial i\Delta \mathbf{x}^i} 
	\int_0^\infty \, \frac{d|\mathbf{p}| \mathbf{p}^2}{2\omega_\mathbf{p}} \,
	\frac{1}{i|\mathbf{p}||\Delta \mathbf{x}|}
	\left(
	e^{-i|\mathbf{p}||\Delta \mathbf{x}|}
	-
	e^{i|\mathbf{p}||\Delta \mathbf{x}|}
	\right)
	\left[
	i\lambda_i
	\frac{1}{\kappa}
	\left(
	\frac{3m^2}{2\omega_\mathbf{p}^2}
	-
	\frac{9}{2}
	\right)
	\right]
\end{align}
which can be equivalently written as
\begin{align}
	\frac{\pi}{|\Delta \mathbf{x}|} 
	\frac{\lambda^i}{\kappa}
	\frac{\partial}{\partial i\Delta \mathbf{x}^i} 
	\int_{-\infty}^\infty \, \frac{d|\mathbf{p}| |\mathbf{p}|}{2\omega_\mathbf{p}} \,
	e^{i|\mathbf{p}||\Delta \mathbf{x}|}
	\left(
	\frac{3m^2}{\omega_\mathbf{p}^2}
	-
	9
	\right)
\end{align}
This means that we only need to compute the two Fourier transforms
\begin{align}\label{ft1}
	\frac{3m^2}{2}
	\int_{-\infty}^\infty \, d|\mathbf{p}|  \,
	e^{i|\mathbf{p}||\Delta \mathbf{x}|}
	\frac{|\mathbf{p}|}{(m^2 + \mathbf{p}^2)^{3/2}}
\end{align}
\begin{align}\label{ft2}
	-\frac{9}{2}
	\int_{-\infty}^\infty \, d|\mathbf{p}|  \,
	e^{i|\mathbf{p}||\Delta \mathbf{x}|}
	\frac{|\mathbf{p}|}{(m^2 + \mathbf{p}^2)^{1/2}}
\end{align}
One can easily check that, since $m>0, |\Delta \mathbf{x}|>0$, we have
\begin{align}
	\frac{3m^2}{2}
	\int_{-\infty}^\infty \, d|\mathbf{p}|  \,
	e^{i|\mathbf{p}||\Delta \mathbf{x}|}
	\frac{|\mathbf{p}|}{(m^2 + \mathbf{p}^2)^{3/2}}
	=
	3i m^2
	|\Delta \mathbf{x}| K_0(m|\Delta \mathbf{x}|)
\end{align}
\begin{align}\label{ft2}
	-\frac{9}{2}
	\int_{-\infty}^\infty \, d|\mathbf{p}|  \,
	e^{i|\mathbf{p}||\Delta \mathbf{x}|}
	\frac{|\mathbf{p}|}{(m^2 + \mathbf{p}^2)^{1/2}}
	=
	-9 i m
	K_1(m|\Delta \mathbf{x}|)
\end{align}
where $K_0(x), K_1(x)$ are the modified Bessel functions of the second kind. The additional term in eq. \eqref{nnlstep} therefore becomes
\begin{align}
	\frac{\pi}{|\Delta \mathbf{x}|} 
	\frac{\lambda^i}{\kappa}
	\frac{\partial m|\Delta \mathbf{x}|}{\partial \Delta \mathbf{x}^i} 
	\frac{\partial}{\partial m|\Delta \mathbf{x}|}
	\left(
	3m^2
	|\Delta \mathbf{x}| K_0(m|\Delta \mathbf{x}|)
	-9
	m
	K_1(m|\Delta \mathbf{x}|)
	\right)
\end{align}
Using the fact that
\begin{align}
	\frac{\partial}{\partial x}
	K_0(x) = -K_1(x)
	\qquad
	\frac{\partial}{\partial x}
	K_1(x)
	=
	-\frac{1}{2}
	(K_0(x)+K_2(x))
\end{align}
we end up with
\begin{align}
	2\pi
	\frac{m}{\kappa}
	\lambda_i  
	\frac{\Delta \mathbf{x}^i}{|\Delta \mathbf{x}|} 
	\left(
	\frac{15 m}{2} 
	K_0(m|\Delta \mathbf{x}|)
	-
	3 m^2
	|\Delta \mathbf{x}| 
	K_1(m|\Delta \mathbf{x}|)
	+
	\frac{9m}{2}
	K_2(m|\Delta \mathbf{x}|))
	\right)
\end{align}
For small values of $|\Delta \mathbf{x}|$ one can show that 
\begin{align}
	\frac{15 m}{2} 
	K_0(m|\Delta \mathbf{x}|)
	-
	3 m^2
	|\Delta \mathbf{x}| 
	K_1(m|\Delta \mathbf{x}|)
	+
	\frac{9m}{2}
	K_2(m|\Delta \mathbf{x}|))
	\approx
	\frac{9}{m |\Delta \mathbf{x}|^2}
\end{align}
while for large values of $|\Delta \mathbf{x}|$ one has
\begin{align}
	\frac{15 m}{2} 
	K_0(m|\Delta \mathbf{x}|)
	-
	3 m^2
	|\Delta \mathbf{x}| 
	K_1(m|\Delta \mathbf{x}|)
	+
	\frac{9m}{2}
	K_2(m|\Delta \mathbf{x}|))
	\approx
	-3 m^{\frac{1}{2}}
	\sqrt{\frac{\pi}{2}}
	e^{-m|\Delta \mathbf{x}|}
	\sqrt{|\Delta \mathbf{x}|}.
\end{align}
As a final comment, let us note that if we used the commutation relation in eq. \eqref{kappamink-differentpoints} instead of eq. \eqref{ultraloc}, then we would have a non-trivial co-product acting on the star product. However, the additional contributions that one gets can be shown not to cancel the extra factor in eq. \eqref{nnlstep}.

\section{Phenomenological consequences of deformed $CPT$ transformations}\label{def-cpt-pheno}

We are now at point $xi)$ of our roadmap in section \ref{roadmap2}. We have already seen the behaviour of infinitesimal boosts on states in eq. \eqref{Naction}, from which one can get the finite (canonical) action given by
\begin{align}
	|p_1,p_2,p_3\rangle_a &\mapsto |\cosh\xi\, p_1 +\sinh\xi\, \omega_\mathbf{p},p_2,p_3\rangle_a \label{finiteboost-a} \\
	|p_1,p_2,p_3\rangle_b &\mapsto |\cosh\xi\, p_1 +\sinh\xi\, \omega_\mathbf{p},p_2,p_3\rangle_b.\label{finiteboost-b}
\end{align}
We now consider the case of a particle and an antiparticle originally at rest, and we put ourselves in their center of mass frame. Notice that we are not considering the two as being part of a single tensor state, but as two separate single states. As such, we are ignoring contributions coming from the finite boost of two-particle states described in section \ref{twopartkin} of chapter \ref{Chapter1}. If we boost their initial single-particle states at rest with the boosts \eqref{finiteboost-a}, \eqref{finiteboost-b}  we have
\begin{align}\label{finiteboost0}
	|\mathbf{p}=0\rangle_a &\mapsto |M\sinh\xi,0,0\rangle_a\nonumber\\
	|\mathbf{p}=0\rangle_b &\mapsto |M\sinh\xi,0,0\rangle_b
\end{align}
where
\begin{align}
	\mathcal{P}_1 |M\sinh\xi,0,0\rangle_a
	&	=
	-S(M\sinh\xi)_1|M\sinh\xi,0,0\rangle_a
	\\
	\mathcal{P}_0 |M\sinh\xi,0,0\rangle_a
	&	=
	-S(M\cosh\xi) |M\sinh\xi,0,0\rangle_a
\end{align}
\begin{align}
	\mathcal{P}_1 |M\sinh\xi,0,0\rangle_b
	&	=
	M\sinh\xi|M\sinh\xi,0,0\rangle_b
	\\
	\mathcal{P}_0 |M\sinh\xi,0,0\rangle_b
	&	=
	M\cosh\xi |M\sinh\xi,0,0\rangle_b 
\end{align}
with
\begin{align}
	-S(M\sinh\xi)_1 &= \frac{\kappa M\sinh\xi}{M\cosh\xi  +\sqrt{\kappa^2 +M^2}}\,,\label{antifen1}\\
	-S(M\cosh\xi)&=M \cosh\xi - \frac{M^2\sinh^2\xi}{M\cosh\xi  +\sqrt{\kappa^2 +M^2}}\label{antifen2}
\end{align}
It is clear that the $C$ operator switches a particle for an antiparticle of a different momentum due to its non-commutativity with the boost. In order to derive phenomenological consequences, it is helpful to first expand at first order in $1/\kappa$ eq. \eqref{antifen1} and \eqref{antifen2}, obtaining
\begin{eqnarray}
	-S(\omega_\mathbf{p}) & = & \omega_\mathbf{p}-\frac{\mathbf p^2}{\kappa}
	+{\mathcal O}(1/\kappa^2), \label{antifen1-esp}  \\
	-S(\mathbf p)_i 
	& = & (\mathbf p)_i-\frac{(\mathbf p)_i \omega_\mathbf{p}}{\kappa} + {\mathcal O}(1/\kappa^2) \label{antifen2-esp}.
\end{eqnarray}
The $\gamma$ factor corresponding to the two boosts will be different because of the presence of the antipode, and in particular it will be $\omega_\mathbf{p}/M$ for the particle and $-S(\omega_\mathbf{p})/M$ for the antiparticle. 

We now consider an unstable particle/antiparticle couple. Their evolution is dictated by a complex energy, where the imaginary part is responsible for the description of the decay (since it corresponds to an exponential decay of both the amplitude and the probability density function). In particular, using proper time $t>0$ as a parameter, the amplitudes will be given by
\begin{eqnarray}
	\psi_{\mbox{\scriptsize part}}(t) & = & A(M,\Gamma,E) \, \exp \, \left[-i(E-\frac{i}{2}\Gamma) \frac{E}{m} \,t\right], \label{psi1} \\
	\psi_{\mbox{\scriptsize apart}}(t) & = & A[M,\Gamma,S(E)] \, \exp \,\left[-i\left(S(E)-\frac{i}{2}\Gamma\right) \frac{S(E)}{m} \,t\right], \label{psi2}
\end{eqnarray}
where $A$ is a normalization factor. Notice that we used the notation $\omega_\mathbf{p} = E - \frac{i}{2}\Gamma$, where $\Gamma = \frac{1}{\tau}$ describes the reciprocal of the decay time and is called decay width. Furthermore, we used the fact that 
\begin{align}
	S(\omega_\mathbf{p}) = S(E) - \frac{i}{2}S(\Gamma)
	\approx
	S(E) - \frac{i}{2}\Gamma
\end{align}
i.e. we used the assumption that the decay width is the same for particles and antiparticles (we are only considering total decay width, so there are no corrections on considering only some of the total decay channels), so that it is unaffected by the antipode. This is an a-priori well defined assumption since any correction to $\Gamma$ (which is computed in the center of mass frame for both particles and antiparticles) can only be proportional to $\frac{m}{\kappa}$ or $\frac{M}{\kappa}$, where $M$ is the mass of the decaying particle and $m$ stands for the mass of some decay product. Because we expect $\kappa$ to be of the order of magnitude of the Planck energy, such contributions (which are also not enhanced by boosting since both $m$ and $M$ are the invariant rest masses) are negligible. We explicitly verify such a-priori estimate by computing the correction to $\Gamma$ for the decay of a single particle $\phi$ of mass $M$ into two particles $\chi$ of mass $m$ in the next section \ref{Appendix}. 

From the amplitudes, one can get the probability density functions
\begin{eqnarray}
	{\cal P}_{\mbox{\scriptsize part}}(t) & = & \frac{\Gamma E}m\exp \,\left(-\Gamma \,\frac {E}{m}\, t \right),  \label{decay}\\
	{\cal P}_{\mbox{\scriptsize apart}}(t) & = & \frac{\Gamma S(E)}{m}\, \exp \, \left[-\Gamma \, \frac{ S(E)}{m}\, t\right] \nonumber \\
	& = & \Gamma\left(\frac Em - \frac{\mathbf p^2}{\kappa m}\right) \,\exp\,\left [-\Gamma \, \left(\frac Em - \frac{\mathbf p^2}{\kappa m}\right)\, t\,\right ].\label{decaya}
\end{eqnarray}
For the moment, we will not discuss possible contributions coming from non-zero mass distribution width coming from loop corrections to the propagator of the decaying particle. We will return to this issue at the end of the next chapter \ref{Chapter3}, section \ref{mdw}. 

In order to measure the effects on decay times due to deformation, the best candidate are particle/antiparticle pairs with small mass and high momentum, so that the quantity $\frac{\mathbf{p}^2}{M\kappa}$ is the biggest possible. Natural candidates are therefore $\mu^+, \mu^-$, for which lifetimes are also known with high accuracy. A detailed discussion in this case can be found in \cite{Arzano:2019toz}. One can also highlight the phenomenological consequences for the next best candidates \cite{Arzano:2020jro}.

We consider a particle decaying into a particle/antiparticle pair in the center of mass, which will produce two particles moving back to back (any correction to the modulus of the spatial momentum due to deformation, and in particular the antipode, will contribute higher order corrections to eq. \eqref{decay}, \eqref{decaya}, and so it can be ignored for the moment). Boosting in the lab frame, and choosing only particles which (in the center of mass frame) decay orthogonally to the boost direction, we will obtain particles whose momenta are pointing approximately along the boost direction, with deviation angle $\theta$ of the order of $10^{-4}$ rad to $10^{-6}$ rad. One can then rotate one momentum over the other, aligning them. This is equivalent to assuming that any effect which is not $CPT$-deformation related is absent. Indeed, for example, possible experimental signatures of anisotropy-induced corrections to CPT and Lorentz symmetries have been extensively studied, confirming the absence of a preferential direction in spacetime with better accuracy than we are considering in this section \cite{Babusci:2013gda}, \cite{Babusci:2021}, \cite{lhcb_cpt}, \cite{babar_cpt}, \cite{d0_cpt}, \cite{focus_cpt}. Notice that, in general, anisotropy-induced corrections to CPT and Lorentz symmetries require working with angles which are much larger than the $\theta$ we are considering here. With big enough data samples, one could obtain a high enough accuracy on the average value of $\theta$. Such a control over $\theta$ is however outside current experimental techniques. The candidate decaying particles can be produced in current particle accelerators (like LHC) or future one (FCC, see \cite{fcc}), and in Table \ref{Table:tab1} we report several possible decay channels, together with the possible values of Lorentz boosts, experimental errors, and limits on the value of $\kappa$. 

\subsection{Estimation of deformed decay width}\label{Appendix}

In this section we present a simple calculation showing that corrections to the decay width resulting from $\kappa$-deformation are neglectable. For simplicity we consider only the $\phi\chi^2$ interaction. We consider decay of a single $\phi$ particle of mass $M$ to two $\chi$ particles of the mass $m$. We have
\begin{align}
	d\Gamma = \frac{1}{2M} d\text{LIPS}_2 |\mathcal{T}|^2
\end{align}
where at tree level $\mathcal{T}=g$ where $g$ is the coupling constant. 
Since $\mathcal{T}$ is computed in the rest frame, the corrections might be only of the form $m/\kappa$ and $M/\kappa$. In the non-deformed theory the Lorentz-invariant phase space factor has the form
\begin{align}
	d\text{LIPS}_2
	&=
	(2\pi)^4 \delta^4(k_1+k_2 - k)
	\frac{d^3 k_1}{(2\pi)^3 2\omega_{k_1}}
	\frac{d^3 k_2}{(2\pi)^3 2\omega_{k_2}} \nonumber\\
	&=
	\frac{1}{4(2\pi)^2\omega_{k_1}\omega_{k_2}} \delta^4(k_1+k_2 - k)
	d^3 k_1
	d^3 k_2\label{dGamma},
\end{align}
where $k$ is the incoming momentum, while $k_1, k_2$ are the outgoing ones.

In the deformed case the only difference comes from the fact that instead of the ordinary delta function we have the one of deformed momentum composition rule
\begin{align}
	\delta^4(k_1 \oplus k_2 \oplus S(k))
\end{align}
Since here we want to find just the leading order deviation from the  non-deformed theory, we consider only one ordering here; all other orderings will give the same first-order result. 

Recall that to the leading in $1/\kappa$ 
we have
\begin{align}
	(p\oplus q)_0 &= p_0+q_0 +\frac{\mathbf{p}\mathbf{q}}\kappa\,,\quad (p\oplus q)_i= p_i+q_i +\frac{p_i\, q_0}\kappa \nonumber\\
	S(p)_0 &= - p_0 + \frac{\mathbf{p}^2}\kappa\,,\quad S(p)_i = - p_i +\frac{p_ip_0}\kappa
\end{align}

We compute $\Gamma$ in the reference frame in which the initial particle is at rest so that $\mathbf{k}=0$. Then
\begin{align}
	&\delta^4(k_1\oplus k_2 \oplus S(k)) \nonumber \\
	&= \delta\left(\omega_{\mathbf{k}_1}\oplus \omega_{\mathbf{k}_2}\oplus S(\omega_{\mathbf{k}})\right)\delta^3\left(\mathbf{k}_1\oplus \mathbf{k}_2\right)\nonumber\\
	&= \delta\left(\omega_{\mathbf{k}_1}+ \omega_{\mathbf{k}_2}- \omega_{\mathbf{k}} + \frac{\mathbf{k}_1\mathbf{k}_2}\kappa\right)\delta^3\left(\mathbf{k}_1+ \mathbf{k}_2+\frac{\mathbf{k}_1\omega_{\mathbf{k}_2}}\kappa \right) \nonumber \\
	&=
	\delta\left(\omega_{\mathbf{k}_1}+ \omega_{\mathbf{k}_2}- \omega_{\mathbf{k}} + \frac{\mathbf{k}_1\mathbf{k}_2}\kappa\right)
	\delta^3\left(
	\mathbf{k}_1
	+ \mathbf{k}_2
	\left(
	1
	-
	\frac{\omega_{\mathbf{k}_2}}{\kappa}
	\right) \right)
	(1 - 3\frac{\omega_{\mathbf{k}_2}}{\kappa})
\end{align}
Notice that 
\begin{align}
	\mathbf{k}_1
	=
	-\mathbf{k}_2
	\left(
	1
	-
	\frac{\omega_{\mathbf{k}_2}}{\kappa}
	\right)
	=
	-\mathbf{k}_2
	+
	\frac{\mathbf{k}_2 \omega_{\mathbf{k}_2}}{\kappa}
\end{align}
and therefore
\begin{align}
	\omega_{\mathbf{k}_1}
	=
	\sqrt{m^2 + \mathbf{k}_2^2 - 2\frac{\mathbf{k}_2^2 \omega_{\mathbf{k}_2}}{\kappa}}
	=
	\omega_{\mathbf{k}_2}
	\sqrt{1 - 2\frac{\mathbf{k}_2^2}{\kappa\omega_{\mathbf{k}_2}}}
	\approx
	\omega_{\mathbf{k}_2}
	\left(
	1 - \frac{\mathbf{k}_2^2}{\kappa\omega_{\mathbf{k}_2}}
	\right)
\end{align}
which means that 
\begin{align}
	\frac{1}{\omega_{\mathbf{k}_1}}
	\approx
	\frac{1}{\omega_{\mathbf{k}_2}}
	\left(
	1 + \frac{\mathbf{k}_2^2}{\kappa\omega_{\mathbf{k}_2}}
	\right)
\end{align}
Substituting this to \eqref{dGamma} and integrating we get
\begin{align}
	\Gamma
	&=\frac{g^2}{2M}\int
	\frac{d^3 k_1
		d^3 k_2}{4(2\pi)^2\omega_{\mathbf{k_1}}\omega_{\mathbf{k_2}}} 
	\, \delta\left(\omega_{\mathbf{k}_1}+ \omega_{\mathbf{k}_2}- \omega_{\mathbf{k}} + \frac{\mathbf{k}_1\mathbf{k}_2}\kappa\right)\delta^3\left(\mathbf{k}_1+ \mathbf{k}_2+\frac{\mathbf{k}_1\omega_{\mathbf{k}_2}}\kappa \right)\nonumber\\
	&=
	-\frac{g^2}{2M}\int
	\frac{\mathbf{k}_2 d^3 k_2}{4(2\pi)^2\omega_{\mathbf{k_2}}^2} 
	\left(
	1 + \frac{\mathbf{k}_2^2}{\kappa\omega_{\mathbf{k}_2}}
	\right)
	\left(
	1
	-
	3
	\frac{\omega_{\mathbf{k}_2}}{\kappa}
	\right)
	\, \delta\left(\omega_{\mathbf{k}_1}+ \omega_{\mathbf{k}_2}- \omega_{\mathbf{k}} + \frac{\mathbf{k}_1\mathbf{k}_2}\kappa\right)
	\nonumber\\
	&=
	-\frac{g^2}{2M}\int
	\frac{\mathbf{k}_2^2 d k_2
	}{4\pi\omega^2_{\mathbf{k_2}}} 
	\,
	\left(
	1
	-
	\frac{3m^2 + 2\mathbf{k}_2^2}{\kappa\omega_{\mathbf{k}_2}}
	\right)
	\delta\left(
	2 \omega_{\mathbf{k}_2}- M - \frac{2\mathbf{k}^2_2}\kappa
	\right)
	\label{Gammaint}.
\end{align}
The argument of the delta function is zero for
\begin{align}
	\mathbf{k}^2_2 + m^2 = \left(\frac{\mathbf{k}^2_2}\kappa+\frac{M}{2}\right)^2
\end{align}
which means that
\begin{equation}
	\mathbf  k^2_2 = \left(\frac{M^2} 4 - m^2\right)\left(1-\frac{M}\kappa\right)
\end{equation}
Furthermore, notice that
\begin{align}
	\omega_{\mathbf{k}_2}^2 
	= 
	m^2 + \mathbf  k^2_2 
	&= 
	m^2 + \left(\frac{M^2} 4 - m^2\right)\left(1-\frac{M}\kappa\right) \\
	&=
	\frac{M^2}{4}
	-
	\frac{M}{\kappa}
	\left(\frac{M^2} 4 - m^2\right)
\end{align}
and therefore we have
\begin{align}
	\omega_{\mathbf{k}_2}^{-1} 
	&=
	\frac{2}{M}
	\left(
	1
	+
	\frac{M}{2\kappa}
	\left(1 - 4\frac{m^2}{M^2}\right)
	\right)
\end{align}
Substituting this result in the above steps we get
\begin{align}
	\Gamma
	&=
	-\frac{g^2}{2M}\int
	\frac{\mathbf{k}_2^2 d k_2
	}{4\pi\omega^2_{\mathbf{k_2}}} 
	\,
	\left(
	1
	-
	\frac{3m^2 + 2\mathbf{k}_2^2}{\kappa\omega_{\mathbf{k}_2}}
	\right)
	\delta\left(
	2 \omega_{\mathbf{k}_2}- M - \frac{2\mathbf{k}^2_2}\kappa
	\right) \\
	&=
	-\frac{g^2}{2M}
	\frac{1}{4\pi}
	\frac{4}{M^2}
	\left(
	1
	+
	\frac{M}{\kappa}
	\left(1 - 4\frac{m^2}{M^2}\right)
	\right) 
	\left(\frac{M^2} 4 - m^2\right)\left(1-\frac{M}\kappa\right)\times \nonumber \\
	&\times
	\Big[
	1
	-
	\frac{3m^2}{\kappa}
	\frac{2}{M}
	\left(
	1
	+
	\frac{M}{2\kappa}
	\left(1 - 4\frac{m^2}{M^2}\right)
	\right) \nonumber \\
	&-
	\frac{2}{\kappa}
	\frac{2}{M}
	\left(
	1
	+
	\frac{M}{2\kappa}
	\left(1 - 4\frac{m^2}{M^2}\right)
	\right)
	\left(\frac{M^2} 4 - m^2\right)\left(1-\frac{M}\kappa\right)
	\Big]
\end{align}
and keeping only the terms up to leading order in $1/\kappa$ and using the fact that
\begin{align}
	\left(
	1
	+
	\frac{M}{\kappa}
	\left(1 - 4\frac{m^2}{M^2}\right)
	\right) 
	\left(\frac{M^2} 4 - m^2\right)
	\left(1-\frac{M}\kappa\right)
	&\approx
	\left[
	\left(\frac{M^2} 4 - m^2\right)
	+
	\frac{M}{\kappa}
	\left(
	\frac{4 m^4}{M^2}
	\right)
	\right]
\end{align}
we get
\begin{align}
	\Gamma
	&\approx
	-\frac{g^2}{2M}\int
	\frac{\mathbf{k}_2^2 d k_2
	}{4\pi\omega^2_{\mathbf{k_2}}} 
	\,
	\left(
	1
	-
	\frac{3m^2 + 2\mathbf{k}_2^2}{\kappa\omega_{\mathbf{k}_2}}
	\right)
	\delta\left(
	2 \omega_{\mathbf{k}_2}- M - \frac{2\mathbf{k}^2_2}\kappa
	\right) \\
	&=
	-\frac{g^2}{2M}
	\frac{1}{4\pi}
	\frac{4}{M^2}
	\left[
	\left(\frac{M^2} 4 - m^2\right)
	+
	\frac{M}{\kappa}
	\left(
	\frac{4 m^4}{M^2}
	\right)
	\right]
	\left[
	1
	-
	\frac{3m^2}{\kappa}
	\frac{2}{M}
	-
	\frac{2}{\kappa}
	\frac{2}{M}
	\left(\frac{M^2} 4 - m^2\right)
	\right] 
\end{align}
To simplify things a bit, notice that
\begin{align}
	&\left[
	\left(\frac{M^2} 4 - m^2\right)
	+
	\frac{M}{\kappa}
	\left(
	\frac{4 m^4}{M^2}
	\right)
	\right]
	\left[
	1
	-
	\frac{3m^2}{\kappa}
	\frac{2}{M}
	-
	\frac{2}{\kappa}
	\frac{2}{M}
	\left(\frac{M^2} 4 - m^2\right)
	\right] \nonumber \\
	&=
	\left(\frac{M^2} 4 - m^2\right)
	+
	\frac{m^2M}{2\kappa}
	+
	\frac{6m^4}{M\kappa}
	-
	\frac{M^3}{4\kappa}
\end{align}
so that, calling $\Gamma^U$ the  non-deformed $\Gamma$ we see that
\begin{align}
	\Gamma
	=
	\Gamma^U
	\left[
	1 + \frac{M}{\kappa}
	\left(
	2\frac{m^2}{M^2}
	+
	\frac{24 m^4}{M^4}
	-
	1
	\right)
	\right]
\end{align}
and we see that the kinematical corrections to the integral \eqref{Gammaint} resulting from the deformation are of the form $m/\kappa$ or $M/\kappa$ . 
Together with possible corrections to the coupling constant $g$, which are of the same order, we conclude that overall corrections to the decay width $\Gamma$ are at most $m/\kappa$ or $M/\kappa$, i.e. of order $10^{-19}$, and therefore completely negligible. 
Moreover, since even in deformed case the masses of particles and antiparticles are identical, the corrections to the decay width $\Gamma$ are the same for particles and antiparticles.

\begin{sidewaystable}[!htbp]
	\caption{\em The following table is taken from \cite{Bevilacqua:2022fbz}. It represents the limits on the deformation parameter $\kappa$ for sets of particle-antiparticle pairs and energies which may be (or are) produced at LHC and FCC. All the values of decay times with their respective errors, as well as the particle masses, are taken from \cite{pdg}. The assumed lifetime accuracies were $\frac{\sigma_\tau}{\tau}=10^{-6}$ everywhere. The Lorentz boosts $\gamma$ were obtained for the assumed energies 6.5 TeV (LHC) and 50 TeV (FCC).}
	\label{Table:tab1}
			\begin{tabular}{ p{12mm} p{20mm} p{20mm} p{20mm} p{24mm} p{20mm} p{20mm} p{20mm} p{20mm} p{20mm}}
				\hline
				\addlinespace[2mm]
				\scriptsize Particle & \makecell{\scriptsize Parent \\ \scriptsize resonance} & $\hskip 10pt \tau$[s] & $M$ [GeV] & $\hskip 10pt\frac{\Gamma}{M}$ & \makecell{$\frac{\sigma_\tau}{\tau}$ \\ \scriptsize (from PDG)} & \makecell{$\gamma$ \\ \scriptsize (LHC)}   & \makecell{$\gamma$ \\ \scriptsize (FCC)}  & \makecell{$\kappa=\frac{p^2}{M\delta_\tau}$ \\ \scriptsize(LHC)} & \makecell{$\kappa=\frac{p^2}{M\delta_\tau}$ \\ \scriptsize (FCC)} \\
				\addlinespace[2mm]
				\hline
				
				\addlinespace[2mm]
				$\mu^\pm$ & $J/\psi, \Upsilon$ & $2.2\times 10^{-6}$ & $\hskip 10pt 0.11$ & $2.8\times 10^{-18}$ & $1\times 10^{-6}$ & $6.1\times 10^4$ & $4.7\times 10^5$ & $4\times 10^{14}$ & $2\times 10^{16}$\\
				\addlinespace[2mm]
				
				\addlinespace[2mm]
				$\tau^\pm$ & $J/\psi, \Upsilon$ & $2.9\times 10^{-13}$ & $\hskip 10pt 1.8$ & $1.3\times 10^{-12}$ & $1.7\times 10^{-3}$ & $3.6\times 10^3$ & $2.8\times 10^4$ & $2.5\times 10^{13}$ & $1.5\times 10^{15}$\\
				\addlinespace[2mm]
				
				\addlinespace[2mm]
				$\pi^\pm$ & \makecell{$K_S, \rho^0, \omega^0$ \\ $D^0, B^0$} & $2.6\times 10^{-8}$ & $\hskip 10pt 0.14$ & $1.8\times 10^{-16}$ & $1.9\times 10^{-4}$ & $4.6\times 10^4$ & $3.6\times 10^5$ & $3\times 10^{14}$ & $1.8\times 10^{16}$\\
				\addlinespace[2mm]
				
				\addlinespace[2mm]
				$K^\pm$ & $\phi^0, D^0, B^0$ & $1.2\times 10^{-8}$ & $\hskip 10pt 0.49$ & $1.1\times 10^{-12}$ & $1.6\times 10^{-3}$ & $1.3\times 10^4$ & $1.0\times 10^5$ & $8.5\times 10^{13}$ & $5.1\times 10^{15}$\\
				\addlinespace[2mm]
				
				\addlinespace[2mm]
				$D^\pm$ & $\psi, B^0$ & $1.0\times 10^{-12}$ & $\hskip 10pt 1.9$ & $3.4\times 10^{-13}$ & $6.7\times 10^{-3}$ & $3.5\times 10^3$ & $2.7\times 10^4$ & $2.2\times 10^{13}$ & $1.3\times 10^{15}$\\
				\addlinespace[2mm]
				
				\addlinespace[2mm]
				$B^\pm$ & $\Upsilon$ & $1.6\times 10^{-15}$ & $\hskip 10pt 5.3$ & $0.8\times 10^{-13}$ & $2.4\times 10^{-3}$ & $1.2\times 10^3$ & $0.9\times 10^4$ & $0.8\times 10^{13}$ & $0.5\times 10^{15}$\\
				\addlinespace[2mm]
				\hline
				
			\end{tabular}
	\end{sidewaystable}

\chapter{$\kappa$-deformed propagator, and 1-loop correction to it} 

\label{Chapter3} 

\section{Introduction}

In the previous chapter we studied in details the properties of the complex scalar field, its charges under continuous transformations and the behaviour of discrete symmetries, and the phenomenological consequences of the deformation of $CPT$ transformations. 

In this chapter we will compute the Feynman propagator of our model, as well as the imaginary part to the 1-loop correction to the propagator. We will begin by computing the propagator in two ways in section \ref{defeypro}. We will then give a standard example of computation of the imaginary part of the 1-loop correction to the propagator in section \ref{nondefloopex}. We do this to highlight that each of the tools which is used is independent from the presence of $\kappa$-deformation. As such, we will immediately go to the deformed case in section \ref{defloop}. Because of the peculiar feature of the momentum sum, we will need to consider 4 different cases, and each of them is treated separately. Finally, in section \ref{mdw} we discuss possible experimental signatures coming from the non trivial mass distribution width contribution to the experimental signatures of deformed $CPT$ discussed in section \ref{def-cpt-pheno}. Everything presented in this chapter (except for the definitions for the functional derivative in the $\kappa$-deformed context, which have been taken from \cite{Arzano:2018gii}) is original work.

\section{Feynman propagator}\label{defeypro}

We now have a well-defined formalism of a complex scalar field in $\kappa$-Minkowski spacetime, with a well defined action invariant both under discrete symmetries ($C,P,T$), and under continuous symmetries. We now exploit this model to define the propagator, and to compute the imaginary part of the 1-loop correction to it.

\subsection{From the generating functional}

The off-shell action (in normal ordering) is given by  eq. \eqref{Stot-momentumspace}. Furthermore, following \cite{Arzano:2018gii}, we can define the generating functional as
\begin{align}
	Z[J, J^\dag] 
	=
	\frac{1}{Z[0,0]}
	\int\mathcal{D}[\phi] \mathcal[\phi^\dag]
	e^{i S_{\text{free}}[\phi, \phi^\dag] 
		+ 
		\frac{i}{2}\int
		[
		\phi^\dag \star J + J \star \phi^\dag + J^\dag \star \phi + \phi \star J^\dag
		]
	}
\end{align}
Notice that, differently from \cite{Arzano:2018gii}, we include all the possible orderings for the star product between the sources $J, J^\dag$ and the fields $\phi, \phi^\dag$. 

The first step that we have to do is to rewrite the exponent in the generating functional in momentum space, because in this way we will be able to complete the squares and then easily compute the propagator. The action has already been written in eq. \eqref{Stot-momentumspace}. Furthermore, in analogy with the fields in eq. \eqref{fa-off-simp}, \eqref{fb-off-simp}, we define $J, J^\dag$ as follows 
\begin{equation}
	J(x) = \int \frac{d^4p}{p_4/\kappa}\, \theta(p_0)\, 
	\zeta(p)\, 
	J_\mathbf{p}\, e^{-i(\omega_pt-\mathbf{p}\mathbf{x})} + \int \frac{d^4p}{p_4/\kappa}\, \theta(p_0)\,
	\zeta(p)\,
	J^\dag_\mathbf{p }\, e^{-i(S(\omega_p)t-S(\mathbf{p})\mathbf{x})} 
\end{equation}
\begin{equation}
	J^\dag(x) = \int \frac{d^4p}{p_4/\kappa}\, \theta(p_0)\, \zeta(p)\, 
	J^\dag_\mathbf{p}\, e^{-i(S(\omega_p)t-S(\mathbf{p})\mathbf{x})} + \int \frac{d^4p}{p_4/\kappa}\, \theta(p_0)\, 
	\zeta(p)\, 
	J_\mathbf{p}\, e^{-i(\omega_pt-\mathbf{p}\mathbf{x})}.
\end{equation}
If we perform the same steps as for the computation of the momentum space action, everything remains the same except for the absence of the terms $(p_\mu p^\mu - m^2)$ and $(S(p)_\mu S(p)^\mu - m^2)$ (recall that the mixed terms containing $c a $ and $d^\dag b^\dag$ all go away because they are multiplied by $\delta(\omega\oplus \omega)$ or $\delta(S(\omega)\oplus S(\omega))$ whose arguments can never be zero).
Therefore, the final result (applying normal ordering) would be 
\begin{align}
	\phi^\dag \star J
	+
	J \star \phi^\dag 
	=
	\int \frac{d^4p}{p_4^2/\kappa^2}\, \theta(p_0)\,
	\zeta^2(p)
	\left(
	1 + \frac{|p_+|^3}{\kappa^3}
	\right)
	[
	a^\dag_\mathbf{p} J_\mathbf{p}
	+
	J^\dag_{\mathbf{p}} b_{\mathbf{p} }
	] \\
	\phi \star J^\dag
	+
	J^\dag \star \phi 
	=
	\int \frac{d^4p}{p_4^2/\kappa^2}\, \theta(p_0)\,
	\zeta^2(p)
	\left(
	1 + \frac{|p_+|^3}{\kappa^3}
	\right) 
	[
	J^\dag_\mathbf{p} a_\mathbf{p}
	+
	b^\dag_{\mathbf{p} } J_{\mathbf{p}}
	]
\end{align}

For simplicity, we now concentrate only on the $a, a^\dag$ part of the above action, the computations for the $b, b^\dag$ parts are the same. Therefore, we only consider
\begin{align}
	\int \frac{d^4p}{p_4^2/\kappa^2}\, \theta(p_0)\,
	\zeta^2(p)
	\left(
	1 + \frac{|p_+|^3}{\kappa^3}
	\right)
	[
	(p_\mu p^\mu - m^2)
	a^\dag_\mathbf{p} a_\mathbf{p}
	+
	a^\dag_\mathbf{p} J_\mathbf{p}
	+
	J^\dag_\mathbf{p} a_\mathbf{p}
	]
\end{align}
and we complete the squares using the fact that
\begin{align}
	[
	(p_\mu p^\mu - m^2)
	a^\dag_\mathbf{p} a_\mathbf{p}
	+
	a^\dag_\mathbf{p} J_\mathbf{p}
	+
	J^\dag_\mathbf{p} a_\mathbf{p}
	]
	&=
	\Bigg|\left(
	\sqrt{p_\mu p^\mu - m^2}
	a_\mathbf{p}
	+
	\frac{J_\mathbf{p}}{\sqrt{p_\mu p^\mu - m^2}}
	\right)\Bigg|^2 \nonumber  \\
	&-
	\frac{J_\mathbf{p}^\dag J_\mathbf{p}}{p_\mu p^\mu - m^2}
\end{align}
The generating functional in momentum space can be written as 
\begin{align}
	Z[J, J^\dag] 
	=
	\frac{1}{Z[0,0]}
	\int\mathcal{D}[\tilde\phi] \mathcal{D}[\tilde\phi^\dag]
	e^{i S_{\text{free}}[\tilde\phi, \tilde\phi^\dag] 
		+ 
		\frac{i}{2}\int \frac{d^4p}{p_4^2/\kappa^2}\, \theta(p_0)\, 
		\zeta^2(p)
		\left(
		1 + \frac{|p_+|^3}{\kappa^3}
		\right)
		[
		a^\dag_\mathbf{p} J_\mathbf{p}
		+
		J^\dag_{\mathbf{p}} b_{\mathbf{p} }
		+
		J^\dag_\mathbf{p} a_\mathbf{p}
		+
		b^\dag_{\mathbf{p} } J_{\mathbf{p}}
		]
	}
\end{align}
where $\mathcal{D}[\tilde\phi] \mathcal{D}[\tilde\phi^\dag]$ is a shorthand notation for $\mathcal{D}[a_\mathbf{p}]\mathcal{D}[a_\mathbf{p}^\dag]\mathcal{D}[b_{\mathbf{p}*}]\mathcal{D}[b_{\mathbf{p}*}^\dag]$. Using the fact that the measure is invariant under constant shifts like 
\begin{align}
	a_\mathbf{p} \rightarrow a_\mathbf{p} - \frac{J_\mathbf{p}}{p_\mu p^\mu - m^2}
\end{align}
and recalling that the same exact computations can be reproduced for the $b, b^\dag$ operators, the generating functional reduces to 
\begin{align}\label{generatingfunctsimpl}
	Z[J, J^\dag]
	=
	\exp
	\left[
	-i
	\int \frac{d^4p}{p_4^2/\kappa^2}\, \theta(p_0)\,
	\zeta^2(p)
	\left(
	1 + \frac{|p_+|^3}{\kappa^3}
	\right)
	\frac{J_\mathbf{p}^\dag J_\mathbf{p}}{p_\mu p^\mu - m^2 + i\epsilon}
	\right]
\end{align}
where the remaining contribution
\begin{align}
	\int\mathcal{D}[\tilde\phi] \mathcal{D}[\tilde\phi^\dag]
	e^{\frac{i}{2} 
	\int \frac{d^4p}{p_4^2/\kappa^2}\, \theta(p_0)\,
	\zeta^2(p)
	\left(
	1 + \frac{|p_+|^3}{\kappa^3}
	\right)
	(p_\mu p^\mu - m^2)
	[
	a^\dag_\mathbf{p} a_\mathbf{p}
	+
	b^\dag_{\mathbf{p} } b_{\mathbf{p} }
	]
	}
\end{align}
gets cancelled by the $\frac{1}{Z[0,0]}$ factor. Notice that there is no prefactor $\frac{1}{2}$ in eq. \eqref{generatingfunctsimpl} because one has two equal factors coming from the computations of $a, a^\dag$ and $b, b^\dag$. At this point, following \cite{Arzano:2018gii} we can define the the functional derivative as follows
\begin{align}
	\frac{\delta Z[J, J^\dag]}{\delta J_\mathbf{q}}
	&=
	\lim_{\varepsilon\rightarrow 0}
	\frac{Z[J_\mathbf{p} + \varepsilon\delta(S(p)\oplus q), J_\mathbf{p}^\dag] - Z[J,J^\dag]}{\varepsilon} \\
	\frac{\delta Z[J, J^\dag]}{\delta J_\mathbf{q}^\dag}
	&=
	\lim_{\varepsilon\rightarrow 0}
	\frac{Z[J_\mathbf{p}, J_\mathbf{p}^\dag + \varepsilon\delta(S(p)\oplus S(q))] - Z[J,J^\dag]}{\varepsilon}
\end{align}
Notice that these expressions reduce to the canonical ones in the limit $\kappa\rightarrow \infty$. Using these we can compute the Feynman propagator $\tilde\Delta_\kappa(p,q)$ as follows. 
\begin{align}\label{Feypropdef}
	i\tilde{\Delta}_\kappa(p,q)
	=
	\left(
	i 
	\frac{\delta}{\delta J_\mathbf{l}^\dag}
	\right)
	\left(
	-i 
	\frac{\delta}{\delta J_\mathbf{q}}
	\right)
	Z[J, J^\dag] 
	\Big|_{J,J^\dag = 0}
	=
	\left(
	\frac{\delta}{\delta J_\mathbf{l}^\dag}
	\right)
	\left(
	\frac{\delta}{\delta J_\mathbf{q}}
	\right)
	Z[J, J^\dag] 
	\Big|_{J,J^\dag = 0}
\end{align}
We have\footnote{Notice that we are now dealing with canonical expressions of the variables $p, q$, and we don't need to use the star product because there are no mixed functions of spacetime and momentum space. Furthermore $J, J^\dag$ are not dynamical quantities, and therefore their Poisson brackets are trivial.}
\begin{align}
	\left(
	\frac{\delta}{\delta J_\mathbf{q}}
	\right)
	Z[J, J^\dag]
	&=
	\lim_{\varepsilon\rightarrow 0}
	\Bigg\{
	\exp
	\left[
	-i
	\int \frac{d^4p}{p_4^2/\kappa^2}\, \theta(p_0)\, 
	\zeta^2(p)
	\left(
	1 + \frac{|p_+|^3}{\kappa^3}
	\right)
	\frac{J_\mathbf{p}^\dag J_\mathbf{p}}{p_\mu p^\mu - m^2 + i\epsilon}
	\right] \times \nonumber \\
	&\times \frac{\exp
	\left[
	-i\varepsilon
	\int \frac{d^4p}{p_4^2/\kappa^2}\, \theta(p_0)\,
	\zeta^2(p)
	\left(
	1 + \frac{|p_+|^3}{\kappa^3}
	\right)
	\frac{J_\mathbf{p}^\dag \delta(S(p)\oplus q)}{p_\mu p^\mu - m^2 + i\epsilon}
	\right]}{\varepsilon} \nonumber \\
	&-
	\frac{\exp
		\left[
		-i
		\int \frac{d^4p}{p_4^2/\kappa^2}\, \theta(p_0)\,
		\zeta^2(p)
		\left(
		1 + \frac{|p_+|^3}{\kappa^3}
		\right)
		\frac{J_\mathbf{p}^\dag J_\mathbf{p}}{p_\mu p^\mu - m^2 + i\epsilon}
		\right]}{\varepsilon}
	\Bigg\} \\ \nonumber \\
	&=
	\exp
	\left[
	-i
	\int \frac{d^4p}{p_4^2/\kappa^2}\, \theta(p_0)\,
	\zeta^2(p)
	\left(
	1 + \frac{|p_+|^3}{\kappa^3}
	\right)
	\frac{J_\mathbf{p}^\dag J_\mathbf{p}}{p_\mu p^\mu - m^2 + i\epsilon}
	\right] \times \nonumber \\
	&\times \left\{
	-i
	\int \frac{d^4p}{p_4^2/\kappa^2}\, \theta(p_0)\,
	\zeta^2(p)
	\left(
	1 + \frac{|p_+|^3}{\kappa^3}
	\right)
	\frac{J_\mathbf{p}^\dag \delta(S(p)\oplus q)}{p_\mu p^\mu - m^2 + i\epsilon}
	\right\} \\ \nonumber \\
	&=
	\exp
	\left[
	-i
	\int \frac{d^4p}{p_4^2/\kappa^2}\, \theta(p_0)\,
	\zeta^2(p)
	\left(
	1 + \frac{|p_+|^3}{\kappa^3}
	\right)
	\frac{J_\mathbf{p}^\dag J_\mathbf{p}}{p_\mu p^\mu - m^2 + i\epsilon}
	\right] \times \nonumber \\
	&\times \left\{
	-i
	\frac{\kappa}{p_4} 
	\zeta^2(q)
	\left(
	1 + \frac{|q_+|^3}{\kappa^3}
	\right)
	\frac{J_\mathbf{q}^\dag}{q_\mu q^\mu - m^2 + i\epsilon}
	\right\}
\end{align}
where we expanded the exponential containing $\epsilon$ and in the last passage we used the fact that \cite{Arzano:2018gii}\footnote{In particular ibid. eq. (59),(60), and discussion below eq. (60). Notice that the measure $d\bar{\mu}(q)$ in eq. (59) in \cite{Arzano:2018gii} is the left-invariant measure, which can be equivalently written as $\frac{d^4p}{p_4/\kappa}\theta(p_0)$ (see eq. (48) and (51) in \cite{Arzano:2018gii}). }
\begin{align}\label{diracdeltaprop}
	\delta(S(p)\oplus q) = \delta(S(q)\oplus p)
	\qquad
	\int \frac{d^4p}{p_4/\kappa} \delta(S(q)\oplus p) f(p) = f(q).
\end{align}
Hence (considering only the relevant term which do not go away when $J, J^\dag = 0$)
\begin{align}
	\left(
	\frac{\delta}{\delta J_\mathbf{l}^\dag}
	\right)
	\left(
	\frac{\delta}{\delta J_\mathbf{q}}
	\right)
	Z[J, J^\dag] \Bigg|_{J, J^\dag = 0}
	&=
	\lim_{\varepsilon\rightarrow 0}
	\frac{1}{\varepsilon}
	\left\{
	-i
	\frac{\kappa}{p_4} 
	\zeta^2(q)
	\left(
	1 + \frac{|q_+|^3}{\kappa^3}
	\right)
	\frac{\varepsilon\delta(S(q)\oplus S(p))}{q_\mu q^\mu - m^2 + i\epsilon}
	\right\} \\ \nonumber \\
	&=
	-i
	\frac{\kappa}{p_4} 
	\zeta^2(q)
	\left(
	1 + \frac{|q_+|^3}{\kappa^3}
	\right)
	\frac{\delta(S(q)\oplus S(l))}{q_\mu q^\mu - m^2 + i\epsilon}
\end{align}
Notice that, if we were to invert the order of the functional derivatives in eq. \eqref{Feypropdef} we will not be able to immediately apply the first Dirac delta, but this does not matter because at the end of the computations of both derivatives we are left with an integral containing the product of Dirac deltas $\delta(S(p)\oplus q)\delta(S(p)\oplus S(l))$, which using eq. \eqref{diracdeltaprop} can be rewritten as $\delta(S(q)\oplus p)\delta(S(p)\oplus S(l))$, and applying the integration in $p$ to the first Dirac delta we still get a final $\delta(S(q)\oplus S(l))$, like in the previous order of the functional derivatives. 

The only remaining fact which is left to clarify is the presence of the additional factor 
\begin{align}
	\frac{\kappa}{p_4} 
	\zeta^2(q)
	\left(
	1 + \frac{|q_+|^3}{\kappa^3}
	\right)
\end{align}
in front of the propagator. We already discussed about this factor in section \ref{role-of-zeta}, and for simplicity we will be using the convention expressed in eq. \eqref{convention1} from now onwards.

\subsection{As inverse of the field operator}

We know that the equation of motion for the field in momentum space is given by the inverse of the field operator, which in momentum space is given by $p^2-m^2$ (any additional factor goes away because of the convention in eq. \eqref{convention1}). Therefore, we have
\begin{align}
	(p^2-m^2)(-i\Delta_\kappa) = 1
	\qquad
	\implies
	\qquad
	-i\Delta_\kappa
	=
	\frac{1}{p^2-m^2+i\epsilon}
\end{align}

\section{Example of loop correction to the propagator: non-deformed case}\label{nondefloopex}

\subsection{Using renormalization}

In the  non-deformed context, we consider the following diagram 
\begin{center}
	\begin{tikzpicture}
		\begin{feynman}
			\vertex (a) at (-0.6,0) {\(i \Pi(p^2) = \)};
			\vertex (b) at (1,0);
			\vertex (c) at (2,0);
			\vertex (d) at (3,0);

			\diagram* 
			{(a)-- [ultra thick] (b),
			(b)-- [scalar,half left,looseness=1.7,momentum=\(k\)] (c)-- [scalar,half left,looseness=1.7,momentum=\(k-p\)] (b),
			(c)-- [ultra thick] (d),};
		\end{feynman}
	\end{tikzpicture}
\end{center}
Assuming a $\psi\phi^2$ theory with interaction vertex $g\psi\frac{\phi^2}{2!}$, the vertex is given by $i g$ (times a Dirac delta enforcing momentum conservation) and the propagator is given by
\begin{align}
	\frac{i}{p^2 - m^2 +i\epsilon}
\end{align}
We have therefore
\begin{align}\label{initialamp}
	i\Pi(p^2)
	=
	(ig)^2
	\int \frac{d^4 k}{(2\pi)^4} 
	\underbrace{\frac{i}{k^2 - m^2 +i\epsilon}}_{B}
	\underbrace{\frac{i}{(p-k)^2 - m^2 +i\epsilon}}_{A}
\end{align}
We now use the so called Feynman trick, which says that 
\begin{align}\label{feytrickn}
	\frac{1}{A_1 \dots A_n}
	=
	\int_0^1 
	\prod_{i=1}^{n} dx_i \,\,
	\delta\left(
	\sum_{j=1}^{n} x_j - 1
	\right)
	\frac{(n-1)!}{\left(
		\sum_{l=1}^{n} x_l A_l
		\right)^n}
\end{align}

and we have
\begin{align}
	A+ (B-A)x 
	&=
	[k-p(1-x)]^2
	+
	\underbrace{p^2x(1-x)- m^2}_{-\Delta(x)} + i\epsilon
\end{align}
Notice that $\Delta(x)\geq 0$. Thus, after using the Feynman parameters, we have
\begin{align}
	i\Pi(p^2)
	=
	-\frac{(ig)^2}{(2\pi)^4}
	\int d^4k
	\int_0^1 dx \,
	\frac{1}{\{k^2-\Delta(x) + i\epsilon\}^2}
\end{align}
where we performed the constant linear shift $k \mapsto k + p(1-x)$ which leave the metric invariant. 
The poles can be represented as follows in the plane below. 
We can therefore use the integration contour in red in the same picture. 
\begin{center}
	\begin{tikzpicture}
		
		\draw[color=black, -{Stealth[length=2mm]}] (-2,0) -- (2,0);
		\draw[color=black, -{Stealth[length=2mm]}] (0,-2) -- (0,2);
		
		\filldraw[color=blue] (1,-0.5) circle (0.6mm);
		\filldraw[color=blue] (-1,0.5) circle (0.6mm);
		
		\draw[color=red, -{Stealth[length=2mm]}] (-1,0) -- (1,0);
		\draw[color=red, -{Stealth[length=2mm]}] (1,0) arc (0:90:1cm);
		\draw[color=red, -{Stealth[length=2mm]}] (0,1) -- (0,-1);
		\draw[color=red, -{Stealth[length=2mm]}] (0,-1) arc (270:180:1cm);
		
	\end{tikzpicture}
\end{center}
The integral computed in the red circuit $\Gamma$ is zero, and if $R\rightarrow \infty$ the contribution from the circular sections goes to zero, meaning that we can use the change of variables
\begin{align}
	k^0 \mapsto i k^0_E
\end{align}
which is called Wick rotation. In this way one has
\begin{align}
	k^2 = (k^0)^2 - \mathbf{k}^2 \mapsto -k_E^2 = -(k^0_E)^2 - \mathbf{k}^2.
\end{align}
The amplitude reduces to
\begin{align}
	i\Pi(p^2)
	=
	-i(ig)^2
	\int_0^1 dx
	\int
	\frac{d^4 k}{(2\pi)^4}
	\frac{1}{(k_E^2 + \Delta)^2}
\end{align}
where there is no additional $i\epsilon$ because there are no more singularities in the integrand. The momentum space integral is a known integral since 
\begin{align}\label{dimregint}
	\mu^{4-d}
	\int
	\frac{d^d k}{(2\pi)^4}
	\frac{(k^2)^a}{(k^2 + \Delta)^b}
	=
	\frac{\Gamma(b-a-\frac{1}{2}d)\Gamma(a+\frac{1}{2}d)}{(4\pi)^{d/2}\Gamma(b)\Gamma(\frac{1}{2}d)}
	\Delta^{-(b-a-d/2)}\mu^{4-d}
\end{align}
Notice that $[\mu]=1$, and that the term $\mu^{4-d}$ has been added so that the integral still has the same mass dimension as before dimensional regularization (i.e. 4), regardless of or choice of $d$. 
In our case, using dimensional regularization, we can assume $d=4-\epsilon$, and in our case we have $a=0$ and $b=2$. Therefore the momentum space integral is
\begin{align}
	\frac{\Gamma(\frac{\epsilon}{2})}{(4\pi)^2 \Gamma(2)}
	\Delta^{-\frac{\epsilon}{2}}
	(\mu^2)^{\frac{\epsilon}{2}}
\end{align}
One can use the expansions
\begin{align}
	\Delta^{-\frac{\epsilon}{2}}(\mu^2)^{\frac{\epsilon}{2}} = 1 - \frac{\epsilon}{2}\log \frac{\Delta}{\mu^2} + O(\epsilon^2)
\end{align}
Furthermore, 
we have
\begin{align}
	\Gamma(z)
	&=
	\frac{1}{z}
	-
	\gamma
	+
	\frac{z}{2}
	\left(
	\gamma^2 + \frac{\pi^2}{6}
	\right)
	-
	\frac{z^2}{6}
	\left(
	\gamma^3 +\frac{\gamma\pi^2}{2} +2\zeta(3)
	\right)
	+
	\dots
\end{align}
Therefore we have
\begin{align}\label{useful}
	\Gamma(\frac{\epsilon}{2})
	\Delta^{-\frac{\epsilon}{2}}
	(\mu^2)^{\frac{\epsilon}{2}}
	&=
	\left[
	\frac{2}{\epsilon}
	-
	\gamma
	+
	\frac{\epsilon}{4}
	\left(
	\gamma^2 + \frac{\pi^2}{6}
	\right)
	-
	\frac{\epsilon^2}{24}
	\left(
	\gamma^3 +\frac{\gamma\pi^2}{2} +2\zeta(3)
	\right)
	\right]
	(1 - \frac{\epsilon}{2}\log \frac{\Delta}{\mu^2}) \\
	& \approx
	\frac{2}{\epsilon} - \log \frac{\Delta}{\mu^2} -\gamma + O(\epsilon)
\end{align}
Putting everything together we end up with 
\begin{align}\label{undefresbeforeren}
	\Pi(p^2)
	&=
	-\frac{(ig)^2}{(4\pi)^2 \Gamma(2)}
	\int_0^1 dx \,
	\left[
	\frac{2}{\epsilon} - \log \frac{\Delta}{\mu^2} -\gamma + O(\epsilon)
	\right] \\
	&=
	-\frac{(ig)^2}{(4\pi)^2}
	\int_0^1 dx \,
	\left[
	\frac{2}{\epsilon} - \log \frac{\Delta}{\mu^2} -\gamma + O(\epsilon)
	\right] 
\end{align}
To renormalize this amplitude we add the counterterms $Ak^2 + B m^2$ to the amplitude, we then chose $A, B$ in such a way that the divergence is eliminated leaving behind only a finite contribution, and finally we fix these finite contributions by imposing the canonical restrictions $\Pi(m^2)=\Pi'(m^2)=0$ where $\Pi' = \frac{d\Pi}{dp^2}$. We have
\begin{align}
	\Pi(p^2)
	&=
	-\frac{(ig)^2}{(4\pi)^2}
	\int_0^1 dx \,
	\left[
	\frac{2}{\epsilon} - \log \frac{\Delta}{\mu^2} -\gamma + O(\epsilon)
	\right] 
	+
	Ap^2 + B m^2
\end{align}
so that choosing $A = \kappa_A $, $B=\kappa_B + \frac{(ig)^2}{(4\pi)^2} \frac{2}{m^2\epsilon}$ we have
\begin{align}
	\Pi(p^2)
	&=
	\frac{(ig)^2}{(4\pi)^2}
	\int_0^1 dx \,
	\log \frac{\Delta}{\mu^2} 
	+
	\kappa_A p^2
	+
	\kappa_B m^2
\end{align}
At this point, notice that the logarithm may have an imaginary part. It is sufficient to notice that the argument of the logarithm is positive unless $p^2 x(1-x)> m^2$ (recall that $\Delta = m^2 - p^2 x(1-x)$), in which case the numerator becomes negative, and therefore the whole argument of the logarithm becomes negative. Notice that this is indeed possible since $p$ refers to the momentum of the external particle, meaning that $p^2 = m_\psi^2>m^2$ where $m = m_\phi$ is the mass appearing in the propagators in the loop. Therefore, as long as $|x| < \frac{1}{2} \pm \frac{1}{2} \sqrt{1-4\frac{m^2}{p^2}}$ and $p^2>4m^2$, the amplitude gains a small imaginary part. However, notice that this imaginary part cannot be cancelled (or in general dealt with) by the counterterms, since they are real parameters coming from a Hermitian Lagrangian. Therefore, we can at best deal with the real part of the amplitude. In other words, instead of simply imposing $\Pi(p^2=m_\psi^2)=0$, we are actually imposing $\Re(\Pi(p^2=m_\psi^2)) = 0$, and the same for $\Pi'$.

The renormalized amplitude reads
\begin{align}\label{finalresundef1}
	\Pi(p^2)
	&=
	\frac{(ig)^2}{(4\pi)^2}
	\int_0^1 dx \,
	\log \frac{\Delta}{|\Delta^0|} 
	+
	\frac{(ig)^2}{m^2(4\pi)^2}
	\left(
	\frac{2\pi}{3\sqrt{3}} - 1
	\right)
	(p^2 - m^2)
\end{align}
Notice that this final expression is dimensionally correct. 

We can now deal with the imaginary part of the amplitude. We can obtaining it by simply assuming that the argument of the logarithm becomes negative. 

In this case $\Im \log \frac{\Delta}{|\Delta^0|} = -\pi$, where the minus arises from the fact that the quantity $\Delta-i\epsilon$ (which is found in the denominator $(k^2 - (\Delta(x)-i\epsilon))^2$ after the Feynman trick) has a negative imaginary part. Therefore, recalling that the only part of the $x$ integration which contributes is the one such that 
\begin{align}
	|x| < \frac{1}{2} \pm \frac{1}{2} \sqrt{1-4\frac{m^2}{p^2}}
\end{align}
we have
\begin{align}\label{undefim}
	\Im \Pi(p^2) = \frac{g^2\pi}{(4\pi)^2} \int dx = \frac{g^2}{16\pi}\sqrt{1-4\frac{m^2}{p^2}}.
\end{align}
Notice the very important fact that the only real role of the imposition of the conditions $\Pi(m^2) = \Pi'(m^2)=0$ (for what concerns the imaginary part contribution) is to substitute $\mu^2 \leftrightarrow |\Delta^0|$ inside the argument of the logarithm, but both $\mu^2$ and $|\Delta^0|$ are strictly positive, which means that they are irrelevant in the determination of the imaginary part of the logarithm. In other words, one can get the imaginary contribution of the 1-loop correction directly from eq. \eqref{undefresbeforeren}.

Furthermore, the reason we showed in such detail the non-deformed computation was to show that all the tools that are conventionally employed in non-deformed QFT can be used as well in the deformed context. In other words, we have all the tools to approach the same calculations in the $\kappa$-deformed context.

\section{Example of loop correction to the propagator: $\kappa$- deformed case}\label{defloop}

We can consider the case of one incoming particle (with momentum $S(p)$) and two outgoing ones (let us call them $p_2$ and $p_3$). The conservation of momentum then implies that
\begin{align}
	S(p) \oplus p_2 \oplus p_3 = 0
\end{align}
This identity is immediately satisfied if $p_2 \oplus p_3 = p$, and we can use one of the following parametrizations for $p_2$ and $p_3$
\begin{align}\label{Skp}
	p_2 = k
	\qquad
	p_3 = S(k)\oplus p
\end{align}
\begin{align}\label{pSk}
	p_2 = p\oplus S(k)
	\qquad
	p_3 = k
\end{align}
\begin{align}\label{kp}
	p_2 = S(k)
	\qquad
	p_3 = k \oplus p
\end{align}
\begin{align}\label{pk}
	p_2 = p\oplus k
	\qquad
	p_3 = S(k)
\end{align}
We can now perform the computation as in the  non-deformed case, keeping in mind to add all the above contributions.

\subsection{Dimensional regularization and cut-off in the $\kappa$-deformed context}

We start from the same amplitude as before, namely 
\begin{center}
	\begin{tikzpicture}
		\begin{feynman}
			\vertex (a) at (-0.6,0) {\(i \Pi(p^2) = \)};
			\vertex (b) at (1,0);
			\vertex (c) at (2,0);
			\vertex (d) at (3,0);

			\diagram* 
			{(a)-- [ultra thick] (b),
				(b)-- [scalar,half left,looseness=1.7,momentum=\(k\)] (c) -- [scalar,half left,looseness=1.7,reversed momentum=\(S(k)\oplus p\)] (b),
				(c)-- [ultra thick] (d),};
		\end{feynman}
	\end{tikzpicture}
\end{center}
\begin{align}\label{initialamp-def1-DR}
	i\Pi(p^2)
	=
	(ig)^2
	\int \frac{d^4 k}{(2\pi)^4} 
	\frac{i}{k^2 - m^2 +i\epsilon}
	\frac{i}{(S(k)\oplus p)^2 - m^2 +i\epsilon}
\end{align}
Notice that, apart from the deformed conservation of momentum, we are assuming that each vertex only contributes $ig$.

From the relations
\begin{align}
	(P \oplus Q)_0 
	&= \frac{1}{\kappa} P_0 Q_+ + \kappa \frac{Q_0}{P_+}+ \frac{\vec{P}\cdot \vec{Q}}{P_+} \\
	(P \oplus Q)_i
	&=
	\frac{1}{\kappa} P_i Q_+ + Q_i \\
	(P \oplus Q)_4
	&=
	\frac{1}{\kappa} P_4 Q_+ - \kappa \frac{Q_0}{P_+}- \frac{\vec{P}\cdot \vec{Q}}{P_+} \\
	P_+ = P_0 + &P_4 = P_0 + \sqrt{\kappa^2 + P_0^2  - \vec{P}^2} 
\end{align}
we have
\begin{align}
	(S(k)\oplus p)_0
	&=
	\frac{1}{\kappa}S(k)_0 p_+ 
	+ 
	\kappa
	\frac{p_0}{S(k)_+}
	+
	\frac{S(\mathbf{k})\mathbf{p}}{S(k)_+} \\
	&=
	\frac{p_+}{\kappa}
	\left(
	-k_0 + \frac{\mathbf{k}^2}{k_+}
	\right)
	+
	\frac{p_0k_+}{\kappa}
	-
	\frac{\kappa\mathbf{k}\mathbf{p}}{k_+}
	\frac{k_+}{\kappa^2} \\
	&=
	-\frac{k_0 p_+}{\kappa}
	+
	\frac{p_+\mathbf{k}^2}{\kappa k_+}
	+
	\frac{p_0k_+}{\kappa}
	-
	\frac{\mathbf{k}\mathbf{p}}{\kappa}
\end{align}
\begin{align}
	(S(\mathbf{k})\oplus \mathbf{p})_i
	&=
	\frac{1}{\kappa}S(\mathbf{k})_i p_+ + \mathbf{p}_i \\
	&=
	-
	\frac{\mathbf{k}_i}{k_+}p_+ + \mathbf{p}_i
\end{align}
from which we get 
\begin{align}\label{Skplusp0-BPHZ}
	(S(k)\oplus p)_0^2
	&=
	\left(
	-\frac{k_0 p_+}{\kappa}
	+
	\frac{p_+\mathbf{k}^2}{\kappa k_+}
	+
	\frac{p_0k_+}{\kappa}
	-
	\frac{\mathbf{k}\mathbf{p}}{\kappa}
	\right)^2 
	\\
	&+
	2
	\frac{p_+\mathbf{k}^2p_0}{\kappa^2}
	-
	2
	\frac{p_+\mathbf{k}^2(\mathbf{k}\mathbf{p})}{\kappa^2 k_+}
	-
	2
	\frac{p_0k_+(\mathbf{k}\mathbf{p})}{\kappa^2}
\end{align}
\begin{align}\label{Skpluspi-BPHZ}
	(S(\mathbf{k})\oplus \mathbf{p})^2
	=
	\frac{\mathbf{k}^2p_+^2}{k_+^2}
	+
	\mathbf{p}^2
	-
	2\frac{(\mathbf{k}\mathbf{p})p_+}{k_+}
\end{align}
It is useful to obtain the following approximated relations, valid up to second order in $\frac{1}{\kappa}$.
\begin{align}
	(p\oplus q)_0 
	\approx 
	p_0+q_0 
	+ 
	\frac{\mathbf{p}\mathbf{q}}{\kappa}
	+
	\frac{1}{2\kappa^2}
	\left(
	p_0^2 q_0 + p_0 q_0^2 + q_0 \mathbf{p}^2 - p_0\mathbf{q}^2 - 2p_0 \mathbf{p}\mathbf{q}
	\right)
\end{align}
\begin{align}
	(\mathbf{p}\oplus \mathbf{q})_i
	=
	\mathbf{p}_i + \mathbf{q}_i
	+
	\frac{\mathbf{p}_i q_0}{\kappa}
	+
	\frac{\mathbf{p}_i}{2\kappa^2}
	\left(
	q_0^2 - \mathbf{q}^2
	\right)
\end{align}
\begin{align}\label{t-SN-2nd}
	(S(p)\oplus q)_0 
	\approx 
	-p_0 + q_0 
	- \frac{\mathbf{p}\mathbf{q}}{\kappa} + \frac{\mathbf{p}^2}{\kappa}
	+
	\frac{1}{2\kappa^2}
	\left(
	p_0^2 q_0 - p_0 q_0^2 + q_0 \mathbf{p}^2 + p_0\mathbf{q}^2 
	\right)
\end{align}
\begin{align}\label{s-SN-2nd}
	(S(\mathbf{p})\oplus \mathbf{q})_i
	=
	-\mathbf{p}_i + \mathbf{q}_i
	-
	\frac{\mathbf{p}_i q_0}{\kappa}
	+
	\frac{\mathbf{p}_i p_0}{\kappa}
	-
	\frac{\mathbf{p}_i}{2\kappa^2}
	\left(
	p_0^2 - 2p_0q_0 + q_0^2 + \mathbf{p}^2 - \mathbf{q}^2
	\right)
\end{align}
\begin{align}
	(p\oplus S(q))_0
	&\approx
	p_0-q_0
	-
	\frac{\mathbf{p}\mathbf{q}}{\kappa} 
	+\frac{\mathbf{q}^2}{\kappa} \nonumber \\
	&+
	\frac{1}{2\kappa^2}
	\left(
	-p_0^2q_0 + p_0q_0^2 - q_0\mathbf{p}^2 + 2 p_0 \mathbf{p}\mathbf{q} + 2 q_0 \mathbf{p}\mathbf{q} - p_0\mathbf{q}^2
	\right)
\end{align}
\begin{align}
	(\mathbf{p}\oplus S(\mathbf{q}))_i
	=
	\mathbf{p}_i - \mathbf{q}_i
	-
	\frac{\mathbf{p}_i q_0}{\kappa}
	+
	\frac{\mathbf{q}_iq_0}{\kappa}
	+
	\frac{1}{2\kappa^2}
	\left(
	q_0^2 \mathbf{p}_i
	-
	q_0^2 \mathbf{q}_i
	+
	\mathbf{q}^2\mathbf{p}_i
	-
	\mathbf{q}^2\mathbf{q}_i
	\right)
\end{align}
Notice that we are going up to second order because, as will be shown below, the first order contribution to the propagator identically vanish, so that we are reduced to the non-deformed case. Since we want to study the effects of deformation, we need therefore to consider terms up to second order.

We first show that the first order contribution in $\frac{1}{\kappa}$ to $\frac{i}{(S(k)\oplus p)^2 - m^2 +i\epsilon}$ vanish.
\begin{align}
	(S(k)\oplus q)^2 
	&=
	(S(k)\oplus q)_0^2
	-
	(S(\mathbf{k})\oplus \mathbf{q})^2  \\
	&\approx
	\left(
	-k_0 + q_0 - \frac{\mathbf{k}\mathbf{q}}{\kappa} + \frac{\mathbf{k}^2}{\kappa}
	\right)^2
	-
	\left(
	-\mathbf{k}_i + \mathbf{q}_i
	-
	\frac{\mathbf{k}_i q_0}{\kappa}
	+
	\frac{\mathbf{k}_i k_0}{\kappa}
	\right)^2 \\
	&\approx
	(q-k)^2
	+
	2(-k_0+q_0)
	\left(
	- \frac{\mathbf{k}\mathbf{q}}{\kappa} + \frac{\mathbf{k}^2}{\kappa}
	\right) \nonumber \\
	&-
	2
	(-\mathbf{k}_i + \mathbf{q}_i)
	\left(
	-
	\frac{\mathbf{k}_i q_0}{\kappa}
	+
	\frac{\mathbf{k}_i k_0}{\kappa}
	\right) \\
	&=
	(q-k)^2
	+
	2k_0
	\frac{\mathbf{k}\mathbf{q}}{\kappa}
	-
	2k_0
	\frac{\mathbf{k}^2}{\kappa}
	-
	2q_0
	\frac{\mathbf{k}\mathbf{q}}{\kappa}
	+
	2q_0
	\frac{\mathbf{k}^2}{\kappa} \nonumber \\
	&-
	2k_0
	\frac{\mathbf{k}\mathbf{q}}{\kappa}
	+
	2k_0
	\frac{\mathbf{k}^2}{\kappa}
	+
	2q_0
	\frac{\mathbf{k}\mathbf{q}}{\kappa}
	-
	2q_0
	\frac{\mathbf{k}^2}{\kappa} \\
	&=
	(q-k)^2.
\end{align}
We need therefore to go to second order contributions. We have (once again using Mathematica) using eq. \eqref{t-SN-2nd} and \eqref{s-SN-2nd}
\begin{align}
	(S(k)\oplus q)^2 
	&=
	(S(k)\oplus q)_0^2
	-
	(S(\mathbf{k})\oplus \mathbf{q})^2  \\ \nonumber \\
	&\approx
	(q-k)^2
	+\frac{1}{\kappa^2}
	\Big(
	-k_0^3 q_0 
	+ 
	(- \mathbf{k}^2 + \mathbf{k}\mathbf{q})(q_0^2 + \mathbf{k}\mathbf{q} - \mathbf{q}^2) \nonumber \\
	&+ 
	k_0^2(2q_0^2 - 2 \mathbf{k}^2 + \mathbf{k}\mathbf{q} - \mathbf{q}^2)
	+
	k_0q_0
	(-q_0^2 + 3 \mathbf{k}^2 - 2\mathbf{k}\mathbf{q} + \mathbf{q}^2)
	\Big)
	\\ \nonumber \\
	&:=
	(q-k)^2 
	+
	\frac{\Delta_2[(S(k)\oplus q)^2]}{\kappa^2}
\end{align}
where we introduce the short-hand notation
\begin{align}\label{delta2-BPHZ}
	\Delta_2[(S(k)\oplus p)^2]
	&:=
	\left\{
	\frac{1}{2}
	\frac{d^2}{d(1/\kappa)^2}
	[(S(k)\oplus p)^2
	-
	(p-k)^2]
	\right\}_{1/\kappa=0}
	\\
	&=
	\Big(
	-k_0^3 p_0 
	+ 
	(- \mathbf{k}^2 + \mathbf{k}\mathbf{p})(p_0^2 + \mathbf{k}\mathbf{p} - \mathbf{p}^2) \nonumber \\ 
	&+ 
	k_0^2(2p_0^2 - 2 \mathbf{k}^2 + \mathbf{k}\mathbf{p} - \mathbf{p}^2) 
	+
	k_0p_0
	(-p_0^2 + 3 \mathbf{k}^2 - 2\mathbf{k}\mathbf{p} + \mathbf{p}^2)
	\Big)
\end{align}
where the index ${}_2$ in $\Delta_2$ means that it is the coefficient of the second order of the expansion of the quantity inside square brackets. Therefore, we have
\begin{align}
	\frac{i}{(S(k)\oplus p)^2 - m^2 +i\epsilon}
	&\approx
	\frac{i}{(p-k)^2 
		+
		\frac{\Delta_2[(S(k)\oplus p)^2]}{\kappa^2} - m^2 +i\epsilon} \\
	&=
	\frac{i}{ 1
		+
		\frac{\Delta_2[(S(k)\oplus p)^2]}{\kappa^2}
		\frac{1}{(p-k)^2 - m^2+i\epsilon}}
	\frac{1}{(p-k)^2 - m^2 +i\epsilon} \\ 
	&=
	\frac{i}{(p-k)^2 - m^2 +i\epsilon}
	-
	\frac{1}{\kappa^2}
	\frac{i\Delta_2[(S(k)\oplus p)^2]}{[(p-k)^2 - m^2 +i\epsilon]^2}
\end{align}
The leading contributions in the $1/\kappa$ expansion of the amplitude in eq. \eqref{initialamp-def1-DR} is given by
\begin{align}
	i\Pi(p^2) 
	&= (ig)^2
	\int \frac{d^4k}{(2\pi)^4}
	\frac{i}{k^2-m^2+i\epsilon}
	\frac{i}{(p-k)^2 - m^2 +i\epsilon} \\
	&-
	\frac{(ig)^2}{\kappa^2}
	\int \frac{d^4k}{(2\pi)^4}
	\frac{i}{k^2-m^2+i\epsilon}
	\frac{i\Delta_2[(S(k)\oplus p)^2]}{[(p-k)^2 - m^2 +i\epsilon]^2} \\ \nonumber \\
	&=
	i\Pi^U(p^2)
	-
	\frac{g^2}{\kappa^2}
	\int \frac{d^4k}{(2\pi)^4}
	\frac{1}{k^2-m^2+i\epsilon}
	\frac{\Delta_2[(S(k)\oplus p)^2]}{[(p-k)^2 - m^2 +i\epsilon]^2}.
\end{align}
Considering only the integral, after treating the denominator in the same as the non-deformed case using eq. \eqref{feytrickn} we are reduced to 
\begin{align}
	-\frac{g^2}{\kappa^2}
	\int_0^1 dx 
	\int \frac{d^4k}{(2\pi)^4}
	\frac{\Delta_2^{shift}[(S(k)\oplus p)^2]}
	{[k^2 - \Delta(x) +i\epsilon]^3}
\end{align}
where we have used the generalized Feynman trick
\begin{align}
	\frac{1}{A^m B^n}
	=
	\frac{\Gamma(m+n)}{\Gamma(m)\Gamma(n)}
	\int_0^\infty
	\frac{\lambda^{m-1}d\lambda}{[\lambda A + B]^{n+m}}
\end{align}
which can be rewritten in our particular case as
\begin{align}
	\frac{1}{A B^2}
	&=
	\int_0^1 \, dx \, dy\,
	\frac{2\delta(x+y-1)}{[xA + yB]^3}
	=
	\int_0^1 \, dy \,
	\frac{2}{[A + (B-A)y]^3} \nonumber \\
	&=
	\int_0^1 \, dx \,
	\frac{2}{[A + (B-A)x]^3}
\end{align}
and therefore
\begin{align}
	\frac{1}{k^2-m^2+i\epsilon}
	\frac{1}{[(p-k)^2-m^2+i\epsilon]^2}
	=
	\int_0^1 dx \,\,
	\frac{2}{\{[k-p(1-x)]^2 - \Delta(x) + i\epsilon\}^3}
\end{align}
\begin{align}
	-\Delta(x)
	=
	p^2 x (1-x) - m^2.
\end{align}

Since the poles are in the same position as before, we can now do a Wick rotation obtaining
\begin{align}
	i\frac{g^2}{\kappa^2}
	\int_0^1 dx 
	\int \frac{d^4k}{(2\pi)^4}
	\frac{\Delta_2^{shift+Wick}[(S(k)\oplus p)^2]}
	{[k^2 + \Delta(x)]^3}
\end{align}
where the $i\epsilon$ has gone away and where, as we computed before, we have
\begin{align}
	&\Delta_2^{shif+Wick}[(S(k)\oplus p)^2] \nonumber \\
	&=
	i k_0^3 p_0+ k_0^2 \left(-3 p_0^2 x+p_0^2+2 \mathbf{k}^2-4 \mathbf{k} \mathbf{p} x+3 \mathbf{k} \mathbf{p}+2 \mathbf{p}^2 x^2-3 \mathbf{p}^2 x+2 \mathbf{p}^2\right) \nonumber \\
	&+k_0 
	\Big(-3 i p_0^3 x^2+2 i p_0^3 x+4 i p_0 \mathbf{k}^2 x-i p_0 \mathbf{k}^2-8 i p_0 \mathbf{k} \mathbf{p} x^2+8 i p_0 \mathbf{k} \mathbf{p} x \nonumber \\
	&-2 i p_0 \mathbf{k} \mathbf{p}+4 i p_0 \mathbf{p}^2 x^3-7 i p_0 \mathbf{p}^2 x^2+6 i p_0 \mathbf{p}^2 x-2 i p_0 \mathbf{p}^2\Big) \nonumber \\
	&+
	\Big(p_0^4 x^3-p_0^4 x^2-2 p_0^2 \mathbf{k}^2 x^2+p_0^2 \mathbf{k}^2 x+4 p_0^2 \mathbf{k} \mathbf{p} x^3-5 p_0^2 \mathbf{k} \mathbf{p} x^2 \nonumber \\
	&+2 p_0^2 \mathbf{k} \mathbf{p} x-2 p_0^2 \mathbf{p}^2 x^4+4 p_0^2 \mathbf{p}^2 x^3-4 p_0^2 \mathbf{p}^2 x^2+2 p_0^2 \mathbf{p}^2 x-\mathbf{k}^3 \mathbf{p} \nonumber \\
	&+\mathbf{k}^2\mathbf{p}^2 x
	+
	(\mathbf{k}\mathbf{p})^2(2x-1)-3 \mathbf{k} \mathbf{p}^3 x^2+2 \mathbf{k} \mathbf{p}^3 x+\mathbf{p}^4 x^3-\mathbf{p}^4 x^2\Big)
\end{align}
Terms proportional to $k_0$ or $\mathbf{k}$ elevated to an odd power do not contribute to the final integral. The numerator can therefore be simplified as
\begin{align}
	\Delta_2^{shif+Wick}[(S(k)\oplus p)^2]
	&=
	k_0^2 \left(-3 p_0^2 x+p_0^2+2 \mathbf{k}^2+2 \mathbf{p}^2 x^2-3 \mathbf{p}^2 x+2 \mathbf{p}^2\right) \nonumber \\
	&+
	\Big(p_0^4 x^3-p_0^4 x^2-2 p_0^2 \mathbf{k}^2 x^2+p_0^2 \mathbf{k}^2 x
	-2 p_0^2 \mathbf{p}^2 x^4 \nonumber \\
	&+4 p_0^2 \mathbf{p}^2 x^3-4 p_0^2 \mathbf{p}^2 x^2+2 p_0^2 \mathbf{p}^2 x \nonumber\\
	&+\mathbf{k}^2\mathbf{p}^2 x
	+
	(\mathbf{k}\mathbf{p})^2(2x-1)+\mathbf{p}^4 x^3-\mathbf{p}^4 x^2\Big)
\end{align}
Since we are integrating in the Euclidean space, each time we encounter $\mathbf{k}^2$ alone we can substitute it with $3k_0^2$. Furthermore, since the terms linear in $k$ do not contribute, the mixed terms in $(\mathbf{k}\mathbf{p})^2$ do not contribute, so that we also have
\begin{align}
	(\mathbf{k}\mathbf{p})^2
	=
	\sum_{i,j} \mathbf{k}_i \mathbf{p}^i \mathbf{k}_j \mathbf{p}^j
	\mapsto
	\sum_i \mathbf{k}_i^2 (\mathbf{p}^i)^2
	\mapsto
	k_0^2 \mathbf{p}^2
\end{align}
Therefore, the above numerator can be simplified to 
\begin{align}
	\Delta_2^{shif+Wick}[(S(k)\oplus p)^2]
	&=
	k_0^2 \left(p_0^2 \left(1-6 x^2\right)+\mathbf{p}^2 (2 x (x+1)+1)\right)+2 k_0^2\mathbf{k}^2 \\
	&+(x-1) x \left(p_0^4 x-2 p_0^2 \mathbf{p}^2 ((x-1) x+1)+\mathbf{p}^4 x\right) \nonumber 
\end{align}
Notice that there is one obvious exception to the substitution done above, and this exception is obtained once we have terms like $k_0^2\mathbf{k}^2$, which explain the remaining $\mathbf{k}^2$ in the expression above. 

We now have three types of integral:
\begin{itemize}
	\item[1)]
	\begin{align}
		\int \frac{d^4k}{(2\pi)^4}\,\,
		\frac{k_0^2}{(k^2+\Delta)^3}
	\end{align}
	This integral can be computed with the help of eq.  \eqref{dimregint}
	Considering the case $a=1$, the above integral reduces to
	\begin{align}\label{onedimint-DR}
		\sum_{i=1}^{d}
		\mu^{4-d}
		\int
		\frac{d^d k}{(2\pi)^4}
		\frac{k_i^2}{(k^2 + \Delta)^b}
		&=
		d
		\int
		\frac{d^d k}{(2\pi)^4}
		\frac{k_i^2}{(k^2 + \Delta)^b} \nonumber \\
		&=
		\frac{\Gamma(b-1-\frac{1}{2}d)\Gamma(1+\frac{1}{2}d)}{(4\pi)^{d/2}\Gamma(b)\Gamma(\frac{1}{2}d)}
		\Delta^{-(b-1-d/2)}\mu^{4-d}
	\end{align}
	which means that
	\begin{align}\label{intkii-DR}
		\mu^{4-d}
		\int
		\frac{d^d k}{(2\pi)^4}
		\frac{k_0^2}{(k^2 + \Delta)^b}
		=
		\frac{1}{d}
		\frac{\Gamma(b-1-\frac{1}{2}d)\Gamma(1+\frac{1}{2}d)}{(4\pi)^{d/2}\Gamma(b)\Gamma(\frac{1}{2}d)}
		\Delta^{-(b-1-d/2)}\mu^{4-d};
	\end{align}

\item[2)]
 \begin{align}
 	\int \frac{d^4k}{(2\pi)^4}\,\,
 	\frac{1}{(k^2+\Delta)^3}
 \end{align}
This integral is just an example of eq. \eqref{dimregint} with $a=0$. We have therefore
\begin{align}
	\mu^{4-d}
	\int
	\frac{d^d k}{(2\pi)^4}
	\frac{1}{(k^2 + \Delta)^b}
	=
	\frac{\Gamma(b-\frac{1}{2}d)\Gamma(\frac{1}{2}d)}{(4\pi)^{d/2}\Gamma(b)\Gamma(\frac{1}{2}d)}
	\Delta^{-(b-d/2)}\mu^{4-d};
\end{align}

\item[3)] \begin{align}
	\int \frac{d^4k}{(2\pi)^4}
	\,\,
	\frac{k_i^2 k_j^2}{(k^2+m^2)^3}
	\qquad
	i\neq j
\end{align}
To compute this in dimensional regularization we cannot simply use the relation \eqref{dimregint}, but we need to compute it from scratch. We use $d$ dimensions and we go in spherical coordinates. Using the relations $k_i = |k| \cos(\alpha_i)$ (where $\alpha_i$ is the angle which the vector $k$ forms with respect to the $i$-th axis) we have
\begin{align}
	\int 
	\frac{d^dk}{(2\pi)^d}
	\,\,
	\frac{k_i^2 k_j^2}{(k^2+\Delta)^b}
	=
	\frac{1}{(2\pi)^d}
	\int d\Omega \,\, \cos^2\alpha_i \cos^2\alpha_j
	\int_0^\infty
	dk
	\frac{k^{d-1}k^4}{(k^2 + \Delta)^b}
\end{align}
We can now can just define the constant
\begin{align}
	\Theta(i,2,j,2) := \int d\Omega \,\, \cos^2\alpha_i \cos^2\alpha_j
\end{align}
which is a strictly positive quantity bounded from above by the surface $S_d$ of a $d$-dimensional sphere. The remaining integral can be evaluated with the help of the beta functions
\begin{align}
	B(\alpha, \gamma)
	=
	\frac{\Gamma(\alpha)\Gamma(\gamma)}{\Gamma(\alpha+\gamma)}
	=
	\int_0^\infty
	dy \,\, y^{\alpha-1}(1+y)^{-\alpha-\gamma}
\end{align}
In our case we have
\begin{align}
	\int_0^\infty
	dk
	\,\,
	k^{d+3}(k^2 + \Delta)^{-b}
	&=
	\Delta^{-b}
	\int_0^\infty
	dk
	\,\,
	k^{d+3}\left(\frac{k^2}{\Delta} + 1\right)^{-b} \\
	&=
	\Delta^{-b}
	\int_0^\infty
	\frac{1}{2}\Delta^{\frac{1}{2}} y^{-\frac{1}{2}} dy
	\,\,
	(\sqrt{\Delta y})^{d+3}\left(y + 1\right)^{-b} \\
	&=
	\frac{1}{2}
	\Delta^{-b+\frac{1}{2} + \frac{d}{2}+\frac{3}{2}}
	\int_0^\infty
	dy \,\,
	y^{\frac{d}{2}+2-1}
	(1+y)^{\frac{d}{2}+2-b -(\frac{d}{2}+2)} \\
	&=
	\frac{\Gamma(2+\frac{1}{2}d)\Gamma(b - 2 - \frac{1}{2}d)}{2\Gamma(b)}
	\Delta^{-b + 2 + \frac{d}{2}}
\end{align}
where we used the change of coordinates 
\begin{align}
	y = \frac{k^2}{\Delta}
	\quad
	\implies 
	\quad
	dy
	=
	2\frac{k}{\Delta} dk
	\quad
	\implies 
	\quad
	dk = \frac{\Delta}{2 \sqrt{y\Delta}} dy
	=
	\frac{1}{2}\Delta^{\frac{1}{2}} y^{-\frac{1}{2}} dy
\end{align}
which corresponds to the result we would have gotten from eq. \eqref{dimregint} but with the factors $\frac{1}{(4\pi)^{d/2}\Gamma(\frac{1}{2}d)}$ exchanged with the factor $\frac{\Theta(i,j)}{2(2\pi)^d}$. Therefore we have
\begin{align}\label{intkikj-DR}
	\mu^{4-d}
	\int 
	\frac{d^dk}{(2\pi)^d}
	\,\,
	\frac{k_i^2 k_j^2}{(k^2+\Delta)^b}
	=
	\frac{\Theta(i,2,j,2)}{(2\pi)^d}
	\frac{\Gamma(2+\frac{1}{2}d)\Gamma(b - 2 - \frac{1}{2}d)}{2\Gamma(b)}
	\Delta^{-b + 2 + \frac{d}{2}}\mu^{4-d}
\end{align}
where at the end we wrote once again the term $\mu^{d-4}$ for dimensional reasons.
\end{itemize}

We can now explicitly perform the integral. We have to compute 
\begin{align}
	i\frac{g^2}{\kappa^2}
	\Bigg\{
	&\int_0^1 dx 
	\underbrace{\left(p_0^2 \left(1-6 x^2\right)+\mathbf{p}^2 (2 x (x+1)+1)\right)}_{:=F(p,x)}
	\int \frac{d^4k}{(2\pi)^4}
	\frac{k_0^2 }
	{[k^2 + \Delta(x)]^3}\label{F}
	 \\
	+&
	2\int_0^1 dx 
	\int \frac{d^4k}{(2\pi)^4}
	\frac{k_0^2 \mathbf{k}^2}
	{[k^2 + \Delta(x)]^3} \\
	+&
	\int_0^1 dx 
	\underbrace{(x-1) x \left(p_0^4 x-2 p_0^2 \mathbf{p}^2 ((x-1) x+1)+\mathbf{p}^4 x\right)}_{:=G(p,x)}
	\int \frac{d^4k}{(2\pi)^4}
	\frac{1}
	{[k^2 + \Delta(x)]^3}
	\Bigg\}\label{G}
\end{align}
Using the equations obtained above, and using the canonical choice $d=4-\epsilon$ which is used in dimensional regularization, we have
\begin{align}
	\int \frac{d^4k}{(2\pi)^4}
	\frac{k_0^2 }
	{[k^2 + \Delta(x)]^3}
	&=
	\frac{1}{d}
	\frac{\Gamma(b-1-\frac{1}{2}d)\Gamma(1+\frac{1}{2}d)}{(4\pi)^{d/2}\Gamma(b)\Gamma(\frac{1}{2}d)}
	\Delta^{-(b-1-d/2)}\mu^{4-d} \Bigg|_{b=3} \nonumber \\
	&=
	\frac{1}{4}
	\frac{\Gamma(3-1-2+\frac{\epsilon}{2})\Gamma(1+2)}{(4\pi)^{2}\Gamma(3)\Gamma(2)}
	\Delta^{-(3-1-2+\frac{\epsilon}{2})}\mu^{\epsilon} \nonumber \\
	&=
	\frac{\Gamma(\frac{\epsilon}{2})}{4(4\pi)^2}
	\left(
	\frac{\Delta}{\mu^2}
	\right)^{-\frac{\epsilon}{2}}
\end{align}
\begin{align}
	\int \frac{d^4k}{(2\pi)^4}
	\frac{k_0^2 \mathbf{k}^2}
	{[k^2 + \Delta(x)]^3}
	&=
	3
	\frac{\Theta(0,2,j,2)}{(2\pi)^d}
	\frac{\Gamma(2+\frac{1}{2}d)\Gamma(b - 2 - \frac{1}{2}d)}{2\Gamma(b)}
	\Delta^{-b + 2 + \frac{d}{2}}\mu^{4-d}\Bigg|_{b=3} \nonumber \\
	&=
	3
	\frac{\Theta(0,2,j,2)}{(2\pi)^4}
	\frac{\Gamma(4)\Gamma(3 - 2 - 2+\frac{\epsilon}{2})}{2\Gamma(3)}
	\Delta^{-3 + 2 + 2}
	\left(
	\frac{\Delta}{\mu^2}
	\right)^{-\frac{\epsilon}{2}} \nonumber \\
	&=
	9
	\frac{\Theta(0,2,j,2)}{2(2\pi)^4}
	\Delta
	\Gamma(-1+\frac{\epsilon}{2})
	\left(
	\frac{\Delta}{\mu^2}
	\right)^{-\frac{\epsilon}{2}}
\end{align}
\begin{align}
	\int \frac{d^4k}{(2\pi)^4}
	\frac{1}
	{[k^2 + \Delta(x)]^3}
	&=
	\frac{\Gamma(b-\frac{1}{2}d)\Gamma(\frac{1}{2}d)}{(4\pi)^{d/2}\Gamma(b)\Gamma(\frac{1}{2}d)}
	\Delta^{-(b-d/2)}\mu^{4-d} \Bigg|_{b=3} \nonumber \\
	&=
	\frac{\Gamma(1)}{(4\pi)^{2}\Gamma(3)}
	\Delta^{-1} \nonumber \\
	&=
	\frac{1}{2(4\pi)^2 \Delta}
\end{align}
We now need the following relations 
\begin{align}
	\Gamma(-n+\epsilon)
	=
	\frac{(-1)^n}{n!}
	\left(
	\frac{1}{\epsilon} - \gamma + \sum_{k=1}^{n}\frac{1}{k} + O(\epsilon)
	\right)
\end{align}
\begin{align}
	\eqref{useful}\rightarrow \quad \Gamma(\frac{\epsilon}{2})
	\Delta^{-\frac{\epsilon}{2}}
	(\mu^2)^{\frac{\epsilon}{2}}
	& \approx
	\frac{2}{\epsilon} - \log \frac{\Delta}{\mu^2} -\gamma + O(\epsilon)
\end{align}
In this way, the amplitude becomes
\begin{align}
	\Pi(p)
	=
	\Pi^U(p)
	+
	\frac{g^2}{\kappa^2}
	\Bigg\{
	&
	\frac{1}{4(4\pi)^2}\int_0^1 dx 
	F(p,x)
	\left(
	\frac{2}{\epsilon} - \log \frac{\Delta}{\mu^2} -\gamma
	\right)
	\\
	-&
	9
	\frac{\Theta(0,2,j,2)}{(2\pi)^4}
	\int_0^1 dx 
	\Delta
	\left(
	\frac{2}{\epsilon} - \gamma + 1
	\right)
	\left(
	1-\frac{\epsilon}{2}
	\log \frac{\Delta}{\mu^2}
	\right)
	\\
	+&
	\frac{1}{2(4\pi)^2}
	\int_0^1 dx 
	G(p,x)
	\frac{1}{\Delta}
	\Bigg\}
\end{align}
Once again, notice that the amplitude contains a real and an imaginary part, because the $\Delta$ is exactly the same as the one in the  non-deformed case. As such, all the renormalization procedure is intended to be valid only for the real part $\Re(\Pi)$ of it. 

Like in the non-deformed case, the only contribution comes when the argument of the logarithm becomes negative (notice, after imposing the renormalization conditions, the $\mu^2$ in the argument of the logarithms will be replaced by $\Delta_0$ as previously defined, which is always positive for $x\in[0,1]$), and therefore $\Im \log\frac{\Delta}{\Delta_0} = -\pi$. Hence, recalling that the integration domain on the $x$ for the imaginary part is
\begin{align}
	|x| < \frac{1}{2} \pm \frac{1}{2} \sqrt{1-4\frac{m^2}{p^2}}
\end{align}
the imaginary part of the integral in the amplitude is given by
\begin{align}\label{step}
	&\frac{g^2}{\kappa^2}
	\Bigg\{
	\frac{\pi}{4(4\pi)^2}\int_{L_1}^{L_2} dx 
	F(p,x)
	-
	9\pi
	\frac{\Theta(0,2,j,2)}{(2\pi)^4}
	\int_{L_1}^{L_2} dx 
	\Delta
	\Bigg\} 
\end{align}
where we called 
\begin{align}
	L_1 = \frac{1}{2} - \frac{1}{2}\sqrt{1-4\frac{m^2}{p^2}}
	\qquad
	L_2 = \frac{1}{2} + \frac{1}{2}\sqrt{1-4\frac{m^2}{p^2}}.
\end{align}
Computing everything explicitly, and recalling that $F(p,x)$ and $G(p,x)$ are defined respectively in eq. \eqref{F} and \eqref{G}, we get (recall that we are using the metric convention $+---$)
\begin{align}
	\int_{L_1}^{L_2} F(p,x)dx
	&=
	\int_{L_1}^{L_2} 
	\left(p_0^2 \left(1-6 x^2\right)+\mathbf{p}^2 (2 x (x+1)+1)\right) \\
	&=
	\sqrt{1-4\frac{m^2}{p^2}}
	\left(
	\frac{8}{3}p^2 - \frac{5}{3} p_0^2 -\frac{2}{3}m^2 -4\frac{m^2}{p^2}p_0^2
	\right)
\end{align}
\begin{align}
	\int_{L_1}^{L_2}\Delta dx
	&=
	\int_{L_1}^{L_2} (m^2-p^2 x (1-x)) dx \\
	&=
	\sqrt{1-4\frac{m^2}{p^2}}
	\left(-\frac{2}{3}m^2 +\frac{1}{6}p^2\right)
\end{align}
(notice that we switched $\mathbf{p}^2 = p_0^2 - p^2$). 
To get a fully numerical computation, we need to compute
\begin{align}
	I=\Theta(0,2,j,2)
	=
	\int \cos^2 \alpha_0 \cos^2 \alpha_j d\Omega
\end{align}
The cosines of $\alpha_j$ are called directional cosines. We have 
\begin{align}
	\cos \alpha_j = \frac{k_j}{\sqrt{k_0^2+k_1^2+k_2^2+k_3^2}}
\end{align}
so that by definition  we have
\begin{align}\label{directionalcosprop}
	\sum_{i=0}^3 \cos^2 \alpha_j = 1
\end{align}
Since $j$ is fixed, we can assume without loss of generality that $j=1$. Using the above relation we can substitute $\cos^2\alpha_0$ by $1 - \sum_{i=1}^3\cos^2\alpha_i$ so that 
\begin{align}
	I = \underbrace{\int \cos^2\alpha_1d\Omega}_{I_2} - \int \cos^4\alpha_1 d\Omega
	-
	2I
\end{align}
where we used the fact that the integral of the product of any two different directional cosines does not depend on which pair we choose, because Euclidean space is isotropic. In the same way, we can write 
\begin{align}
	I_2 =\int \cos^2\alpha_1d\Omega
	=
	\int d\Omega - 3 I_2
	=
	S_4 - 3I_2
\end{align}
so that 
\begin{align}
	I_2 = \frac{S_4}{4}
\end{align}
where $S_4$ is the surface of a $n=4$ dimensional sphere. We are therefore reduced to 
\begin{align}
	I
	=
	\frac{S_4}{12}
	-
	\frac{1}{3}
	\int \cos^4\alpha_1 d\Omega
\end{align}
Now we only have one integral left, and this we can compute directly. In fact, the sperical surface element in $n$ dimensions in Euclidean space and in hyperspherical coordinates is given by (we are considering a fixed radius $r=1$, since we already integrated over the radius in the formula where $\Theta$ was introduced in the first place)
\begin{align}
	d\Omega
	=
	\sin^{n-2}(\phi_1) \sin^{n-3}(\phi_2) \dots \sin(\phi_{n-2}) d\phi_1 \dots d\phi_{n-1}
\end{align}
where $\phi_1, \dots , \phi_{n-2}\in [0,\pi]$ and $\phi_{n-1}\in[0,2\pi]$. In our case $n=4$  and therefore we have 
\begin{align}
	d\Omega = \sin^2(\omega)\sin(\theta) d\omega d\theta d\phi
\end{align}
Such an explicit choice of coordinate brakes the symmetry of Euclidean space, since we are now choosing a preferred axis. In particular, the directional cosine should be taken with respect to this preferred axis, in the same way in which in three dimension the only directional cosine is the one taken with respect to the preferential axis, which in three dimension is the $z$ axis. Therefore we have
\begin{align}
	\int
	\cos^4(\omega) \sin^2(\omega)\sin(\theta) d\omega d\theta d\phi
	=
	\frac{\pi^2}{4}
\end{align}
Furthermore, recalling that 
\begin{align}
	S_d
	=
	\frac{2\pi^{\frac{d}{2}}}{\Gamma(\frac{d}{2})}
	\quad \implies \quad
	S_4
	=
	2\pi^2
\end{align}
we have
\begin{align}\label{directcos}
	I=
	\frac{\pi^2}{6} - \frac{\pi^2}{12}
	=
	\frac{\pi^2}{12}
\end{align}
Therefore, the imaginary part of the integral becomes
\begin{align}
&
	\frac{g^2}{16\pi\kappa^2}
	\sqrt{1-4\frac{m^2}{p^2}}
	\Bigg\{
	p_0^2
	\left(
	-
	\frac{5}{12}
	-
	\frac{m^2}{p^2}
	\right)
	+
	p^2
	\left(
	\frac{2}{3}
	-
	\frac{1}{8}
	\right)
	+
	m^2
	\left(
	-\frac{1}{6}
	+
	\frac{1}{2}
	\right)
	\Bigg\} \\
	=&
	\frac{g^2}{16\pi\kappa^2}
	\sqrt{1-4\frac{m^2}{p^2}}
	\Bigg\{
	p_0^2
	\left(
	-
	\frac{5}{12}
	-
	\frac{m^2}{p^2}
	\right)
	+
	\frac{13}{24}
	p^2
	+
	\frac{1}{3}
	m^2
	\Bigg\}
\end{align}
Recalling from eq. \eqref{undefim} that $\Im \Pi^U(p^2) =  \frac{g^2}{16\pi}\sqrt{1-4\frac{m^2}{p^2}}$ we can write
\begin{align}\label{Pi1}
	\boxed{
	\Im \Pi^{(1)}(p^2)
	=
	\Im \Pi^U(p^2)
	\left\{
	1
	+
	\frac{1}{\kappa^2}
	\left[
	p_0^2
	\left(
	-
	\frac{5}{12}
	-
	\frac{m^2}{p^2}
	\right)
	+
	\frac{13}{24}
	p^2
	+
	\frac{1}{3}
	m^2
	\right]
	\right\}}
\end{align}
where the index ${}^{(1)}$ just means that this is the amplitude corresponding to the first possibility of the choice of momenta.

We could perform the same integrals using a hard cutoff instead of dimensional regularization. In fact, from a principled point of view, we are expanding in powers of $1/\kappa$, which means that momenta should have a momentum smaller than $\kappa$. This in turn implies that we should not be able to integrate up to infinity in momentum space when integrating virtual momenta. Of course, the finite part of any integrand does not depend on the choice of regularization, but limiting ourself to a hard cutoff makes the approximations used in the expansion in powers of $1/\kappa$ explicit at all steps. In particular, if we assume that our momenta are organized according to the following relation
\begin{align}
	\boxed{p,k \ll \Lambda \ll \kappa}
\end{align}
where $\Lambda$ is the cutoff scale, then our expansion in powers of $1/\kappa$ is fully justified. 


Therefore, we compute the same integral using a cutoff in momentum space. We have what follows:
\begin{itemize}
	\item[1)] 
	\begin{align}
		\int \frac{d^4k}{(2\pi)^4}\,\,
		\frac{k_0^2}{(k^2+\Delta)^3}
	\end{align}
If we go to spherical coordinates this integral becomes (using Mathematica)
\begin{align}
	\frac{1}{4}
	\frac{S_4}{(2\pi)^4}
	\int_0^\Lambda dk \,
	\frac{k^5}{(k^2+\Delta)^3}
	&=
	\frac{1}{32\pi^2}
	\Big[
	\frac{1}{4}
	(-3-2\log \Delta) \nonumber \\
	&+
	\frac{1}{4}
	\left(
	\frac{\Delta(3\Delta + 4\Lambda^2)}{(\Delta+\Lambda^2)^2}
	+
	2\log(\Delta+\Lambda^2)
	\right)
	\Big]
\end{align}
\item[2)]
 \begin{align}
	\int \frac{d^4k}{(2\pi)^4}\,\,
	\frac{1}{(k^2+\Delta)^3}
\end{align}
In this case going to spherical coordinates we have (again with Mathematica)
\begin{align}
	\frac{S_4}{(2\pi)^4}
	\int_0^\Lambda dk \,
	\frac{k^3}{(k^2+\Delta)^3}
	=
	\frac{1}{8\pi^2}
	\frac{\Lambda^4}{4\Delta(\Delta+\Lambda^2)^2}
\end{align}
\item[3)]
\begin{align}
	\int \frac{d^4k}{(2\pi)^4}
	\,\,
	\frac{k_i^2 k_j^2}{(k^2+m^2)^3}
	\qquad
	i\neq j
\end{align}
To compute this with a cutoff we must proceed as before, and we go in spherical coordinates. Using the relations $k_i = |k| \cos(\alpha_i)$ (where $\alpha_i$ is the angle which the vector $k$ forms with respect to the $i$-th axis) we have
\begin{align}
	\int 
	\frac{d^4k}{(2\pi)^4}
	\,\,
	\frac{k_i^2 k_j^2}{(k^2+\Delta)^3}
	=
	\frac{1}{(2\pi)^4}
	\int d\Omega \,\, \cos^2\alpha_i \cos^2\alpha_j
	\int_0^\Lambda
	dk
	\frac{k^7}{(k^2 + \Delta)^3}
\end{align}
We know from eq. \eqref{directcos} that the angular integral corresponds to $\Theta(0,2,j,2)=\frac{\pi^2}{12}$, and the radial integral can be computed explicitly obtaining 
\begin{align}
	\int_0^\Lambda
	dk
	\frac{k^7}{(k^2 + \Delta)^3}
	&=
	\frac{1}{4}
	\Delta
	(5 + 6\log \Delta) \nonumber \\
	+
	&\frac{-5\Delta^3 - 4 \Delta^2 \Lambda^2 + 4 \Delta\Lambda^4 + 2 \Lambda^6 - 6\Delta(\Delta+\Lambda^2)^2\log(\Delta+\Lambda^2)}{4(\Delta+\Lambda^2)^2}
\end{align}
\end{itemize}
As in the dimensional regularization case, the divergencies can be absorbed in the counterterms, but here we are only interested in the parts which can contribute an imaginary part to the integral, which are once again the logarithms. Notice here that the logarithms of the form $\log(\Delta+\Lambda^2)$ cannot contribute any immaginary part because, although $\Delta$ can be negative, the sum $\Delta+\Lambda^2$ cannot ($\Lambda^2$ is positive and very large). Therefore, limiting ourselves to the logarithms, the contribution from the above integrals become respectively
\begin{align}\label{cutoff-log-1}
	\int \frac{d^4k}{(2\pi)^4}\,\,
	\frac{k_0^2}{(k^2+\Delta)^3}
	\rightarrow
	-\frac{1}{4(4\pi)^2}
	\log\Delta
\end{align}
\begin{align}\label{cutoff-log-2}
	\int \frac{d^4k}{(2\pi)^4}\,\,
	\frac{1}{(k^2+\Delta)^3}
	\rightarrow
	0
\end{align}
\begin{align}\label{cutoff-log-3}
	2\int \frac{d^4k}{(2\pi)^4}
	\,\,
	\frac{k_0^2 \mathbf{k}^2}{(k^2+m^2)^3}
	\rightarrow
	\frac{\pi^2}{12}
	\frac{1}{(2\pi)^4}
	9
	\Delta\log\Delta
\end{align}
We know from the previous equations, and in particular from eq. \eqref{step} and \eqref{directcos}, that these are the same coefficients as before. We obtain therefore the same imaginary part, as was expected. 

\bigskip

We now pass to the second possibility, namely 
\begin{center}
	\begin{tikzpicture}
		\begin{feynman}
			\vertex (a) at (-0.6,0) {\(i \Pi(p^2) = \)};
			\vertex (b) at (1,0);
			\vertex (c) at (2,0);
			\vertex (d) at (3,0);

			\diagram* 
			{(a)-- [ultra thick] (b),
				(b)-- [scalar,half left,looseness=1.7,momentum=\(k\)] (c) -- [scalar,half left,looseness=1.7,reversed momentum=\(p\oplus S(k)\)] (b),
				(c)-- [ultra thick] (d),};
		\end{feynman}
	\end{tikzpicture}
\end{center}
\begin{align}\label{initialamp-def1-DR-2}
	i\Pi(p^2)
	=
	(ig)^2
	\int \frac{d^4 k}{(2\pi)^4} 
	\frac{i}{k^2 - m^2 +i\epsilon}
	\frac{i}{(p\oplus S(k))^2 - m^2 +i\epsilon}
\end{align}
Every consideration goes exactly as before, but this time we need to compute 
\begin{align}
	\Delta_1^{shif+Wick}[(S(k)\oplus p)^2],
	\qquad
	\Delta_2^{shif+Wick}[(S(k)\oplus p)^2]
\end{align}
In fact, contrary to the previous case, this time there is a $1/\kappa$ contribution to the expansion of the denominator. The explicit expression is given by 
\begin{align}
	i\Pi(p^2) 
	&=
	i\Pi^U(p^2) \\
	&-
	\frac{g^2}{\kappa}
	\int \frac{d^4k}{(2\pi)^4}
	\frac{1}{k^2-m^2+i\epsilon}
	\frac{\Delta_1[(p\oplus S(k))^2]}{[(p-k)^2 - m^2 +i\epsilon]^2}\\
	&-
	\frac{g^2}{\kappa^2}
	\int \frac{d^4k}{(2\pi)^4}
	\frac{1}{k^2-m^2+i\epsilon}
	\frac{\Delta_2[(p\oplus S(k))^2]}{[(p-k)^2 - m^2 +i\epsilon]^2}
\end{align}

We follow the same steps as before. We have
\begin{align}
	\Delta_1[(p\oplus S(k))^2]
	=
	2(\mathbf{k}-\mathbf{p}) (p_0 \mathbf{k}-k_0 \mathbf{p})
\end{align}
\begin{align}
	\Delta_2[(p\oplus S(k))^2]
	&=
	\Big(k_0^3 (-p_0)+k_0^2 \left(2 p_0^2-2 \mathbf{k}^2+2 \mathbf{k} \mathbf{p}-\mathbf{p}^2\right) \nonumber \\
	&-k_0 p_0 \left(p_0^2-\mathbf{k}^2+\mathbf{p}^2\right)-p_0^2 \mathbf{k} (\mathbf{k}-2 \mathbf{p})\Big)
\end{align}
Since the denominators are always the same as before, we can treat them in the same way, obtaining
\begin{align}
	&-\frac{g^2}{\kappa}
	\int_0^1 dx 
	\int \frac{d^4k}{(2\pi)^4}
	\frac{\Delta_1^{shift}[(p\oplus S(k))^2]}
	{[k^2 - \Delta(x)+i\epsilon]^3} \\
	&-
	\frac{g^2}{\kappa^2}
	\int_0^1 dx 
	\int \frac{d^4k}{(2\pi)^4}
	\frac{\Delta_2^{shift}[(p\oplus S(k))^2]}
	{[k^2 - \Delta(x)+i\epsilon]^3}
\end{align}
The shift is again $k\mapsto k+p(1-x)$ and the numerators become
\begin{align}\label{secondDshift1}
	\Delta_1^{shift}[(p\oplus S(k))^2]
	&=
	2 (\mathbf{k}+\mathbf{p} (1-x)-\mathbf{p}) (p_0 (\mathbf{k}+\mathbf{p} (1-x)) \nonumber \\
	&-\mathbf{p} (k_0+p_0 (1-x)))
\end{align}
\begin{align}\label{secondDshift2}
	&\Delta_2^{shift}[(p\oplus S(k))^2] \nonumber \\
	&=
	(k_0+p_0 (1-x))^2 \left(2 p_0^2-2 (\mathbf{k}+\mathbf{p} (1-x))^2+2 \mathbf{p} (\mathbf{k}+\mathbf{p} (1-x))-\mathbf{p}^2\right) \nonumber \\
	&-p_0 (k_0+p_0 (1-x)) \left(p_0^2-(\mathbf{k}+\mathbf{p} (1-x))^2+\mathbf{p}^2\right) \nonumber \\
	&-p_0 (k_0+p_0 (1-x))^3-p_0^2 (\mathbf{k}+\mathbf{p} (1-x)) (\mathbf{k}+\mathbf{p} (1-x)-2 \mathbf{p})
\end{align}
We can now perform the Wick rotation. The deformed part of the amplitude becomes
\begin{align}
	&i\frac{g^2}{\kappa}
	\int_0^1 dx 
	\int \frac{d^4k}{(2\pi)^4}
	\frac{\Delta_1^{shift+Wick}[(p\oplus S(k))^2]}
	{[k^2 + \Delta(x)]^3} \\
	&+
	i
	\frac{g^2}{\kappa^2}
	\int_0^1 dx 
	\int \frac{d^4k}{(2\pi)^4}
	\frac{\Delta_2^{shift+Wick}[(p\oplus S(k))^2]}
	{[k^2 + \Delta(x)]^3}
\end{align}
Notice that the $i\epsilon$ has gone away and there is a global sign change coming from $(-k^2-\Delta)^3 = -(k^2+\Delta)^3$. The additional $i$ comes from the metric. 
We also remove from both $\Delta_1^{shift+Wick}[(p\oplus S(k))^2]$ and $\Delta_2^{shift+Wick}[(p\oplus S(k))^2]$ the irrelevant factors (proportional to an odd power component of $k$), and using the isotropy of Euclidean space. We have
\begin{align}
	\Delta_1^{shift+Wick}[(p\oplus S(k))^2]
	=
	+6 p_0 k_0^2
\end{align}
\begin{align}
	&\Delta_2^{shift+Wick}[(p\oplus S(k))^2] \nonumber \\
	&=
	k_0^2[
	p_0^2(-6x^2+6x-5) + \mathbf{p}^2 (2x^2-2x + 1)
	]
	+2 k_0^2 \mathbf{k}^2 \nonumber
	\\
	&+p_0^4 x^3-p_0^4 x^2 
	-2 p_0^2 \mathbf{p}^2 x^4+5 p_0^2 \mathbf{p}^2 x^3 -5 p_0^2 \mathbf{p}^2 x^2+2 p_0^2 \mathbf{p}^2 x
\end{align}
We now use the same integrals as before in the cutoff regularization, eq. \eqref{cutoff-log-1}, \eqref{cutoff-log-2}, \eqref{cutoff-log-3}. 
Applying these we have
\begin{align}
	&i\frac{g^2}{\kappa}
	\int_0^1 dx 
	\int \frac{d^4k}{(2\pi)^4}
	\frac{\Delta_1^{shift=Wick}[(p\oplus S(k))^2]}
	{[k^2 + \Delta(x)]^3} \nonumber \\
	&=
	i\frac{g^2}{\kappa}
	\int_0^1 dx 
	6p_0
	\int \frac{d^4k}{(2\pi)^4}
	\frac{k_0^2}
	{[k^2 + \Delta(x)]^3} \\
	&=
	-
	i\frac{g^2}{\kappa}
	\frac{1}{64\pi^2}
	\int_{L_1}^{L_2} dx \, 
	6p_0
	\log \frac{\Delta}{|\Delta^0|}
\end{align}
\begin{align}
	&i
	\frac{g^2}{\kappa^2}
	\int_0^1 dx 
	\int \frac{d^4k}{(2\pi)^4}
	\frac{\Delta_2^{shift+Wick}[(p\oplus S(k))^2]}
	{[k^2 + \Delta(x)]^3} \\ \nonumber \\
	&=
	0 
	-
	\frac{1}{64\pi^2}
	i
	\frac{g^2}{\kappa^2}
	\int_{L_1}^{L_2} dx 
	[p_0^2(-6x^2+6x-5) + \mathbf{p}^2 (2x^2-2x + 1)]
	\log \frac{\Delta}{|\Delta^0|} \\
	&+
	i
	\frac{\pi^2}{12}
	\frac{9}{(2\pi)^4}
	\frac{g^2}{\kappa^2}
	\int_{L_1}^{L_2} dx \,
	\Delta \log\frac{\Delta}{|\Delta^0|}
\end{align}
This of course means that 
\begin{align}
	\Pi^{(2)}(p^2)
	&=
	\Pi^U(p^2)
	-
	\frac{g^2}{\kappa}
	\frac{1}{64\pi^2}
	\int_{L_1}^{L_2} dx \, 
	6p_0
	\log \frac{\Delta}{|\Delta^0|} \\
	&-
	\frac{1}{64\pi^2}
	\frac{g^2}{\kappa^2}
	\int_{L_1}^{L_2} dx 
	[p_0^2(-6x^2+6x-5) + \mathbf{p}^2 (2x^2-2x + 1)]
	\log \frac{\Delta}{|\Delta^0|} \\
	&+
	\frac{3}{64\pi^2}
	\frac{g^2}{\kappa^2}
	\int_{L_1}^{L_2} dx \,
	\Delta \log\frac{\Delta}{|\Delta^0|}
\end{align}
Once again, each logarithm will contribute $-\pi$ to the imaginary part, and computing the integral explicitly we end up with
\begin{align}
	\Im \Pi^{(2)}(p^2)
	&=
	\Im \Pi^U(p^2)
	+
	\frac{g^2}{\pi \kappa}
	\frac{3 p_0}{32}
	\sqrt{1-4\frac{m^2}{p^2}} \nonumber \\
	&+
	\frac{g^2}{16\pi \kappa^2}
	\sqrt{1-4\frac{m^2}{p^2}}
	\left(
	p_0
	\left(
	-\frac{1}{6} + \frac{2}{3}\frac{m^2}{p^2}
	\right)
	+
	\frac{1}{24}
	p^2
	+
	\frac{1}{2} m^2
	\right)
\end{align}
Recalling from eq. \eqref{undefim} that $\Im \Pi^U(p^2) =  \frac{g^2}{16\pi} \sqrt{1-4\frac{m^2}{p^2}}$ we finally obtain
\begin{align}\label{Pi2}
	\boxed{
	\Im \Pi^{(2)}(p^2)
	=
	\Im \Pi^U(p^2)
	\left\{
	1
	+
	\frac{1}{\kappa}
	\left(
	\frac{3}{2}p_0
	\right)
	+
	\frac{1}{\kappa^2}
	\left[
	p_0
	\left(
	-\frac{1}{6} + \frac{2}{3}\frac{m^2}{p^2}
	\right)
	+
	\frac{1}{24}
	p^2
	+
	\frac{1}{2} m^2
	\right]
	\right\}
}
\end{align}

We now pass to the third option. In this case we have
\begin{center}
	\begin{tikzpicture}
		\begin{feynman}
			\vertex (a) at (-0.6,0) {\(i \Pi(p^2) = \)};
			\vertex (b) at (1,0);
			\vertex (c) at (2,0);
			\vertex (d) at (3,0);

			\diagram* 
			{(a)-- [ultra thick] (b),
				(b)-- [scalar,half left,looseness=1.7,momentum=\(S(k)\)] (c) -- [scalar,half left,looseness=1.7,reversed momentum=\(k\oplus p\)] (b),
				(c)-- [ultra thick] (d),};
		\end{feynman}
	\end{tikzpicture}
\end{center}
\begin{align}\label{initialamp-def1-DR-3}
	i\Pi(p^2)
	=
	(ig)^2
	\int \frac{d^4 k}{(2\pi)^4} 
	\frac{i}{k^2 - m^2 +i\epsilon}
	\frac{i}{(k\oplus p)^2 - m^2 +i\epsilon}
\end{align}
which expanded to the first two non-trivial orders in powers of $1/\kappa$ is 
\begin{align}
	i\Pi(p^2) 
	&= 
	i\Pi^U(p^2) \\
	&-
	\frac{g^2}{\kappa}
	\int \frac{d^4k}{(2\pi)^4}
	\frac{1}{k^2-m^2+i\epsilon}
	\frac{\Delta_1[(k\oplus p)^2]}{[(p+k)^2 - m^2 +i\epsilon]^2}\\
	&-
	\frac{g^2}{\kappa^2}
	\int \frac{d^4k}{(2\pi)^4}
	\frac{1}{k^2-m^2+i\epsilon}
	\frac{\Delta_2[(k\oplus p)^2]}{[(p+k)^2 - m^2 +i\epsilon]^2}
\end{align}
Notice the presence of $(p+k)^2$ in the denominator. This has no impact on the  non-deformed amplitude, because we can simply switch $k\mapsto - k$, the metric remains invariant as well as the integration domain, and we recover the usual integral. In the deformed terms, however, we have to keep in mind that we must send $k\mapsto-k$ also in the numerator before doing the Feynman trick, so that we can once again use the same steps af before. 

Apart from this, every consideration goes exactly as before, and this time we need to compute 
\begin{align}
	\Delta_1^{shif+Wick}[(k\oplus p)^2],
	\qquad
	\Delta_2^{shif+Wick}[(k\oplus p)^2]
\end{align}
once again. 
We have
\begin{align}
	\Delta_1[(k\oplus p)^2]
	=
	2 \mathbf{k} (k_0 \mathbf{p}-p_0 \mathbf{k})
\end{align}
\begin{align}
	\Delta_2[(k\oplus p)^2]
	&=
	(k_0+p_0) \left(p_0 \left(k_0 (k_0+p_0)+\mathbf{k}^2\right)-2 k_0 \mathbf{k} \mathbf{p}-k_0 \mathbf{p}^2\right) \nonumber \\
	&-p_0^2 \mathbf{k}^2+\mathbf{k} \left(\mathbf{p}^2-p_0^2\right) (\mathbf{k}+\mathbf{p})+\frac{\mathbf{k}^2 \mathbf{p}^2}{3}
\end{align}
where we used the identity
\begin{align}
	(\mathbf{k}\mathbf{p})^2
	=
	\sum_{i,j} \mathbf{k}_i \mathbf{p}^i \mathbf{k}_j \mathbf{p}^j
	\mapsto
	\sum_i \mathbf{k}_i^2 (\mathbf{p}^i)^2
	\mapsto
	k_0^2 \mathbf{p}^2
	\mapsto
	\frac{1}{3} \mathbf{k}^2 \mathbf{p}^2.
\end{align}

We now send $k\mapsto -k$. It is obvious that $\Delta_1[(k\oplus p)^2]$ is not affected by such a change. On the other hand $\Delta_2[(k\oplus p)^2]$ is modified. We have
\begin{align}
	\Delta_1[(k\oplus p)^2]_{new}
	=
	2 \mathbf{k} (k_0 \mathbf{p}-p_0 \mathbf{k})
\end{align}
\begin{align}
	\Delta_2[(k\oplus p)^2]_{new}
	&=
	(p_0-k_0) \left(p_0 \left(\mathbf{k}^2-k_0 (p_0-k_0)\right)-2 k_0 \mathbf{k} \mathbf{p}+k_0 \mathbf{p}^2\right) \nonumber \\
	&-p_0^2 \mathbf{k}^2-\mathbf{k} \left(\mathbf{p}^2-p_0^2\right) (\mathbf{p}-\mathbf{k})+\frac{\mathbf{k}^2 \mathbf{p}^2}{3}
\end{align}
Since now the denominators are again the same as before, we can treat them in the same way, obtaining
\begin{align}
	&-\frac{g^2}{\kappa}
	\int_0^1 dx 
	\int \frac{d^4k}{(2\pi)^4}
	\frac{\Delta_1^{shift}[(k\oplus p)^2]_{new}}
	{[k^2 - \Delta(x)+i\epsilon]^3} \\
	&-
	\frac{g^2}{\kappa^2}
	\int_0^1 dx 
	\int \frac{d^4k}{(2\pi)^4}
	\frac{\Delta_2^{shift}[(k\oplus p)^2]_{new}}
	{[k^2 - \Delta(x)+i\epsilon]^3}
\end{align}
The shift is again $k\mapsto k+p(1-x)$, and 
performing the Wick rotation the amplitude becomes 
\begin{align}
	&i\frac{g^2}{\kappa}
	\int_0^1 dx 
	\int \frac{d^4k}{(2\pi)^4}
	\frac{\Delta_1^{shift+Wick}[(k\oplus p)^2]_{new}}
	{[k^2 + \Delta(x)+i\epsilon]^3} \nonumber \\
	&+
	i
	\frac{g^2}{\kappa^2}
	\int_0^1 dx 
	\int \frac{d^4k}{(2\pi)^4}
	\frac{\Delta_2^{shift+Wick}[(k\oplus p)^2]_{new}}
	{[k^2 + \Delta(x)+i\epsilon]^3}
\end{align}
The numerators are
\begin{align}
	\Delta_1^{shift+Wick}[(k\oplus p)^2]_{new}
	=
	-2 p_0 \mathbf{k}^2
\end{align}
\begin{align}
	\Delta_2^{shift+Wick}[(k\oplus p)^2]_{new}
	&=
	-3 k_0^2 p_0^2 x+k_0^2 p_0^2 +2 k_0^2 \mathbf{p}^2 x-k_0^2 \mathbf{p}^2 \nonumber \\
	&+p_0^4 x^3-p_0^4 x^2+p_0^2 \mathbf{k}^2 x-2 p_0^2 \mathbf{k}^2 \nonumber \\
	&-p_0^2 \mathbf{p}^2 x^3-p_0^2 \mathbf{p}^2 x^2+3 p_0^2 \mathbf{p}^2 x-p_0^2 \mathbf{p}^2+\frac{4 \mathbf{k}^2 \mathbf{p}^2}{3} \nonumber \\
	+\frac{4 \mathbf{p}^4 x^2}{3}-\frac{5 \mathbf{p}^4 x}{3}+\frac{\mathbf{p}^4}{3}
\end{align}
Once again, being in the Eulidean space, we can switch any isolated $\mathbf{k}^2$ with $3k_0^2$, obtaining 
\begin{align}
	\Delta_1^{shift+Wick}[(k\oplus p)^2]_{new}
	=
	-6 p_0 k_0^2
\end{align}
\begin{align}
	\Delta_2^{shift+Wick}[(k\oplus p)^2]_{new}
	&=
	k_0^2
	[-5 p_0^2 + \mathbf{p}^2 (2x+3)] \nonumber \\
	&+p_0^4 x^3-p_0^4 x^2
	-p_0^2 \mathbf{p}^2 x^3-p_0^2 \mathbf{p}^2 x^2+3 p_0^2 \mathbf{p}^2 x-p_0^2 \mathbf{p}^2 \nonumber \\
	&+\frac{4 \mathbf{p}^4 x^2}{3}-\frac{5 \mathbf{p}^4 x}{3}+\frac{\mathbf{p}^4}{3}
\end{align}
Notice that the $1/\kappa$ contribution is numerically equivalent to the previous case, but with a sign difference, so we don't need to compute it. For what remains, since we are only interested in the logarithmic contribution, using eq. \eqref{cutoff-log-1}, \eqref{cutoff-log-2}, \eqref{cutoff-log-3}, since each $\log$ contributes a $-\pi$, and since 
\begin{align}
	\int_{L_1}^{L_2} dx 
	[-5 p_0^2 + \mathbf{p}^2 (2x+3)]
	=
	\sqrt{1-4\frac{m^2}{p^2}}
	\left(
	p_0^2 +4p^2
	\right)
\end{align}
we end up with 
\begin{align}
	\Im \Pi^{(3)} = \Im \Pi^U(p^2) 
	-
	\frac{g^2}{\pi\kappa}
	\sqrt{1-4\frac{m^2}{p^2}}
	\frac{3p_0}{32}
	+
	\frac{g^2}{64\pi \kappa^2} 
	\sqrt{1-4\frac{m^2}{p^2}}
	\left(
	p_0^2 +4p^2
	\right)
\end{align}
Recalling from eq. \eqref{undefim} that $\Im \Pi^U(p^2) =  \frac{g^2}{16\pi}\sqrt{1-4\frac{m^2}{p^2}}$ we finally obtain
\begin{align}\label{Pi3}
	\boxed{
	\Im \Pi^{(3)} = \Im \Pi^U(p^2)
	\left[
	1 - \frac{1}{\kappa}
	\left(\frac{3}{2} p_0\right)
	+
	\frac{1}{\kappa^2}
	\left(\frac{1}{4} p_0^2 + p^2 \right)
	\right]
}
\end{align}

Finally, we consider the last graph.
\begin{center}
	\begin{tikzpicture}
		\begin{feynman}
			\vertex (a) at (-0.6,0) {\(i \Pi(p^2) = \)};
			\vertex (b) at (1,0);
			\vertex (c) at (2,0);
			\vertex (d) at (3,0);

			\diagram* 
			{(a)-- [ultra thick] (b),
				(b)-- [scalar,half left,looseness=1.7,momentum=\(S(k)\)] (c) -- [scalar,half left,looseness=1.7,reversed momentum=\(p\oplus k\)] (b),
				(c)-- [ultra thick] (d),};
		\end{feynman}
	\end{tikzpicture}
\end{center}
\begin{align}\label{initialamp-def1-DR-4}
	i\Pi(p^2)
	=
	(ig)^2
	\int \frac{d^4 k}{(2\pi)^4} 
	\frac{i}{k^2 - m^2 +i\epsilon}
	\frac{i}{(p\oplus k)^2 - m^2 +i\epsilon}
\end{align}
The computations for this graph are exactly the same with the previous with $k \leftrightarrow p$ up to the shift, so we have
\begin{align}
	i\Pi(p^2) 
	&=
	i\Pi^U(p^2) \\
	&-
	\frac{g^2}{\kappa}
	\int \frac{d^4k}{(2\pi)^4}
	\frac{1}{k^2-m^2+i\epsilon}
	\frac{\Delta_1[(p\oplus k)^2]}{[(p+k)^2 - m^2 +i\epsilon]^2}\\
	&-
	\frac{g^2}{\kappa^2}
	\int \frac{d^4k}{(2\pi)^4}
	\frac{1}{k^2-m^2+i\epsilon}
	\frac{\Delta_2[(p\oplus k)^2]}{[(p+k)^2 - m^2 +i\epsilon]^2}
\end{align}
and 
\begin{align}
	\Delta_1[(k\oplus p)^2]_{new}
	=
	-2 \mathbf{p} (k_0 \mathbf{p}-p_0 \mathbf{k})
\end{align}
\begin{align}
	\Delta_2[(k\oplus p)^2]_{new}
	&=
	-\mathbf{p} \left(\mathbf{k}^2-k_0^2\right) (\mathbf{k}-\mathbf{p})-k_0^2 \mathbf{p}^2 \nonumber \\
	&+(k_0-p_0) \left(k_0 \left(\mathbf{p}^2-p_0 (k_0-p_0)\right)+p_0 \mathbf{k}^2-2 p_0 \mathbf{k} \mathbf{p}\right)+\frac{\mathbf{k}^2 \mathbf{p}^2}{3}
\end{align}
Since now the denominators are again the same as before, we can treat them in the same way, obtaining
\begin{align}
	&-\frac{g^2}{\kappa}
	\int_0^1 dx 
	\int \frac{d^4k}{(2\pi)^4}
	\frac{\Delta_1^{shift}[(k\oplus p)^2]_{new}}
	{[k^2 - \Delta(x)+i\epsilon]^3} \\
	&-
	\frac{g^2}{\kappa^2}
	\int_0^1 dx 
	\int \frac{d^4k}{(2\pi)^4}
	\frac{\Delta_2^{shift}[(k\oplus p)^2]_{new}}
	{[k^2 - \Delta(x)+i\epsilon]^3}
\end{align}
The shift is again $k\mapsto k+p(1-x)$, and performing then the Wick rotations the amplitude becomes 
\begin{align}
	&i\frac{g^2}{\kappa}
	\int_0^1 dx 
	\int \frac{d^4k}{(2\pi)^4}
	\frac{\Delta_1^{shift+Wick}[(p\oplus k)^2]_{new}}
	{[k^2 + \Delta(x)+i\epsilon]^3} \\
	&+
	i
	\frac{g^2}{\kappa^2}
	\int_0^1 dx 
	\int \frac{d^4k}{(2\pi)^4}
	\frac{\Delta_2^{shift+Wick}[(p\oplus k)^2]_{new}}
	{[k^2 + \Delta(x)+i\epsilon]^3}
\end{align}
and the numerators (after the same simplifications) become
\begin{align}
	\Delta_1^{shift+Wick}[(k\oplus p)^2]_{new}
	=
	0
\end{align}
\begin{align}
	\Delta_2^{shift+Wick}[(k\oplus p)^2]_{new}
	&=
	k_0^2
	[p_0^2(-6x+1)+ \mathbf{p}^2 (10x-5)]\nonumber \\
	&+p_0^4 x^3-p_0^4 x^2 -2 p_0^2 \mathbf{p}^2 x^3+2 p_0^2 \mathbf{p}^2 x^2+p_0^2 \mathbf{p}^2 x \nonumber \\
	&-p_0^2 \mathbf{p}^2  +\mathbf{p}^4 x^3-\frac{5 \mathbf{p}^4 x^2}{3}+\frac{\mathbf{p}^4 x}{3}+\frac{\mathbf{p}^4}{3}
\end{align}
Recalling once more eq. \eqref{cutoff-log-1}, \eqref{cutoff-log-2}, \eqref{cutoff-log-3}, since each $\log$ contributes a $-\pi$ and since 
\begin{align}
	\int_{L_1}^{L_2} dx 
	[p_0^2(-6x+1)+ \mathbf{p}^2 (10x-5)]
	=
	2 \sqrt{1-4\frac{m^2}{p^2}} p_0^2
\end{align}
we end up with 
\begin{align}
	\Im\Pi^{(4)}(p^2)
	=
	\Im\Pi^U(p^2)
	+
	\frac{g^2}{\pi\kappa^2}
	\sqrt{1-4\frac{m^2}{p^2}}
	\left(
	\frac{1}{32}p_0
	\right)
\end{align}
Recalling from eq. \eqref{undefim} that $\Im \Pi^U(p^2) =  \frac{g^2}{16\pi}\sqrt{1-4\frac{m^2}{p^2}}$ we finally obtain
\begin{align}\label{Pi4}
	\boxed{
		\Im \Pi^{(4)} = \Im \Pi^U(p^2)
		\left[
		1 
		+
		\frac{1}{\kappa^2}
		\left(\frac{1}{2} p_0^2 \right)
		\right]
	}
\end{align}

\subsubsection{Summary of the individual diagrams and total 1-loop amplitude}

We considered the following diagrams
\begin{center}
	\begin{tikzpicture}
		\begin{feynman}
			\vertex (a) at (-0.6,0) {\(i \Pi^{(1)}(p^2) = \)};
			\vertex (b) at (1,0);
			\vertex (c) at (2,0);
			\vertex (d) at (3,0);

			\diagram* 
			{(a)-- [ultra thick] (b),
				(b)-- [scalar,half left,looseness=1.7,momentum=\(k\)] (c) -- [scalar,half left,looseness=1.7,reversed momentum=\(S(k)\oplus p\)] (b),
				(c)-- [ultra thick] (d),};
		\end{feynman}
	\end{tikzpicture}
\end{center}
\begin{center}
	\begin{tikzpicture}
		\begin{feynman}
			\vertex (a) at (-0.6,0) {\(i \Pi^{(2)}(p^2) = \)};
			\vertex (b) at (1,0);
			\vertex (c) at (2,0);
			\vertex (d) at (3,0);

			\diagram* 
			{(a)-- [ultra thick] (b),
				(b)-- [scalar,half left,looseness=1.7,momentum=\(k\)] (c) -- [scalar,half left,looseness=1.7,reversed momentum=\(p\oplus S(k)\)] (b),
				(c)-- [ultra thick] (d),};
		\end{feynman}
	\end{tikzpicture}
\end{center}
\begin{center}
	\begin{tikzpicture}
		\begin{feynman}
			\vertex (a) at (-0.6,0) {\(i \Pi^{(3)}(p^2) = \)};
			\vertex (b) at (1,0);
			\vertex (c) at (2,0);
			\vertex (d) at (3,0);

			\diagram* 
			{(a)-- [ultra thick] (b),
				(b)-- [scalar,half left,looseness=1.7,momentum=\(S(k)\)] (c) -- [scalar,half left,looseness=1.7,reversed momentum=\(k\oplus p\)] (b),
				(c)-- [ultra thick] (d),};
		\end{feynman}
	\end{tikzpicture}
\end{center}
\begin{center}
	\begin{tikzpicture}
		\begin{feynman}
			\vertex (a) at (-0.6,0) {\(i \Pi^{(4)}(p^2) = \)};
			\vertex (b) at (1,0);
			\vertex (c) at (2,0);
			\vertex (d) at (3,0);

			\diagram* 
			{(a)-- [ultra thick] (b),
				(b)-- [scalar,half left,looseness=1.7,momentum=\(S(k)\)] (c) -- [scalar,half left,looseness=1.7,reversed momentum=\(p\oplus k\)] (b),
				(c)-- [ultra thick] (d),};
		\end{feynman}
	\end{tikzpicture}
\end{center}
Each of them gives the imaginary part to the amplitude respectively in eq. \eqref{Pi1}, \eqref{Pi2}, \eqref{Pi3}, \eqref{Pi4}. To get the complete contribution, we just sum the four different results
\begin{align}
	\Im \Pi^{TOT}(p^2)
	=
	\frac{1}{4}
	\left(
	\Im \Pi^{(1)}(p^2)
	+
	\Im \Pi^{(2)}(p^2)
	+
	\Im \Pi^{(3)}(p^2)
	+
	\Im \Pi^{(4)}(p^2)
	\right)
\end{align}
where the factor $1/4$ is necessary to get the correct $\kappa\rightarrow\infty$ limit. We get
\begin{align}\label{Pitot}
	\boxed{
	\Im \Pi^{TOT}(p^2)
	=
	\Im\Pi^U(p^2)
	\left\{
	1
	-
	\frac{1}{\kappa^2}
	\left[
	p_0^2
	\left(
	\frac{1}{48}
	+
	\frac{1}{12}
	\frac{m^2}{p^2}
	\right)
	-
	p^2
	\frac{19}{48}
	-
	\frac{5}{24}m^2
	\right]
	\right\}
}
\end{align}
Notice the remarkable fact that, although some of the diagrams contain a contribution proportional to $1/\kappa$, this contribution goes away after we sum all the contributions from every diagram, leaving a leading contribution proportional to $1/\kappa^2$. 

Notice the important point that we chose our initial particle to have momentum $S(p)$, but we may just as well have decided to take a particle with initial momentum $-p$. In this case momentum conservation would amount to
\begin{align}
	(-p)\oplus p_2\oplus p_3
\end{align}
which is immediately satisfied if $p_2\oplus p_3 = S(-p)$ (in order to avoid confusion, it is understood that the equivalent computations using the momenta $p^*$ would amount to having a particle with momentum $p^*$, so that $p_2\oplus p_3 = S(p^*)$). It is therefore sufficient to switch every $p$ in eq. \eqref{Skp}, \eqref{pSk}, \eqref{kp}, \eqref{pk} with $S(-p)$ and perform the same computations again. Since we are expanding up to second order in $1/\kappa$, the final result will be the same as before plus some additional contribution. Since we have already shown in detail the previous computations, here we limit ourselves to present the additional contribution to the numerator of the sum of the four diagrams after all the shifts, Wick rotation, and Euclidean simplifications have been performed. We call $\delta\Delta_2^{shift+Wick}$ the additional contribution to the sum of the numerators. We have
\begin{align}
	\delta\Delta_2^{shift+Wick}
	=
	\frac{8 k_0^2 \mathbf{p}^2}{\kappa^2}+\frac{8 \mathbf{p}^2 \left(p_0^2 \left(-x^2+x-2\right)+\mathbf{p}^2 (x-1) x\right)}{\kappa^2}
\end{align}
where we included that $1/\kappa$ powers to show that also in this case we have no $1/\kappa$ contribution after summing all four diagrams contributions. Notice that, because of eq. \eqref{cutoff-log-2}, the term with no $k$ in the numerator does not contribute to the imaginary part, since it does not generate a $\log$ term upon integrating over $k$. The only term which contributes is the term $\frac{8 k_0^2 \mathbf{p}^2}{\kappa^2}$, and we know from eq. \eqref{cutoff-log-1} that it contributes to the total amplitude as
\begin{align}
	-8\mathbf{p}^2 \frac{1}{64\pi^2}\log \Delta
\end{align}
and since the $\log$ contributes as $-\pi$ to the imaginary part we only need to compute the quantity
\begin{align}
	-\frac{g^2}{\kappa^2}
	\frac{1}{64\pi}\int_{L_1}^{L_2} 8\mathbf{p}^2 dx
	=
	-
	\Im \Pi^U(p^2)
	\left(2\frac{p_0^2}{\kappa^2} - 2\frac{p^2}{\kappa^2}\right).
\end{align}
In other words, calling $\Im\Pi_{S(p)}$ the total amplitude $\Im\Pi^{TOT}(p^2)$ in eq. \eqref{Pitot} corresponding to an initial on-shell particle of momentum $S(p)$, and calling $\Im\Pi_{-p}$ the equivalent quantity with momentum $-p$, we have
\begin{align}
	\Im\Pi_{S(p)}
	-
	\Im\Pi_{-p}
	=
	\Im \Pi^U(p^2)
	\left(2\frac{p_0^2}{\kappa^2} - 2\frac{p^2}{\kappa^2}\right).
\end{align}
The quantity $\frac{p^2}{\kappa^2} = \frac{m_\psi^2}{\kappa^2}$ is invariant, but the quantity $\frac{p_0^2}{\kappa^2}$ is sensible to the particles energy, which means that this difference can be experimentally highlighted by going to higher energies.

\section{Mass distribution width}\label{mdw}

In section \ref{def-cpt-pheno} in chapter \ref{Chapter2} we addressed possible phenomenological consequences of the deformation of $CPT$ symmetry, but we did not consider for the moment possible contributions coming from the higher order corrections to the propagator of the decaying particle. Such corrections would induce a widening of the mass distribution, and such contributions need to be taken into account. 

More formally, given a mass distribution function $\omega(m; M, \Gamma)$, where $m$ is the resonant mass, and with mean mass $M$ and decay width $\Gamma$, one can describe the time and momentum dependent decay amplitude with the expression
\begin{eqnarray}\label{a1}
	a(t,\mathbf p)=\int_{-\infty}^\infty dm\,\omega(m; M,\Gamma)\,e^{-it\sqrt{m^2+\mathbf p^2}}.
\end{eqnarray} 
The decay probability is then given by the usual formula $\mathcal{P} = |a(t, \mathbf{p})|^2$. Since we are interested in ultra-relativistic particles (we need high momenta and energies in order to counterbalance the large value of $\kappa$ in the denominator), we consider the relativistic Breit-Wigner distribution
\begin{align} \label{a3}
	\omega(m; M,\Gamma) & =  \frac{f(M,\Gamma)}{(m^2-M^2)^2+M^2\Gamma^2},
	\\
	f(M,\Gamma)&=\frac{2\sqrt{2}}{\pi}\frac{M\Gamma\sqrt{M^2(M^2+\Gamma^2)}}{\big[M^2+\sqrt{M^2(M^2+\Gamma^2)}\big]^{1/2}}.
\end{align}
Notice that $f(M,\Gamma)$ is independent of $m$. Since we want to understand the effect of deformation, we now switch the energy $E = \sqrt{m^2 + \mathbf{p}^2}$ at the exponent for its antipode $S(E)$ (see eq. \eqref{antifen1-esp}). Furthermore, one can show that
\begin{eqnarray} \label{a4}
	a(t,\mathbf p) & = & \int dm\,\frac{f(M,\Gamma)}{(m^2-M^2)^2+M^2\Gamma^2} e^{-it(\sqrt{m^2+\mathbf p^2}-\mathbf p^2/\kappa)} \nonumber \\
	& = & e^{-it(\sqrt{M^2+\mathbf p^2+iM\Gamma}-\mathbf p^2/\kappa)}.
\end{eqnarray}
The new, corrected decay width $\tilde{\Gamma}$ (which takes into consideration the non-trivial mass distribution width) can be now computed by taking into account the real part of the exponent in eq. \eqref{a4} (which will describe the exponential decay of the amplitude). However, the only effect of deformation comes into play in the term $-it\frac{\mathbf{p}^2}{\kappa}$, i.e. in the imaginary part, so no correction to $\Gamma$ can be obtained from a non-trivial mass distribution width. Indeed, one explicitly has 
\begin{eqnarray} \label{a5}
	\tilde\Gamma & = & 2\,\Im \Big(\sqrt{M^2+\mathbf p^2+iM\Gamma}-\frac{\mathbf p^2}{\kappa}\Big) \nonumber \\
	& = & \sqrt{2} \Big[\sqrt{(M^2+\mathbf p^2)^2+M^2\Gamma^2}-(M^2+\Gamma^2) \Big]^{1/2}.
\end{eqnarray}
so that the only corrections are the same that would be present in the canonical, non-deformed context. 


\section{Summary}

In this work, we have presented our model of scalar field theory, and analysed its features and the possible phenomenological consequences coming from our results. 

We first introduced the framework of $\kappa$-Minkowski spacetime and $\kappa$-Poincar\'e Hopf algebra, which are the fundamental ingredients for any other construction. Before going to the complex scalar field, we tackled the interesting problem of the boost of a two-particle state system. The determination of the finite boost of a two-particle system allows us to draw interesting phenomenological consequences for any type of particle, since the results are purely of kinematical nature.

We then passed to the complex scalar field. After some preliminary mathematical work concerning the Weyl maps, the action of derivatives on star products, and the integration-by-parts relations, we introduced the complex scalar field action and computed the EoM (which turned out to be the canonical Klein-Gordon equations). This allowed us to derive an explicit formula for the on-shel fields $\phi$, $\phi^\dag$. Furthermore, using the surface terms obtained during the integration-by-parts, we obtained the Noether charges coming from the continuous translation symmetry of the action. 

To compute the remaining Noether charges we used the symplectic form, which was also important for the fact that it allowed us to obtain the creation/annihilation operators commutation relation. Together with the charges, these allowed us to check that the algebra satisfied by the charges is the canonical Poincar\'e algebra. Notice that, despite the fact that the EoM are the canonical Klein-Gordon equations, and the fact that the charges satisfy the canonical Poincar\'e algebra, there are still effect of deformation on single-particle states which are made manifest by looking at the non-trivial interaction between discrete and continuous transformations. Here we explicitly checked that, in the deformed case, we have $[\mathcal{N}_i, C]\neq 0$. Furthermore, $\kappa$-deformation affects in non-trivial way the features vacuum expectation values of fields. In this work, we discussed in detail Greenberg's theorem, showing that it does not hold in general in the $\kappa$-deformed context. We then explored some phenomenological consequences of the fact that $[\mathcal{N}_i, C]\neq 0$, by boosting a particle and an antiparticle, both originally at rest.

We then proceeded to compute the propagator and the imaginary part of the 1-loop correction to it. To do so, we exploit the path integral formalism for the determination of the tree-level propagation, and we assume a deformed conservation of momenta at each vertex. Because of this, a single diagram in the non-deformed context is translated into four different diagrams in the deformed context. We computed all of them for initial momentum of a particle given by $S(p)$ and $-p$, obtaining a difference in the imaginary part proportional to $p_0^2/\kappa^2$. We conclude with some comments on the relevance of a non-trivial mass distribution width on the decay times of boosted particles and antiparticles.






\appendix 



\end{document}